# Basic Paradoxes of Statistical Classical Physics and Quantum Mechanics

Oleg Kupervasser

Scientific Research Computer Center, Moscow State University, 119992 Moscow, Russia

olegkup@yahoo.com

**Abstract** Statistical classical mechanics and quantum mechanics are developed and well-known theories that represent a basis for modern physics. Statistical classical mechanics enable the derivation of the properties of large bodies by investigating the movements of small atoms and molecules which comprise these bodies using Newton's classical laws. Quantum mechanics defines the laws of movement of small particles at small atomic distances by considering them as probability waves. The laws of quantum mechanics are described by the Schrödinger equation. The laws of such movements are significantly different from the laws of movement of large bodies, such as planets or stones. The two described theories are well known and have been well studied. As these theories contain numerous paradoxes, many scientists doubt their internal consistencies. However, these paradoxes can be resolved within the framework of the existing physics without the introduction of new laws. To clarify the paper for the inexperienced reader, we include certain necessary basic concepts of statistical physics and quantum mechanics in this paper without the use of formulas. Exact formulas and explanations are included in the Appendices. The text is supplemented by illustrations to enhance the understanding of the paper. The paradoxes underlying thermodynamics and quantum mechanics are also discussed. The approaches to the solutions of these paradoxes are suggested. The first approach is dependent on the influence of the external observer (environment), which disrupts the correlations in the system. The second approach is based on the limits of the self-knowledge of the system for the case in which both the external observer and the environment are included in the considered system. The concepts of observable dynamics, ideal dynamics, and unpredictable dynamics are introduced. The phenomenon of complex (living) systems is contemplated from the point of view of these dynamics.

## 1. Introduction

A number of important points should be noted.

1) In contrast with other papers regarding paradoxes of quantum mechanics, this paper is not a philosophical paper on physics. We employ scientific methods to consider a solution of these paradoxes. We also construct the physics by excluding these paradoxes and obtain requirements that facilitate their feasibility. The misunderstanding of physics causes these paradoxes and produces physical errors instead of philosophical errors.
2) This paper does not constitute an attempt to provide a new interpretation of quantum mechanics. All interpretations (for example, multi-world interpretation and Copenhagen) aim to provide an evident explanation of quantum mechanics. These interpretations neither solve any paradoxes nor introduce any new appearance in the physics. The author considers all existing reasonable and admissible interpretations. In this paper, a paradox solution is not related to an interpretation but is based on general physics.
3) This paper does not constitute a popular scientific paper and includes original ideas. The paper is designed for an extensive set of specialists, including biologists, physicists (in quantum mechanics, statistical physics, thermodynamics, and non-linear dynamics), and computer science specialists. Therefore, we have provided a popular review of physics that may be trivial for one expert and useful to another expert. Formulas are not included in this paper; this paper is composed of figures and text. All formulas are contained in the appendices. The author is not a pioneer of this writing style. Examples of this writing style are books by Penrose [1,2], Hofstadter [3], Mensky [4], and Licata [85]. These books are not popular despite their "simplistic" style.
4) This paper includes not only a review of completed papers (although many references are provided) but also includes original ideas of the author.
5) The author does not attempt to discover new laws of physics[1]. All reviews are conducted within the framework

---

of previously existing physics. The motivation to write this paper was the fact (paradox) that the author has not encountered any paper or physics textbook in which a complete and clear explanation of these paradoxes of physics (Fig. 1) and its consequences is provided. These paradoxes are disregarded in numerous papers, whereas the explanations are incomplete or inaccurate in other papers. In numerous papers, the solution is based on certain interpretations of physics (usually multiworld). New (but not necessary) laws of physics are sometimes used as explanations.

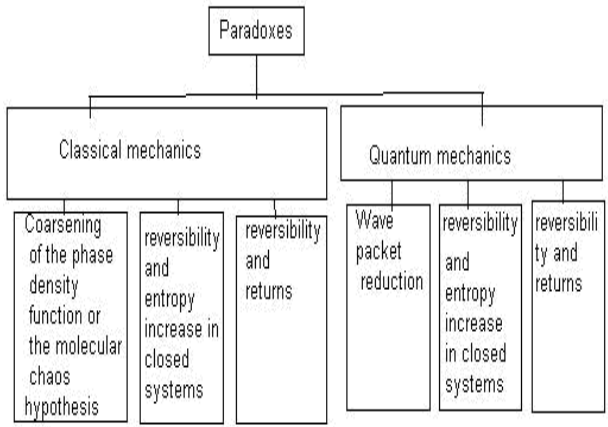

**Figure 1**   Paradoxes in classical and quantum mechanics

# 2. Principal Paradoxes of Classical Statistical Physics

## 2.1. Macroscopic and Microscopic Parameters of Physical Systems [5, 6]

We begin our discussion from the viewpoint of statistical physics. We examine the gas outflow from a jet engine nozzle (Fig. 2).

We observe the distribution of density and velocity of flowing gas for large volumes. These volumes include enormous quantities of invisible molecules. These distinct density and velocity distribution of flowing gas are defined as macroscopic parameters of the system. They provide incomplete descriptions of the system. The complete set of the parameters is includes the velocities and positions of all gas molecules. These parameters are defined as microscopic parameters. Flowing gas is defined as an observable system. The system is considered isolated if it does not interact with its environment. The internal energy of the system is the sum of its molecular energies.

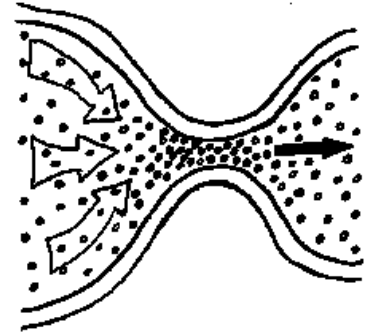

**Figure 2**   The outflow of gas from a nozzle.   Molecules of gas, which are invisible by the naked eye, are magnified.

We subsequently consider isolated systems and define internal energy and finite volume (unless the contrary is stated).

## 2.2. Phase Spaces and Phase Trajectories [5, 6, 12, 17]

Here, we introduce the multidimensional space. Coordinates and velocities of all molecules of the system will define the axes of this space. The system will be denoted by a point of this space. The position of this point will provide the complete microscopic description of the system. This space is defined as a system phase space. The system state change, which is featured by the point moving in this space, is defined as a phase trajectory (Fig. 3).

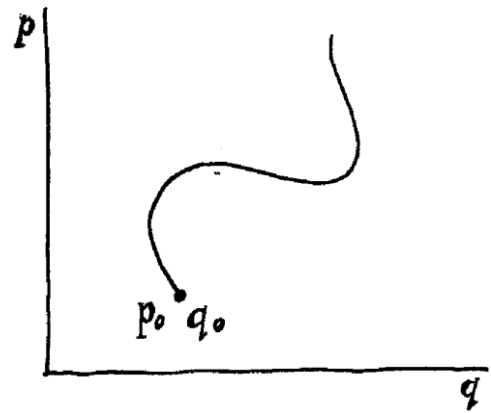

**Figure 3**   Trajectory in the phase space. A current state of the system described by a   point in phase space p, q. The time evolution is described by a trajectory beginning at the initial state point $p_0$, $q_0$.

If we assume that macroscopic parameters are known and microscopic parameters are unknown, the system can be described in phase space by a continuous set of points that correspond to these macroscopic parameters. It is the phase volume ("cloud") of the system or the ensemble of Gibbs (Fig. 4). All points of this volume have equal probability and correspond to different microscopic (but identical macroscopic) parameters. (Appendix A) [5, 6]

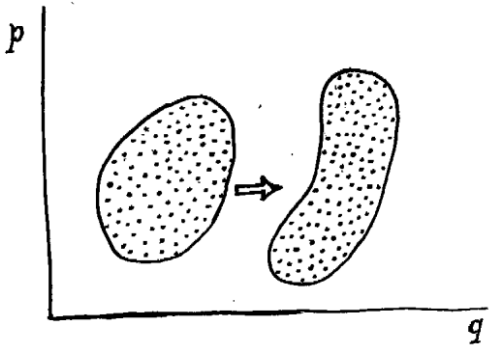

**Figure 4** Ensembles in the phase space. The Gibbs ensemble is described by a cloud of points with different initial conditions. The form of the cloud changes during evolution.

For each set of macroscopic parameters (a macroscopic state), the correspondent ensemble of microscopic parameter sets can be obtained. To produce a finite ensemble, we divide the phase space into separate small meshes. This method is defined as discretisation of the continuous space. In this review, the system with a finite volume and a given internal energy can be featured by a major but finite ensemble of states. The corresponding major but finite ensemble of microscopic states can be obtained for each macroscopic state. The properties of the majority of systems are such that a significant part of their possible microscopic states correspond to a principal macroscopic state, which is the equilibrium state. For example, for gas in a given volume, the equilibrium state corresponds to the uniform distribution of molecules in the volume (Appendix E) [5, 6].

## 2.3. Ergodicity and Mixing [12-14]

The majority of real systems possess the property of ergodicity [**13, 14**]: almost any phase trajectory should eventually visit all meshes of microstates, which are possible for the given energy of a system. The system should remain for approximately equivalent periods in each microstate mesh. Ergodic systems possess a remarkable property. The average value on time of any macroparametre over a trajectory will be identical for all trajectories. It coincides with the average value over the ensemble of systems featuring thermodynamic equilibrium. This ensemble (referred to as microcanonical) is distributed over a constant energy surface.

The majority of real systems possess a property that is termed chaos or mixing (the first step in describing this property is the so-called KAM (Kolmogorov-Arnold-Moser) theorem [**13, 14**], followed by all subsequent theorems). I.e., in the neighbourhood of many points in phase space, there is nearly always another point at which the phase trajectories of these two points diverge exponentially quickly [**12- 14**] (Figure 5).

Indeed, the KAM theorem states that there are persistent portions of phase space in which systems continue to behave like harmonic oscillators. However, the KAM theorem also states that there are noticeable portions of phase space in which harmonic-oscillator systems are destroyed and become chaotic. The KAM theorem describes the mechanism of this destruction. The KAM theorem considers small perturbations of integrable systems. However, for large perturbations (which are applicable for most real systems) and with more degrees of freedom, the portion of the phase space in which the systems continue to behave like harmonic oscillators becomes small: (http://en.wikipedia.org/wiki/Kolmogorov%E2%80%93Arn old%E2%80%93Moser_theorem): "The non-resonance and non-degeneracy conditions of the KAM theorem become increasingly difficult to satisfy for systems with more degrees of freedom. As the number of dimensions of the system increases, the volume occupied by the tori decreases. Those KAM tori that are not destroyed by perturbation become invariant Cantor sets, named Cantori by Ian C. Percival in 1979. As the perturbation increases and the smooth curves disintegrate we move from KAM theory to Aubry-Mather theory which requires less stringent hypotheses and works with the Cantor-like sets."

Pure ergodic or mixing systems are rare (e.g., Sinai's hard spheres), but the majority of systems are very close to chaotic (ergodic or mixing) systems. For example, weather is predictable only on short time scales because of chaos (the butterfly effect).

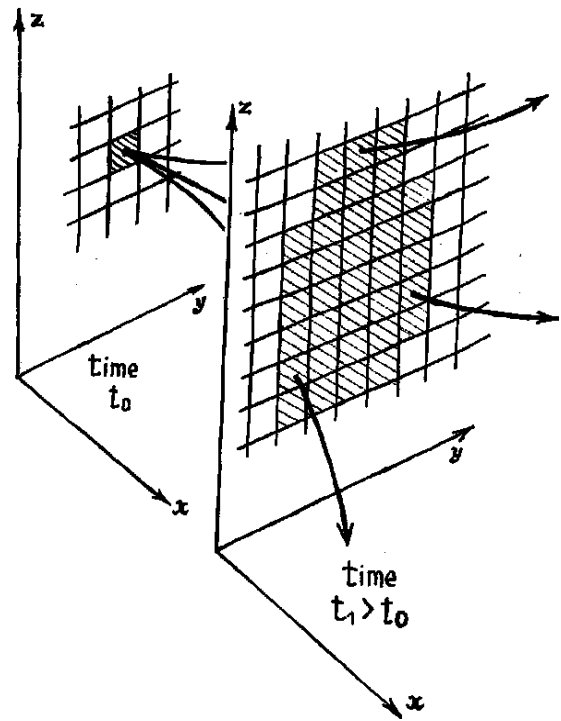

**Figure 5** Illustration of increased uncertainty increasing or loss of information in a dynamic system. The shaded square at time $t_0$ describes uncertainty in knowledge of the initial condition.

Exponential speed is defined as follows: if the trajectories diverge two times from the initial magnitude after one second then they will diverge four times from the initial magnitude after the two seconds. After the third second, they will diverge 8 times from the initial magnitude. It is a rapid type of divergence (Fig. 5). Ergodic systems always possess the property of mixing (Fig. 6,7, and 8).

## 2.4. Reversibility and Poincare's Theorem

Microstate evolution is reversible. For each trajectory in phase space, the inverse trajectory is obtained by the inversion of all velocities of molecules into opposite values. It is equivalent to the reverse demonstration about the process for a film. After a specific duration (most likely large), almost any trajectory returns to its initial microstate. This statement is named Poincare's theorem (Appendix C) [6]. The majority of real systems are chaotic and unstable, and phase trajectories from previously neighbouring microstates are rapid and divergent. Therefore, the return time for such systems is unequal for previously neighbouring microstates. It is dependent on the exact position of the initial trajectory point in a mesh that contains a divided phase space. However, for a small class named integrable systems, this return time is approximately identical for all initial points of phase mesh. These returns occur periodically or almost periodically.

## 2.5. Entropy. [5, 6, 13- 15]

We introduce the basis concept for statistical mechanics—macroscopic entropy. Assume that a certain macroscopic state corresponds to 16 microstates. Using «yes» or «no» answers, how many questions are required to determine which of these 16 microscopic states contains the system? If we consider each microstate, 15 questions are required. A smarter method involves dividing all microscopic states into two groups, with 8 microstates in each group. The first question is to which group does the microstate apply? The specified group will be divided into two subgroups with 4 microstates in each subgroup; we will then ask the same question. We will continue this procedure until a single microstate of the system is obtained. Only four questions, which is the minimal number of questions for the current case, are required. This minimal number of questions can be defined as macroscopic entropy of macrostate [5, 6, 15] (Appendix B). Entropy is simply calculated as the logarithm with basis 2 of the microstate number;the entropy increases with an increase in microstate number. The equilibrium state has maximum entropy as it corresponds to a maximal number of microstates. Entropy is frequently considered a measure of disorder. The greater the disorder in a system, the more questions are necessary to determine the microstate of a system. Therefore, the entropy also increases. Why do we need to introduce an "abstruse" conception as entropy? It would be easier to use the number of microstates instead. However, entropy possesses a remarkable property. We assume that we have a system that consists of two disconnected subsystems. The entropy of a complete system is the sum of two entropies of its subsystems. (The number of questions that correspond to a system is the sum of questions for each of the subsystems). However, the number of microstates are multiplied. Summing the numbers is easier than multiplying the numbers.

Statistical mechanics describes a number of important properties of physical systems:

Let the initial macroscopic state correspond to a certain volume in phase space. A theorem for the reversible Newtonian evolution of the system exists, in which the value of this phase volume is conserved (Appendix F) [6]. Therefore, the number of microstates that correspond to it is also conserved. The entropy that corresponds to this set of microstates is defined as the entropy of ensemble. Based on the conservation of phase volume, the entropy of ensemble is constant over time.

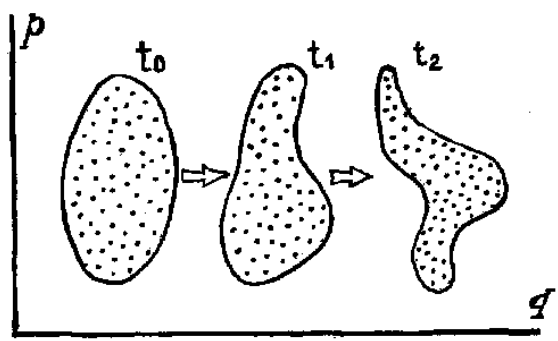

**Figure 6**   Conservation of volume in phase space.

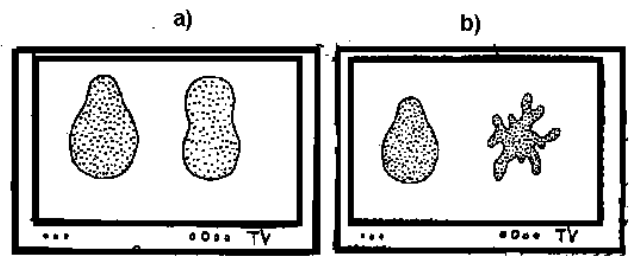

**Figure 7**   Change in phase volume element in a) stable and b) unstable cases.

a)

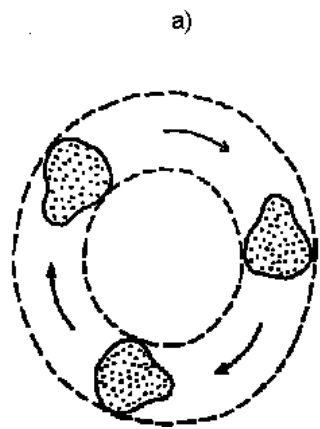

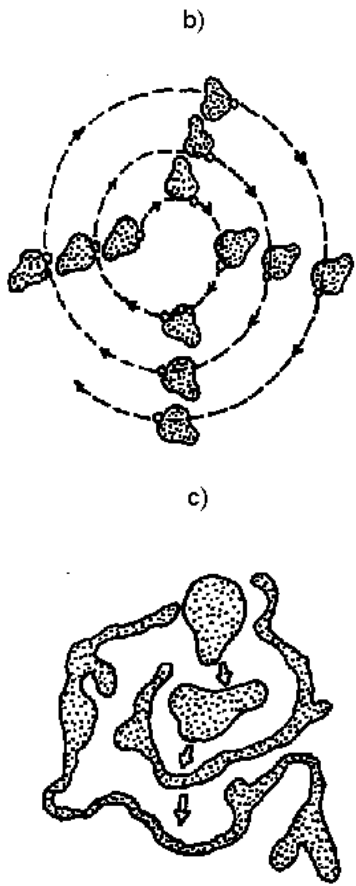

b)

c)

**Figure 8.** Different types of flow in phase space: a) nonergodic flow, b) ergodic flow without mixing, and c) ergodic flow with mixing.

## 2.6. Evolution of Macroscopic Entropy for Chaotic Systems

Based on the properties of ergodicity, a system will eventually transfer from almost any initial microstate to an equilibrium state and will remain in this state for the majority of time. During the evolution of a system, the majority of transiting microstates correspond to an equilibrium state. The equilibrium state exhibits maximum macroscopic entropy. If the macroscopic entropy of an initial state was small, it would increase substantially after convergence to thermodynamic equilibrium. This property is opposite of the property of ensemble entropy, which remains constant during evolution. The entropy of ensemble is defined by numerous microstates that do not change during evolution. It is constant and equivalent to the initial number of microstates, whereas macroscopic entropy is defined by a number of microstates that corresponds to a current macroscopic state. Thermodynamic equilibrium involves a large number of microstates.

For chaotic systems (systems with mixing), the following theorem is valid:

Processes of the evolution of macroparametres with macroscopic entropy decreasing are highly unstable with respect to small external noise. In contrast, in the processes of evolution of macroparametres with macroscopic entropy increasing are stable[2].

We prove above formulated theorem by considering a process with entropy growth. The initial state of the system is featured by a macroscopic state that is far from thermodynamic equilibrium. This state is characterised with a compact (closed and limited) and convex (containing a straight segment connected by any two points) phase volume. As the system is chaotic, another point is present at an exponentially diverging distance in the neighbourhood of each point. Due to phase volume conservation (Appendix F), another point is always present in the neighbourhood of each phase point, such that these two points exponentially converge and diverge. As a result of mixing, an initially compact phase volume, which is not convex, will completely spread over the constant energy surface in phase. The phase volume possesses a large quantity of "sleeves" or "branches". The full volume of phase "drop" is conserved. "Sleeves" are exponentially expended over their lengths and exponentially shrink over their widths. The number of "sleeves" or "branches" expands; they are incurvate and completely covered by its "net" at the phase energy surface. This process is named as spreading of phase "drop" **[13, 14]**. Assume that a small external noise has thrown out a phase point from the "sleeve" of a phase drop. However shrinking occurs perpendicularly to the "sleeve," and the phase point moves closer to the "sleeve" and not further from it, which indicates that the process of phase drop spreading is stable with respect to noise.

Therefore, noise can significantly influence a microstate but does not influence a macrostate. A macrostate corresponds to a substantial number of molecular microstates. Although external to each individual molecule, the complete contribution of all molecules to a macrostate remains unchanged. It is related to the "law of large numbers" in probability theory **[16].** The majority of microstates, which correspond to a current macrostate, evolve in entropy growth as the probability of this evolution is significant. When the phase drop almost spreads along the entire constant energy surface, its macrostate corresponds to thermodynamic equilibrium. Thus, a small noise cannot distinctly affect its macrostate as the majority of microstates in a system correspond to equilibrium.

We consider an inverse process with decreasing entropy. The initial state is defined by points set in phase space, which are obtained from a direct process («a phase drop» spreading), and the final state is defined by the reversion of molecular velocities. In a reversion of velocities, the initial shape of «a phase drop» does not change. However, the direction of shrinkage due to the reversion of velocities occurs in parallel with its "branches". Shrinkage occurs instead of phase drop spreading. Assume that a small external noise has thrown out a phase point from a "sleeve" of a phase drop. Shrinking occurs in parallel and the spreading occurs perpendicularly to the "sleeve", which causes the phase point to move away from the "sleeve" instead of approaching it. This condition indicates that the process of phase drop shrinkage is unstable with respect to noise (Fig. 9).

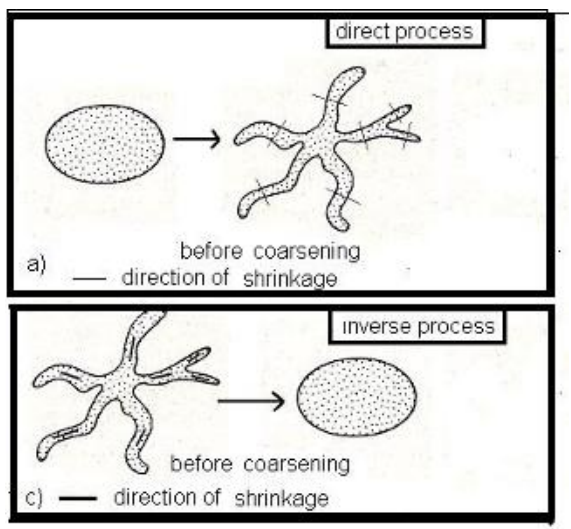

**Figure 9.** Direct process with an increase in macroscopic entropy and its inverse process. The direction of shrinkage is denoted.

## 2.7. The Second Law of Thermodynamics and Related Paradoxes

We discuss the second law of thermodynamics and related paradoxes. The second law states the following principles:

In isolated finite volume systems, macroscopic entropy does not decrease but increases or remains constant. Macroscopic entropy reaches a maximum in a thermodynamic equilibrium state [5, 6].

The reversibility is inconsistent with this entropy growth law according to the previously mentioned basic properties of statistical physics.

Due to the reversibility of each process with increasing entropy, an inverse process with decreasing entropy exists. It is paradox of Loschmidt.

Poincare's theorem of returns states that the system must return to an initial state. Its entropy will also return to an initial value. This is a paradox of Poincare.

The concept of the **correlations** of the velocities and positions of molecules is related to these two paradoxes.

## 2.8. Additional Unstable Microscopic Correlations and Their Connection with Paradoxes of Statistical Physics[17])

Correlation is a measure of the mutual dependence of variables. (In our case, it is a measure of the mutual dependence of the velocities and positions of molecules). Pearson's correlation, which is the most well-known correlation, is a measure of the linear relation of two variables (Appendix D). More complex dependencies and corresponding more complex correlations exist. Correlations between various variables produce restrictions on the possibility of selection of certain values of these variables.

The knowledge of a macroscopic state of a system is a source of correlations. Certain microstates correspond to the given macroscopic state. Thus, their set is already restricted, which will generate to a restriction for possible velocities and positions of molecules, i.e., restriction on the possible microstate of the system. Note that all such correlations are macroscopic and are manifested in dependence between macroscopic parameters of a system. For macroscopic states with small entropy, the restriction on the selection of possible microstates is significant and the quantity of macroscopic parameters and their correlations is also significant. A system at thermodynamic equilibrium entropy reaches the maximum, and the quantity of macroscopic parameters and correlations is small.

Additional or microscopic unstable correlations [17] are defined not only by a knowledge of the current macroscopic state but also by the knowledge of a previous macroscopic history of a system. Assume that the physical system evolved from an initial macroscopic state into another current macroscopic state. Thus, all microscopic states that conform to a current macroscopic state are not possible. Only such states in which the reversion of velocities of molecules produces an initial state can be considered (property of reversibility of motion.): this additional restriction on the states superimposes additional restrictions (correlations) on a set of the microstates that correspond to a current macroscopic state. Additional unstable correlations can be identified in another way; not via knowledge of the past but via knowledge of the future. According to Poincare's theorem, the system should return to a known initial macroscopic initial state after a specific period. By knowing a certain current macroscopic state and when it will return, we can superimpose additional restrictions (correlation) on a set of microstates that correspond to this current macroscopic state. These correlations are considered unstable as they are extremely unstable with respect to external noise (as will be discussed subsequently). Based on the definition of additional unstable correlations it can be seen that these correlations are close related to Poincare's and Loschmidt's paradoxes.

These correlations are additional to the macroscopic correlations. (Macroscopic correlations correspond to macrostate representations). The existence of these additional correlations produce a violation of the second law of thermodynamics and ensures a possibility of returns and reversibility, i.e., appearances observed in Poincare's and Loschmidt's paradoxes.

A basic property of additional or microscopic correlations is instability. The interaction of different parts within an observable system or the interaction of the system with environmental systems (including the observer) causes additional disappearances of correlations. Specifically, these additional correlations "spread" between parts of the system and/or between the system and surrounding systems. Assume that an initial state with small entropy exists. After a brief period of time, an initial collision occurs between molecules. Their positions and velocities become correlated (They can be verified by converting velocities. We can then obtain the initial state). Only close pairs of the molecules during collision are correlated. However, as collisions increase, correlations will include greater quantities of molecules. Correlations will be spread over the entire increasing volume of the system. "Spreading" of correlations in the system occurs (Fig. 10).

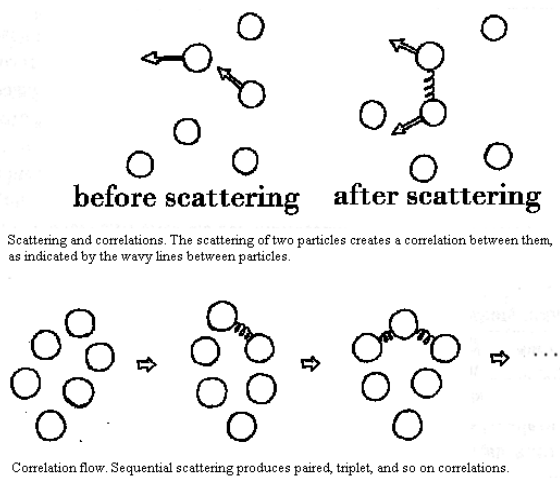

**before scattering    after scattering**

Scattering and correlations. The scattering of two particles creates a correlation between them, as indicated by the wavy lines between particles.

Correlation flow. Sequential scattering produces paired, triplet, and so on correlations.

**Figure 10.** Scattering and correlations and correlation flow. Similarly, if the system consists of two non-interacting systems, correlations will only exist inside each system. Returns and reversibility are possible for each of these subsystems. Assume that a small interaction between these subsystems exists. Correlations "will flow" from one subsystem to another, which will create two dependent systems. Only their joint return or reversibility will be possible.

# 3. Principal Paradoxes of Quantum Mechanics

### 3.1. Basic Concepts of Quantum Mechanics—The Wave Function, Schrodinger Equations, Probability Amplitude, Observables, and Indeterminacy Principle of Heisenberg [18, 19]

For clarity, we provide the basic concepts of quantum mechanics.

Motion in quantum mechanics is represented by a wave function and not by a trajectory. Motion is a probability wave; specifically, it is a "probability amplitude" wave, which signifies that the quadrate of the amplitude module of the wave function at some point provides the probability of detecting a particle at this point. The variation in the time of this probability wave is defined by the Schrodinger equation [19]. This is a linear equation, i.e., the sum of its two solutions comprises the solution. Thus, the amplitudes of the probabilities are summed; however, the probabilities are not summed. The probability is defined by the quadrate of amplitude. It imports nonlinearity to the evolution process of a wave function.

Any observable (for example, momentum) is featured by an orthonormal and complete set of functions (a set of eigenfunctions of a observable). The wave function can be expanded to this set of eigenfunctions. Each eigenfunctions set corresponds to a specific value of a observable (eigenvalue). Expansion coefficients provide a probability amplitude for each value. If the wave function is equivalent to an eigenfunction of the observable set, the value of the observable in this case is equivalent to the correspondent eigenvalue. If this is not the case, we can only specify probabilities for various eigenvalues.

The concept of particle velocity has no explicit physical explanation because no well-defined trajectory of a particle exists; only a probability wave exists [19]. Momentum is defined not via a product of velocity and mass but through wave function expansion coefficients over momentum eigenfunctions. This set of eigenfunctions is similar to a complete orthonormal set of Fourier functions used in Fourier analysis.

Coordinate eigenfunctions are proportional to Dirac delta functions. The coefficients of wave function expansion over Dirac delta functions are given by the value of a wave function in an infinite point of a Dirac delta function. The value of a wave function corresponds to the previously defined sense of a wave function as probability amplitudes.

Both momentum and coordinates correspond to various sets of eigenfunctions [19]. Therefore, no wave function, which is in contrast with can simultaneously correspond to both a single momentum eigenvalue and a single coordinate eigenvalue classical mechanics. The explanation for the well-known uncertainty of Heisenberg has been defined [19] (See Appendix G) and is related to the variation in the definitions of quantum and classical mechanics.

### 3.2. Pure and Mixed States. Density Matrix [15,18, 20]

Quantum mechanics is best described by its wave function, which is the pure state. For classical mechanics, a point in a phase space has similar sense. What is analogous to a classical and statistical ensemble of systems (a cloud of points in a phase space) in quantum mechanics? It is a set of wave functions in which each function corresponds with its probability (instead of the "probability amplitude" for the expansion of a pure state over eigenfunctions). This is the definition of the mixed state.

Assume that a system is part of a larger system. Even if the larger system is represented by a pure state, the smaller subsystem must be generally represented by a mixed state, with the exception of the case in which the pure state of the large system can be described as a product of the small system wave function and its environment wave function. Assume, for example, that the small quantum system interacts with the device that is in a pure state. Although the large system (including the device and the small quantum system) can be represented by a pure state, the small quantum system after measurement is generally already represented by a mixed state.

For equivalent representations of mixed and pure states, the density matrix is used [20]. We select a observable and a corresponding set of eigenfunctions as the density matrix representation based on these eigenfunctions is represented by a square matrix. Every function corresponds to a diagonal element of a density matrix. The value of the element is equivalent to the probability of detecting the corresponding eigenvalue during measurement of the observable.

Nondiagonal elements of a density matrix define the correlations between corresponding pairs of eigenfunctions. Nondiagonal elements possess a maximum value in a pure state; however, their magnitude decreases for the mixed states and can attain a value of zero. The density matrix can always be rewritten over a different set of eigenfunctions that correspond to a different observable. The density matrix provides a maximally complete description of the state of the system. Consequently, the evolution of the density matrix provides a full description of the evolution of the system. (Appendix I.)

### 3.3. Properties of the Isolated Quantum System with Finite Volume and a Finite Number of Particles [15]

Similar to classical systems, we consider the properties of the isolated (closed) quantum system with finite volume and a finite number of particles.
1) These quantum systems evolve for reversible equations of motion (Schrodinger's equation)
2) Poincare's theorem is also accurate for these systems. The properties of quantum systems are similar to the properties of classical integrable systems. (Integrable systems are a small part of all possible classical systems.) Their returns occur in a nearly periodic fashion. The period of these returns is slightly dependent on initial conditions.
3) For quantum systems, it is also possible to define the entropy of ensemble. Entropy is a measure of uncertainty about the state of a system. A pure state provides a maximally complete description of a quantum system. Therefore, any pure state entropy is zero by definition. For the mixed-state case, the system corresponds to a set of pure states. Therefore, entropy already exceeds zero. Assume that the probability of a pure state is near 1. This mixed state is almost pure and its entropy is almost zero. On the other hand, when all pure states of the mixed state have equivalent probabilities, entropy reaches a maximum.
4) During the evolution of a quantum system, the pure state can evolve to a pure state only. In the mixed state, the probabilities of pure states also remain unchanged. Thus, the entropy of ensemble does not change during the evolution of a quantum system.
5) We can represent a large quantum system by a small number of parameters named macroscopic parameters. A large set of pure states defined by microscopic parameters corresponds to this mixed macroscopic state. The entropy of a macroscopic state can be calculated based on this pure set.

We define this entropy as macroscopic entropy. In contrast with the entropy of ensemble, the macroscopic entropy should not be conserved during the evolution of a quantum system.
6) A quantum system will not be considered as an isolated system due to its interaction with the measuring device. Its initial pure state evolves to a mixed state and its microscopic entropy increases. This evolution cannot be reversed by inversion of the measured system as inversion of the measuring device is also necessary.

### 3.4. Theory of Measurement in Quantum Mechanics [15, 18]  (Appendix J, O, P)

To verify a scientific theory, measurements must be performed with measuring devices. A minimum of two measurements are required: a measurements for the initial and a measurements for the final state. If we know the initial state, we can compare the measured final state with the state predicted by theory. Thus, we can verify the accuracy of the theory using this approach.

In classical mechanics, measurement is a simple process of finding current parameters of the system that does not influence its dynamics. In this case, a complete description of the system, which is provided by all microparametres, yields the unique result of measurement.

This situation is substantially much more complicated in quantum mechanics. Measurement influences the dynamics of a quantum system. For a general case of quantum mechanics, we can predict only the probability of a measurement result despite even the full knowledge of its state (The full knowledge of measured system correspond to a pure state).

We explore the measurement process in quantum mechanics. Let the initial system be represented by a wave function. The measurement of some observables causes the situation in which the wave function transfers to an eigenfunction of a observable with a certain probability. This eigenfunction corresponds to a measured value of a observable, which is equivalent to its eigenvalue. As previously stated, the probability of this measurements is proportional to the quadrate of the amplitude of the wave function. The probability is obtained by expansion to eigenfunctions. After measurement , the system transfers from a pure state to a mixed state and is an ensemble of these possible measurement results with corresponding probabilities. This process is named the reduction of wave function. It is not described by Schrodinger's equation. The Schrodinger equation describes the evolution from a pure state in the pure one. The result of a reduction is a mixed state obtained from an initial pure state. The Schrodinger equation is reversible. The process of reduction is nonreversible. The second type of quantum evolution is possible as the quantum system is not isolated during measurement —it interacts with the macroscopic classical device.

To be consistently represented by quantum mechanics, the macroscopic device should be ideally macroscopic, i.e., either to exist in infinite space or to consist of an infinite number of particles. The ideal macroscopic device does not obey Poincare's theorem of returns and exhibits a certain macroscopic state during all moments of measurement. For the ideal macroscopic device, quantum laws yield the same results as classical laws during any finite period. Note that the real measuring device (i.e., in a finite volume with a finite number of particles) is approximately macroscopic. This condition is essential to our future analysis as it is the main source of the paradoxes considered in the subsequent section. Thus, the evolution of a quantum system is divided into two aspects. The first aspect is reversible—Schrodinger's evolution. The second aspect is the nonreversible reduction of the wave function, which occurs at the interaction with the macroscopic classical device.

We only observe classical devices and not small quantum systems. Thus, a representation of these "mysterious" quantum objects is not necessary. We can consider quantum objects as mathematical abstracts, which enables the establishment of connections between the results obtained by the measuring devices. Measuring devices are classical and representative. They do not have, for example, parameters that cannot be measured simultaneously, such as coordinates and momentum in quantum mechanics. "Evident", "physical", and "intuitive" representations of quantum mechanics are necessary to simplify the understanding of the most complex mathematical models of quantum mechanics. This understanding cannot be completely possible as our intuition is based on the classical world around us. However, as previously stated, no such practical necessity exists. This impossibility is a real source of the well-known "magic" and "mysteriousness" of a quantum mechanics. In reality, "mysteriousness" does not exist.

### 3.5. Complexity of an Attempt of a "Classical" Interpretation of Quantum Mechanics: Introduction of Hidden Parameters and Paradox EPR [18, 103]

The laws of quantum mechanics are probabilistic, and many observables cannot be measured simultaneously. However, many laws of classical statistical mechanics are also probabilistic. Their probabilistic nature can be attributed to hidden microscopic parameters (Appendix U): the velocities and positions of all molecules. Any classical macroscopic state is featured by a set of possible corresponding microstates. Similarly, we can try to interpret a quantum probability by introducing hidden parameters. The knowledge of these hidden parameters enables the unique determination of all variables in the quantum system. Similarly, regarding the macroscopic classical state, an observed state in quantum mechanics corresponds to a set of possible values for the hidden parameters. However, the existence of these hidden parameters is possible in quantum mechanics only under the following assumptions:
1) Measurement (with the exception of special cases in which one of the observable eigenfunctions is equivalent to

the wave function of the observed system) changes the state of the observed system. In a classical case, it is possible (at least in principle) to perform any measurement without the perturbation of the observed system.
2) All hidden parameters cannot be measured simultaneously. Note the Heisenberg uncertainty principle. The measurement changes a system state (a wave function reduction) and thus all hidden parameters cannot also be measured by a set of sequential measurements. All hidden parameters possess well-defined values. However, there is no such *real and observable* physical state in which all hidden parameters possess these well-defined values; a probability distribution exists. In any real experiment, we can measure only a part of these parameters. This measurement will simultaneously produce uncontrollable perturbation for the remainder of the parameters.
3) Existence of the hidden parameters in quantum theory is impossible without introduction of long-range interactions [18]. This long-range interaction acts instantaneously across an infinite distance. However, there is no contradiction with the relativity theory maximal velocity limit as this interaction cannot transfer any information or mass. As the parameters are hidden; the interaction between them is also hidden and not observed. The *observed* appearances can be explained by the correlation of random values.

This necessity of introduction of unobserved long-range interaction is a high price for classical "presentation". The interpretation of hidden parameters is not used in the literature about quantum mechanics. It is easier to consider the laws of quantum mechanics as a mathematical method for the calculation of typical random correlations of observable macroscopic parameters for measuring devices.

This necessity of introduction of long-range interactions of the hidden parameters is illustrated by the well-known Einstein-Podolsky-Rozen "Paradox" (EPR) [18] (Appendix R). This "paradox" is fictitious and is derived from a classical interpretation (i.e., interpretation of hidden parameter) of quantum mechanics by the use of "the small price" (i.e. without long-range interaction).

This paradox is based on an analysis of electron-positron pair states. Initially three particles were in proximity; they subsequently scattered over a large distance.

An electron (or positron) exhibits intrinsic angular momentum, which is defined as spin. The classical analogue of intrinsic angular momentum is the angular momentum of rotation around an intrinsic axis. In contrast with classical intrinsic angular momentum, the absolute value of a spin projection has invariable magnitude (1/2). and its projection to any axis has only two possible values: along the axis and across the axis (+1/2 and-1/2). If we choose the other axis, it will possess the same property. However, projections to two different axes cannot be measured simultaneously. There is no quantum state in which spin projections to two different axes have certain values. Assume that the electron-positron pair is conceived with a total spin of zero. The electron and the positron will move in opposite directions to a large distance between them. The exact values of their spins are unknown. Assume that we have measured a spin +1/2 of an electron along an axis (we designate it axis Z). From the law of conservation of the complete spin, the projection of the positron spin projection on the same axis is -1/2. We can also measure the positron spin projection along any other axis. If this axis is perpendicular to the Z axis, we can measure both +1/2 and-1/2 with equal probability. For a different axis, the provisions of quantum mechanics also enable the precise calculation of the mutual probabilities of positron-electron spin projections (Fig. 11).

Assume that spin projections have "classical" interpretations as hidden parameters. Measurement yields previously hidden values. Why does the dependence of measured spin projections of an electron and a positron exist? Are these common random correlations? Does a long-distance interaction between the hidden parameters exist? Using Bell's inequality (Appendix T), it is possible to prove the following theorem: There is no set of hidden parameters such that the probability distributions of these hidden parameters can explain the mutual probabilities calculated from quantum mechanics without introducing a long-distance interaction between these hidden parameters. The interaction of long-distance spins must exist in hidden spins theory.

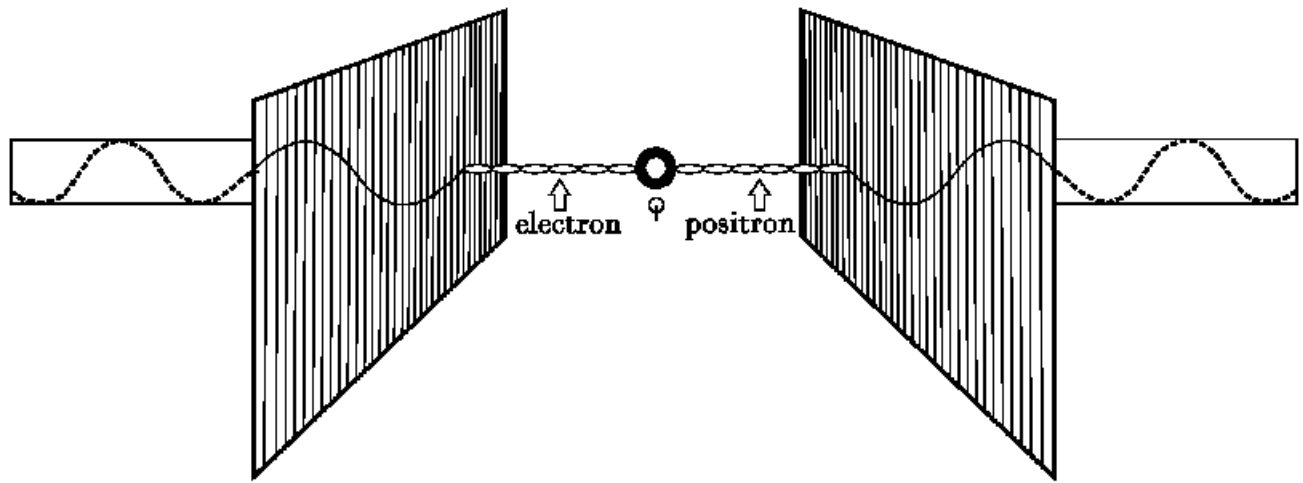

electron    positron

**Figure 11.** EPR experiment

However, it is possible to not consider all of the hidden parameters but only the results of measurement. In this case, the dependence between the electron and positron spins can be easily explained using typical random correlations between the measured parameters. We can simultaneously measure only two projections of a spin (one projection for an electron and one projection for a proton) from numerous possible hidden parameters.

The dependence between long-range objects is defined as a quantum correlation. This correlation cannot be explained by hidden parameters without the long-range interaction of these hidden parameters. This explanation is always possible for classical correlations. After measurement, the quantum correlations are transformed to classical correlations.

The introduction of hidden parameters in quantum mechanics is impossible without long-range interactions between these parameters. It is possible to disregard hidden parameters and to consider quantum mechanics, such as a mathematical apparatus provides random dependence between the measured properties of large classical devices. In this case, long-range interaction is not required. Dependence between observed values can be explained by random correlation. Correlation exists as the measured quantum objects initially co-existed.

### 3.6. Problem of Two Slots As an Illustration of the Complexity of Quantum Mechanics [96,97,104]

Due to the impossibility of an "easy" classical interpretation of quantum mechanics, a well-known American physicist named Richard Feynman assumed that nobody understands quantum mechanics. He noted that «the single secret of quantum mechanics can be expressed by just one experiment using a double slot and electrons. It is a modified version of a simple classical experience in 1801 by an English scientist Thomas Young, who demonstrated the wave nature of light. In Young's experiment, light from a source (a narrow slot $S$) illuminates a screen with two closely positioned slots $S_1$ and $S_2$. While transiting through each of the slots, the light is scattered by diffraction; therefore, the light beams that have transited through slots $S_1$ and $S_2$, were overlapped on a white screen E. In the field of overlapping light beams, numerous alternating light and dark bands formed—we define this overlapping as an interference pattern. Young interpreted the dark lines as places in which "crests" of light waves from one slot meet "troughs" of waves from the other slot and quench each other. The bright lines occur in places in which crests or troughs from both slots coincide, which amplified the light. During almost two hundred years, variants of the two-slot-hole Young experiments were considered to be proof of the nature of waves on water, radio signals, X-rays, sound and thermal radiation (Fig. 12).

We define the concept of the path difference of waves from slots. Assume that a point exists on the final screen. The difference in the distances from the two slots to this point, which is measured in wave length units, is named the path difference for this point. If it is an integer, the maximum wave occurs at this point. If it is an integer and a half, the minimum occurs at this point.

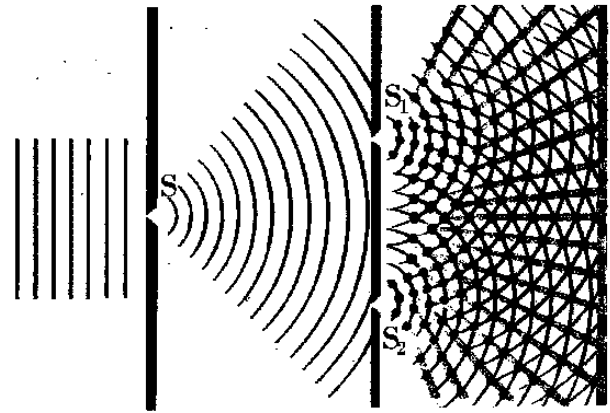

**Figure 12.** Young's experiment with light.

It is remarkable that Young's experiment can also be conducted with electrons (Fig. 13,14). Instead of a sunlight beam, a beam of electrons transits through parallel slots. The screen plate is coated with a luminophor (similar to the screen of a television tube). Each electron colliding with the luminophor produces an illuminating point, which registers its arrival in the form of a common particle. However, the image generated by all electrons yields a surprising impression. The image gives an interference pattern similar to the pattern obtained in the case of light. The image contrasts with the pattern that we would obtain by throwing balls at a fence with two boards removed (similar to two slots). The two-slot-hole experiment with electrons demonstrates that these particles can behave as a wave.

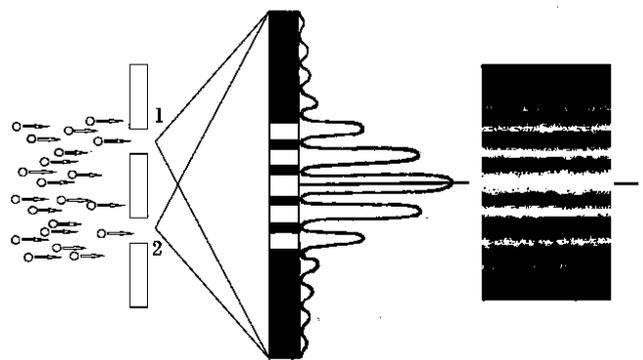

**Figure 13.** Young's experiment with electrons.

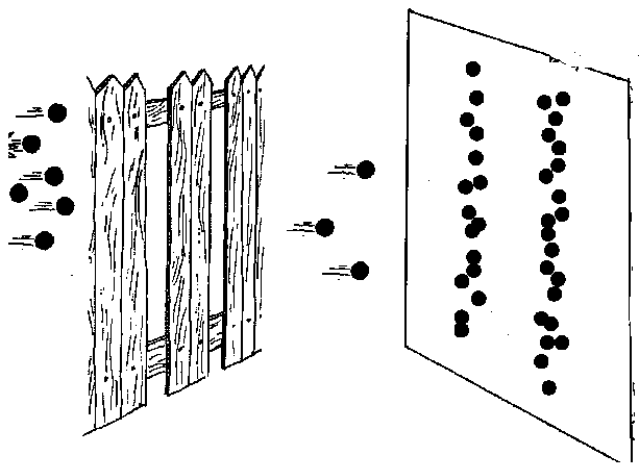

**Figure 14.** Young's experiment with balls and a fence.

If one of the slots is closed in the two-slot electron diffraction experiment, an interference pattern disappears. Instead, the band of the electrons is registered. We open the second slot and close the first slot. In this manner, we obtain the second band. The final pattern is similar to the pattern obtained by the ball game described previously, i.e., a simple sum of these two bands. However, if both slots are opened simultaneously, we observe a complex interference pattern. The results of this experiment cannot be explained by the interaction of electrons—the same result is obtained by emitting electrons one at a time—as an electron position is not defined by a specific trajectory but by a wave of probabilities. The two waves from the two slots are summed to yield an interference pattern. The quadrate of the amplitude of sum of these two waves on the screen yields the probability of locating an electron in this area.

Assume that we arrange a detector that shows the slots through which an electron transits. In this case, the final pattern is similar to the results of the experiment with alternately closed slots, i.e., the interference pattern disappears. This result is explained by the influence of the measuring device—the detector. A reduction of wave function occurs, and it's pure state transfers to the mixed state. Thus, instead of the sum of wave amplitudes from two slots, the probabilities are condensed and the interference pattern disappears.

This experiment demonstrates two main properties of quantum mechanics. First, we cannot predict the exact final position of an electron on the screen but we can discover the probabilities of all points. Only a large number of electrons produces a certain and predictable distribution pattern on the screen. In the classical case, the result was predictable for a single particle. Second, we cannot perform *any* measurement of the intermediate state of an electron without perturbation of this intermediate state that causes variation in the measurement results. After observing which slots the electron has transited, we destroy the additional interference pattern. In classical mechanics, it is always possible, at least in principle, to perform a measurement without the perturbation of the system dynamics. In quantum mechanics, this measurement is possible only if the wave function of the measured system is identical to an eigenfunction of an observable.

The experiment with two slots also facilitates an explanation of the mechanism of the vanishing of quantum interference effects for macroscopic systems, which occurs under the following three requirements:

1) A coherent "monochromatic" wave, which is considered in the experiment, interacts with its environment or its source. This interaction causes the transformation of its pure state to the mixed state. Consequently, the probability wave does not consist of an infinite sine curve but of a set of sine curve segments. This segment of the sine curve is named a wave packet. A phase of a wave packet possesses a random value. The length of a wave packet is approximately 10–20 wave lengths with the order of wave-atom interaction radius. The atoms correspond to the surrounding medium or the source.

2) Consider a system with a macroscopic size. The distances between the slots (D) are significantly larger than the lengths of a wave packet (n $\lambda$) and the distances from the slots to the screen (L). Precisely, D>> $(L \cdot n\lambda)^{1/2}$, where $\lambda$ is a wave length.

3) The introduction of macroscopic parameters (the average wave intensity for a length is considerably larger than the lengths of a wave packet and occurs during a period that is considerably longer than the time of the wave packet transiting through a point).

As the distance between slots increases (at constant value L), the path difference becomes much larger than the length of a wave packet for the majority of points on the screen. As result, the phases of the waves that originate at the slots become random. Thus, coarsened macroscopic intensities are condensed rather than the amplitudes of the waves. The interference disappears for the majority of points on the screen. As the distance between the slots becomes larger than the distance to the screen, the interference remains in the small neighbourhood of a screen point, which would be precisely between the slots. The size of an interference range becomes equivalent to a wave packet. As the distances between slots increase, the wave intensity in this interference range begins to decrease and converges to zero without a decrease in size[3]. Thus, small interference effects are not observed for coarsened (i.e., macroscopic) descriptions.

These effects of loss of interference are caused by the macroscopic nature of the system and its parameters and by the mixed nature of the initial state. This transformation of a wave from a pure coherent state to a mixed state due to interaction (entangling) with its environment is named decoherence (from the Latin *cohaerentio*—connection) [21-25] (Appendix P). The system is mixed or entangled with a surrounding medium. For macroscopic (i.e., very large) systems, decoherence causes a loss of quantum interference, as discussed previously in the experiments with two slots. The decoherence theory has an important consequence: for macrostate quantum theory, the predictions almost coincide with the predictions of classical theory. However, the price for this coincidence is irreversibility, as will be subsequently discussed.

---

[3] Note that this system has an infinite size in the wave propagation direction and the screen with interference (its distance to the screen with slots is constant and equal to L) does not reflect the wave. Therefore, in contrast with the finite systems considered in the subsequent section, the interference does not reappear after it disappears. Conversely, when the distance between the slots converges to infinity (at the constant value L), the effects of quantum interference converge to zero (for any finite wave packet length and for any finite degree of macroparameter coarsening).

### 3.7. Schrodinger's Cat Paradox [26] and Spontaneous Reduction [18, 98]

The complete violation of the wave superposition principle (i.e., the complete vanishing of interference) and the wave function reduction occurs only during the interaction of a quantum system with an ideal macroscopic object or device. The ideal macroscopic object either possesses infinite volume or consists of an infinite number of particles. Such an ideal macroscopic object can be consistently described by quantum and classical mechanics[4].

We consider (unless the other is assumed) only systems with finite volume with a finite number of particles, which is similar to the classical case. Such devices or objects can be considered as only approximately macroscopic[5].

A real experiment demonstrates that, for nonideal macroscopic objects, the destruction of superposition and corresponding wave function reduction may occur. We define this reduction of imperfect macroscopic objects as spontaneous reduction. Despite its tremendous successes, spontaneous reduction produces paradoxes that produce doubt in the completeness of quantum mechanics. We reduce the most impressive paradox from this series—Schrodinger's cat paradox (Schrodinger 1935) [26] (Fig. 15).

Schrodinger's cat paradox is a thought experiment that clarifies the principles of superposition and wave function reductions. A cat is placed in a box. With the exception of the cat, there is a capsule with poisonous gas (or a bomb) in the box, which can detonate with a 50 percent probability due to the radioactive decay of a plutonium atom or a casually illuminated light quantum. After awhile, the box is opened and it is revealed whether the cat is still alive. Until the box is opened (measurement is not performed), the cat stays in a strange superposition of two states: "alive" and "dead". For macroobjects, this situation appears to be mysterious[6]. (Therefore, for quantum particles, the superposition of two different states is natural.) No basic prohibition for the quantum superposition of macrostates exists.

The reduction of these states by an external observer at the moment the box is opened does not produce any inconsistency in quantum mechanics. It is easily explained by the interaction of the external observer with the cat during the evaluation of the cat's state.

However, the paradox arises in the closed box when the observer is the cat. The cat possesses consciousness and is capable of observing both itself and the environment. After introspection, the cat cannot be simultaneously alive and dead but must exist in one of these two states. Experience demonstrates that any conscious creature feels alive, otherwise it is dead. These situations do not exist simultaneously. Therefore, spontaneous reduction to two possible states (alive and dead) occurs[7]. The cat, along with the contents of the box, does not represent an *ideal* macroscopic object. Thus, an observable and nonreversible spontaneous reduction contradicts reversible Schrodinger quantum dynamics. In this case, it cannot be explained by an external influence, as the system is isolated. [18, 27, 7].

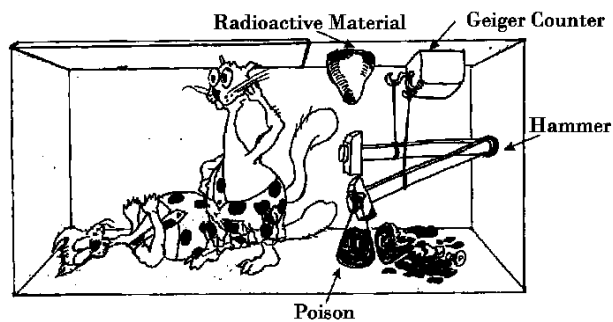

**Figure 15.** Experiment with Schrodinger' cat.

Numerous problems are related to spontaneous reduction. Does spontaneous reduction actually contradict Schrodinger quantum dynamics? When the system is substantially macroscopic, does spontaneous reduction occur? Must a macroscopic system have a consciousness similar to the consciousness of the cat? When does spontaneous reduction occur?

### 3.8. Zeno Paradox or the Paradox of a Kettle that Never Boils

The "the paradox of a kettle that never boils" is related to the last previously mentioned problem. Two paradoxes are discussed.

Assume a quantum process. For example, the decay of a particle or transmission of a particle from one energy level to another level.

The first paradox is as follows: If intervals between acts of registering converge to zero, then the specified process never occurs during any chosen finite interval. This paradox can be explained by the influence of quantum measurement. Measurement causes a reduction to the mixed state (broken and unbroken particles). The relative process rate (over one particle) converges to zero when the measurement interval converges to zero. These two facts cause an end to the process of frequent measurements.

---

[4] Thus, by observing the light of a remote star, we study the light but do not influence the light as the observation may have been expected on the basis of quantum measurement theory. We only change the state of star photons that reach us as we consider the Universe space as infinite. Thus, illuminated photons have no chance of returning to a star and changing its state. In the case of the finite Universe, observable photons can return to a star and influence it. However, for an extremely large Universe, an extremely long period may be required.
[5] For example, this star-observer system in a large finite Universe would behave similarly to an infinite Universe only during a long finite period.
[6] These situations can be described with art, such as the use of using "paradoxical" images [3, 28], (Appendix V).

[7] Certain attempts have been made to understand how the consciousness can perceive these exotic states of macro-objects [3, 28], (Appendix V, and the previous references).

The second paradox is as follows: In real life, the decay of a substance that contains numerous particles is always represented by exponential law. The decay is not casual. The relative velocity of this decay is constant over time. It is impossible to determine the observational "age" of this substance if we do not know the initial quantity of unbroken particles and the quantity of particles that are removed from the system. However, quantum decay, according to the quantum mechanical equations, is not described by exponential law. Therefore, the relative rate of decay is zero at the beginning of the process and, subsequently, increases. We paradoxically conclude that it is possible to introduce a nonphysical concept of the system—"age". "Age" can be easily determined through the current relative rate of system decay.

We will resolve this second paradox in the part of the paper concerned with observable dynamics.

### 3.9. Quantum Correlations of System States and their Connection with the Paradox of Schrodinger's Cat

The concept of the quantum correlation of system states is closely related to the paradox of Schrodinger's cat. Assume a spontaneous reduction of states of the living or dead cat. Any additional measurement will be dependent on the previous state of the cat. The cat is either "living" or "dead". The observed data can be divided into two nonoverlapping groups: one group corresponds to a "living cat" and the other group corresponds to a "dead cat". However, if the cat is in a quantum superposition of these two states, the results of the additional measurements will be dependent on both states of the cat. It cannot be divided into two nonoverlapping groups. This connection between initial states, for which the division of additional results of measurement into independent nonoverlapping groups corresponds to these initial states, is named the "quantum correlation of system states".

In mathematical language, this fact is explained by the nonlinearity of the connection between the probability of observed data and a wave function. The quadrate of the sum is not equivalent to the sum of the quadrates. Additional terms (or the interference terms) are the measures of quantum correlations.

Quantum correlations also correspond to nondiagonal elements of a density matrix. For the mixed state obtained from measurement, all nondiagonal terms are zero.

We express the paradox of Schrodinger's cat in the language of quantum correlations. Introspection about the cat provides only one of two possible results: a "living cat" or a "dead cat". Thus, spontaneous reduction exists, and a quantum correlation between these states disappears, which indicates that additional results of measurement **can be** divided into two independent nonoverlapping groups that correspond to the initial states.

According to Schrodinger's equations, a quantum correlation cannot disappear by itself without the presence of external forces. Additional measurement results can*not* be divided into two independent nonoverlapping groups that correspond to the initial states.

This inconsistency between Schrodinger dynamics and the observable spontaneous reduction produces the paradox.

## 4. Quantum Mechanics Interpretations: Their Failure to Solve Paradoxes

One of the problems that we discussed previously is the difficulty of understanding quantum mechanics based on our classical intuition of real word experience. Various interpretations of quantum mechanics, which **[18]** can facilitate this understanding, exist. Note that none of the interpretations of quantum mechanics can solve the previously mentioned paradoxes; they only enable a visual, distinct and intuitive understanding of quantum mechanics. From an extensive list of possible interpretations, we restrict this discussion to three interpretations. The multiworld interpretation is currently the most popular interpretation.

### 4.1. Multiworld Interpretation. [29, 30, 18, 99]

We describe the multiword interpretation in more detail. Based on the example of Schrodinger's cat, we discover that quantum evolution can generate various and macroscopically distinguishable conditions. We observe only one of these conditions. A multiworld interpretation states that although these states simultaneously exist in certain "parallel worlds", we (or a cat in our mental experiment) can only observe one macroscopic alternative.

A similar approach illustrates the concept of spontaneous reduction: As all worlds exist simultaneously, they all have the capability of influencing measurement results. Generally, measurement results cannot be divided into two disconnected groups related to the living and dead cat. This observation signifies that these worlds are correlated and each influences the measurement results. The presence of spontaneous reduction in measurement causes a loss of this correlation. The measurement results separate into independent groups that correspond to various worlds (Fig. 16).

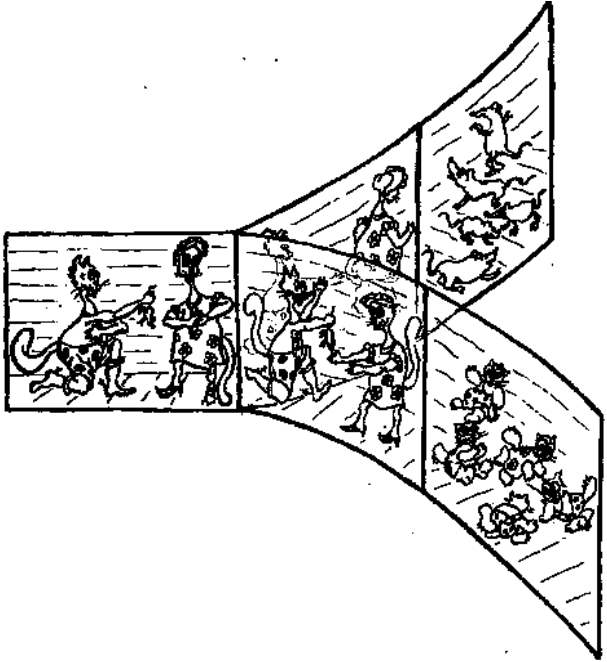

**Figure 16.** Multiworld interpretation.

The multiworld interpretation does not explain the paradox of Schrodinger's cat. The cat observes only one of the existing worlds. The results of additional measurements are dependent on correlations that exist between the worlds. However, neither these worlds nor these correlations are observed. «Parallel worlds» that we know nothing about can exist. These worlds can significantly affect the results of future experiments, i.e., knowledge of only the current state (in our "world") and the laws of quantum mechanics does not enable us to probabilistically predict the future. However, quantum mechanics have been developed for these predictions. Nevertheless, based on a spontaneous reduction, which destroys the quantum correlations between the worlds, we can predict the future using knowledge of only the current

(and observed) state of our "world". We see that the paradox of Schrodinger's cat returns, but in a different form.

The multiworld interpretation does not solve some problems. For example, the definition of macroscopic states that correspond to "separation" of "the parallel worlds" is unclear. (The wave function expansion is ambiguous, and different sets of orthogonal functions can be employed for this purpose.) The approach for obtaining the exact moments of time when this "separation" occurs is unclear. However, obtaining solutions for the paradoxes (in contrast with a common error) is not the purpose of interpreting quantum mechanics.

### 4.2. Copenhagen Interpretation [100]

The Copenhagen interpretation, which has been referenced in existing studies, is a standard for the majority of books and papers in the field of quantum mechanics. It states that at the moment of the observation of macroscopic states, spontaneous reduction occurs and quantum correlations disappear. Thus, the paradoxes described previously are generated.

Note that the reduction in the Copenhagen interpretation occurs only for a chosen *final* observer in the sequence of measurements. The reduction can be different for different observers. Indeed, the experiment should be described from the observer's point of view. The reduction, similar to the velocity of the system, is dependent on the choice of the observation system.

Assume that an external observer investigates another observer, for example, the external observer (scientist) investigates Schrodinger's cat. No spontaneous reduction, which is observed by the cat, occurs for the external observer (Fig. 17).

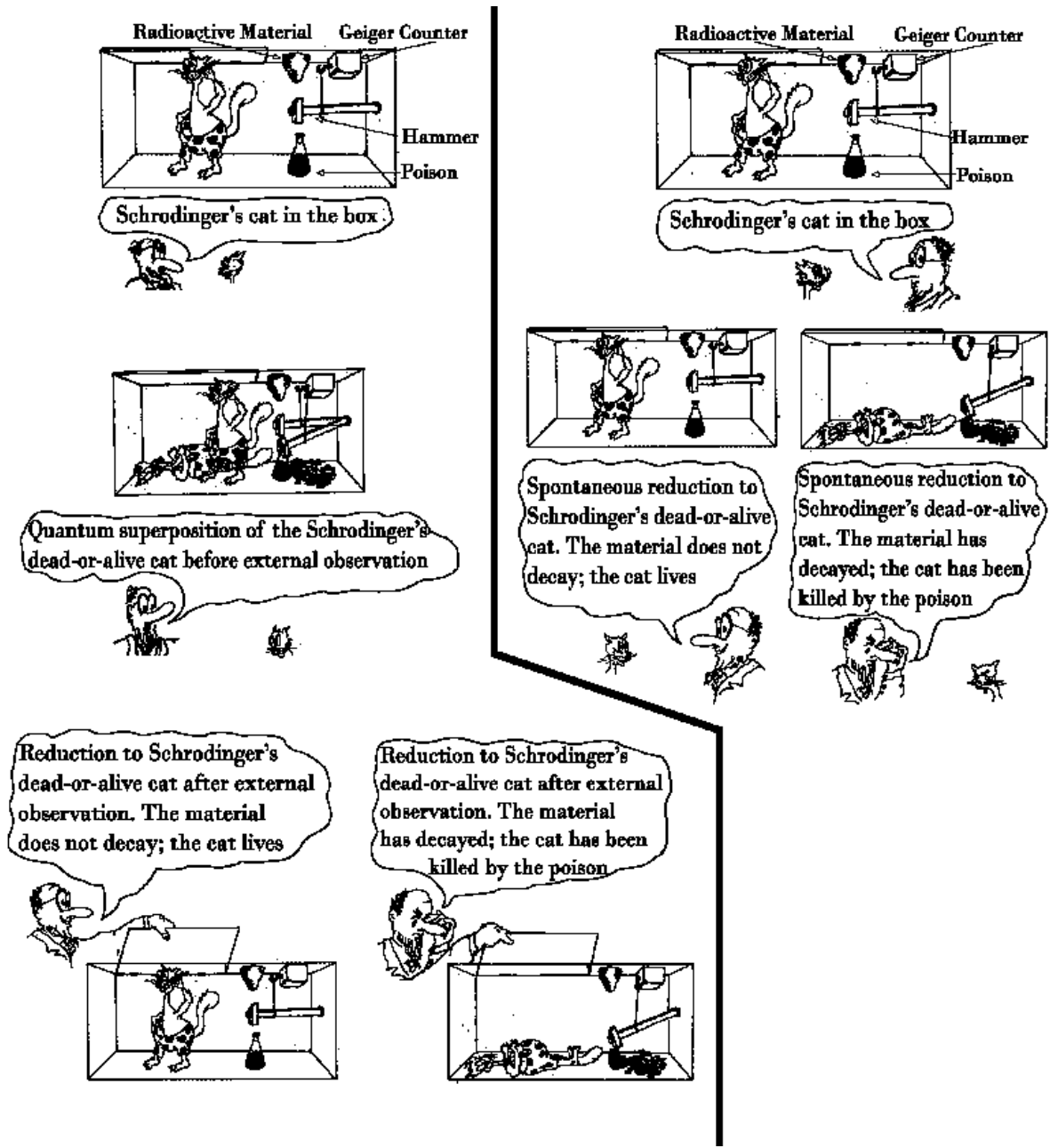

**Figure 17.** Schrodinger's cat experiment from the viewpoint of an external observer (the experimenter) and from self-observation (the cat)

From the point of view of an external observer, the reduction occurs when the external experimenter opens the box and interacts with the cat, which results in the reduction. No paradox exists for the external observer.

Only when the cat is regarded as the final observer does spontaneous reduction occur and, consequently, the paradox described previously exists. The cat can only feel alive or dead but cannot simultaneously exist in these two states.

This observation is very important as its misunderstanding can produce erroneous statements [29, 30], such as the Copenhagen interpretation, which is incompatible with the multiworld interpretation. The difference between these interpretations are not observable; thus, both interpretations can be employed.

### 4.3. Interpretation Via Hidden Parameters. [18], (Appendix S,T,U)

Introduction of the hidden parameters defines another interpretation related to the EPR paradox, such as the wave-pilot theory of de Broglie-Bohm [18]. This theory includes coordinates, velocities, spins and wave functions (wave- pilot) that change over time, according to Schrodinger equations, as hidden parameters. Thus, quantum correlations (as  presented during the discussion of the EPR paradox) produce a locality violation, i.e., long-range interactions among the hidden parameters. For an explanation of the connection among actual measured (not hidden) parameters, such long-range interactions are not necessary. These connections are perfectly described by the typical correlations of variables. Thus, the reduction of a macroscopic state (or a state that occurs at measurement or that is spontaneous) causes the quantum correlations, which become classical correlations, to vanish.

The difference between quantum correlations and classical correlations, which occur after reduction, is expressed not only by existing long-range interactions. Let the correlated long-distance parts of the system (these parts that initially existed in a pure state) appear together after a considerable period of time. Thus, in a quantum case, we again obtain a pure state, but in the classical case, reduction (or a reduction that occurs at measurement or that is spontaneous)—the mixed reduction—also occurs. Spontaneous reduction creates inconsistency in the Schrodinger evolution. Paradoxes do not disappear but only acquire different appearances.

# 5. Definition of the Complete Physical System in the Theory of Measurement

In measurement theory, it is necessary to include both the observer and a surrounding medium in a complete system because, in many cases, even their small influence cannot be neglected. As will be subsequently explained, this necessity of taking into account of the observer's influence is valid not only for quantum mechanics but also for classical mechanics. Generally, a complete system consists of three parts: the observable system, the surrounding medium, and the observer. The observer also consists of three parts: the measuring device, the observer and the memory of the observer. Memory is required for maintaining the sequence of observation. This observation sequence can be employed for the comparison with the theory. Memory must be isolated from its entire environment, with the exception of the channel of receiving information. If certain external factors can influence, change or delete its contents, no experiments for theory verification are possible. This statement, which is critical, helps to resolve many paradoxes, including the paradoxes considered in the subsequent section.

The final point of the complete physical system is the observer's memory. The system includes only one observer. Although many observers can exist, we choose the point of view of only one observer. The remaining observers are considered as parts of the observable system or the environment. Which observer must be chosen? The problem is solved using an approach that is similar to relativity theory—it is possible that any observer can be selected. However, it is important to interpret all facts from the point of view of the single chosen observer. For the case of Schrodinger's cat paradox, the observer can be either the cat or the external observer (but not both).

# 6. Solution of the Paradox of Schrodinger's Cat

Note, the paradox of Schrodinger's cat is inconsistent with regards to the spontaneous reduction observed by a cat and the evolution by Schrodinger forbidding such a reduction. To understand the paradox of Schrodinger's cat, it is necessary to consider the paradox from the point of view of two observers: the external observer and the cat, i.e., introspection.

In the case of the external observer-experimenter, the paradox does not occur. If the experimenter attempts to determine whether the cat is alive, it inevitably influences the observable system (in agreement with quantum mechanics), which causes a reduction. The system is not isolated and thus cannot be represented by a Schrodinger equation. The reducing role of the observer can also be played by the surrounding medium. This case is defined as decoherence. In this case, the role of the observer is more natural and reduced to fixing decoherence. In both cases, the measured system is entangled with the environment or the observer, i.e., correlations of the measured system exist with the environment or the observer.

What will occur if we consider the closed complete physical system, including the observer, observed system and environment? This is the case of the cat introspection. The system includes the cat and his box environment. Note that complete introspection (complete in the sense of quantum mechanics) and the complete verification of the laws of quantum mechanics is impossible in the isolated system, including by the observer. We can precisely measure and analyse a state of an external system in principal. However, if we also include ourselves in the consideration, natural restrictions exist. This occurrence is related to the possibility of retaining memory and the analysis of the states of molecules using these molecules. This assumption produces inconsistencies (Appendix M). Therefore, the possibility of experimentally establishing the inconsistency between Schrodinger evolution and spontaneous reduction using introspection in an isolate system is also restricted.

We discovered mental experiments that cause inconsistency between Schrodinger evolution and spontaneous reduction.
1) The first example is related to the reversibility of quantum evolution. Assume that we have introduced a Hamiltonian that is capable of reversing the quantum evolution of the cat-box system [29, 30]. Although it is almost practically impossible, no theoretical impossibility exists. If spontaneous reduction occurs, the process would be nonreversible. If spontaneous reduction does not occur, the cat-box system will return to an initially pure state. However, only the external observer can construct such a validation. The cat cannot validate this observation through introspection because the cat's memory will be erased after returning to an initial state.
No paradox exists from the point of view of the external observer as he does not observe a spontaneous reduction that can produce a paradox.

2) The second example is related to the necessity of Poincare's return of a quantum system to an initial state. Assume that the initial state was pure. If a spontaneous reduction exists in the case of cat introspection, the reduction produces a mixed state. Return would be impossible—the mixed state cannot transfer to a pure state according to Schrodinger equations. Thus, if the cat has a fixed return, it would result in an inconsistency with spontaneous reduction. However, the cat cannot achieve a fixed return (in the case of the fidelity of quantum mechanics) because a return will erase the cat's memory. Therefore, no paradox exists.

The external observer can observe this return by measurement of the initial and final state of this system. No paradox exists as the observer does not observe a spontaneous reduction that can generate a paradox (Fig. 18).

Note that the inconsistency between spontaneous reduction and Schrodinger evolution is experimentally observable only when the spontaneous reduction is retained in the memory of the observer and this memory is not erased or damaged. All of the previously described do not include this requirement. Thus, these examples distinctly show that although a spontaneous reduction can cause a violation of Schrodinger evolution, this violation is not experimentally observed.

3) The third example concerns the notion that quantum mechanics yields a superposition of a living and dead cat in a box. Theoretically, an exterior observer can always measure this superposition *exactly* if it was one of the measurement eigenfunctions. In contrast with the case in which the living cat and the dead cat are eigenfunctions of a measurement, this measurement would not destroy superposition. Having informed the cat about the result of measurement, we enter into an inconsistency with a spontaneous reduction observed by the cat [31, 32]. This argument possesses a double error.

At first, this experiment is conducted for the *verification* of the existence of the cat's spontaneous reduction when the observer is the Cat. The external observer only influences the cat's memory when spontaneous reduction does not occur, and the cat's state is a superposition of live and dead states. However, the cat's state does influence and can destroy the cat's memory in the case of spontaneous reduction. Thus, this experiment is not legitimate for the verification of the existence of a spontaneous reduction.

Thus, no contradiction with the past exists.

Second, the data transmitted to the cat is retained in his memory. This transmission changes both the state and all further evolutions of the cat, i.e., the system cannot be considered to be isolated in the following section. Thus, no contradiction with the future exists.

The external observer does not observe a spontaneous reduction and therefore does not observe the paradox. From the external observer's point of view, this verification is possible and legitimate. It does not influence the external observer's memory. This verification, which does not break the evolution of the observable system, enables not only the measurement of an initial and final state of a system but also all of the intermediate states, i.e., the verification implements a continuous nonperturbative observation.

Note that the external observer can only theoretically observe the superposition of the live and dead cat. This observation is almost impossible practically. In contrast with small quantum systems, the superposition is observable. Quantum mechanics is usually considered to be the theory of small systems. However, for small macroscopic (**mesoscopic**) objects, such observations are also possible. The large set of particles at low temperatures or certain photon states [**33**] can be employed as an example.

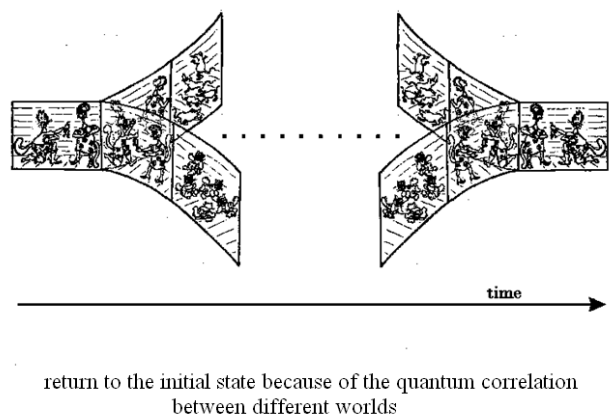

return to the initial state because of the quantum correlation between different worlds

**Figure 18.** Poincare's return close to the initial state due to quantum correlations.

# 7. Solution of the Paradoxes of Loschmidt and Poincare in Classical Mechanics: An Explanation of the Law of Increasing Macroscopic Entropy [105]

Consider two cases: the case in which an observer is included in an observable system and the case in which an observer is external to an observable system.

The basic inconsistency of classical statistical mechanics consists of the inconsistency between the law of increasing entropy and the laws of reversible classical motion. This inconsistency is expressed as the Poincare and Loschmidt paradoxes.

In contrast with quantum mechanics, in the case of classical mechanics, a simpler case is when the observer is included in an observable system. Poincare's return of the system to an initial state causes memory erasing, which is similar to the arguments in the previous chapter. This memory erasing makes experimental observation of the Poincare impossible.

The reversion of velocities produces a decrease in entropy. However, the direction of time is relative and not absolute. We should define the positive direction of the time arrow. It is reasonable to set the time arrow in the direction of entropy growth. We define this time arrow as the proper time arrow of the system. With respect to this proper time arrow, Loschmidt's paradox disappears.

Note that for both of the solutions of the paradoxes there exists a memory loss in the final system state and entropy growth in the direction of the proper system time arrow. The direction of the proper system time arrow reverses with respect to the initial state direction near the final state. The main reason for the impossibility of paradox observation is the impossibility of complete system state knowledge through introspection.

For the external observer, the situation is more difficult. Theoretically, an interaction between the observer and the observable system is arbitrarily small in classical mechanics. Nothing prevents a decrease in entropy. In this case, the direction of the proper time arrow of the observer is opposite of the observable system. Is this possible? The theoretical answer is yes, but the practical answer is that this is almost impossible. The majority of real physical systems consist of mixing (chaotic) systems. Their phase trajectories exponentially diverge in the presence of a small amount of noise. Small interactions between the observer and the observable system or the surrounding medium and the observable system inevitably occur in most cases. Note that the following theorem is valid for chaotic systems:

Processes of *macroparametre evolution* with decreasing macroscopic entropy are highly unstable with respect to small amounts of external noise. Conversely, the process of *macroparametre evolution* with macroscopic entropy growth is stable.

Therefore, inevitable small interactions cause the destruction of the processes of decreasing entropy and the proper time arrow synchronisation with the observer and the observable system. Thus, the system cannot be considered to be isolated, and classical mechanical paradoxes for isolated systems are irrelevant. Small interactions between the surrounding medium and the observer and the observable system will have the same effect as the interaction of the observer with the observable system—the generation of the proper time arrow synchronisation for all subsystems. In this case, the role of the observer is more passive and natural (Fig. 19).

In these studies [1, 2, 6, 34], the synchronism of all of the proper time arrows in the surrounding world is considered to be a mystery. Why are processes with decreasing entropy never satisfied, although their probability is equivalent to the probability of processes of increasing entropy? Attempts have been made to explain the origin of our Universe[8] **[1, 2, 6, 34];** however, these attempts are unnecessary. Simple, inevitable, and small interactions always exist between systems, which cause the visible synchronisation of all proper time arrows.

Thus, the trajectories that cause a decrease in entropy are unstable with respect to small amounts of noise from the external observer (Fig. 20). In the case of quantum mechanics, this noise is theoretically inevitable during measurement if we do not know the true initial state of the measured system. State measurement produces the inevitable violation of this measured state. In classical mechanics, accurate measurement can theoretically be performed. Therefore, we should "manually" introduce a small amount of external noise and/or initial state errors to explain entropy growth for the observer.

In actual measurement, small amounts of external noise are always present. Substantial efforts, which cause environmental entropy growth, are necessary to prevent the influence of noise. This environmental entropy growth is considerably larger than the decrease in entropy obtained by the isolation of this observable system. Thus, the law of increasing entropy is satisfied. An analogy to the paradox resolution regarding Maxwell's Demon [37-38], which employs the sorting of molecules for decreasing entropy, exists. The acquisition of information for sorting leads to a greater entropy increase compared to the correspondent entropy decrease.

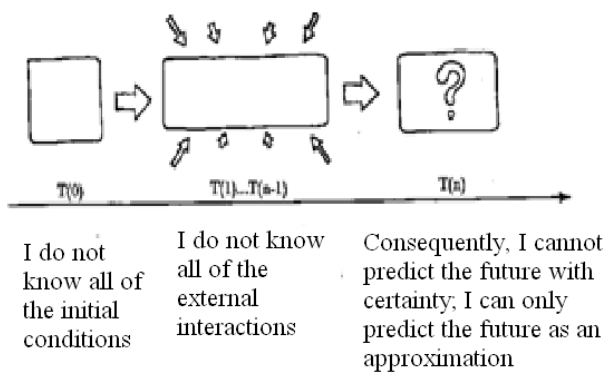

| $T(0)$ | $T(1)...T(n-1)$ | $T(n)$ |
|---|---|---|
| I do not know all of the initial conditions | I do not know all of the external interactions | Consequently, I cannot predict the future with certainty; I can only predict the future as an approximation |

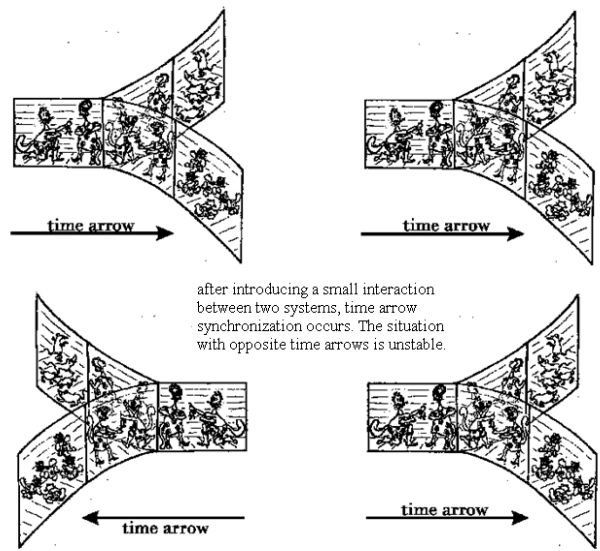

**Figure 20.** Synchronisation of time arrows.

Although the prevention of small and pervasive interactions between systems is a problem for macroscopic systems, it is easily executable for small systems. We observe small fluctuations that correspond to violations from the law of increasing entropy. If we neglect minor friction, reversible processes are also observable in gravitational astronomy.

## 8. Deep Analogies between Quantum Mechanics (QM) and Statistical Classical Mechanics (CM) (Appendix N)

Based on the previously described explanations, it is possible to predict that there is almost a complete analogy between the properties and paradoxes of CM and QM and also between their solution methods. We detail these analogies as follows:

1) Both types of mechanics are reversible in time.

2) Poincare's theorem of returns is applicable to both types of mechanics. In CM, almost-periodic systems constitute a small class of systems, whereas all systems in a finite volume are considered almost-periodic in CM.

3) In CM, correlations with macrostate knowledge and additional hidden microscopic correlations related to the knowledge of its "history" are evident. In QM, two types of correlation are also possible: the classical correlations described by diagonal elements of a density matrix, which are conserved during measurement reduction, and hidden quantum correlations, which are described by nondiagonal elements of a density matrix, leading to paradoxes. Small amounts of external noise from the observer or the environment destroy additional correlations in CM and cause the coarsening of the phase density function. Similarly, in QM, entangling between the observable system and the observer (an inevitable interaction during measurement) or entangling between the observable system with the environment (decoherence) causes the loss of quantum correlations (the loss of the density matrix nondiagonal elements of the density matrix) and wave function reduction.

4) In the case of the introspective observations of Poincare or Loschmidt, returns are impossible due to memory loss. As a result, introspective observations of additional correlations (in CM) or quantum correlations (in QM), which produce paradoxes, is also impossible.

5) Two types of entropy can be defined in QM and CM—entropy of ensembles (phase density function) and macroscopic entropy. Entropy of ensembles is conserved during reversible evolution, whereas macroscopic entropy can either increase or decrease. At introspection, a decrease in entropy becomes unobservable. A small interaction by the external observer with an observable system or by observable systems with the environment also prevents an observation of the decrease in entropy (or makes it hard to achieve). We attempt to isolate an observable system from environmental noise to observe the system of decreasing entropy. A significant increase in entropy, which is considerably larger than the decrease in entropy obtained by isolation, is necessary. Although completely isolated and impenetrable cavities or ideal infinitely light-weight particles do not exist in our real world, small and pervasive interactions do exist.

6) The process of spontaneous reduction in QM is related to the disregard for quantum correlations and transition from a pure state to a mixed state, which causes an increase in macroscopic entropy. An increase in macroscopic entropy can be achieved in a manner similar to the Boltzmann equation in CM. An increase in macroscopic entropy is achieved by the introduction of the "molecular chaos hypothesis", which is related to the disregard for correlations between particles (i.e., their momentums and coordinates are assumed to be independent). Thus, the distribution function of the two particles is considered to be the products of one-particle functions. The introduction of a spontaneous reduction in QM equations is similar to the introduction of laws of increasing entropy in CM equations.

7) The laws of QM are statistical. In QM, observations of the unknown state inevitably result in changes in the evolution of the observable system. As the majority of systems in CM are mixing systems, their behaviour is casual. First, this casual behaviour is related to the small interaction of an observable system with an observer or environment. Second, our knowledge of the initial state is incomplete. However, theoretically, to obtain an accurate initial state measurement and complete isolation, the behaviour of the system can be predicted as precisely as we desire. However, a vast increase in the entropy of the environment is required to attain such accuracy in reality.

8) In both cases, paradoxes arise only for macrosystems. The laws of the behaviour of microsystems are not applicable due to small amounts of external noise and a finite accurate knowledge of the initial state. The single critical difference between QM and CM is described as follows: In CM, small but finite interaction during measurement (observation) or small errors of the initial state knowledge are "manually" introduced, whereas errors arises naturally in QM due to the theoretically inevitable interaction that occurs during the measurement of an unknown quantum state[9].

In conclusion, we make an unexpected deduction: the paradox of Schrodinger's cat in QM is the quantum analogue of the law of increasing entropy in CM. In QM, microsystems are usually investigated, whereas macrosystems are usually investigated in CM. Thus, these equivalent paradoxes possess different forms but possess the same premise (Fig. 21).

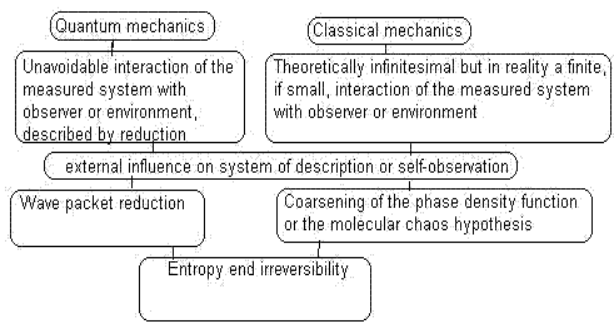

**Figure 21.** Sources of irreversibility and entropy in physics.

# 9. Time Arrows Synchronisation/Decoherence [88-94]

Assume that a process in which entropy decreases exists. Specifically, we consider this process to be a spontaneous reconstruction of a house (previously destroyed in an earthquake).

We also consider a simple example of gas expanding from a small region of space into a large volume. After a specific period of time, the velocities reverse, and the gas is now located in the initial small volume. If we use a camera to take a series of snapshots that record the different stages of a spontaneous house reconstruction/(gas shrinking), we expect that the camera will record this spontaneous process. Why is the camera not able to record it? What prevents the camera from recording these snapshots?

The answer to these questions are as follows: even a small interaction between the camera and the observed system destroys the process of inverse decreasing entropy and results in the synchronisation of the time arrow directions of the observer and the observed system. (A time arrow direction is defined as the direction of the increase in entropy.) This small interaction appears because of light, which is eliminated by the observed object and reflected by the camera (and also due to light eliminated by the camera). In the absence of the observer with the camera, the environment can eliminate and reflect the light. (Any process without an observer is nonsensical. Although the observer's influence is substantially smaller than the influence of the environment, the observer must appear at a stage of the process.) External amounts of noise (interaction) from the observer/the environment destroy the correlation among molecules of the observed system, which results in the prevention of the inverse process with decreasing entropy. In quantum mechanics, this process is defined as "decoherence". The house reconstruction/(gas shrinking) ceases, i.e., the house is not reconstructed/(the gas will not shrink). Conversely, the processes of increasing entropy are stable.

Decoherence (synchronisation of time arrows and "entangling") and relaxation (a system achieves its equilibrium during relaxation) are absolutely different processes. During relaxation, macroscopic variables (entropy, temperature, and pressure) change significantly to their equilibrium values, and invisible microscopic correlations between parts of the system increase. During decoherence, the macroscopic variables (entropy, temperature, and pressure) are almost constant. Invisible microscopic correlations inside the subsystems (environment, observer, and observed system) are destroyed, but new correlations between the subsystems are evident, which are named "entangling" correlations in quantum mechanics. During this process, synchronisation of the time arrows also occurs. The period of relaxation is much longer than the period of decoherence.

Assume a simple example of a gas that expands from a small region of space into a large volume. In this entropy-increasing process, the time evolution of macroscopic parameters is stable for small external perturbations. If all molecular velocities are eventually reversed, the gas will be located in the initial small volume; this prediction is valid in the absence of any perturbation. This entropy-decreasing process is distinctly unstable, and a small external perturbation would trigger a continuous entropy growth. Thus, entropy-increasing processes are stable, whereas entropy-decreasing processes are unstable.

The following example is a citation from Maccone's paper [39]:

"However, an observer is macroscopic by definition, and

[9] Numerous examples of "purely quantum paradoxes" exist, which ostensibly are not analogous to classical statistical mechanics. An example is the Elitzur-Vaidman paradox [36] regarding the bomb that can be discovered without an explosion:
1) Assume that the wave function of one photon branches into two possible channels of certain devices. These channels eventually unite to produce an interference of two probability waves. Introducing a bomb into one of the channels will break the process of interference and will enable the discovery of the bomb even if the photon does not detonate it. (The photon is capable of detonating the bomb)
2) The complete analogy is the following experiment in CM. In one of the channels, that does not contain a bomb, we add a macroscopic beam of numerous lightweight particles. In the channel that houses the bomb, only one lightweight particle travels. Since this particle is not capable of detonating the bomb, the bomb can be added into the channel. This particle is macroscopically undetectable due to the finite sensitivity of the devices. If the beam of particles at the end of the channel exhibits unstable dynamics, the presence of one undetectable additional particle can significantly change the dynamics (named the «butterfly effect»). It will enable a new particle to transit through the second channel if bombs are not present. This explanation constitutes the complete analogy between QM and CM paradoxes.

all remotely interacting macroscopic systems become correlated very rapidly (e.g., Borel famously calculated that moving a gram of material on the star Sirius by 1 m can influence the trajectories of the particles in a gas on earth on a time scale of s [106])"

No problem exists to reverse the observer (the camera), and the observed system simultaneously. Due to Poincaré's return theorem for a closed system (including the observer and the observed system), the return must automatically occur after a considerable period. However, the memory loss of the observer does not facilitate to observe this process.

The majority of real systems are *chaotic*—a weak perturbation may lead to an exponential divergence of trajectories, and a non-negligible interaction always exists between an observed system and an observer/environment. *In principle,* in both quantum mechanics and classical mechanics, we can make unperturbative observations of the process of decreasing entropy. An example of such a mesoscopic device is a quantum computer: the law of increasing entropy does not exist for such a system. This device is isolated from the environment and the observer. *In practice,* unperturbative observations are almost impossible for macroscopic systems. We can conclude that the law of increasing entropy is the *FAPP* (for all practical purposes) *law.*

Consider the synchronisation of time arrows for two noninteracting (prior to an initial moment) systems. Initially, the systems exhibit opposite time arrows, which indicate that there are two noninteracting systems: in one system, time flows (i.e., entropy increases) in one direction, whereas time flows in another (opposite) direction in the second system. However, when the systems interact with each other, one system (the "stronger" one) will cause the other ("weaker") system to flow in its ("stronger") direction; eventually, time flows in the same direction ion both systems.

What does it mean to be "stronger? Is it something that increases with the number of degrees of freedom of the system? This definition is incorrect. The concepts of "stronger" or "weaker" are not dependent on the number of degrees of freedom of the system. For the first system, the interaction occurred in the *future* after an initial moment (in the initial moment, the systems have opposite time arrows). For the second system, the interaction occurred in the *past.* The situation is *not symmetric with time*, and the first system is always "stronger", which can be attributed to the instability of the processes of decreasing entropy and the stability of the processes of increasing entropy, as described previously.

Assume that we have two initially isolated vessels that contain a gas. In the first vessel, the gas expands (increase in entropy). In the second vessel, the gas shrinks (decrease in entropy).

In the first vessel, the gas expands from a small volume in the centre of the vessel. The velocities of the molecules are directed from the centre of the vessel to its boundary. Physically, a small perturbation of the velocities cannot stop gas from expanding. After a random small perturbation, the velocities will continue to be directed from the centre of the vessel to its boundary. Noise increases with the expansion. Thus, the expansion process is stable.

In the second vessel, the gas shrinks from the full volume of a vessel to its centre. The velocities of all of the molecules are directed to the centre of the vessel. Physically, a small random perturbation of the velocities can easily stop the gas from shrinking. After a small perturbation, the velocities will not be directed to the centre of the vessel. Thus, the shrinking process is stopped. We can conclude that the shrinking process is unstable. This shrinking process can be obtained by reversing the expansion of the gas. If we reverse the molecular velocities of the expanding gas *before* the molecules collide with each other and the vessel boundary, this instability is linear and not strong. However, if we reverse the molecular velocities of the expanding gas *after* the molecules collide with each other, this instability is exponential and considerably stronger.

Both directions of time have equivalent roles; a small random noisy interaction can break this symmetry for these two systems due to the instability of the decreasing entropy processes. The symmetry of time exists only for a *complete* system, including the two previously defined subsystems. However, the time arrows of the interacting subsystems must be equivalent.

In reality, the interaction with infinite time can be replaced by the large finite time T, which is chosen to be much smaller than Poincaré's return time. In the first system, an interaction occurs during [0, T]; in the second system, an interaction occurs during [-T, 0]. Can our argument still apply? Instead of the asymmetry of the forces in this case, we obtain a asymmetry of the initial conditions: At initial moment 0 for the first coordinate system [0,T], the two vessels have different eigen time arrows. However, at initial moment -T for the second coordinate system [-T,0], the two vessels have the same eigen time arrows in a negative direction. The situation will be symmetric only if T is equivalent to Poincaré's return time. For this situation, the two eigen time arrows are also different in moment T but are the opposite of their initial directions at time 0. The "stronger" system has future interacting forces with respect to its eigen time arrow.

This theory explains why entropy growth in the Universe occurs in the same direction but cannot explain the low-entropy initial condition of the Universe. The low-entropy initial condition of the Universe is most likely a result of the anthropic principle [42].

# 10. The Law of Increasing Entropy and the "Synchronisation of Time Arrows"/Decoherence in Gravitation Theory

In Einstein's general relativity theory, motion is reversible, which is similar to classical mechanics. An important difference also exists between general relativity and classical mechanics. General relativity is an ambiguous theory. In general relativity, two various initial states can yield infinitesimally close states after a *finite* interval. This process of converging occurs, for example, during the formation of a black hole as a result of a collapse. Consider the inverse process, which is described as a white hole. In this process, the infinitesimally close initial states after the *finite* interval can yield different final states. This divergence process signifies that an observer/environment can considerably affect its evolution during a *finite* interval even when the observer/environment infinitesimally and weakly interacts with the white hole. As a result of this property, the law of increasing entropy becomes an exact law but not FAPP (for all practical purposes). Thus, entropy becomes a fundamental concept. The entropy of a black hole is a fundamental concept. The existence of this entropy can be explained by the perturbation created by the observer. This perturbation may be infinitesimally weak compared with classical mechanics. During the formation of the black hole, entropy increases. Time reversion produces the appearance of the white hole and entropy subsequently decreases.

The white hole cannot exist in reality due to the decrease in entropy. The decrease in entropy is prohibited in general relativity due to the same reasons that it is prohibited in classical mechanics. However, the instability of the decrease in entropy processes is significantly stronger for general relativity than for classical mechanics. This instability results in the synchronisation of the eigen time arrows of the white hole and the observer/environment. The direction of the eigen time arrow of the white hole changes to the opposite direction, which coincides with the eigen time arrow of the observer/environment. The white hole transforms into a black hole.

The well-known black hole information paradox [43] is as follows: information (which in classical and quantum mechanics is conserved) disappears in a black hole forever. This situation should not pose a problem: information is most likely stored inside of the black hole in some form. However, chaotic Hawking radiation explicitly defines this process of information loss: the black hole evaporates but the information is not recovered.

Hawking radiation concerns semiclassical gravitation. However, the paradox can also be formulated within the framework of the general relativity theory. The spherical black hole can be "changed" into a white hole at some moment (It is difficult to realise in reality, but a similar transformation is considered in [95]. The reversion in time is changed by a "wormhole" from a black hole to a white hole that is located in another Universe). Thus, the process is converted over time. However, information cannot be recovered due to the ambiguity (the infinitely high instability) of the evolution of the white hole.

Typically, only two solutions are considered for this problem: information disappears or, due to interior correlations of Hawking radiation (or exact reversion of the black hole process after its transmutation into a white hole), the information is conserved. The third solution is most likely the true solution. Due to the inevitable influence of the observer/environment, it is impossible to experimentally distinguish these two situations. If a subject cannot be experimentally verified, it is not a subject of science.

The paradox of the general relativity theory and semiclassical gravitation can be resolved using the influence of the observer/environment. Assume that Hawking radiation is correlated and not chaotic (or the white hole would be exactly inversed into a black hole). The infinitesimal influence of the observer/environment causes the inevitable loss of these correlations (and the corresponding information) during a finite interval. The inclusion of the observer in the described system is nonsensical: complete self-description and introspection is not impossible. The information conservation law cannot be verified experimentally for this case, even if it is correct.

We have no general theory of quantum gravitation. However, for a special case of a 5-dimentional anti-de-Sitter space, this paradox is considered to be resolved by many scientists. Information is supposed to be conserved, as a hypothesis regarding AdS/CFT dualities, i.e., the hypotheses that quantum gravitation in the 5-dimensional anti-de-Sitter space (that is, with the negative cosmological term) is mathematically equivalent to a conformal field theory on a 4-surface of this world. This equivalence was verified for special cases but not proven for the general case. Assume that if this hypothesis is true, it automatically solves the information problem. The conformal field theory is unitary. If it is dual to quantum gravitation, then the corresponding quantum gravitation theory is also unitary. Thus, the information in this case is not lost. Assume that it is incorrect. The process of the formation of a black hole and its subsequent evaporation occurs on *all surfaces* of the anti-de-Sitter space (described by the conformal quantum theory). Although such process includes the observer/environment, the observer cannot precisely know an initial state and analyse the system behaviour as he is a part of this system. However, his influence on the system cannot be neglected. Thus, the experimental verification of the information paradox becomes impossible.

Consider the law of increasing entropy from the point of view of a paradoxical object of general relativity theory, such as a wormhole [44]. We consider the Morris-Thorne wormhole [45]. Using a simple procedure (we place one of the wormhole mouths on a spaceship; the spaceship moves with relativistic velocity over a closed loop and the mouth returns to its initial location), the wormhole traversing space can be transformed into a traverse period. After this transformation, the wormhole can be used as a time machine, which indicates the well-known paradox of a grandfather. How can this paradox be resolved?

For macroscopic wormholes, the solution can be determined using the law of increasing entropy. The realisation of this law is ensured by the instability of the processes of decreasing entropy, which results in the synchronisation of time arrows.

The wormhole traversing space does not produce the paradox. Assume an object enters one mouth at a certain moment and exits from the other mouth after a later moment. Thus, the object travels from an initial high-order low-entropy environment to a future low-order high-entropy environment. During the trip along the wormhole, the entropy of the object also increases. Thus, the directions of the time arrows of the object and the environment are equivalent. The same conclusions are correct for travelling from the past to the future into a wormhole traversing time.

However, for travelling from the future to the past, the directions of the time arrows of the object and the environment will be opposite. The object travels from the initial low-order high-entropy environment to the future high-order low-entropy environment. Its entropy increases and does not decrease. As we stated previously, this process is unstable and will be prevented or forcedly converted by the process of synchronisation of the time arrows. This process must occur at the moment that the mouth of the wormhole returns to its initial state.

"Free will" enables us to initiate only irreversible processes with an increase in entropy but not with a decrease in entropy. Thus, we cannot send the object from the future to the past. The process of synchronisation of the time arrows (and the corresponding entropy growth law) forbids the initial conditions that are necessary for macroscopic objects to travel to the past (and the realisation of conditions for the paradox of a grandfather).

Paper [46] demonstrates that for the thermodynamic time arrow, it is impossible to have an identical orientation with the coordinate time arrow over a closed time-like curve due to the entropy growth law. The process of synchronisation of the time arrows (concerned with infinitely large instability and ambiguity of the processes of decreasing entropy) considers the *physical mechanism*, which ensures this impossibility and the realisation of the entropy growth law for the same thermodynamic time arrow.

For microscopic wormholes, the situation is absolutely different. If the initial conditions are compatible to travelling to the past through a wormhole, no reasons exist that can prevent it. If a small (even infinitesimally small) perturbation of initial conditions produces an inconsistency with the wormhole existence, the wormhole can be easily destroyed [47]. The property of general relativity is evident: infinitely large instability (ambiguity) exists, which indicates that the infinitesimal perturbation of initial conditions can result in finite changes to the final state during a finite period.

However, this is not a solution of the grandfather paradox, which is a macroscopic phenomenon. Assume that processes with opposite time arrow directions exist: a cosmonaut and the surrounding Universe. The cosmonaut travels over a wormhole from the Universe's future to the Universe's past. For the eigen time arrow direction of the cosmonaut, the cosmonaut will be travelling from his past into his future. For the theory of general relativity, this situation is impossible, even in principle (in contrast with classical mechanics): even an infinitesimal interaction leads to synchronisation of the time arrow directions as an infinitely large instability

(ambiguity) of processes with decreasing entropy (in this case the "process with decreasing entropy" consists of the cosmonaut travelling from the future to the past). This synchronisation of the time arrow directions can be accompanied by the destruction of the wormholes [47], conservation of the wormhole and a modification of initial conditions [46]. The entropy growth law (and the corresponding synchronisation of time arrow directions) *does not allow the occurrence* of these situations with an inconsistency between macroscopic initial conditions and an initially defined (unchanging and invariable) macroscopic space-time topology (including a set of wormholes) [46]. We formulate a final conclusion: *for macroscopic processes,* the infinitely large instability (ambiguity) of processes with a decrease in entropy (and the corresponding synchronisation of time arrow directions) represents an impossible occurrence of initial conditions, which is incompatible with the existence of the given wormholes. This instability also prevents the destruction of both wormholes and the transportation of macroscopic objects to the past (resulting in "the grandfather paradox"). In sum, the same reasons (which enabled us to resolve the reduction paradox and the Loshmidt and Poincare paradoxes) also enable the resolution of the information paradox of black holes and the grandfather paradox for wormholes. This is a remarkable universality.

# 11. Ideal and Observable Dynamics

## 11.1. Definition of Ideal and Observable Dynamics. Why is Observable Dynamics Necessary?

We determine that the exact equations of quantum and classical mechanics describe IDEAL dynamics, which are reversible and lead to Poincare's returns. Conversely, the equations of physics describe OBSERVABLE dynamics, for example, of a hydrodynamic equation of viscous fluid, of the Boltzmann equation in thermodynamics, and of the law of growth of entropy in isolated systems (*master equations*), are nonreversible and are excluded by Poincare's returns to an initial state (Fig. 22).

We provide the definition of observable and ideal dynamics and also explain the introduction of observable dynamics. Ideal Dynamics is described by exact laws of quantum or classical mechanics. Why do we use the word "ideal" in our definition? We employ this word because observed laws (the laws of increasing entropy and spontaneous reduction) contradict ideal dynamics laws. This violation of ideal dynamics, which is explained by the interaction of measured systems with the environment or/and an observer or by the limits of self-knowledge of the system obtained by introspection in the case of the external observer and the environment, is included in the considered system. Real systems are either open or not completely described, i.e., ideal dynamics are impossible.

Can we conclude that using physics laws is impossible in these cases? Absolutely not. The majority of these systems can be described by equations of exact (or probability) dynamics, despite their nonisolation or incompleteness. We define these dynamics as observable dynamics. The majority of equations in physics, which are named *master equations* (such as the hydrodynamic equations of viscous fluid, the Boltzmann equation in thermodynamics, and the law of growth of entropy for isolated systems), are equations of observable dynamics.

To possess this property, observable dynamics should satisfy certain requirements. Observable dynamics cannot be operated with a complete set of microvariables. In observable dynamics, we use a considerably smaller number of macrovariables, which are functions of microvariables. Thus, observable dynamics are significantly more stable with respect to initial condition errors and to external noise. The change in the microstate does not inevitably cause a change in the macrostate because one macrostate corresponds to a large set of microstates. For example, the macrovariables for gas consist of density, pressure, temperature and entropy. Microvariables are the velocities and coordinates of all molecules.

How can observable dynamics be derived from ideal dynamics? They are derived either by the introduction of the ideal dynamics of small but finite amounts external noise or by the introduction of errors in the initial state. These amounts of noise or initial state errors always exist in real experiments but do not occur in ideal models. What is the meaning of small but finite amounts of noise? The amount of noise or errors should be large enough to prevent nonobserved appearances (inverse processes with decreasing entropy or Poincare's returns). Conversely, the amount of noise or initial state errors should be sufficiently small enable the dynamics of macroparametres to be unchanged for the processes of entropy growth.

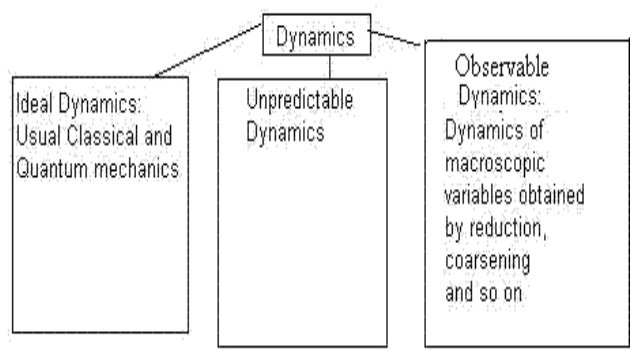

**Figure 22.** Three types of dynamics.

Observable dynamics for processes with entropy growth yield results that coincide with ideal dynamics. However, inverse processes with decreasing entropy and Poincare's returns are impossible for observable dynamics.

The possibility of observable dynamics is related to the stability of the processes with entropy growth with respect to initial condition errors and external noise. Conversely, the processes with decreasing entropy and Poincare's returns are unstable for small noise and small initial condition errors. However, these appearances are not observed in real experiments.

What is an explanation for the stability of entropy growth processes in observable dynamics with respect to noise (under the appropriate selection of macrovariables)? Two main explanations were defined by Schrodinger [48].

The first explanation has been described previously. The macrostate corresponds to a vast quantity of molecular states. Although the amount of external noise can significantly change the state for each molecule, a complete contribution of all molecules to their macrostates remains unchanged. Such stability to the large change of the molecules state is related to the "law of large numbers" in probability theory [16]. Examples of the corresponding law are macroscopic fluid or gas motion laws.

The second explanation is related to the discretisation of states in quantum mechanics. As different quantum states are discrete and have different energies, small amounts of noise cannot influence them. This is the reason for the stability of a chemical bond and this fact facilitates the consideration of macromolecule thermodynamics.

Is it necessary to use observed dynamics instead of ideal dynamics if the two types of dynamics yield identical results in all important observational situations related to entropy growth? The description of observable dynamics is much simpler than ideal dynamics. Observable dynamics eliminates unobservable processes (such as inverse processes with decreasing entropy or Poincare's returns) and uses a smaller quantity of variables and simpler equations. It enables the abstraction of small amounts of external noise or the incompleteness of initial conditions of the system, which produce an exact or probable description of the system.

We know that the true theory consists of ideal dynamics. Observable dynamics yield different results. Is it possible to experimentally discover the difference between these two theories provided that ideal dynamics is true? A theory can be considered as either correct or incorrect if a real experiment can reject the theory. This theory was termed falsifiable, according to Karl Popper [49]. Assume that ideal dynamics are correct. Are observable dynamics falsifiable?

For the complete physical system, including the observer, the observable system and the surrounding medium, observable dynamics is not falsifiable according to Popper (under the condition of *fidelity of ideal dynamics*), i.e., the difference between ideal and observable dynamics in this case cannot be experimentally observed[10].

---

[10] The difference can be useful. Some calculations are easier in a different framework. The calculations can be simpler for understanding. For example, in the framework of Galileo, which is related to the sun, we can calculate the dynamics of planets easier than with the framework of Ptolemy, which is related to the Earth, although we may choose framework of Ptolemy for calculation also. The selection between «the Earth rotates around the Sun» and «the Sun rotates around the Earth» remains free and arbitrary. This relationship is defined only by the beauty of the description and our convenience.
Similarly, in mathematical science, the selection of a number of definitions

For the system that is only measured, without the inclusion of an observer and the environment, observable dynamics is essentially falsifiable. It is necessary to exclude any influence from the surrounding medium or observer of the measured system to prepare a known initial state, to measure a final state and to compare the obtained result with the theory. Are some observable dynamics true? [1, 2, 8]. Assume that spontaneous reduction occurs for macroscopic systems that are sufficiently large, i.e., the spontaneous reduction is observed not only by introspection but for the external observer at the complete isolation of a macrosystem from environment noise. However, in the case of small isolated systems, when ideal dynamics can be observationally verified, this ideal dynamics always appears to be true.[11]

However, for macroscopic systems, the exclusion of any perturbation from a surrounding medium or observer on measured system is a difficult problem; thus, the practical unfalsifiability of observable dynamics exists, i.e., it is theoretically possible to reject but difficult to construct in a real experiment.

Note that certain cases are possible when observable dynamics cannot be correctly formulated. Thus, the system remains unpredictable due to nonisolation or initial state incompleteness. This is the case of unpredictable Dynamics, which are examined in the following chapter.

## 11.2. What Restrict the Selection of Observable Dynamics Macroscopic Variables?

It is of great importance that the selection of macroparametres cannot be arbitrary (Fig. 23). Observable dynamics should lead to the law of increasing entropy and irreversibility. For an accurate definition of macrovariables, processes with increasing entropy should be stable and equivalent for both ideal and observable dynamics. Conversely, processes of decreasing entropy should be unstable in ideal dynamics and impossible in observable dynamics. This requirement superimposes crucial restrictions on the selection of possible macroscopic states. In classical statistical mechanics, a set of microstates that corresponds to a macrostate resembles a compact convex drop in phase space.

This macrostate and its dynamics of increasing entropy are stable with respect to small external noise. Consider a set of phase space points that correspond to the spread phase drop with a set of narrow branches ("sleeves") and the reverse velocities of all molecules. This ensemble cannot correspond to any macrostate, although the ensemble corresponds to initial states of processes with decreasing entropy. The impossibility of the correspondent macrostate is explained by the instability of this set with respect to small noise. Similarly, it is impossible to select microparametres, i.e., velocities and coordinates of all molecules for system decryption (even in the limiting case). In a chaotic system, this microstate is highly unstable. Another reason is the enormous quantity of these parameters.

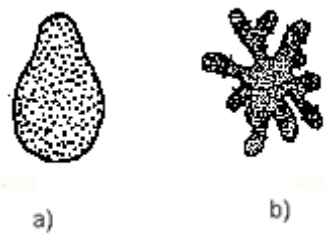

**Figure 23.** a) Possible and b) impossible distributions of a macrostate.

Similarly, the following problem exists in QM: for example, for the basic macrostates of Schrodinger's cat, why do we choose two basic states defined by a living or dead cat but not two basic states defined by their difference and sum [54]? From the point of view of QM, such difference and sum of the states are also possible states. The reason is that a living or dead cat is stable with respect to the environmental perturbations. In contrast, their sum or difference are entangled with an environment and transfer a mixture of a living cat and a dead cat during a short period. We previously named this process decoherence. Decoherence occurs much faster than all other thermodynamic relaxational processes [20, 23-25], i.e., the selection of two states, which are exactly the living cat or the dead cat, is dictated by the necessity of stability to external noise.

Which property of ideal dynamics equations causes the prioritisation of these states? This property is named the locality of interaction. Only molecules close in space are highly interactive. States of a living cat and a dead cat are separated in space. Thus, their superposition (sum or difference) is easily reduced to their mixture by interaction with the molecules of the surrounding medium. The definition of priority macroscopic states (named *pointer states*) in the case of quantum mechanics is featured in studies by Zurek [21, 22]. Assume that the interaction between molecules is defined not by proximity of the positions of molecules but by the proximity of its momentums, for example. In this case, the priority macroscopic states (pointer states) would be different.

---

and axioms is bounded only by our convenience and the requirement of the consistency of axioms. The theory that explains how to make a selection is absent (unlike Göde's theorem of incompleteness). Usually, arguments of "beauty" and theorems of "generality" are used to explain a choice. However, these items require exact definitions. [50]

[11] These experiments do exist and are performed for systems of intermediate size between macro- and micro-, which are named mesoscopic-, systems. These experiments confirm ideal dynamics. In these experiments, quantum interference is observed (in the absence of a spontaneous reduction) and entropy fluctuations exist [3, 8]. However, for extremely large macroscopic systems, similar experiments will not be possible in the foreseeable future. In fundamental physics, a similar situation exists for string theories and great unifications. Experiments that can confirm or reject these theories will not be possible in the foreseeable future, with the exception of an unexpected miracle. In Einstein's theory of gravitation, which is precisely verified only for small gravitation forces, the situation is similar. (We note, for example, mysterious dark substances, energy, and new gravitation theories of Milgrom [51] and Logunov [52-53]).

For systems closed to thermodynamic equilibrium, priority macroscopic states (pointer states) correspond to energy eigenfunctions. In an energy representation of thermodynamic equilibrium, a density matrix is diagonal.

Note that the selection of the set of macrovariables set and the corresponding observable dynamics equations are ambiguous. A large set of possible observable dynamics exists. All thermodynamics master equations (for example, the hydrodynamic equations of viscous fluid, the Boltzmann equation in thermodynamics, and the entropy growth law for isolated systems) are observable dynamics equations. Various observable dynamics differ by their degree of «macroscopicity» and by their choice of a proper set of macrovariables.

Consider an ensemble that is in equilibrium with a thermostat. In QM, this ensemble is represented by an energy representation of a density matrix. Its nondiagonal elements, which correspond to correlations, are zero. Similarly, in CM, no correlations between the molecules exist in equilibrium. Two types of nonequilibrium exist. The first type is defined by macroscopic correlations of macroparametres. They are expressed in QM by diagonal elements, with the values different from equilibrium, for a density matrix in the energy representation. These correlations, which disappear during the process, are named relaxation to an equilibrium state. The second type is defined by microscopic correlations (this type consists of quantum correlations in QM or additional unstable correlations in CM). Quantum correlations correspond to nonzero nondiagonal elements of the density matrix in an energy representation. These microscopic correlations are significantly more unstable and faster than macroscopic correlations. Their vanishing process is named decoherence. The period of decoherence is substantially less than the period of relaxation.

### 11.3. Two Methods for the Derivation of Observable Dynamics (Master Equations)

Consider the methods for the derivation of observable dynamics. According to two principal reasons for the application of observable dynamics (i.e., external noise for the external observer or incompleteness of system state knowledge for introspection), we can divide all methods for derivation of observable dynamics into two groups.

The first method is related to the introduction of small amounts uncontrollable noise from a large external thermostat (for example, the vacuum is a thermostat with zero temperature). [20] This noise destroys the additional microscopic correlations that produce returns and reversibility.

The second method is related to incompleteness of the complete system state knowledge for introspection. This method enables the implementation of the coarsening ("spreading") of the function, which features system states [6, 13] (Appendix K). In CM, this function is the phase density function, whereas it is a density matrix in QM. If we consider the case of CM, coarsening involves the flattening (averaging) of the phase density function in a neighbourhood of each point over some period of time. Flattening evolution is represented by common equations of ideal dynamics. In QM, a similar procedure is related to a periodic reduction of a wave function [18], and Schrodinger's equations uses reductions. Similar methods of coarsening destroy the additional correlations that produce returns and convertibility. These correlations are not experimentally observed for an introspection case.

Observable dynamics (at introspection) should feature the behaviour of the system (with entropy growth) and should produce results that correspond with ideal dynamics during a finite time interval. This time is much smaller than Poincare's return time. The system cannot be experimentally observed during a larger interval due to the loss of observer memory at returns.

### 11.4. Solution of the Zeno Paradox from the Point of View of Observable Dynamics. Exponential Particle Decay is a Law of Observed Instead of Ideal Dynamics

Due to the necessity of the concept of observable dynamics, we note the quantum Zeno paradox. We provide a solution for the second part of this paradox (regarding nonexponential decay) within the framework of observable dynamics.

Choose a number of undecayed particles, N (with a small possible error), which is the measured macroparameter. During the initial moment, $t_0$, there were $N_0$ undecayed particles. The ideal quantum dynamics of the additional decay are not exponential. However, the quantum measurement of decay inevitably imports perturbations to ideal dynamics; the ideal dynamics becomes inapplicable. We can reduce perturbations that are imported by measurement by increasing the intervals between measurements. The time interval between measurements (reductions) should be sufficiently large to prevent the influence of the dynamics of decay ($\Delta t \gg \hbar/\Delta E$, where $\Delta t$ denotes between reductions, $\hbar$ denotes a constant lath, $\Delta E$ is the difference of energies between states of the broken and unbroken particles). Conversely, the interval must be less than an average lifetime of the decayed particle ($\hbar/\Delta E \ll \Delta t < \tau$, where $\tau$ is a medial lifetime of the decayed particle). The complete time of the decay process observation should be considerably less than Poincare's return time. For isolated finite volume systems, including the observer, these returns are not observed due to the loss of observer memory ($n \cdot \Delta t \ll T_{return}$, where n is the number of observations [reductions] and $T_{return}$ is Poincare's return time). Assume that the interval between measurements and the complete time of observation are correctly chosen, i.e., satisfy all of the requirements. In this case, the law of decay is strictly exponential and is not dependent on the concrete exact value of the chosen interval between measurements (reductions). This exponential law of decay ($N = N_0 \cdot \exp(-(t-t_0)/\tau)$) is already the law of observable dynamics according to the definition of observable dynamics.

## 11.5. Examples of Various Methods of Deriving Observed Dynamics by «Coarsening»: The Boltzmann Equation and Prigogine's New Dynamics

An example of observable dynamics is the Boltzmann equation [5, 6]. Coarsening ("spreading") of the phase density function (Appendix K) is produced over two stages. In the first stage, the phase density function is replaced by one particle function that corresponds to the phase density function averaged over all particles, with the exception of one particle. The equation for a one-particle function is a reversible equation of ideal dynamics that is dependent on a two-particle function. The equation of observable dynamics is obtained by the next coarsening step. «Molecular chaos hypothesis» is introduced. This hypothesis signifies that the correlation between any two particles is assumed to be zero; thus, a two-particle function is replaced by the product of two one-particle functions. By substituting this two-partial function in the equation described previously, we obtain a nonreversible and nonlinear Boltzmann equation. This equation is similar to a QM reduction, in which all correlations between possible measurement results (that is, nondiagonal density matrix elements) are forced to be zero.

Another example of the derivation of observable dynamics is a coarsening method used in Prigogine's «New Dynamics» [14, 55], (Appendix L). It is a wonderful method. Both the coarsening ("spreading") procedure and the equations of motion (obtained by simple substitution of inverse to coarsened function in the equations of ideal dynamics) are linear. In CM, the nonisotropic coarsening of the phase density function is employed. As described previously, for chaotic systems in a neighbourhood of each phase-space point there exists a direction along which trajectories diverge exponentially (a spreading direction). There also exists a direction along which trajectories converge exponentially (a shrinking direction). Coarsening ("spreading") of a phase function is yielded in the shrinking direction.

Consider a macroscopic state corresponding to a compact convex «phase drop». This «phase drop», spreading on a phase space, provides many "branches". The shrinking direction is perpendicular to these "branches". Therefore, function coarsening along the branches causes an increase in the «phase drop» square, and, correspondingly, causes an increase in the number of microstates and entropy.

Consider an inverse process. The initial state is described by a set of phase space points obtained from the final state of the direct process («phase drop» spreading) by the reversion of all molecular velocities. Velocity reversion does not change the shape of the spreading «phase drop». The shrinking direction becomes parallel to its "branches". Therefore, for this case, coarsening along the spreading direction almost does not change the «phase drop» square. The number of microstates and amount of entropy do not vary, which is in contrast with the magnification of entropy spread «a phase drop» in the direct process (Fig. 24).

Thus, nonisotropic coarsening breaks the symmetry of time direction. Consequently, the corresponding equations appear nonreversible.

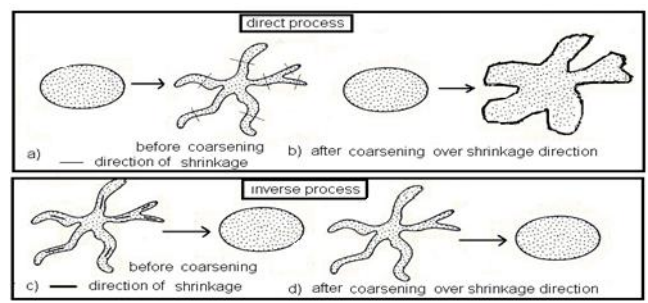

**Figure 24.** Direct process with increasing macroscopic entropy **a)** prior to coarsening and **b)** after coarsening. Inverse process **c)** before coarsening and **d)** after coarsening. The directions of shrinkage are denoted. Anisotropic coarsening is produced along the direction of shrinkage in Prigogine's "New Dynamics".

Consider «New Dynamics» for quantum systems. A similar linear nonisotropic coarsening procedure can occur in the case of QM but only for infinitely large quantum systems (an infinite volume or infinite number of particles). Finite quantum systems with a finite number of particles are almost-periodic and have a finite return time. For these cases, the «New Dynamics» are not applicable[12].

Prigogine suggests that all real quantum systems be considered as infinitely large. The solution, however, can be much easier. Observable dynamics can be used during time interval much smaller than time interval of Poincare's return. Indeed, a system cannot be observed experimentally during a large interval due to memory loss at returns during introspection. However, for small intervals, the behaviours of large-volume systems in a thermodynamic limit are similar to the behaviours of infinite systems. The thermodynamic limit is a limit in which the system volume moves to infinity but the ratio of particle number to complete system volume remains constant. If we have a system for which the thermodynamic limit does not exist (for example, the macromolecule), it can be considered as surrounded by the environment, for which such a limit does exist. These methods enable us to use the equations of "New Dynamics" for finite QM systems.

«New Dynamics» is often subjected to criticism [34]. The basic argument against this theory is as follows: "We can explain all paradoxes and problems of QM/CM without «New Dynamics»". Critics conclude that «New Dynamics» is not necessary and redundant. However, we must consider «New Dynamics» not only as a replacement for QM/CM (unfortunately, Prigogine did this) but also as one of the *useful and simple forms* of its observable dynamics. It allows us to describe physical systems by *simpler and irreversible* laws, with the exception of unobservable experimentally

---

[12] New Dynamics is not applicable to the almost-periodic systems of CM. However, the majority of real CM systems are systems with mixing (chaotic). Therefore, in CM, this problem is less important.

reversibility and returns. It is a real and significant advantage.

Consider the situation in which it is not possible to locate any observable dynamics. This case was defined previously as unpredictable dynamics.

# 12. Unpredictable Dynamics [40, 102]

Ideal dynamics, when broken by exterior noise (or by incompleteness at introspection), cannot always be replaced by predictable observable dynamics. For some systems, the dynamics become unpredictable in principle. Thus, we define the dynamics of such a system as unpredictable dynamics. Based on the definition, for such systems, it is impossible to introduce macroparametres that are typical of observable dynamics and to predict their behaviour. Their dynamics are not featured or predicted by scientific methods. Thus, science places boundaries on the applicability of dynamics.

We do not doubt the fidelity and universality of the basic laws of physics. However, the impossibility of the complete knowledge of (system states)/(system dynamics laws) exists due to the interaction with the observer/environment or incompleteness at introspection. This impossibility renders a complete experimental verification of these basic physical laws impossible in some cases. It enables the freedom to change these laws without any inconsistency in experimentation. When these modifications lead to the predicted dynamics of a system, they are named observable dynamics. For the case in which predicted dynamics are not possible, the predicted dynamics is named unpredictable dynamics.

We provide examples of unpredictable dynamics.
1) Phase transition or bifurcation points. These examples consist of places or *points* of the branching or forking of a single macroscopic state and are represented by observable dynamics into a set of possible macroscopic states. This branching occurs by changing a regulated external parameter or through time evolution. For these points, observable dynamics loses unambiguity. Vast macroscopic fluctuations exist, and the use of macroparametres becomes senseless. Evolution also becomes unpredictable for these points, i.e., unpredictable dynamics are evident.
2) Additional appearances are included in modern cosmological models, with the exception of appearances that have been previously introduced. They are related to a loss of information and the corresponding incompleteness of system state knowledge. The examples are a Black Hole or unobservable space of the Universe that exist out of an observer light cone.
3) The case of uncontrollable and unstable microscopic quantum correlations for a system isolated from decoherence of the environment. Assume that an observer fixes the initial state of a quantum system. He can predict and measure any future system state providing there is no interaction between the observed system and the environment;

otherwise, the observer exists prior to this measurement. Consider another external observer who does not know the initial system state. Unlike the first observer, system behaviour is primarily unpredictable for the second observer. Any attempt by the second observer to unpredictably measure the system state destroys the quantum correlations of the system and the corresponding evolution, i.e., from the point of view of the second observer, unpredictable dynamics exists. Well-known examples of these systems include quantum computers (Fig. 25) and quantum cryptographic transmitting systems (Fig. 26) [23-25].

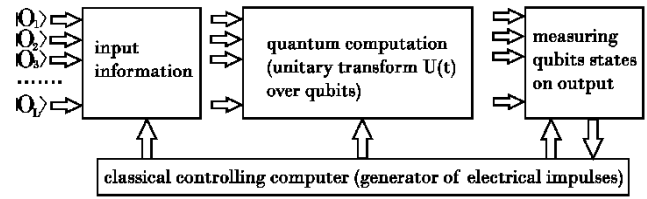

**Figure 25.** Quantum computer.

A quantum computer is unpredictable for any observer who does not know its state at the beginning of its calculations. Any attempt by an observer to measure the intermediate state of a quantum computer during calculation destroys the calculation process in an unpredictable manner. Another important property is the significant parallelism in calculation, which is a consequence of the laws of linearity in QM. The initial state can be considered as the sum of many possible initial states of "quantum bits of information". Due to the laws of linearity in QM, all components of this sum can evolve independently. This parallelism enables numerous important problems, which usual cannot be solved by computers in real time, to be solved rapidly. It generates the potential for future applications of quantum computers.

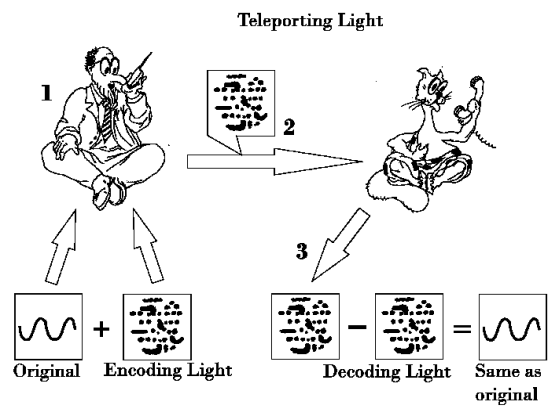

**Figure 26.** Quantum encoding and decoding.

Quantum cryptographic transmitting systems use the properties of unpredictability and unobservability of "messages" that cannot be read during transmission by any external observer. These "messages" are common quantum systems that are governed by quantum laws and quantum correlations. An external observer, who has no information about its initial states and attempts to perform a measurement (reading) of the "message" over transmission, inevitably destroys this transmission. Thus, message interception appears *principally* impossible under the laws of physics.

In contrast to prevailing opinions, both quantum computers and quantum cryptography [23-25] have classical analogues. In contrast with quantum systems, precise measurement in classical systems can be performed in principle without any distortion of measured states. However, in classical chaotic systems, uncontrollable and unstable microscopic additional correlations result in reversibility and Poincare's returns. The introduction "by hands" of small finite perturbations or initial state errors destroys these correlations and eliminates this principal difference between classical and quantum system behaviours. This small environmental noise always exists in any real system. By isolating chaotic classical systems from this external noise, we obtain classical analogues of isolated quantum devices with quantum correlations [86-87].

Analogues of quantum computers consist of molecular computers [56, 86-87]. The vast quantity of molecules ensures the parallelism of evaluations. The unstable additional microscopic correlations (which produce reversibility and returns) ensure that the dynamics of intermediate states are unpredictable for the external observer, who is not informed about the initial state of the computer. The observer would destroy the computer calculations during an attempt to measure an intermediate state.    .

Similar arguments can be used for classical cryptographic transmitting systems based on these classical unstable microscopic additional correlations for information transition. A "message" comprises a classical system that is chaotic in intermediate states. As in the case of QM, any attempt to intercept the "message" inevitably destroys it.

4) Conservation of unstable microscopic correlations can be ensured not only by passive isolation from an environment and the observer but also by active dynamic mechanism of perturbation cancelling. Such active dynamic mechanism occurs in physical stationary systems, in which the steady state is supported by a continuous stream of energy or substance through a system. An example is a micromaser [57]—a small and well-conducting cavity that contains electromagnetic radiation. The size of a cavity is so small that radiation is required with the help of QM. Radiation damps due to interaction with conducting cavity walls. This system is well-represented by a density matrix of base energy eigenfunctions. These eigenfunctions are Zurek's "pointer states" (similar to any system closed to equilibrium). This set is the best choice for observable dynamics. Microscopic correlations correspond to nondiagonal elements of the density matrix. Nondiagonal elements converge to zero much faster than diagonal elements during radiation damping. The period of decoherence time is significantly less than the period of relaxation. However, a beam of excited particles passing through a micromaser generates the strong damping deceleration of nondiagonal elements of a density matrix (microcorrelations). It also leads to nonzero radiation in the steady state.

5) An example of complex stationary systems are living systems. Their states are far from thermodynamic equilibrium and extremely complex. Although these systems are highly ordered, their order is significantly different from an order of lifeless periodical crystals. Low entropy disequilibrium of a living system is supported by entropy growth in the environment [13]. It is metabolism—the continuous stream of substance and energy through a living organism. Conversely, metabolism not only supports disequilibrium, this disequilibrium is also a catalytic agent of metabolic process, i.e., this disequilibrium performs and supports metabolism on a necessary level. As the state of living systems exhibit nonequilibrium, this disequilibrium can support existing unstable microcorrelations, which usually disrupted by decoherence. These correlations can exist between the parts of the living system and between different living systems (or living systems with lifeless systems). If these correlations occur, the dynamics of a living system can be referred to as unpredictable dynamics. Successes in molecular biology enable descriptions of superior dynamics of living systems. However, no proof that we are capable of completely representing all complex processes of a living system exists.

It is difficult to analyse real living systems within the framework of ideal, observed and unpredictable dynamics due to their vast complexity. It is possible to construct mathematical models with nonequilibrium stationary systems with a metabolism that are considerably less difficult. We can analyse these systems. This is an important problem for the future work of physicists and mathematicians. Future steps in this direction are described in the chapter about synergetic systems.

Unstable microcorrelations exist not only in quantum mechanics but also in classical mechanics. For example, they exist in classical chaotic systems (with mixing). Thus, these models should not exist only in quantum mechanics. They can also be classical. The assumption that only quantum mechanics can feature similar appearances is a common error [1, 2]. We previously specified that the introduction "by hands" of small but finite interactions or initial state errors eliminate unstable microcorrelations. The principal difference between quantum and classical mechanics disappears.

6) The previous cases do not describe all possible types of unpredictable dynamics. The exact requirements for which ideal dynamics transfer to observable and unpredictable dynamics is a problem that has not been solved completely by mathematics and physics. Another problem that has not been solved completely (and, apparently, is related to the previous problem) is the important role of these three types of dynamics in complex (living) systems. The solution of this problem will enable an understanding that is more profound

---

[13] For example, the entropy of the sun increases. It is an infinite energy source for life on Earth.

than the physical principles of life. We devote the following chapter to this problem.

# 13. Complex (living) Systems

We mention that although the previous statements were strict and exact, the statements in this chapter are hypothetical. Assume that life completely corresponds with the laws of physics. The following problems should be considered:

What is life and death from the point of view of physics?
Are certain properties of living organisms incompatible with physics?
Which properties of living systems are different from lifeless systems from the point of view of physics?
Do living systems have consciousness and free will from the point of view of physics?

Life is usually defined as a special high-ordered form of existence of organic molecules, which are capable of metabolism, reproduction, adaptation, motion, response to external irritants, and self-preservation; they can also be self-organising. This definition is valid but narrow: many living systems possess only some of these properties; some of these properties also can exist in lifeless substances and inorganic forms of life.

## 13.1. Life from the Point of View of Physics—Previous Reports

The first attempt to describe life from the point of view of physics was attributed to Schrodinger [48]. In this book, life is defined as an aperiodic crystal, i.e., the highly ordered[14] form of a substance that is not based on simple repetition, such as normal crystals. It also presents two reasons that observable dynamics of living systems are stable to interior and exterior noise: the statistical "law of large numbers" and the step-type behaviour of quantum transitions ensures the stability of chemical bonds. There exists a marked similarity between a living organism and a clock: both exhibit «an order from the order» despite high-temperature noise.

The following steps for understanding life were developed by Bohr [58]. He paid attention to the fact that according to QM, complete measurement of a living system state inevitably perturbs its behaviour. It can results in unpredictability and incognisable characteristics of life. The criticism of the viewpoint of Bohr by Schrodinger [59] is not well-grounded. He states that the complete knowledge of the quantum system state is probabilistic. Thus (unlike the classical case), the complete quantum system state knowledge most likely enables the prediction of the future. The problem, however, is that measurement creates *stronger* ambiguity than the probabilistic character defined by QM, i.e., it is impossible to predict its behaviour even probabilistically. It means that in the absence of measurement, a different behaviour would be evident [41]. Measurement eliminates unstable classic (or quantum) correlations between system parts and changes its behaviour. Thus, measurement causes the behaviour of a system to be principally unpredictable and not only probabilistic. This occurrence is evident in QM and CM, in which a small finite interaction between real systems exists; these interactions destroy additional unstable microcorrelations.

Bauer's[15] book [60] indicates that high ordering (low entropy) is defined not only by a nonequilibrium distribution of material in a living organism but also by the high-ordered (significantly unstable) structure of a living substance. This significantly unstable structure is not only supported by a metabolism process but is a catalytic agent for metabolism. Proteins or viruses have ordered structures in a crystalline form. However, it is possible to satisfy the high-ordered and low-entropy modifications of living substances. The degradation of structures will eventually occur over time, which will inevitability cause death and the necessity of reproduction for the conservation of life phenomena, i.e. a metabolic process only significantly decelerates the decay of the complex structure of a living substance, instead of supporting them to be unchanged. The experiments described by Bauer confirm energy production and the respective increase in entropy in autolysis. Autolysis is the decay of a living substance due to the lack of a supporting metabolism. In the first stage of the process, energy is produced due to destruction of highly unstable initial structure of a living substance; in the second stage of the process, energy is produced due to the appearance or release of protolytic (decomposing) enzymes. Bauer considered the existence of this *structural* energy as an essential feature of life.

The majority of the previously mentioned papers considered only individual living organisms, whereas it is possible to define and describe life as the totality of all living organisms (biosphere). The issue of the origins of life arises. The most complete answer to these questions from the point of view of modern physics is given in the paper by Elitzur [38]. The origin of life is considered as an ensemble of self-replicating molecules. Based on Darwinian natural selection, life accumulates genetic information (or rather *useful* information (knowledge), in terms suggested by Elitzur) about the environment. This process of information accumulation increases the level of the system's organisation (negentropy) that corresponds with the second law of thermodynamics. Lamarck's views contradicted this law of physics. An extensive range of issues is discussed in this paper. However, Elitzur's arguments also possesses the following drawbacks:

The description is valid for life as a whole, i.e., as a phenomenon but not for an individual live organism.

His suggested reasoning against Lamarck's theory

---

disproves only the most simple and straightforward version of Lamarck's theory, whereas numerous hypotheses and experiments illustrate the possibility of realisation of Lamarck's viewpoints in real life [61].

From Elitsur's viewpoint, the self-organising dissipating systems suggested by Prigogine (e.g., Benard cells [62]), in contrast with living organisms, are denied the capability of adaptation. The adaptability of Benard cells is incomparable with the adaptability of live systems; however, the adaptability of Benard cells exists, even if in a primitive form. Thus, Benard cells change their geometry or even disappear depending on the temperature difference between the upper and lower layers of the fluid. This finding can be considered a primitive form of adaptation.

### 13.2. Life as a Process of Preventing Relaxation and Decoherence: Conservation of both Macroscopic Correlations and Unstable Classical (or Quantum) Microscopic Correlations

Bauer considers life as a highly unstable system of self-maintenance due to motion and metabolism. Living systems resist the transition from an unstable state to a stable state. We can assume that this instability is primarily due to highly unstable (additional in CM or quantum in QM) microscopic correlations. Living systems tend to maintain these correlations and preserve them due to resistance to the decoherence process. Note that living systems can support these correlations between their internal parts and the environment.

We define two types of correlations in physical systems: the first type of correlation is stable to small amounts of external noise (macroscopic correlations between system parameters). For example, it is the connection between pressure, density and temperature for ideal gas. The second type of correlation consists of unstable microscopic correlations that produce reversibility and Poincare's returns in quantum and classical systems. Decoherence destroys these correlations by breaking reversibility and preventing returns. Decoherence leads to the entropy increasing law. We assume that living systems possess the ability to conserve these unstable correlations by slowing down or preventing decoherence.

We compare the properties of living and lifeless systems for the inhibition of the destruction and transformation to thermodynamic equilibrium. Systems can actively inhibit the relaxation process in cases of living or lifeless stationary systems exchanging energy or substances with an environment. However, the system can inhibit not only relaxation but also decoherence. In lifeless systems, such inhibition can be achieved in a passive manner, i.e., by isolation of the system from the environment. In living open systems, it is attained through active interaction with the environment, external and internal motion, and metabolism.

The ability of life to support not only macroscopic but additional classic (or quantum) microscopic correlations renders life unpredictable, as assumed by Bohr. It is important that quantum mechanics is not required; similar correlations also exist in classical mechanics. Analogues of quantum correlations are additional microscopic correlations in CM.

Achievements in molecular genetics do not contradict the influence of unstable and unpredictable correlations, which may be essential for life. It is possible to create a type of observable dynamics for life. Living systems are open systems that actively interact with a casual environment. The external observer interacts with living systems less frequently and cannot change their behaviour considerably. However, an attempt to understand and predict life too explicitly and in too many details can break the complex and thin correlations preserved by life and can lead to unpredictable dynamics of living systems, which was predicted by Bohr.

### 13.3. Synergetic Systems—Models of the Physical Properties of Complex (Living) Systems

We introduce the concept of synergetic physical systems. We define them as simple physical or mathematical systems by illustrating some properties of complex (living) systems. First, we are interested in synergetic models of systems, which are capable of inhibiting external noise (decoherence in QM). They conserve system correlations (quantum or classical), which produce reversibility or Poincare's returns.

Three methods exist for constructing these systems:
1) The passive method: a creation of "walls" those are impenetrable to noise. Examples are models of modern quantum computers.
2) The active method, which are the inverse method to the passive method: a type of dissipative or living systems that conserves disequilibrium with the help of active interaction environment and by interchanging energy and substance with the environment (metabolism). Future models of quantum computers are included in this field.
3) When correlations are considered for systems, including the entire Universe, external noise is impossible. The source of correlations for this system is the Big Bang. We define these correlations, which correspond to the entire Universe, as global correlations.

Two factors are noted:
1) During their evolution, numerous complex systems pass dynamic bifurcation points (time moments). A set of possible approaches to future evolution exists after this moment of time. The selection of a concrete approach to future evolution is dependent on a small perturbation of the system state in the bifurcation time moment [63-65]. In this moment, even small correlations (which can be conserved using the methods described previously) can exert a significant influence on the evolution of the system. The existence of these correlations restricts the predictive capability of physical science but does not restrict personal intuition. As we constitute an integral part of this Universe, we are capable of "anticipating" these correlations on a subjective level in principle. As these correlations are not feasible for experimental scientific observation, no contradiction with scientific laws exists.

2) Powerful sources of negentropy in an environment are necessary for both passive and active correlation conservation methods. Therefore, complete entropy of a system and its environment can only increase. The law of increasing entropy continues to be accurate for complete systems, including observable systems, the environment and the observer; however, it is not valid for isolated observable systems. Although a reduction in entropy in a complete system can occur in principle (according to ideal dynamics), it is unobservable, as explained previously. We do not need to consider such unobservable situations.

We demonstrate examples of synergetic models of physical processes.

Crystal growth models the ability of reproduction in live systems. The analysis of these systems enables the discovery of doubt [66-67] in Wigner's argumentation [27] regarding the inconsistency between the ability of a living organism to reproduce and QM. Assume that the interaction between a reproducing quantum system and its environment is random. For this interaction, Wigner proves that the probability of reproduction is near zero. However, this interaction is not casual but defined by a crystalline lattice of existing crystals. In the protein synthesis process in living organisms, the crystalline lattice is replaced by existing DNA nucleotide sequences; thus, the interaction is not random.

The active deceleration of decoherence, which is similar to active correlation conservation methods in living systems, exists in open systems, such as micromasers; these systems were previously described [57]. They comprise another example of synergetic systems.

Another active mechanism of decoherence inhibition is «quantum teleportation» [24]. This process can reproduce the exact copy of any initial quantum state by destroying its initial state. (The creation of a copy of any quantum state by conservation of this state is impossible in QM [24].) We can reproduce this copy multiple times with small time intervals. It is possible to conserve any initial quantum system state for a long period with a process that inhibits decoherence. This process is equivalent to conservation of the initial measured state in the previously described paradox «a kettle that never boils». It is observed for multiple and frequent measurements of the current state. A difference between these two cases also exists. In the «quantum teleportation» case, the initial state remains not only unchanged, as in the case of the "kettle paradox", but also remains unknown.

Dissipative systems are also active synergetic systems. They illustrate properties of open living systems, such as relaxation deceleration, conservation of low entropy and a primitive form of adaptation to surrounding medium change.

Another example is the quantum isolated system (for example, modern quantum computers at low temperatures). They demonstrate the property to conserve unstable quantum correlations. This property is similar to the conservation of the instability in living systems, which is also related to the conservation of similar quantum or classical correlations.

However, this conservation is passive compared with living systems. Penrose provides an example of this system, which is most likely used by the brain for thinking [1, 2]. This example is system of tubulin dimers that serve as a basis for cytoskeleton microtubules of neurons (main cells of the brain). The system of tubulin dimers is considered by Penrose as a quantum computer [23-25]. Even if this hypothetical model is not accurate, it would illustrate the principal possibility of the existence of quantum correlations in the brain. Analogues of quantum correlations in CM are additional unstable microcorrelations, and an analogue of the quantum computer is the molecular computer with additional correlations between molecules. Similar correlations appear in chaotic or almost chaotic classical systems (with mixing). The construction of a model of the brain that is based on classical (not quantum) chaotic systems is likely possible. Thus, they would have exhibit properties of quantum computers—unpredictability and parallelism of calculations. As Penrose incorrectly considers classical chaotic systems as unsuitable for modelling living systems, he does not consider this possibility.

Another example of synergetic systems that illustrates the properties of quantum correlations is quantum-oscillating systems, which are almost isolated from an environment [8]. Assume that a superconducting ring exists. State "A" corresponds to a clockwise current and state "B" corresponds to counter-clockwise current. This oscillating quantum and almost isolated system can change its states as follows: A -> A+B -> B -> A-B -> A. In this example, "A+B" and "A-B" are quantum superpositions of states A and B. Assume that we would like to measure the current directions within the ring. This measurement can destroy the superposition state if the system exists in this state at the moment of measurement. Thus, it can change the dynamics of a system and destroy the quantum correlations between states of superposition [8].

It is possible to construct another example of an oscillation system that is sensitive to measurement, which may interest biologists and chemists. This example contains active protection from external noise influence. We describe a process that consists of three stages. In the first stage, an enzyme with unstable conformation "A" appears. In the second stage, it catalyses a chemical process. This process inhibits the destruction of unstable conformation A. Consider a more complex situation. Assume the existence of two unstable conformations, "A" and "B", which are capable of a catalysing process in the second stage. The enzyme has no fixed conformation, "A," but a sequential set of unstable transitions from A to B and from B to A. In the third stage, the enzyme is involved in the chemical reaction. If the initial moment enzyme was in conformation A, it would catalyse this reaction. Otherwise, no catalysis occurs. Thus, the third stage is dependent on a finite enzyme conformation state at the end of the second stage. During the second stage (at a certain time moment), we decided to measure the enzyme conformation state (by nuclear magnetic resonance, for example). As the transition process from A to B (or B to A) is unstable, measurement can break the phase of this process.

As a result, the initial enzyme is evident at conformation B instead of conformation A in the third stage. Thus, the third stage reaction cannot commence. Thus, measurement of the intermediate enzyme state can destroy the process by changing its resulting products. This scenario can occur for both the quantum and the classical mechanics processes.

With the help of synergetic "toy" models, it is possible to understand synchronicity (simultaneity) that is not coupled casually with processes and global correlations phenomena.

Examples are nonstationary systems with "peaking" (blow-up) [63-65], which were considered by Kurdyumov. In these processes, a function is defined on a plane. Its dynamics are featured by nonlinear equations that are similar to the burning equation. Blowing-up a solution function can converge to infinity for a *finite* period in a single or several closed points on a plane. The function attains infinity in these points at the same moment of time, i.e., synchronously.

Using these models, we can illustrate population growth (or engineering level of civilisations) in the megacities of our planet [68]. Points of infinite growth comprise megacities and population density is a function value.

We complicate the problem by assuming that during some initial time moment there is rapid expansion ("inflation") of the plane in the blowing-up process. Processes of converging to infinity in set points remain synchronous despite the fact that these points are not closed and are located at significant distances.

This complex model can qualitatively explain the synchronism of processes in far reaches of our Universe after the rapid expansion caused by the Big Bang ("inflation"). These blowing-up processes appear only at some narrow set of burning equation coefficients. Such he narrow diapason of the appropriate values enables the construction of an analogy based on the «anthropic principle» [69]. The anthropic principle states that fundamental constants have values that enable our observed Universe to appear with anthropic entities, which are capable of observation.

The complex processes can be illustrated using "cellular" models. A good example is the discrete Hopfield model [70, 71]. This system can be represented as a two-dimensional square lattice of meshes, which can be either black or white. We establish an initial state of a lattice. The coefficients of the linear interaction between the meshes are unequal. They can be chosen in such a way that during its discrete evolution the initial state transfers in one of possible terminating states from the previously defined known set of states (attractors). We denote these attractors by the letters A or B.

Certain initial unstable states differ by one mesh (a critical element). Thus, one state has attractor A and another state has attractor B. This synergetic model is a good illustration of the *global instability of* complex systems. It also demonstrates that this instability represents the entire system but does represent its parts. Only an external observer can change a critical element and, thus, change the evolution of the system evolution. The internal dynamics of the system cannot do such change. The *global correlation* between meshes of unstable initial states unambiguously define which attractor must be chosen by this lattice (either A or B).

This model can be interpreted as a neural network with feedback or as a spin lattice (spin glass) with unequal interactions between spins. The system can be employed for pattern recognition.

It is possible to complicate the model. Let each mesh in the lattice represent itself as a similar sublattice. Assume that the process runs in two stages.

In the first stage, large meshes do not interact; interaction only occurs in sublattices that change under the usual method. The initial states of all sublattices can be designated as unstable. We associate the final state A of the sublattices with a black mesh of the large lattice, and the final state B of sublattices with a white mesh.

The second stage of evolution is defined as a common evolution of this large lattice, without modifications in sublattices. Its initial state, which appeared at the previous stage, can also be unstable.

Let the total state of coarsened lattice and each large mesh be denoted by the letter A. Let us name this state «A-A». The appearance of this exact final state (but not a different state) is capable of explaining the correlations in an unstable initial state by global meshes and by defined values of interaction coefficients between meshes.

Assume that prior to this two-stage process, our coarsened lattice occupied a small field of space, but as a result of expansion ("inflation"), it was extended to vast sizes. The process began after this expansion. Thus, the existence of global correlations in an unstable initial state (results in lattice attractor «A-A») can be explained by the initial closeness of meshes (prior to "inflation"). A specific selection of interaction coefficients between meshes (also results in attractor «A-A») similarly explains the «anthropic principle».

This coarsened lattice can be compared with our "Universe". Its large meshes (sublattice) can be compared with "living organisms" that inhibit (actively or passively) «decoherence». «Decoherence» is an influence of a large mesh "environment" (i.e., influence of other meshes) on existing processes inside this mesh. Then, global correlations of unstable initial lattice states can serve as analogues of possible global correlations of unstable initial states of our Universe, and interaction coefficients of meshes correspond to fundamental constants. The initial process of lattice expansion corresponds to the Big Bang.

## 13.4. Hypothetical Consequences Explaining Life as a Method of Correlations Conservation

The definition of life as the totality of systems that maintains a correlation in contrast with external noise is a reasonable explanation of the mysterious silence of the cosmos, i.e., the absence of signals from other intelligent worlds. All parts of the universe, which possess a unique centre of origin (Big Bang), are correlated, and life maintains these correlations, which comprise the basis of its existence. Therefore, the emergence of life in different parts of the

Universe is correlated; all of the civilisations of the Universe exhibit approximately the same level of development, and no supercivilisations that are capable of reaching the Earth exist.

The effects of long-range correlations can explain a part of the truly wonderful phenomena of human intuition and parapsychological effects. A well-known psychiatrist named Charles Jung discussed this topic in his paper "On Synchronicity" [72-75]. We cite the most interesting excerpts from this paper. The following excerpt defines "synchronicity":

"But I would rather approach the subject the other way and first give you a brief description of the facts which the concept of synchronicity is intended to cover. As its etymology shows, this term has something to do with time or, to be more accurate, with a kind of simultaneity. Instead of simultaneity we could also use the concept of a meaningful coincidence of two or more events, where something other than the probability of chance is involved. A statistical- that is, a probable concurrence of events, such as the "duplication of cases" found in hospitals, falls within the category of chance space and time, and hence causality, are factors that can be eliminated, with the result that acausal phenomena, otherwise called miracles, appear possible. All natural phenomena of this kind are unique and exceedingly curious combinations of chance, held together by the common meaning of their parts to form an unmistakable whole. Although meaningful coincidences are infinitely varied in their phenomenology, as acausal events they nevertheless form an element that is part of the scientific picture of the world. Causality is the way we explain the link between two successive events. Synchronicity designates the parallelism of time and meaning between psychic and psychophysical events, which scientific knowledge so far has been unable to reduce to a common principle. The term explains nothing, it simply formulates the occurrence of meaningful coincidences which, in themselves, are chance happenings, but are so improbable that we must assume them to be based on some kind of principle, or on some property of the empirical world. No reciprocal causal connection can be shown to obtain between parallel events, which are just what gives them their chance character. The only recognizable and demonstrable link between them is a common meaning, or equivalence. The old theory of correspondence, was based on the experience of such connections- a theory that reached its culminating point and also its provisional end in Leibniz' idea pre- established harmony, and was then replaced by causality. Synchronicity is a modern differentiation of the obsolete concept of correspondence, sympathy, and harmony. It is based not on philosophical assumptions but on empirical experience and experimentation."

The following citation provides an example of «synchronicity» based on Jung's personal experience:

"I have therefore directed my attention to certain observations and experiences which, I can fairly say, have forced themselves upon me during the course of my long medical practice. They leave to do with spontaneous meaningful coincidences of so high a degree of probability as to appear flatly unbelievable. I shall therefore describe to you only one case of this kind, simply to give an example characteristic of a whole category of phenomena. It makes no difference whether you refuse to believe this particular case or whether you dispose of it with an ad hoc explanation. I could tell you a great many such stories, which are in principle no more surprising or incredible than the irrefutable result arrived at by Rhine, and you would soon see that almost every case calls for its own explanation. But the causal explanation the only possible one from the standpoint of natural science breaks down owing to the psychic relativisation of space and time which together form the indispensable premises for the cause-and-effect relationship.

My example concerns a young woman patient who, in spite of efforts made on both sides, proved to be psychologically inaccessible. The Difficulty lay in the fact that she always knew better about everything. Her excellent education had provided her with a weapon ideally suited to this purpose, namely a highly polished Cartesian rationalism with an impeccably "geometrical" idea of reality. After several fruitless attempts to sweeten her rationalism with a somewhat more human understanding, I had to confine myself to the hope that something unexpected and irrational would turn up, something that burst the intellectual retort into which she had sealed herself. Well, I was sitting opposite of her one day, with my back to the window, listening to her flow of rhetoric. She had an impressive dream the night before, in which someone had given her a golden scarab-a costly piece of jewellery. While she was still telling me this dream, I heard something behind me gently tapping on the window. I turned round and saw that it was a fairly large flying insect that was knocking against the window from outside in the obvious effort to get into the dark room. This seemed to me very strange. I opened the window and immediately caught the insect in the air as it flew in. It was a scarabaeid beetle, or common rose-chafer, whose golden green colour most nearly resembles that of a golden scarab. I handed the beetle to my patient with the words "Here is your scarab." This broke the ice of her intellectual resistance. The treatment could now be continued with satisfactory results.

This story is meant only as a paradigm of the innumerable cases of meaningful coincidence that have been observed not only by me but by many others, and recorded in large collections. They include everything that goes by the name of clairvoyance, telepathy, etc., from Swedenborg's well-attested vision of the great fire in Stockholm to the recent report by Air Marshal Sir Victor Goddard about the dream of an unknown officer, which predicted the subsequent accident to Goddard's plane.

All the phenomena I have mentioned can be grouped under three categories:

1. The coincidence of a psychic state in the observer with a simultaneous, objective, external event that corresponds to the psychic state or content (e.g., the scarab), where there is no evidence of a causal connection between the psychic state and the external event, and where, considering the psychic relativity of space and time, such a connection is not even conceivable.

2. The coincidence of a psychic state with a corresponding (more or less simultaneous) external event taking place outside the observer's field of perception, i.e., at a distance, and only verifiable afterward (e.g., the Stockholm fire).

3. The coincidence of a psychic state with a corresponding, not yet existent, future event that is distant in time and can likewise only be verified afterward.

In groups 2 and 3 the coinciding events are not yet present in the observer's field of perception, but have been anticipated in time in so far as they can only be verified afterward. For this reason I call such events synchronistic, which is not to be confused with synchronous."

This «synchronicity» described by Jung, from a scientific point of view, can be explained not only by accidental coincidence but also by unstable unobserved correlations that exist between living organisms and environmental objects. As we previously stated, metabolism processes can support these correlations and inhibit their «entangling» with an environment during decoherence. As these correlations are unstable, they are not observed (i.e., correspond with unpredictable dynamics). This explanation does not require the use of quantum mechanics as similar correlations can be found in classical mechanics that have analogues of quantum correlations. Attribution of these correlations exclusively to quantum mechanics is a common mistake.

For the external observer, the unobservability of these correlations does not mean that they cannot be registered by our subjective experience in the form of certain «presentiments». Similarly, the external observer cannot measure or predict the calculation results of a quantum computer as such an attempt would destroy this calculation. Assume that the quantum computer has a "consciousness". The computer is capable of forming a "presentiment" of a future calculation result compared with the external observer who is incapable of this intuition.

This consideration does not "prove" the existence of the unobservable correlations; however, «synchronicity» can be in principle related to unstable correlations. We cannot conclude that this assumption does not contradict physics. Any observational verification of this hypothesis does not seem possible in principle due to the principal unobservability of these correlations.

For example, assume that a person started two initially correlated quantum computers and knew their initial state. Then, the person disappeared. We cannot predict future calculation results for these two computers. Even if we discovered a correlation between their calculated results (for example, they are equivalent), we are not able to choose between the two possibilities: this is caused by a correlation of initial conditions or can be explained by simple accidental coincidence. Any attempt to measure the internal state of a quantum computer during the calculation process inevitably causes the destruction of its correct operation. Similar rationales are valid not only for quantum systems but also for chaotic classical systems with additional unstable microscopic correlations.

Maybe the real world is also a set of correlated computers with unobservable unstable correlations? Maybe the role of living substances consist of the conservation of these correlations? Only God can know their exact initial state if we assume His existence. However, the existence of these correlations is possible because our world appeared from a single point as a result of the «Big Bang». All living organisms on our planet may be a result of unique "protocell" evolution.

Human insight and certain parapsychological effects can exist in a narrow field on the verge of comprehensibility by exact science. This field is the field of unpredictable dynamics. Their basic elusiveness and instability does not enable natural selection to increase these properties [76, 77]. Due to instability and unpredictability, it is not feasible to investigate these appearances using scientific methods, although these appearances do not contradict any physical laws and are possible from the point of view of physics.

In his book, Mensky [4] attempts to justify points of human intuition and parapsychological effects through specific aspects of quantum mechanics. However, he makes a number of typical errors.

1) "Violations" of the laws of quantum mechanics are not required to explain these effects. For example, transitions to other parallel worlds (introduced by the multiworld interpretation) using "consciousness forces" of a "medium" are not necessary. It is sufficient to assume a correlation between the desires of a "medium" and the occurrence of events because these correlations with the environment can "play along" with our desires. Due to these correlations, a consciousness can "have a presentiment" of the future. An attempt "to measure" or "to discover" these unstable quantum correlations will result in their loss and a nonreversible altered evolution. Any "violation" of the common laws of quantum mechanics using a "medium" is unnecessary.

2) Effects that are similar to quantum effects can also occur in classical chaotic systems. Accordingly, these effects can be classically modelled without quantum mechanics.

3) Mensky writes about complexity and the impossibility of validation of the evolution of an individual system evolution from a scientific point of view. During this evolution, we observe nonrepeated events. Their probabilities are described by different probability distributions. In studying the verification of some theory with probabilistic laws, an ensemble of the same events is generally employed with the help of the «law of large numbers» [16]. However, the "law of large numbers" also exists for events of different types. (Generalised Chebyshev's theorem or Markov's theorem) [16]. Thus, using these statements, the consistency in QM laws can also be examined for a complex set of different events, which correspond with the evolution of an individual system.

# 14. Conclusion

This paper does not represent an abstract philosophical discussion. The lack of understanding of the principles discussed in this paper can cause mistakes in solving physical problems. The majority of the real systems that were presented cannot be described with ideal equations of quantum or classical mechanics. The influences of measurements and the environment on the systems (which are inevitable in quantum mechanics and almost always present in classical mechanics) disturb the evolution of the system. The attempt to include the observer and the environment in the system for their descriptions produces the paradox of the self-observing system (Appendix M). This type of system cannot measure and retain complete information of its own state. Even the approximate (self-) description in such a system only has applicability in a time domain that is limited by a time smaller than the return time for this system. After this return time is determined according to Poincare's theorem, all information concerning the previous state of the system is inevitably deleted. However, the description of the system in this case is possible based on observable dynamics. The possibility of this description is explained by the independence of observable dynamics on the type and value of external noise for an extensive range of noise types; thus, observable dynamics are determined not only by the properties of the system. Observable dynamics are based on roughening of the distribution function or density matrix because the initial state of the system is not precisely defined. The difference between observable and ideal dynamics cannot be experimentally verified, even when the observer and the environment are included in the system to be described because self-description is limited in the time domain of precision and observation. Thus, the return of the system to the initial state, as predicted by Poincare's theorem for ideal dynamics, cannot be observed by the self-observing system due to the effect of the deleting of information concerning previous states. The introduction of observable dynamics helps to solve all known paradoxes of classical and quantum mechanics.

The lack of understanding of the principles discussed in this paper can produce mistakes when solving physical problems. Several examples of these mistakes are provided, including the errors in the pole theory for the problems of flame front motion and "finger" growth at the liquid/liquid interface.

Sivashinsky et al. [78] state that the ideal dynamics of poles cause an acceleration of flame front propagation and that this condition is not due to noise as the effect does not disappear with a decrease in noise and is purely dependent on the properties of the system. However, the noise-connected observable dynamics are not dependent on noise over an extensive range of values.

Tanveer *et al.* [79] discovered a discrepancy between the theoretical predictions for "finger" growth in the problem of interfacial fluid flow and the results of numerical experiments. However, no understanding was attained in the paper [79] regarding the connection of this discrepancy with numerical noise, which generates a new observable dynamics.

These examples were discovered during the research of this paper; additional examples can be easily obtained.

The results of this paper are necessary for a thorough understanding of the basics of nonlinear dynamics, thermodynamics, and quantum mechanics.

An extensive analysis, such as the analysis in this paper, is usually not required for typical problems of physics. Physical systems are typically either considered to be systems with a small quantity of particles or systems with numerous particles closed to thermodynamic equilibrium. In these systems, precise ideal dynamics or simplified "approximate" methods can be employed for deriving observable dynamics, for example, a reduction in QM or the Boltzmann equation in CM. Therefore, the interest by physicists to papers similar to this paper is sufficiently small. However, many physical systems are not included in a narrow class that is represented by ideal or simple observable dynamics. Their behaviour is probabilistically unpredictable. We named their behaviour unpredictable dynamics. A quantum computer concerns systems from the point of view of the observer who did not observe its "beginning". Systems with unpredictable dynamics can also include also certain stationary systems far from thermodynamic equilibrium. Living organisms are examples of these stationary systems. Even if these systems could be partially or completely represented by observable dynamics, its derivation is a nontrivial problem. If physics would attempt to represent these complex systems, the understanding of methods stated in this paper becomes necessary. Numerous problems still need to be solved by physicists and mathematicians:

1) Which methods exist for deriving the observable dynamics of complex systems?

2) When a derivation of observable dynamics is impossible, can a system be represented by unpredictable dynamics?

3) Is it possible to create «synergetic» systems «on paper» (for example, similar to Penrose's tubulins) that would illustrate the principal possibility of the appearance and existence of complex stationary systems (both classical, and quantum), which are represented by unpredictable dynamics? We attempt to provide some predictions regarding this possibility:

a) The systems featured by unpredictable dynamics should be capable of inhibiting decoherence and conserving additional unstable classical (or quantum) correlations, both inside complex systems and between complex systems.

b) These systems can have several unstable states and can transfer between these states during evolution. The stream of negentropy, substances or energies (i.e., metabolism) enables the conservation of these unstable states and processes, without destroying these unstable states and processes, and by protecting these unstable states and processes from external noise. Conversely, unstable systems can serve as catalytic agents of this metabolism. Both inverse processes and Poincare's returns are possible in these systems. They are protected from external noise (decoherence) by metabolism. External noise, which would be reduced, is incapable of destroying these processes or states. Any attempt to measure a current state or process in this unstable system would destroy its dynamics. Thus, these dynamics would be unobservable. These systems can be not only quantum but also classical.

c) In physics, a macrostate is generally considered to be a passive function in its microstate. However, assume that a system itself is capable of measurement both its macrostate and its environment of macrostates. In this manner, the feedback of macrostates through a microstate appears [3] (Appendixes M and V).

d) Unpredictable systems can be a type of self-replicated cellular automata [80].

Last year, interesting papers about the construction of these «synergetic» systems, which are most likely similar to live organisms, were published [81-83]. Note that the construction of these models is a problem of physics and mathematics and not philosophy.

## Appendix A. Phase Density Function. [5-6, 14]

The state of a system of N particles, which is identical from the point of view of classical mechanics, can be given by the coordinates $r_1, \ldots, r_N$ and momenta $p_1, \ldots, p_N$ of all particles, N, of the system. For brevity, we use the notation $x_i=(r_i,p_i)$ (i=1, 2, …, N) to designate the set of coordinates and the momentum spatial components of a single particle and the designation $X=(x_1, \ldots, x_N)=(r_1, \ldots, r_N, p_1, \ldots, p_N)$ to denote the set of coordinates and momenta of all particles of the system. The corresponding state of 6 N variables is called 6N-dimensional phase space.

We consider a Gibbs ensemble, i.e., a set of identical macroscopic systems to define the concept of distribution function. The experimental conditions are similar for these systems. However, as these conditions do not unambiguously determine the state of the systems, different values of X correspond to different states of the ensemble at a given time $t$.

We select a volume dX in the vicinity of point X. Assume that at a given time, $t$, this volume contains points that characterise the states of dM systems from the total number, M, of systems in the ensemble. The limit of the ratio of these values

$$\lim_{m \to \infty} \frac{dM}{M} = f_N(X,t)dX$$

(where m=M/dM) defines the density function of the distribution in the phase state at time $t$. This function is distinctly normalised as follows:

$$\int f_N(X,t)dX = 1$$

The Liouville equation for the phase density function can be written in the form

$$i\frac{\partial f_N}{\partial t} = Lf_N$$

where $L$ is the following linear operator

$$L = -i\frac{\partial H}{\partial p}\frac{\partial}{\partial x} + i\frac{\partial H}{\partial x}\frac{\partial}{\partial p}$$

where $H$ is the energy of the system.

## Appendix B. Definitions of Entropy

We provide the following definition of entropy:
S=-k $\int_{(X)} f_N(X,t) \ln f_N(X,t)$
In quantum mechanics, entropy is defined via a density matrix as
S=-k tr $\rho \ln \rho$ [15]
where $tr$ denotes a matrix trace.

This definition of entropy does not change in the case of reversible evolution. Coarsened values of $f_N$ or $\rho$ should be used to obtain the changing entropy.

## Appendix C. Poincare's Proof of the Theorem of Returns

The number of phase points that leave the given phase volume, g, during motion and not returned in motion will be less than any finite portion of the complete number of phase points. We will prove this position.

Consider the system with a finite phase volume $G$. We select a *fixed* surface, σ, inside this volume, which is restricted to the small volume, $g$. We consider the phase points flowing through a surface σ from volume $g$. The velocity of the transition of a phase point on a phase trajectory is only dependent on phase coordinates; therefore, the number of the points flowing in a unit of time through the fixed surface, σ, is not dependent on time. We designate the volume occupied with phase points, which flow in a unit of time from phase volume $g$ and is not being returned in it, by $g'$. During $T g'T$ volumes of a phase, fluid flows from volume $g$, as the flowed-out volume is $g'T$. Under this assumption, it is not returned more in volume $g$ as it should fill a remaining part of the full phase volume $G$. A phase fluid is incompressible; therefore, flowed-out of $g$, the volume $g'T$ should not exceed the volume in which it will flow out, i.e.,

$$Tg' < G-g < G. (1)$$

Volume $G$ is finite; therefore, this inequality can be satisfied only for finite T at finite $g'$. For $T \to \infty$, the inequality (1) is satisfied only at $g' \to 0$, as it must be shown.

# Appendix D. Correlation

Consider the following problem. A series of measurements of two random variables $X$ and $Y$ has been performed, and measurements were performed in a pairwise fashion, i.e., we obtain two values for one measurement - $x_i$ and $y_i$. As the sample consists of pairs $(x_i, y_i)$, we verify whether a dependence exists between these two variables. This dependence is named **correlation**. Correlation can exist not only between two magnitudes but also between larger quantities of magnitudes.

Dependence between random variables can possess a functional character, i.e., to be the strictly functional relation linking their values. However, experimental data dependencies are frequently statistical dependencies. The distinction between two aspects of dependences is that that the functional connection establishes a strict correlation between variables, and statistical dependence only explains why the distribution of random variable $Y$ is dependent on the value accepted by random variable $X$.

### Coefficient of Pearson's Correlation

Various coefficients of correlation exist, to each of which the previously stated correlation is valid. The coefficient of Pearson's correlation, which characterises the degree of a linear relation between variables, is well known and defined as

$$ r = \frac{\sum_i (x_i - \bar{x})(y_i - \bar{y})}{\sqrt{\sum_i (x_i - \bar{x})^2} \sqrt{\sum_i (y_i - \bar{y})^2}} $$

# Appendix E. Thermodynamic Equilibrium of the Isolated System: Microcanonical Distribution[5, 14]

We are interested in an *adiabatic system*—a system that is isolated from external bodies and that contains a specific given energy, E.

### MICROCANONICAL Distribution

Consider an adiabatic system, i.e., a system that cannot exchange energy with external bodies at invariable external parameters. For such a system,

$$ H(X, a) = E = \text{const} \qquad (1) $$

and the phase density function, $\varphi(\varepsilon)$, has an acute maxima at $\varepsilon = E$ because the energy of the system E is fixed and will not change. However, the phase density, $\varphi(\varepsilon)$, in a limit at $\Delta E \to 0$ becomes propositional to a Dirac delta function, $\delta\{\varepsilon - E\}$. Thus, for an adiabatic isolated system, it is possible to assume

$$ \omega(X) = [1 / \Omega(E, \alpha)] \, \delta \{E - H(X, \alpha)\}, \qquad (2) $$

where $1/\Omega(E, \alpha)$ is the norming factor that can be found from the requirement of normalisation, i.e.,

$$ \Omega(E, \alpha) = \int_{(X)} \delta \{E - H(X, \alpha)\} dX \qquad (3) $$

Expression (2) is named the microcanonical Gibbs distribution. Based on basis this distribution, it is possible to calculate the phase averages of any physical quantities for adiabatic isolated system using the formula

$$ \overline{F} = \int_{(X)} F(X) \frac{1}{\Omega(E, a)} \delta\{E - H(X, a)\} dX \qquad (4) $$

The magnitude $\Omega(E, \alpha)$ has visual geometrical meaning. $\Omega(E, \alpha) \, dE$ exhibits a sense of phase volume of lamina concluded between the hypersurfaces $H(X,a) = E$ and $H(X, a) = E + dE$.

# Appendix F. Theorem of the Invariance of Phase "Fluid" Volume

Assume that each mass point of system is represented by the Cartesian coordinates

$$ x_k, y_k, z_k \ (k = 1, 2..., N). $$

We also designate these three coordinates as the vector $\vec{r}_k$. This system of $N$ mass points can also be represented by $3N$ using generalised coordinates:

$$ q_n (x_1..., z_N) \ (n=1, 2..., 3N). $$

The equations of motion of this conservative system consist of the equations of Lagrange

$$ \frac{d}{dt} \frac{\partial L}{\partial \dot{q}_k} - \frac{\partial L}{\partial q_k} = 0 \qquad (k = 1, 2,..., 3N) \qquad (1) $$

where $L = K - U$-the function of Lagrange or a Lagrangian, K denotes the kinetic energy of a system and $U$ denotes the potential energy of a system. However, it is more convenient to use the equations of motion of the Hamilton shape in statistical physics:

$$ \left. \begin{array}{l} \dot{q}_k = \dfrac{\partial H}{\partial p_k} \\[2mm] \dot{p}_k = -\dfrac{\partial H}{\partial q_k} \end{array} \right\} \ k = 1, 2,..., 3N, \\[4mm] H = \sum_{k=1}^{3N} p_k \dot{q}_k - L, \quad p_k = \dfrac{\partial L}{\partial \dot{q}_k} \qquad (2) $$

A Hamiltonian function, or a Hamiltonian $(q_1, q_2..., q_{3N}; p_1, p_2..., p_{3N})$, is *a population of canonical variables*. In blanket deductions, we designate all canonical variables by the letter $X$ and assume:

$$q_k = X_k, \; p_k = X_{k+3N} \quad \{k = 1, 2..., 3\,N\}. \tag{3}$$

Formulas that transform to a compact form of all population of variables ($X_1, X_2..., X_{6N}$) are frequently designated by one letter (X), and the product of all differentials $dX_1\, dX_2\, ...dX_{6N}$ is designated by dX.

The equations of Hamilton represent a system of the differential equations of the first order; thus, all variables X during the moment t are completely defined if the values of these variables, X0, during the moment t = 0 are known. This property of the Hamilton shape mechanically enables the introduction of geometrically evident image of the system and its motion in a phase space. The driving of a phase ensemble in a phase space can be considered as the motion of a phase fluid, which is analogous to the motion of a common fluid in three-dimensional space. Phase space points are identified with points of the imaginary phase fluid filling space.

By satisfying the Hamiltonian equations, it is easy to prove that a phase fluid of a system is incompressible. The denseness of a typical three-dimensional incompressible fluid is constant. Thus, based on the continuity equation

$$-\frac{\partial \rho}{\partial t} = div \; \vec{\upsilon} \; \rho = \rho \; div \; \vec{\upsilon} + \vec{\upsilon} \cdot \nabla \rho \tag{4}$$

and the requirement that $\rho$ = const for an incompressible fluid, we obtain

$$div \; \vec{\upsilon} = \frac{\partial \dot{x}}{\partial x} + \frac{\partial \dot{y}}{\partial y} + \frac{\partial \dot{z}}{\partial z} \tag{5}$$

This theorem can be easily extended for a fluid in a multidimensional space and, consequently, the requirement of equality to zero for multidimensional divergence for an incompressible phase fluid should be satisfied, i.e.,

$$\sum_{k=1}^{6N} \frac{\partial \dot{X}_k}{\partial X_k} = O \tag{6}$$

Based on the equations of Hamilton (2)

$$\sum_{k=1}^{6N} \frac{\partial \dot{X}_k}{\partial X_k} = \sum_{k=1}^{6N} \left( \frac{\partial \dot{q}_k}{\partial q_k} + \frac{\partial \dot{p}_k}{\partial p_k} \right) = \sum_{k=1}^{6N} \left( \frac{\partial^2 H}{\partial q_k \partial p_k} - \frac{\partial^2 H}{\partial p_k \partial q_k} \right) \equiv 0 \tag{7}$$

As a phase fluid is incompressible, the invariable phase volume is occupied during its motion with any part of this fluid.

# Appendix G. The Basic Concepts of Quantum Mechanics[18-20]

## Wave Function.

The basis of a mathematical apparatus of quantum mechanics consists of the statement that the system state can be described by a (complex) function of coordinates, $\Psi\,(q)$, and the quadrate of the module of this function that defines a probability distribution of values of coordinate q: $|\Psi(q)|^2 dq$ is the probability of obtaining values of coordinates in $[q, q+dq]$ space interval for measurement produced over the system. Function $\Psi$ is termed as a system *wave function*.

## Observable Variables

Consider a physical quantity $f$ that characterises the state of a quantum system. In the following discussion, we present not one variable but a complete set. For simplicity, we discuss only one physical quantity.

The values of certain physical quantities are named *eigenvalues*, and their complete set is named a *spectrum* of eigenvalues for this variable. In classical mechanics, a set of all possible values of any variable is generally continuous. In quantum mechanics, physical quantities (for example, coordinates) for which eigenvalues fill the continuous number exist; in these cases, a *continuum spectrum of* eigenvalues exists. If all possible eigenvalues comprise a discrete set, the set is a *discrete spectrum*.

For simplicity, consider that variable $f$ possesses a discrete spectrum. We designate variable eigenvalues $f$ as $f_n$, in which the index $n$ can assume the values 1, 2, 3, …. We designate $\Psi_n$ for a system wave function of state that corresponds to the value $f_n$ of variable $f$. $\Psi_n$ is named the *eigenfunctions* of the given physical quantity $f$. Each of these functions is normalised, i.e.,

$$\int |\Psi_n|^2 dq = 1. \tag{1}$$

If the system exhibits an arbitrary state with a wave function $\Psi$, the measurement of variable $f$ will yield an eigenvalue $f_n$. According to the principle of superposition, the wave function $\psi$ should represent a linear combination of all eigenfunctions $\psi_n$, which corresponds to all values $f_n$ that can be measured with a nonzero probability. Therefore, any state function $\Psi$ can generally be presented in the following form

$$\Psi = \Sigma \; a_n \Psi_n, \tag{2}$$

where the summation is yielded on all $n$, and $a_n$ are constant coefficients.

Thus, we can conclude that any wave function can be expanded to the set of eigenfunctions of a physical quantity. This set is named a *complete set of functions*.

In expansion (2), the module quadrate $|a_n|^2$ defines the probability of measurement the $f_n$ of variable $f$ for a state with a wave function $\Psi$. The sum of all probabilities should be equivalent to one; the following relation should occur

$$\sum_n |a_n|^2 = 1 \tag{3}$$

An observable variable is defined by operators over a function space. The result of an operator action on a function is also a function. Eigenfunctions $\psi k$ and its eigenvalues $\lambda_k$ comprise a solution of the functional equation

$$A \; \psi_k = \lambda_k \psi_k \tag{4}$$

where A is an operator of the corresponding observable variable.

Label H is usually used for Hamiltonian, which is an energy operator.

For one particle in external field $U(x, y, z)$, it is defined by the following formula

$$H \equiv -\frac{\hbar^2}{2m}\Delta + U(x, y, z) \tag{5}$$

where $\Delta = \frac{\partial^2}{\partial x^2} + \frac{\partial^2}{\partial y^2} + \frac{\partial^2}{\partial z^2}$ an operator of Laplace.

The matrix form of quantum mechanics.

Expand a function under operator action and the output function over eigenfunctions of an observable variable. Both of these functions can be noted as columns of these expansion coefficients. The operator of the observable variable can be expressed by the form of a square matrix. The product of this matrix on a column of coefficients of expansion of the first function will yield coefficients of expansion of the second function. This form of operators and functions is named the matrix form of quantum mechanics.

**Schrodinger Equation**

The wave equation of motion for a particle in an external field can be expressed as

$$i\hbar\frac{\partial\Psi}{\partial t} = -\frac{\hbar^2}{2m}\Delta\Psi + U(x, y, z)\Psi. \tag{6}$$

Equation (6), which was proposed by Schrodinger in 1926, is named *a Schrodinger equation.*

**Uncertainty Principle of Heisenberg**

If we characterise the indeterminacies of coordinates and momentums by average quadratic fluctuations

$$\delta x = \sqrt{\overline{(x-\bar{x})^2}}, \ \delta p_x = \sqrt{\overline{(p_x - \bar{p}_x)^2}},$$

the minimal possible value of their product can be obtained. Consider a one-dimensional case—a package with the wave function $\psi(x)$, which is dependent on one coordinate; for simplicity, we will assume that the medial values $x$ and $p_x$ in this state are equivalent to zero. We begin with the distinct inequality

$$\int_{-\infty}^{+\infty}\left|\alpha\ x\varphi + \frac{d\phi}{dx}\right|^2 dx \geq 0$$

where $\alpha$ is any real constant.

Calculate this integral.

Using

$$\int x^2\left|\varphi\right|^2 dx = (\delta x)^2$$

$$\int\left(x\frac{d\phi^*}{dx}\phi + x\phi^*\frac{d\phi}{dx}\right)dx =$$

$$\int x\frac{d\left|\phi\right|^2}{dx}dx = -\int\left|\phi\right|^2 dx = -1$$

,

$$\int\frac{d\phi^*}{dx}\frac{d\phi}{dx}dx =$$

$$-\int\phi^*\frac{d^2\phi}{dx^2}dx = \frac{1}{\hbar^2}\int\phi^*\ \hat{p}_x^2\phi\ dx = \frac{1}{\hbar^2}\left(\delta p_x\right)^2,$$

we obtain

$$\alpha^2(\delta x)^2 - \alpha + \frac{(\delta p_x)^2}{\hbar^2} \geq 0$$

This quadratic trinomial (over $\alpha$) is non-negative for any $\alpha$ if its discriminant is nonpositive. From this condition, we can obtain the following inequality:

$$\delta x\ \delta p_x \geq \hbar/2 \tag{7}$$

The minimal possible value of the product is equivalent to $\hbar/2$.

This uncertainty principle (7) was established by Heisenberg in 1927.

A decrease in the coordinate uncertainty (i.e., *δx) results in an increasing* uncertainty of momentum along the same axis( *δp$_x$)* and vice versa. In this particular case, when the particle is in a certain point in space *(δx = δy = δz = 0)*; thus, *δp$_x$ = δp$_y$ = δp$_z$ = ∞*, which signifies that all values of momentum have an equivalent probability. Conversely, if a particle has a certain momentum p, all positions in space have an equivalent probability.

# Appendix I. Density Matrix

Consider a beam of N$_a$ particles that are prepared in state $|\chi_a\rangle$ and another beam of N$_b$ particles prepared in state $|\chi_b\rangle$, which is independent of the first bean. To describe the combination of beams, we introduce the mixed state operator ρ, which is defined as

ρ=W$_a$|χ$_a$⟩⟨χ$_a$| + W$_b$|χ$_b$⟩⟨χ$_b$|,

where W$_a$=N$_a$/N, W$_b$=N$_b$/N, N=N$_a$+N$_b$

operator ρ is named a density operator or a statistical operator. It describes the way the beams were prepared and contains complete information about the total beam. The mixture is completely defined by the density matrix. In the special case of a pure state $|\chi\rangle$, the density operator is given by the expression

ρ=|χ⟩⟨χ|.

Operator ρ is usually expressed in matrix form. Therefore, we choose a basic set of states (the most commonly used states consist of |+1/2⟩ and |-1/2⟩) and decompose the |χ$_a$⟩ and

$|\chi_b\rangle$ states over this basic set as follows:

$$|\chi_a\rangle = a_1^{(a)}|+1/2\rangle + a_2^{(a)}|-1/2\rangle,$$
$$|\chi_b\rangle = a_1^{(b)}|+1/2\rangle + a_2^{(b)}|-1/2\rangle.$$

In the representation of $|\pm1/2\rangle$ states, we obtain the following relations for these states

$$|\chi_a\rangle = \begin{pmatrix} a_1^{(a)} \\ a_2^{(a)} \end{pmatrix}$$

$$|\chi_b\rangle = \begin{pmatrix} a_1^{(b)} \\ a_2^{(b)} \end{pmatrix},$$

and for the conjugated states

$$\langle\chi_a| = (a_1^{(a)*}, a_2^{(a)*}) \text{ and }$$
$$\langle\chi_b| = (a_1^{(b)*}, a_2^{(b)*}).$$

Using the matrix multiplication rules, we obtain the "external product"

$$|\chi_a\rangle\langle\chi_a| = \begin{pmatrix} a_1^{(a)} \\ a_2^{(a)} \end{pmatrix}\left(a_1^{(a)*}, a_2^{(a)*}\right)$$

$$= \begin{pmatrix} |a_1^{(a)}|^2 & a_{11}^{(a)}a_2^{(a)*} \\ a_1^{(a)*}a_2^{(a)} & |a_2^{(a)}|^2 \end{pmatrix}$$

and a similar expression for the $|\chi_b\rangle\langle\chi_b|$ product. By substituting these expressions into the density operator, we obtain the following density matrix:

$$\rho =$$
$$\begin{pmatrix} W_a|a_1^{(a)}|^2 + W_b|a_1^{(b)}|^2 & W_a a_1^{(a)}a_2^{(a)*} + W_b a_1^{(b)}a_2^{(b)*} \\ W_a a_1^{(a)*}a_2^{(a)} + W_b a_1^{(b)*}a_2^{(b)} & W_a|a_2^{(a)}|^2 + W_b|a_2^{(b)}|^2 \end{pmatrix}$$

As the $|\pm1/2\rangle$ states were employed for the basic state, the obtained expression is named the density matrix in $\{|\pm1/2\rangle\}$ representation.

### Statistical Density Matrix $P_0$.

We develop several concluding remarks concerning the statistical matrix $P_0$, which exhibits remarkable properties. All possible macroscopic states of the system in classical statistical thermodynamics are considered *a priori* equiprobable. The states are considered equally probable; unless information is available concerning the total energy of the system, its contact with the thermostat ensures a constant temperature in the system. Similarly, in wave mechanics, all states of the system that correspond to the functions that form a complete system of orthonormalised functions can be considered *a priori* equiprobable. Let $\varphi_1, \ldots, \varphi_k$ be a system of basic functions. Provided that the system is characterised by a mixture of $\varphi_k$ states, in the absence of other relevant information we can assume that the statistical matrix of the system has the form

$$P_0 = \sum_k p P_{\varphi_k}, \text{ where } \sum_k p = 1,$$

i.e., that $P_0$ is the statistical matrix of a mixed state with all equal statistical weights. As $\varphi_k$ are the basis functions, the matrix $P_0$ can be represented as follows:

$$(P_0)_{kl} = p\delta_{kl}$$

If the matrix $P_0$ characterises the statistical state of the ensemble of systems at the initial moment of time, and the same value A is measured in all systems of the ensemble, the statistical state of the ensemble would be characterised by the $P_0$ matrix.

The equations of motion for the density matrix.

The equations of motion for the density matrix $\rho$ have the form

$$i\frac{\partial\rho_N}{\partial t} = L\rho_N$$

where L is the linear operator
$L\rho = H\rho - \rho H = [H, \rho]$ and
where H is the energy operator of the system.
If A is the operator of a certain observable, then the average value of the observable can be obtained as follows:
$\langle A \rangle = \mathrm{tr} A\rho$

# Appendix J. Reduction of the Density Matrix and the Theory of Measurement

Assume that the states $\sigma^{(1)}$, $\sigma^{(2)}$, ... are "clearly discernible" in measurement of a certain object. The measurement performed over the object in one of these states yields the numbers $\lambda_1$, $\lambda_2$, .... The initial state of the measuring device is designated $a$. If the measured systems were initially in the state $\sigma^{(v)}$, then the state of the complete system—"measured system plus the measuring device" prior to their interaction—is determined by the direct product $a \times \sigma^{(v)}$. After the measurement

$$a \times \sigma^{(v)} \rightarrow a^{(v)} \times \sigma^{(v)}$$

Assume that the initial state of the measured system is not discernible but is an arbitrary mixture: $\alpha_1 \sigma^{(1)} + \alpha_2 \sigma^{(2)} + ...$ of such states. In this case, due to the linearity of the quantum equations, we obtain:

$$a \times [\Sigma\alpha_v\sigma^{(v)}] \rightarrow \Sigma\alpha_v[a^{(v)} \times \sigma^{(v)}]$$

A statistical correlation between the state of the object and the state of the device in the final state results from the measurement. A simultaneous measurement of two values in the system—"measured object and measuring system" (the first value is the measured characteristic of the studied object and the second value is the position of the measuring device indicator)—always produces correlating results. Therefore, one of these measurements is superfluous: a conclusion on the state of the measured object can always be made based on observations of the measuring device.

The state vector obtained as the measurement result cannot be represented as a sum in the right part of the previous relation. It is a mixture, i.e., one of the state vectors of the form

$$a^{(v)} \times \sigma^{(v)},$$

and the probability of this state appears to be the result of the interaction between the measured object and the measuring device is $|\alpha_v|^2$. This transition is named "wave packet reduction," and it corresponds to the transition of the density matrix with the nondiagonal form $\alpha_v \alpha_\mu^*$ to the density matrix with the diagonal form $|\alpha_v|^2 \, \delta_{v\mu}$. This transition is not described by the quantum mechanical equations of motion.

# Appendix K. Coarsening of the Phase Density Function and the Molecular Chaos Hypothesis

Coarsening of the density function is named for its substitution with an approximate value, e.g.,

$f_N^*(X,t)=\int_{(Y)}g(X-Y)f_N(Y,t)dY$

where
$g(X)=1/\Delta \, D(X/\Delta)$
$D(x)= 1$ for $|X|<1$
$D(x)= 0$ for $|X|\geq 1$
Another example of coarsening is the "molecular chaos hypothesis". This hypothesis implies the substitution of a two-particle distribution function with the product of single-particle functions as follows:
$f(x_1,x_2,t) \rightarrow f(x_1,t)f(x_2,t)$

# Appendix L. Prigogine's New Dynamics

The new dynamics introduced by Prigogine is frequently mentioned in this paper. A brief introduction to this theory, which is based on the monographs, is presented below [14, 55].
The linear operator $\Lambda$, which acts on the phase density function or density matrix $\rho$, is introduced such that

$$\tilde{\rho} = \Lambda^{-1} \rho$$

$$\Lambda^{-1} \, 1=1$$

$$\int \tilde{\rho} = \int \rho$$

where $\Lambda^{-1}$ remains positive. The requirement on the operator $\Lambda$ is that the function $\Omega$ is defined via the function $\tilde{\rho}$ as

$$\Omega = tr\tilde{\rho}^+ \, \tilde{\rho} \text{ or } \Omega = tr\tilde{\rho} \, ln \, \tilde{\rho}$$

And complies with the inequality $d\Omega/dt \leq 0$.
The equation of motion for the transformed function $\tilde{\rho}$ is

$$\frac{\partial \tilde{\rho}}{\partial t} = \Phi \tilde{\rho}$$

where $\Phi= \Lambda^{-1} \, L \Lambda$ and
$\Phi$ is a noninvertible Markovian semigroup.
$$\Lambda^{-1}(L)=\Lambda^+(-L)$$
The operator $\Lambda^{-1}$ for the phase density function corresponds to the coarsening in the direction of the phase volume reduction. In quantum mechanics, this operator can only be obtained for an infinite volume or an infinite number of particles. A projection operator, P, which makes all nondiagonal elements of the density matrix zero, is introduced in quantum mechanics. The operator, $\Phi$, and the basis vectors of the density matrix are chosen so that the operators $\Phi$ and P are permutable:

$$\frac{\partial P \tilde{\rho}}{\partial t} = P \, \Phi \tilde{\rho} = \Phi P \tilde{\rho}$$

# Appendix M. Impossibility of Self-Prediction of Evolution of the System [101]

Assume that a powerful computer exists that is capable of predicting its own future and the future of its environment based on the calculation of motion of all molecules. Assume that the prediction involves the rolling of a black or white ball from a certain device, which is an integral part of the computer and is described by the machine. The device rolls out a white ball when the computer predicts a black one and rolls out a black ball when the white one is predicted. The predictions of the computer are always false. As the choice of the environment is arbitrary, this contrary instance proves the impossibility of exact self-observation and self-calculation. As the device always contradicts the predictions of the computer, complete self-prediction of the system, including both the computer and the device, is impossible (Fig. 27).

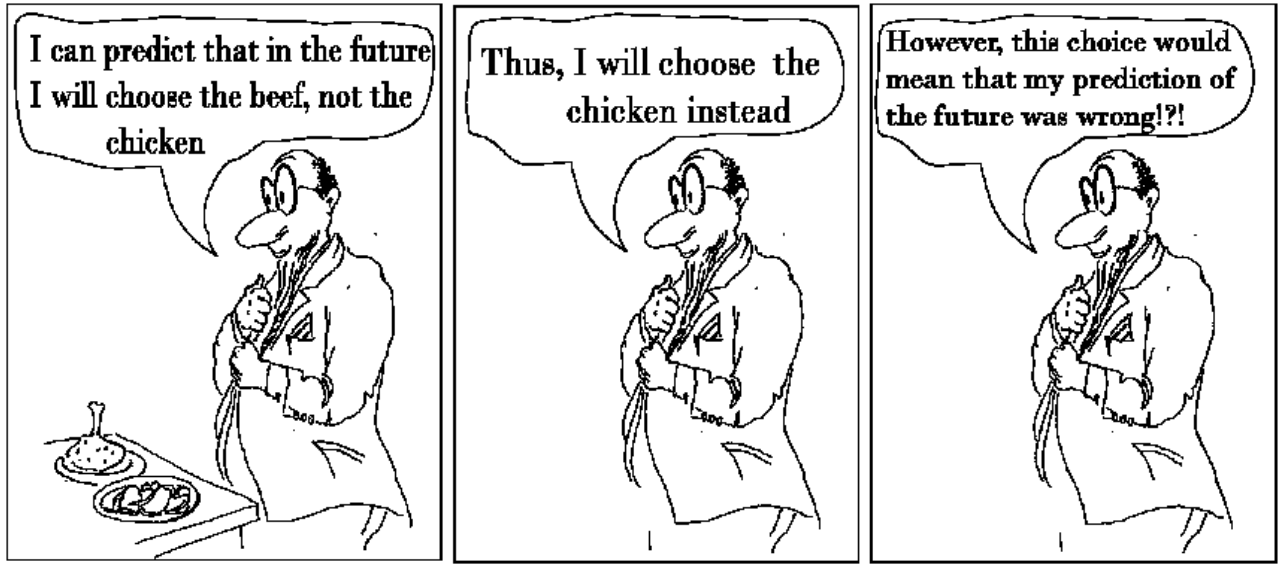

**Figure 27.** Impossibility of self-prediction.

# Appendix N. Tables of Correspondence between Quantum and Classical Mechanics

**Table 1.** Basic Properties of quantum and classical mechanics

| Quantum mechanics | Classical mechanics |
|---|---|
| Density matrix | Phase density function |
| Equation of motion for the density matrix | Liouville equation |
| Wave packet reduction | Coarsening of the phase density function or the molecular chaos hypothesis |
| Unavoidable interaction of the measured system with the observer or the environment as described by reduction | Theoretically infinitesimal but, in reality, a finite small interaction of the measured system with the observer or the environment |
| Nonzero and nondiagonal elements of the density matrix | Correlations between the velocities and positions of particles in different parts of the system |
| Pointer states | Appropriate (i.e., macroscopically stable to small perturbations) macroscopic states |

**Table 2.** Probability formulations in classical and quantum mechanics. [18]

| | Classical mechanics | Quantum mechanics |
|---|---|---|
| Pure state | Point $(q, p)$ of phase space | State vector $|\psi>$ |
| General state | Probability density $\rho(q, p)$ | Positive hermitean operator $\rho$ |
| Normalisation condition | $\int \rho dq dp = 1$ | $\mathrm{tr}\,\rho = 1$ |
| Condition for pure sate | $\rho = \delta$-function | $\rho = |\psi><\psi|$ (operator $\rho$ rank is equal to 1) |
| Equation of motion | $$\frac{\partial \rho}{\partial t} = \{H, \rho\}$$ | $$i\hbar\frac{\partial \rho}{\partial t} = [H, \rho]$$ |
| Observable | Function $A(q, p)$ | Hermitean operator $A$ |
| Average value | $\int A\rho\, dq\, dp$ | $\mathrm{tr}\,(A\rho)$ |

# Appendix O. A System Reduction at Measurement[15, 18]

Consider a situation in which the initial measuring device was in state $|\alpha_0\rangle$ and the object was in a superposition of states $|\psi\rangle = \sum c_i|\psi_i\rangle$, where the $|\psi_i\rangle$ experiment eigenstates. The initial statistical operator is given by expression

$$\rho_0 = |\psi\rangle|\alpha_0\rangle\langle\alpha_0|\langle\psi| \tag{1}$$

The partial track of this operator, which is equal to the statistical operator of the system, including the object, is expressed as

$$tr_A(\rho_0) = \sum_n \langle\varphi_n|\rho_0|\varphi_n\rangle$$

where $|\varphi_n\rangle$- any complete set of device eigenstates. Thus,

$$tr_A(\rho_0) = \sum |\psi\rangle\langle\varphi_n|\alpha_0\rangle\langle\alpha_0|\varphi_n\rangle\langle\psi| = |\psi\rangle\langle\psi|, \tag{2}$$

where the relation $\sum |\varphi_n\rangle\langle\varphi_n| = 1$ and the normalisation condition for $|\alpha_0\rangle$ are used. We have a statistical operator that corresponds to the object state $|\psi\rangle$. After measurement of, a correlation between the device and the object states; thus, the state of the full system including the device and the object is featured by a state vector

$$|\Psi\rangle = \sum c_i e^{i\theta i}|\psi_i\rangle|\alpha_i\rangle. \tag{3}$$

and the statistical operator is given by the expression

$$\rho = |\Psi\rangle\langle\Psi| = \sum c_i c_j^* e^{i(\theta i - \theta j)}|\psi_i\rangle|\alpha_i\rangle\langle\alpha_j|\langle\psi_j|. \tag{4}$$

The partial track of this operator is equal

$$tr_A(\rho) = \sum_n\langle\varphi_n|\rho|\varphi_n\rangle =$$
$$= \sum_{(ij)} c_i c_j^* e^{i(\theta i - \theta j)}|\psi_i\rangle \{\sum_n \langle\varphi_n|\alpha_i\rangle\langle\alpha_j|\varphi_n\rangle\}\langle\psi_j|=$$
$$= \sum_{(ij)} c_i c_j^* \delta_{ij}|\psi_i\rangle\langle\psi_j| \tag{5}$$

(as various states $|\alpha_i\rangle$ of the device are orthogonal to each other); thus,

$$tr_A(\rho) = \sum|c_i|^2|\psi_i\rangle\langle\psi_i|. \tag{6}$$

We have obtained a statistical operator including only the object, which features probabilities $|c_i|^2$ for object states $|\psi_i\rangle$. We formulate the following theorem:

**Theorem 1** (regarding measurement). If two systems, $S$ and $A$, interact in such a manner that to each state $|\psi_i\rangle$ systems $S$, there is a specific corresponding state $|\alpha_i\rangle$ of systems $A$, and the statistical operator $tr_A(\rho)$ over the complete systems *(S and A)* reproduces a wave packet reduction for measurement, yielded over system *S,* which existed prior to measurement the state $|\psi\rangle = \sum c_i|\psi_i\rangle$.

Assume that a subsystem is in a mixed state but that the complete system, including this subsystem, is in a pure state. This mixed state is named an *improper mixed state.*

# Appendix P. The Theorem Regarding Decoherence at the Interaction with a Macroscopic Device. [18, 84]

Consider that the device is a macroscopic system, which indicates that each distinguishable configuration of the device (for example, position of its arrow) is not in a pure quantum state. The device provides no information about the state of each separate arrow molecule. Thus, the initial state of the device, $|\alpha_0\rangle$, should be described by a statistical distribution of microscopic quantum states, $|\alpha_{0,s}\rangle$, in the previously stated reasoning; the initial statistical operator is not given by expression (1) and is equivalent to

$$\rho_0 = \sum_s p_s|\psi\rangle|\alpha_{0,s}\rangle\langle\alpha_{0,s}|\langle\psi|. \tag{7}$$

Each state of the device $|\alpha_{0,s}\rangle$ will interact with each object eigenstate $|\psi_i\rangle$. Thus, it will transform to another state $|\alpha_{i,s}\rangle$, which is one of the quantum states of the set with a macroscopic description that corresponds to an arrow in position i; specifically, we obtain the formula

$$e^{iH\tau/\hbar}(|\psi\rangle|\alpha_{0,s}\rangle) = e^{i\theta i,s}|\psi\rangle|\alpha_{i,s}\rangle. \tag{8}$$

Note the appearance of the phase factor that is dependent on index $s$. Differences in energies for quantum states $|\alpha_{0,s}\rangle$ should have such values phases $\theta_{i,s}$(mod $2\pi$) after time $\tau$ have been randomly distributed between 0 and $2\pi$.

Based on formulas (7) and (8), at $|\psi\rangle = \sum_i c_i|\psi_i\rangle$ the statistical operator after measurement will be given by the following expression:

$$\rho = \sum_{(s,i,j)} p_s c_i c_j^* e^{i(\theta i,s - \theta j,s)}|\psi_i\rangle|\alpha_{i,s}\rangle\langle\alpha_{j,s}|\langle\psi_j| \tag{9}$$

Similar to (9), (6) can be concluded. Thus, the statistical operator (9) reproduces an operation of a reduction applied to a given object. It also practically reproduces an operation of the reduction applied to a device ("practically" in the sense that it is a question about a "macroscopic" observable variable). This observable variable does not distinguish the different quantum states of the device that correspond to the same macroscopic description, i.e., matrix elements of this observable variable correspond to the states $|\psi_i\rangle|\alpha_{i,s}\rangle$ and $|\psi_j\rangle|\alpha_{j,s}\rangle$, which are not dependent on $s$. The average value of this macroscopic observable variable $A$ is equivalent

$$tr(\rho A) = \sum_{(s,i,j)} p_s c_i c_j^* e^{i(\theta i,s - \theta j,s)}\langle\alpha_{j,s}|\langle\psi_j|A|\psi_i\rangle|\alpha_{i,s}\rangle =$$
$$= \sum_{(i,j)} c_i c_j^* a_{i,j}\sum_s p_s e^{i(\theta i,s - \theta j,s)} \tag{10}$$

As phases $\theta_{i,s}$ are distributed randomly, the sum over s is zero at i≠j; thus,

$$tr(\rho A) = \sum |c_i|^2 a_{ii} = tr(\rho'A). \tag{11}$$

where

$$\rho' = \sum |c_i|^2 p_s|\psi_i\rangle|\alpha_{i,s}\rangle\langle\alpha_{j,s}|\langle\psi_j| \tag{12}$$

We obtain a statistical operator that reproduces the operation of reduction on the device. If the device arrow is observed in position i, the device state for a certain s will be $|\alpha_{i,s}\rangle$. The probability of obtaining this state is equal to the probability of obtaining this state prior to measurement of its state at $|\alpha_{i,s}\rangle$. Thus, we propose the following theorem.

Theorem 2 regarding the decoherence of the macroscopic device. Assume that the quantum system interacts with the macroscopic device in such a manner that there is a chaotic distribution of device state phases. Let $\rho$ be a statistical operator of the device after the measurement, which is calculated with the help of Schrodinger equations, and let $\rho'$ be the statistical operator obtained as a result of reduction application to the operator $\rho$. It is impossible to yield such an experiment with the macroscopic device, which would register a difference between $\rho$ and $\rho'$.

For an extensive class of devices, it is proved that the chaotic characteristic of the distribution of phases formulated in theorem 2 occurs if the device evolves nonreversibly at measurement. It is the Daneri-Loinger-Prosperi theorem **[84]**.

# Appendix R. Zeno Paradox: The Theorem of a Continuously Observed Kettle that Does not Boil [18]

Theorem:

Let A be an observable variable of the quantum system, with eigenvalues 0 and 1. Assume that measurements of the observable variable A are yielded in instants $t_0 = 0, t_1..., t_N = T$ over the time interval [0, T] and that the reduction is applied after each such measurement. Let $p_n$ be the probability of that measurement during the moment $t_n$, which yields 0. If $N \to \infty$, then max $(p_{n+1} - p_n) \to 0$,

$$p_N - p_0 \to 0 \qquad (13)$$

(Thus, if the system was a value of observable variable A at instant $t = 0$, it will also have the same value of A at the instant $t = T$).

The proof:

Let $P_0$ be a projection operator on the characteristic space of observable variable A, which corresponds to an eigenvalue 0, and let $P_1 = 1 - P_0$ be the projection operator on the characteristic space with an eigenvalue 1. Let $\rho_n$ be the statistical operator that characterises the state of the system immediately prior to measurement at the instant $t_n$. The statistical operator after measurement is given by the expression

$$\rho_n' = P_0 \rho_n P_0 + P_1 \rho_n P_1 \qquad (14)$$

Thus, the statistical operator that characterises the state of the system state immediately prior to measurement at the instant $t_{n+1}$ is equivalent to

$$\rho_{n+1} = e^{-iHtn} \rho_n' e^{iHtn} \qquad (15)$$

where H is a system Hamiltonian and $\tau_n = t_{n+1} - t_n$. We note that if the operator $\rho_n$ is equivalent to the sum of k terms of type $| \psi> <\psi |$, the operator $\rho_n'$ would comprise the sum of no more than 2k such terms; as $\rho_n = | \psi_0> <\psi_0 |$, it follows that the operator $\rho_n$ is the sum of a finite number of these terms. According to (15), we obtain

$$\rho_{n+1} = \rho_n' - i\tau_n [H, \rho_n'] + O(\tau_n^2). \qquad (16)$$

As $P_0^2 = P_0$ and $P_0 P_1 = 0$, thus

$$P_0 \rho_{n+1} P_0 = P_0 \rho_n P_0 - i\tau_n [P_0 H P_0, P_0 \rho_n P_0] + O(\tau_n^2). \qquad 17)$$

Thus, the probability of that measurement at the instant $t_n$ yields 0, which is equivalent to

$$p_{n+1} = \text{tr} (\rho_{n+1} P_0) = \text{tr} (P_0 \rho_{n+1} P_0) =$$
$$= \text{tr} (P_0 \rho_n P_0) - i\tau_n \text{tr} [P_0 H P_0, P_0 \rho_n P_0] + O(\tau_n^2). \qquad (18)$$

The second equality is valid as $P_0^2 = P_0$. We consider that the operator $P_0 \rho_n P_0$ is equivalent to the sum of a finite number of terms of type $| \psi> <\psi |$ and for any operator X, the following relation is valid:

$$\text{tr} (X| \psi> <\psi |) = <\psi | X | \psi> = \text{tr} (| \psi> <\psi | X). \qquad (19)$$

Thus, the commutator track in (18) is equivalent to zero; therefore,

$$p_{n+1} = p_n + O(\tau_n^2). \qquad (20)$$

Let the maximum value $\tau_n$ be designated by $\tau$ ($\tau$ = max $\tau_n$); a constant value k exists such that

$$p_{n+1} - p_n \leq k\tau_n^2 \leq k\tau\tau_n \qquad (21)$$

Therefore,

$$p_N - p_0 = \sum_{n=0}^{N-1} (p_{n+1} - p_n) \leq k\tau \sum_{n=0}^{N-1} \tau_n = k\tau T \to 0$$

at $\tau \to 0$.

# Appendix S. Einstein-podolsky-rosen Paradox [18]

The point of view that difficulties considered in quantum mechanics are exclusively related by the quantum mechanics vector of a state is feasible. It does not provide complete information about a system state: other variables are obscure, which are named hidden variables. The values of the hidden variables completely characterize a state of system and predict its future behaviour in greater detail than quantum mechanics. A significant argument of existence of these additional hidden variables was advanced by Einstein, Podolsky and Rosen in 1935. We consider an electron and a positron pair, which were born simultaneously, in a state with the complete spin 0. This spin state represents an antisymmetric combination of spin states of two particles with a spin 1/2, i.e., it is expressed as

$$| \Psi \rangle = \frac{1}{\sqrt{2}} \left( |\uparrow\rangle |\downarrow\rangle - |\downarrow\rangle |\uparrow\rangle \right) \qquad (22)$$

where $| \uparrow \rangle$ and $| \downarrow \rangle$ denote the one-particle eigenstates of a component of a spin $s_z$ with eigenvalues +1/2 and -1/2, and the two-particle spin state (22) of an electron is denoted by the first factor.

As the state with the zero angular moment is invariant with respect to rotations, it should resemble (22), independent of axis direction. Thus, it is possible to also note that

$$| \Psi \rangle = \frac{1}{\sqrt{2}} \left( |\rightarrow\rangle |\leftarrow\rangle - |\leftarrow\rangle |\rightarrow\rangle \right) \qquad (23)$$

where $|\leftarrow\rangle$ and $|\rightarrow\rangle$ are the one-particle eigenstates of a component of the spin $s_x$.

Assume that an electron and a positron move in opposite directions and achieve a large distance between each other. The measurement of z component of the electron is yielded. Thus, the measured observable variable is $s_z(e^-)$ of the complete system; after such a measurement, the system state will be projected on a corresponding eigenstate of this observable variable: if measurement yields the value +1/2 after measurement, the system would transfer in a state $|\uparrow\rangle|\downarrow\rangle$, which indicates that the positron will be in a state $|\downarrow\rangle$, and the measured z-component of spin $s_z(e^+)$ will yield the value -1/2. We note that this information about the positron is obtained from the experiment performed on the electron. It is located a substantial distance from the positron and, consequently, cannot influence it. Einstein, Podolsky and Rosen concluded that the result of the experiment on the positron state (namely, that $s_z(e^+) = -1/2$) should be a real fact, which also occurred prior to the experiment with the electron.

Assume that the electron spin is measured not on the z-axis but on the x-axis. Based on (23), it follows that the system state will be projected either on the state $|\leftarrow\rangle|\rightarrow\rangle$ or on the state $|\rightarrow\rangle|\leftarrow\rangle$. Thus, the positron would possess a certain value for its x-component.

$s_x(e^+)$. This positron state should also exist prior to the last experiment. Thus, prior to the experiment, the positron possessed certain values: $s_z(e^+)$ and $s_x(e^+)$. However, they are incompatible observable variables and they do not contain simultaneous eigenstates: no such quantum mechanical state exists in which both contain specific values. Einstein, Podolsky and Rosen concluded that the description of quantum mechanics is incomplete and that "elements of realities", which do not consider quantum mechanics, exist. Consider the explanation of quantum mechanics of the EPR paradox. After the experiment on the electron, the complete system transfers to the eigenstate $|\uparrow\rangle|\downarrow\rangle$ if it was measured with a value of +1/2 for $s_z(e^-)$ or to the eigenstate $|\rightarrow\rangle|\leftarrow\rangle$ if it was measured with a value of +1/2 for $s_x(e^-)$. This finding indicates that after the experiment on the positron is in a certain state, either $|\downarrow\rangle$ or $|\leftarrow\rangle$, and this state differs from the positron state prior to the experiment. However, it does not mean that the positron state has been changed by the experiment on the electron because the positron did not exhibit a specific state prior to the experiment. If the positron is featured separately, it is necessary to converted it into its statistical operator. According to (22) and (23), this operator (prior to the experiment) is expressed as

$$\rho_{r} = tr_{e'} |\Psi\rangle\langle\Psi| = \frac{1}{2}\left(|\uparrow\rangle\langle\uparrow| + |\downarrow\rangle\langle\downarrow|\right) = \quad (24)$$

$$= \frac{1}{2}\left(|\rightarrow\rangle\langle\rightarrow| + |\leftarrow\rangle\langle\leftarrow|\right) \quad (25)$$

i.e., it is equal to a unit operator in a two-dimensional positron spin space multiplied by 1/2. We consider a statistical operator of a positron immediately after the experiment but prior to the information about its result can reach the positron. If the component measured in the experiment with the electron was $s_z$, the positron state would be $|\uparrow\rangle$ or $|\downarrow\rangle$ with equal probability, and the statistical operator would exist (24). If the component measured in the experiment with the electron was $s_x$, the positron state would be $|\rightarrow\rangle$ or $|\leftarrow\rangle$ with equal probability, and the statistical operator would exist (25), i.e., the same as in the previous case and prior to the experiment with an electron. Although these three situations (prior to the experiment, after the experiment on measurement of $s_z$ and after the experiment on measurement of $s_x$) are featured differently regarding the use of positron states, all correspond to the same statistical operator, and no experimentally distinct difference between them exists. Thus, there is no experimentally distinct interaction between the electron and the far positron, i.e., the EPR experiment cannot be used for information transfer with velocity more than light velocity.

# Appendix T. Bell's Inequality [18]

We show that the instantaneous interaction is inevitable for any theory with hidden variables, which produces the same consequences as for quantum mechanics.

Consider a situation in which the experiments are conducted on two particles that are separated in space; we deduce the consequences from an assumption in which the experimental results for one of the particles are dependent on this experiment. They are not dependent on the experimental results for other particles. This property is named *locality*. The locality requirement results for these restrictions on the correlations between the experimental results for different particles, which contradict predictions of quantum mechanics, are discussed in the subsequent section.

No connection exists between a locality and determinism. Assume that probabilities are defined by a set of variables. We designate this set by the symbol $\lambda$ (in the case of two particles separated in space, these variables consist of variables that individually represent both particles and the variables that represent general devices, which simultaneously influence both particles). For each experiment $E$, it is possible to specify the probability $p_E(\alpha\,|\,\lambda)$ of measurement of $\alpha$ when variables have the value $\lambda$. If experiments $E$ and $F$, which are separated in space, are independent in the sense of probability theory, the theory will be *local*. We conclude that

$$p_{E\oplus F}(\alpha\oplus\beta\,/\,\lambda) = p_E(\alpha\,/\,\lambda)p_F(\beta\,/\,\lambda) \quad (26)$$

Any local theory that reproduces all predictions of quantum mechanics concerning the EPR experiment for two separated particles with a spin 1/2 will be equivalent to the deterministic theory. Assume that an electron and a positron are separated by a large distance. In experiment $E$, let the component of electron spin be measured in a certain direction and let the component of positron spin be measured in the same direction in experiment $F$. We denote two possible observed dates by the arrows ↑ and ↓. As the complete spin is equivalent to zero, we know that the experiments $E$ and $F$ will always yield opposite results according to probability theory:

$$p_{E \oplus F}(\uparrow \oplus \uparrow) = p_{E \oplus F}(\downarrow \oplus \downarrow) = 0 \qquad (27)$$

Let $\rho(\lambda)$ be the probability density that characterises the probability that variables have the values $\lambda$; the composite probability (27) is equivalent to

$$p_{E \oplus F}(\uparrow \oplus \uparrow) = \int p_{E \oplus F}(\uparrow \oplus \uparrow)\rho(\lambda)d\lambda =$$
$$= \int p_E(\uparrow | \lambda) p_F(\uparrow | \lambda)\rho(\lambda)d\lambda \qquad (28)$$

As the complete probability is equivalent to zero, an integrand, which is nonnegative, should be zero everywhere. Thus,

either $\rho(\lambda) = 0$ or $p_E(\uparrow | \lambda) = 0$ or $p_F(\uparrow | \lambda) = 0$. (29)

Similarly, we conclude that

either $\rho(\lambda) = 0$ or $p_E(\downarrow | \lambda) = 0$ or $p_F(\downarrow | \lambda) = 0$ (30)

As experiment $E$ has only two results, ↓ and ↑, we obtain the equivalent statements

$$p_E(\uparrow | \lambda) = 0 \Leftrightarrow p_E(\downarrow | \lambda) = 1 \qquad (31)$$

Based on (29) – (31), if $\rho(\lambda) \neq 0$, all four probabilities should be equivalent to either 0 or 1. Thus, for all values of $\lambda$ that are possible, the experimental results are completely defined by the value $\lambda$.

Thus, if we assume that the probability distribution of hidden variables is not influenced by the type of the experiment conducted on the particles, we can conclude that only deterministic theories should be considered.

Assume that each of two separated particles can be subjected by one of three experiments $A$, $B$, $C$, each of which can yield only two results ("yes" or "no"). In the deterministic local theory, the results of experiment $A$ with a particle 1 is defined by the property of system, which we designate $a_1$: it is a variable that can assume the values + and -. We also employ similar variables $b_1$, $c_1$, $a_2$, $b_2$, and $c_2$. We assume that experiment $A$ always yields opposite values for the two particles, $a_1 = -a_2$. We similarly assume that experiments $B$ and $C$ also yield opposite results for both particles, i.e., $b_1 = -b_2$ and $c_1 = -c_2$.

Consider particles that are prepared with the fixed probability of value sets $a$, $b$ and $c$.

Assume $P(a = 1, b = 1)$ denotes the probability that the particle has the specified values $a$ and $b$. Then

$P(b = 1, c = -1) = P(a = 1, b = 1, c = -1) + P(a = -1, b = 1, c = -1)$
$\qquad \leq P(a = 1, b = 1) + P(a = -1, c = -1)$ (32)

Thus, when pairs of particles are prepared with opposite values $a$, $b$ and $c$, we obtain

$P(b_1 = 1, c_2 = -1) \leq P(a_1 = 1, b_2 = -1) + P(a_1 = -1, c_2 = 1)$. (33)

Each item in the right side of this inequality yields the probability of the experimental results for various particles; therefore, the inequality can be verified even for the case in which A, B, C experiments cannot be conducted simultaneously for a single particle.

The probabilities calculated according to the rules of quantum mechanics in the following case do not satisfy inequality (33). We assume that two particles (an electron and a positron) with spin 1/2 are prepared in state with a complete spin that is equivalent to 0; we know that the measurement of a component of a spin in any given direction will yield opposite results for both particles. Let $A$, $B$, and $C$ denote experiments on measurement of the components of a spin along three axes laying in one plane, let the angle between axes $A$ and $B$ be equivalent to $\theta$, and the angle between axes $B$ and $C$ equivalent to $\varphi$. We calculate the probability $P(b_1 = 1, c_2 = 1)$, which is introduced in the left side of the inequality (33); it should be interpreted as the probability that both of the measurements of the components of the spins of particles 1 and 2 along axes in which the angle between the axes is equivalent to 0 will yield the same outcome: +1/2. We assume axis z as the axis for particle 1. At the measurement of a component of a spin of particle 1 along the specified axis, if we obtain a value of 1/2 after measurement, particle 1 will transfer in an eigenstate $| \uparrow \rangle$, and particle 2 – will transfer in an eigenstate $| \downarrow \rangle$. The eigenstates of the measurement yielded over particle 2 are obtained by rotational displacement of states $| \uparrow \rangle$ and $| \downarrow \rangle$ on an angle $\varphi$ (around an axis x); thus, the eigenstate that corresponds to an eigenvalue + 1/2 is expressed as follows:

$$| +(\varphi) \rangle = e^{-i\varphi J_x} | \uparrow \rangle = [\cos(\tfrac{1}{2}\varphi) + 2iJ_x \sin(\tfrac{1}{2}\varphi)] | \uparrow \rangle =$$
$$= \cos(\tfrac{1}{2}\varphi) | \uparrow \rangle + i \sin(\tfrac{1}{2}\varphi) | \downarrow \rangle \qquad (34)$$

The required probability is equivalent to

$P(b_1 = 1, c_2 = 1) = \tfrac{1}{2} | \langle +(\varphi) | \downarrow \rangle |^2 = \tfrac{1}{2} \sin^2(\tfrac{1}{2}\varphi)$ (35)

(As the probability of +1/2 for a particle 1 is equivalent to 1/2). Similarly, it is possible to calculate the probabilities using (33)

$$P(a_1 = 1, b_2 = -1) = \tfrac{1}{2} \cos^2(\tfrac{1}{2}\theta)$$

and

$$P(a_1 = -1, c_2 = 1) = \tfrac{1}{2}\cos^2[\tfrac{1}{2}(\theta + \varphi)].$$

Thus, the inequality (33) is reduced to the inequality

$$\sin^2(\tfrac{1}{2}\varphi) \leq \cos^2(\tfrac{1}{2}\theta) + \cos^2[\tfrac{1}{2}(\theta + \varphi)],$$

or to the inequality

$$\cos \theta + \cos \varphi + \cos(\theta + \varphi) \geq -1, \qquad (36)$$

which are not satisfied for $\theta = \varphi = 3\pi/4$. As a result, we discuss the following theorem.

Theorem 3 (Bell's theorem). Assume that two separated particles can be subjected to one of three two-valued experiments. The same experiment conducted for both particles always yields opposite results. If the particles are represented by local theory and the experiments do not influence the particle property probability distributions, then experimental result probabilities satisfies inequality (33).

This inequality is not satisfied in quantum mechanics for a system of two particles with a spin of 1/2, which contains a complete spin of 0.

# Appendix U. De Broglie - Bohm Theory of Wave-Pilot [18]

Typical features of quantum mechanics (in particular, the interference effects) make the building-up theory of hidden variables a difficult problem. Previously, it was (J. von Neumann) [15] seemingly proved that no theory of this type can reproduce all of the consequences of quantum mechanics. However, this proof seemed erroneous, as the following counterexample shows. We consider a separate simple particle, which moves in the potential of V (*r*). We assume that the particle is represented at the instant $t$ not only as the wave function $\psi$ (*r, t*) but also by the vector q (t) and that the wave function satisfies typical Schrodinger equations

$$i\hbar \frac{\partial \Psi}{\partial t} = -\frac{\hbar^2}{2m}\Delta\Psi + U(x,y,z)\Psi. \qquad (37)$$

and that the vector q satisfies the equation

$$\frac{dq}{dt} = \frac{j(q,t)}{\rho(q,t)}$$

where j and ρ are the density of the probability current and the probability density, respectively, as follows:

$$j = \frac{\hbar}{m}\text{Im}[\bar{\psi}\nabla\psi], \qquad \rho = |\psi|^2. \qquad (38)$$

Assume that in the instant $t = 0$, numerous particles exist, each of which is represented by the same wave function $\psi$ (*r, 0*) but by a different vector q.
Let the portion of particles, for which the value of this vector is in volume dV, contain a point q, which is equivalent to σ(q, 0)dV. Let this portion in instant $t$ be equivalent to σ (q, *t*) *dV*. Considering q as the particle coordinate, it is possible to consider all collective particles as a fluid with density σ and a field of velocities u = j/ρ, according to (37). The last values should satisfy the equation of continuity

$$\frac{\partial \sigma}{\partial t} + \nabla \cdot (\sigma u) = 0 \qquad (39)$$

i.e.,

$$\frac{\partial \sigma}{\partial t} = -\nabla \cdot \left(\frac{\sigma j}{\rho}\right) \qquad (40)$$

The equation of continuity has the single solution $\sigma$ (*r, t*) for a given $\sigma$ (*r, 0*), j (*r, t*) and $\rho(r, t)$. This equation has the solution $\sigma = \rho$. Thus, equation (40) is an equation of continuity. This equation is a consequence of Schrodinger's equations (37). Thus, if the distribution of the values $q$ for particles is represented by function $\rho$ at $t = 0$, it would be characterised by this function at all future instants.

Thus, we can conclude that each particle with a wave function that conforms with Schrodinger's equations (37) has a certain coordinate q in space and that any of our experimental devices, which create particles with the wave function $\psi$, yield particles with a certain distribution of their coordinates; the portion of these particles $| \psi (q) |^2 dV$ is in volume dV near point q. This formulated statement is valid if the experimental device that creates particles with the wave function $\psi$, yields particles in volume $dV$ with probability $| \psi (q) |^2 dV$. As values $\psi$ and q evolve over time according to the deterministic equations (37) and (38), correspondent distribution will be accurate for all instants if it was accurate during the initial moment.

It is possible to also extend this theory to systems from several particles, but one distinct difficulty exists. We consider, for example, a system of two particles. The variables are denoted by $q_1$ and $q_2$, and the two-particle wave function is expressed by $\psi(r_1, r_2)$. As equations of motion, it is necessary to consider a two-particle Schrodinger equation and two equations

$$\frac{dq_1}{dt} = \frac{j_1}{\rho}, \qquad \frac{dq_2}{dt} = \frac{j_2}{\rho},$$

where

$$j = \frac{\hbar}{m}\text{Im}[\bar{\psi}\nabla_1\psi], \quad j = \frac{\hbar}{m}\text{Im}[\bar{\psi}\nabla_2\psi], \rho = |\psi|^2.$$

In this equation, $j_1$ and, consequently, $dq_1/dt$ can be expressed as function $q_2$: the motion of the first particle is dependent on the position of the second particle. Thus, an instantaneous interaction between the two particles exists and when no potential $V (r_1, r_2)$ exists for the interaction between particles. This finding is the result of correlations between the particles, which occurs in the use of quantum mechanics operating with wave functions. In particular, the wave EPR function shows these correlations between separated particles.

# Appendix V. Escher Swirl in Which all Levels are Crossed [3]

Extremely beautiful and at the same time strange disturbing illustration of a cyclone "eye" generated by Entangled Hierarchy is given by Escher in his "Picture gallery" (http://im-possible.info/english/articles/escher_printgallery/).
'The tightest of all strange loops,' says Hofstadter, 'is realized in Print Gallery: a picture of a picture which contains itself. Or is it a picture of a gallery which contains itself? Or of a town which contains itself? Or a young man who contains himself?' "A strikingly beautiful, and yet at the same time disturbingly grotesque, illustration of the cyclonic 'eye' of a Tangled Hierarchy is given to us by Escher in his Print Gallery. What we see is a picture gallery where a young man is standing, looking at a picture of a ship in the harbor of a small town, perhaps a Maltese town, to guess from the architecture, with its little turrets, occasional cupolas, and flat stone roofs, upon one of which sits a boy, relaxing in the heat, while two floors below him a woman - perhaps his mother - gazes out of the window from her apartment which sits directly above a picture gallery where a young man is standing, looking at a picture of a ship in the harbor of a small town, perhaps a Maltese town - What!? We are back on the same level as we began, though all logic dictates that we cannot be. ([3], page 715)"

"... Now are we, the observers of PRINT GALLERY, also sucked into ourselves by virtue of looking at it? Not really. We manage to escape that particular vortex by being outside of the system. And when we look at the picture, we see things which the young man cannot see, such as Escher's signature, "MCE", in the central 'blemish'. Though the blemish seems like a defect, perhaps that defect lies in our expectation, for in fact Escher could not have completed that portion of the picture without being inconsistent with the rules by which he was drawing the picture. That center of the whorl is - and must be - incomplete. Escher could have made it arbitrarily small, but he could not have gotten rid of it. ([3], pages 716-717)"

## Acknowledgments

I thank the painter Gukov Yury Yurjevich for his help in drawing the figures.

# Основные парадоксы статистической классической физики и квантовой механики.

**Купервассер Олег,**

**Московский Государственный Университет**

**E-mail:**
**olegkup@yahoo.com**
**2009 год**

## Аннотация.

Статистическая классическая механика и квантовая механика - это две разработанные и хорошо известные теории. Они представляют собой основу современной физики. Статистическая классическая механика получает свойства больших  тел, исследуя движения мельчайших атомов и молекул, из которых эти тела состоят, используя классические законы Ньютона. Квантовая механика определяет законы движения мельчайших частиц на малых атомных расстояниях, рассматривая их как волны вероятности. Законы квантовой механики описываются уравнением Шредингера. Законы такого движения очень отличаются от законов движения крупных тел, таких как планеты или камни. Описанные две теории давно известны и хорошо изучены. Тем не менее, они содержат ряд парадоксов. Это заставляют многих ученых усомниться во внутренней непротиворечивости этих теорий. **[9]**, **[35]**. Однако данные парадоксы могут быть разрешены в рамках существующей физики, без введения каких либо новых законов. Чтобы сделать изложение понятным даже для неподготовленного читателя, мы вводим в этой статье некоторые необходимые основные понятия статистической физики и квантовой механики без использования формул. Необходимые точные формулы и разъяснения к ним могут быть найдены в Приложениях. Текст снабжен иллюстрациями для лучшего понимания текста. В дальнейшем в статье обсуждаются парадоксы, лежащие в основе термодинамики и квантовой механики. Дано их разрешение с точки зрения влияния внешнего наблюдателя (окружающей среды), разрушающего **дополнительные неустойчивые микроскопические** корреляции между **микроскопическими параметрами** системы, или **ограниченности самопознания** системы (в смысле полного описанная ее состояния и предсказания ее будущего) в случае, когда и наблюдатель, и окружающая среда включены в систему. Введены понятия **Наблюдаемая Динамика**, **Идеальная Динамика** и **Непредсказуемая  Динамика**. Дано рассмотрение явления в сложных системах (в живых системах) с точки зрения этих Динамик.

# Оглавление.

# Введение.

В самом начале этой статьи следует сделать несколько крайне важных замечаний.

1) Эта статья **не является** философской работой на тему основ физики, как некоторые другие работы посвященные парадоксам квантовой механики. Мы рассмариванием разрешение парадоксов физики научными методами и обсуждаем, как можно построить физику, исключающую эти парадоксы, и условия, при которых это возможно. Не понимание основ физики, приводящее к этим парадоксам, приводит к ряду вполне физических, а отнюдь не философских ошибок.

2) Эта работа **не является** попыткой дать очередную новую интерпретацию квантовой механики. Все интерпретации (например, многомировая интерпретация, Копенгагенская и т.д.) лишь пытаются дать более или менее наглядное толкование квантовой механики, не разрешая её парадоксов и не внося ничего нового в саму физику явления. Автор считает все существующие разумные интерпретации допустимыми и дает разрешение парадоксов не связанное с той или иной конкретной интерпретацией, а основанное на самой физике явления.

3) Эта работа **не является** научно-популярной статьей и включает в себя новые, оригинальные идеи. Мы обращаемся в этой статье к очень широкой аудитории, включающей биологов, физиков из самых различных областей этой науки (квантовая механика, статистическая физика, термодинамика и нелинейная динамика), специалистов по компьютерным наукам. Поэтому мы дали популярный обзор основ физики, который хотя может показаться тривиальным для одних специалистов, но, тем не менее, будет очень полезен для других специалистов (в том числе и для физиков из различных областей физики). Автор считает **крайне важным** сохранить эту часть статьи. Кроме того, в тексте нет формул, а лишь поясняющие текст рисунки. Все формулы вынесены в Приложения и необязательны для понимания текста. Следует отметить, что автор не является пионером подобного стиля изложения. Примером служат книги Пенроуза [44], [45] , Хофштадтера [58] Менского [46], Ликата **[85]** которые не являются научно-популярными, несмотря на видимый «легкий» стиль изложения. Автор является сторонником именно такого стиля изложения. Хотя автор и не включает себя сам в число этих очень выдающихся ученых, он крайне надеется, что и ему позволят последовать их заразительному примеру.

4) Эта статья **не является** просто обзором уже сделанных работ (хотя этому уделено много места), но включает в себя и оригинальные идеи автора.

5) Автор **не пытается** открыть новых законов физики[1]. Все рассмотрение идет в рамках уже существующей физики. Мотивацией для написания этой статьи послужил тот факт (и также парадокс!), что автор не встретил **ни одной** статьи или учебника физики, где дается исчерпывающее и полное объяснение этих парадоксов физики (Рис. 1)и следствий из них вытекающих. Более того, во многих работах они или обходятся молчанием, или утверждается их «неважность» для «практической» физики. В других работах их изложение или не полно (например, исчерпывается только лишь одной декогеренцией) или вообще не ясно. Во многих работах их разрешение пытаются увязать неразрывно только с той или иной конкретной интерпретацией физики (как правило, многомировой) или изобрести новые законы физики для их объяснения.

---

[1] Пайерлс **[7],** Менский **[46]** считают, что для разрешения парадокса измерения в квантовой механике необходимо изменение законов квантовой физики путем введения в нее понятие «сознания». Пенроуз **[44], [45]** , Легет **[10]** считают, что законы квантовой механики нарушаются для достаточно больших макроскопических тел. Однако без введения новых законов уже были успешно разрешены многие другие проблемы физики. Такие как, например, парадокс Гиббса **[4]** или интерпретация спина как собственного момента вращения Дираковского электрона **[5]**. Иногда нарушение симметрии Жизни или Вселенной (таких как симметрия направления времени или правого и левого) видят в фундаментальном слабом взаимодействии. Примером может служить работа Элицура **[59]**. Слабое взаимодействие на самом деле нарушает эти симметрии. Однако в данной работе мы пренебрежем этими малыми эффектами и будем искать иные причины асимметрии.

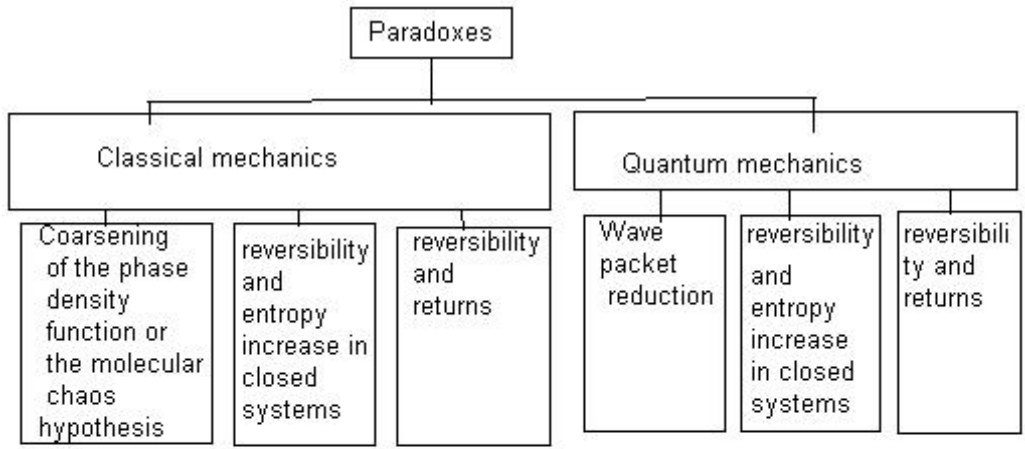

Figure1 Paradoxes in Classical and Quantum mechanics

# 1. Главные парадоксы классической статистической физики.

## 1.1 Макроскопические и микроскопические параметры физических систем. [1],[2].

Начнем со статистической физики. Посмотрим на течение газа из сопла ракетного двигателя (Рис. 2).

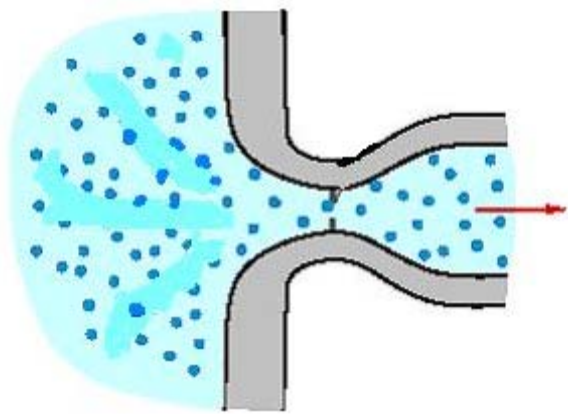

**Рис2**. Истечение газа из сопла. Показаны, в увеличенном размере, на самом деле невидимые невооруженным глазом молекулы газа.

Мы увидим распределение плотности и скорости вытекающего газа, но лишь для больших объемов. Эти объемы включают огромное число невидимых молекул. Подобные легко наблюдаемые распределения скорости и плотности газа в пространстве называются **макроскопическими параметрами** системы. Они дают неполное, частичное описание системы. Полный набор ее параметров определяется скоростями и позицией всех молекул газа. Такие параметры называется **микроскопическими параметрами**. Сам вытекающий газ называется наблюдаемой **системой.** Система называется **замкнутой**, если она не взаимодействует с окружением. **Внутренней энергией** системы называется суммарная энергия всех ее молекул.

   **В дальнейшем, если не оговорено иное, мы будем рассматривать замкнутые системы с заданной внутренней энергией и конечным объемом.**

## 1.2 Фазовое пространство и фазовые траектории.[1],[2]

   Введем многомерное пространство. Осями этого пространства будут координаты и скорости всех молекул системы. Тогда сама система будет изображаться точкой этого пространства. Положение этой точки будет давать полное микроскопическое описание системы. Подобное пространство называется **фазовым пространством** системы. Изменение состояния системы описывается движением точки в этом пространстве и называется **фазовой траекторией** (Рис. 3).

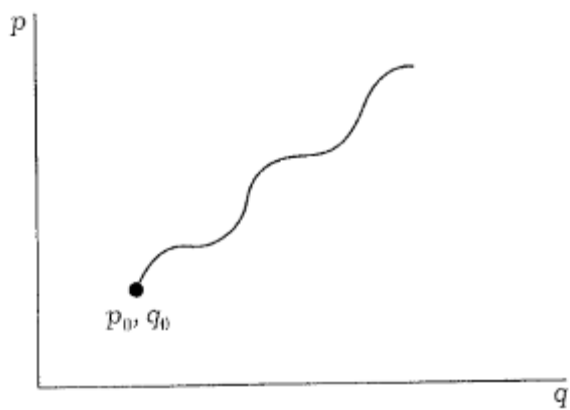

**Figure 3.** Траектория в фазовом пространстве. Динамическое состояние представлено точкой в фазовом пространстве p, q. Эволюцию во времени олисывает траектория, которая исходит из начальной точки $p_0$, $q_0$. (Рис. из [17])

 Пусть известны только макроскопические параметры, а микроскопические параметры неизвестны. Тогда систему можно описать непрерывным набором точек, отвечающим этим макроскопическим параметрам. Это **фазовый объем («облако») системы** или, иначе, **ансамбль Гиббса** (Рис. 4). Все точки этого объема равновероятны и соответствуют разным микроскопическим, но одинаковым макроскопическим параметрам. (Смотри **Приложение А** ) [1], [2]

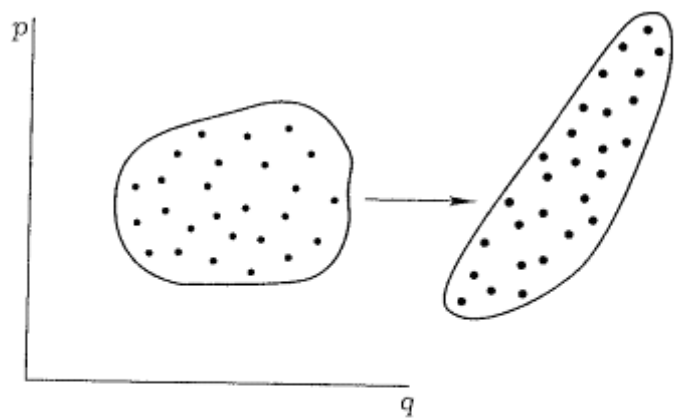

**Figure 4.** Ансамбли в фазовом пространстве. Ансамбль Гиббса представлен облаком частиц, отличающихся начальными условиями. Форма облака изменяется со временем. (Рис. из [17])

Для каждого набора макроскопических параметров (**макроскопическое состояние**) можно найти бесконечное число соответствующих ему наборов микроскопических параметров (**микроскопических состояний**). Для того чтобы сделать это число конечным, разобьем фазовое пространство на отдельные очень малые ячейки. Такой метод называют **дискретизацией** непрерывного пространства. При таком рассмотрении систему с конечным объемом и заданной внутренней энергией можно описать очень большим, но **конечным** числом состояний. Каждому макроскопическому состоянию теперь соответствует большой, но конечный набор микроскопических состояний. Для большинства систем подавляющее число из всех их возможных микроскопических состояний относиться к одному главному макроскопическому состоянию. Оно называется **термодинамически равновесным состоянием**. Например, для газа в заданном объеме, оно соответствует равномерному распределению этого газа по объему. (Смотри **Приложение Е.**) [1], [2]

## 1.3 Эргодичность и перемешивание. [34], [35], [9]

Большинство реальных систем обладает свойством **эргодичности** [9], [35]: почти любая фазовая траектория должна с течением времени побывать во всех микросостояниях, возможных в системе. (Мы здесь имеем в виду микросостояния после дискретизации). В каждом микросостоянии система находиться примерное равное время. Эргодические системы обладают замечательным свойством. Возьмем среднее по времени значение любого макропараметра для траектории. Оно будет одинаково для всех траекторий и совпадает со средним по ансамблю систем, описывающему термодинамическое равновесие.

Подавляющее большинство реальных систем обладает более сильным свойством **хаотичности или перемешивания** (так называемая **теорема КАМ) [9], [35]:** в окрестности любой начальной точки фазового пространства всегда существует другая точка, такая, что фазовые траектории, выходящие из этих двух точек, расходятся экспоненциально быстро **[34], [9]**.

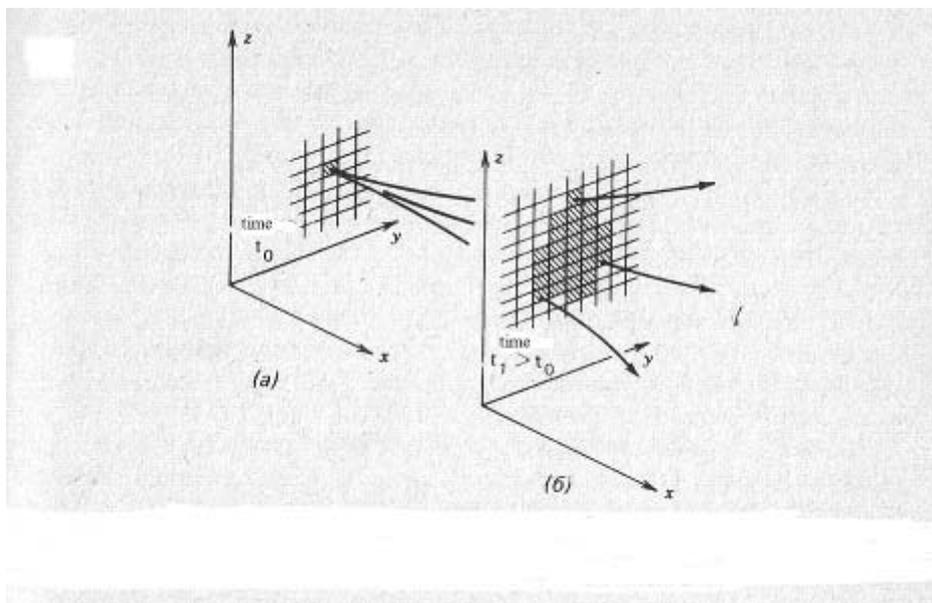

**Figure 5.** Иллюстрация увеличения неопределенности, или потери информации, в динамической системе. Заштрихованный квадрат в момент времени $t=t_0$ показывает неопределенность знания начальных условий. (Рис. из [12])

Экспоненциальная быстрота определяется следующим образом: если за 1 секунду траектории расходятся в два раза от начальной величины, то за следующую секунду они разойдутся уже в 4 раза от начальной величины, ещё за следующую секунду в 8 раз от начальной величины и т.д. Это очень быстрый тип расходимости (Рис. 5). Системы, обладающие свойством перемешивания, всегда являются эргодическими (Рис. 6,7,8).

## 1.4 Обратимость и теорема Пуанкаре.

Изменение микросостояний **обратимо**. Для каждой фазовой **траектории** существует обратная ей траектория, полученная изменением всех скоростей молекул на противоположные значения. Это эквивалентно прокручиванию пленки фильма, снятого о системе, в обратном направлении.

Подавляющая часть фазовых траекторий спустя возможно очень большое время должно вернуться в исходное микросостояние, из которого оно вышло. Это утверждение называется **теоремой Пуанкаре** о возвратах. (Смотри **Приложение С.) [2]** Большинство реальных систем являются хаотическими и обладают неустойчивостью и быстрой расходимостью фазовых траекторий из первоначально близлежащих микросостояний. Поэтому у таких систем время возврата неодинаково и сильно зависит от точного положения начальной точки фазового пространства внутри ячейки, определяющей ее начальное микросостояние. Но для очень узкого класса так называемых **интегрируемых систем** это время возврата примерно одинаково для всех точек начальной фазовой ячейки и эти возвраты происходят периодически или почти периодически.

## 1.5 Энтропия. [1], [2], [29]

Введем базисное для статистической механики понятие **макроскопической энтропии**. Пусть некоторому макроскопическому состоянию соответствует 16 микросостояний. Какое минимальное число вопросов, допускающих только ответы «да» или «нет», нужно задать, чтобы выяснить, в каком из микроскопических состояний находиться система? Если последовательно спрашивать о каждом микросостоянии может потребоваться до 15 вопросов. Но можно поступить и хитрее. Разобьем все микроскопические состояния на две группы, по 8 микросостояний в каждой группе. Первый вопрос будет, к какой группе относиться искомое микросостояние? Затем, указанную группу разобьем на две подгруппы по 4 микросостояния в каждой и зададим тот же вопрос. Продолжим эту процедуру до получения группы из одного искомого микросостояния. Нетрудно подсчитать, что потребуется лишь четыре таких вопроса. Это и будет минимальное число вопросов. Оно носит название **макроскопической энтропии. [1], [2], [29]** ( Смотри **Приложение B )** Ясно, что чем больше микросостояний соответствует макроскопическому состоянию, тем выше его энтропия. Соответственно, термодинамически равновесное состояние имеет максимальную энтропию. Часто говорят, что энтропия является мерой беспорядка. Это естественно. Чем больше беспорядка в системе, тем больше вопросов нужно задать, чтобы понять, в каком точно состоянии система находиться. Зачем нам нужно вводить эту «заумную» энтропию? Можно ведь просто использовать число микросостояний! Но энтропия обладает замечательным свойством. Пусть мы имеем систему, состоящую из двух несвязанных подсистем. Энтропия полной системы тем сумма энтропий двух её подсистем. Действительно, ведь это общее число вопросов об их состояниях, которые просто складываются. А числа микросостояний перемножаются! Суммировать же всегда проще, чем перемножать.

Статистическая механика формулирует несколько важных свойств физических систем: Пусть начальное макроскопическое состояние описывается неким фазовым объемом. Существует теорема, доказывающая, что при обратимом Ньютоновском процессе эволюции системы этот фазовый объем сохраняется (Смотри **Приложение F .) [2]**. Соответственно сохраняется и соответствующее ему число микросостояний. Энтропия, соответствующая этому набору состояний, называется **энтропией ансамбля** систем. Из свойства сохранения фазового объема следует, что энтропия ансамбля постоянна во времени.

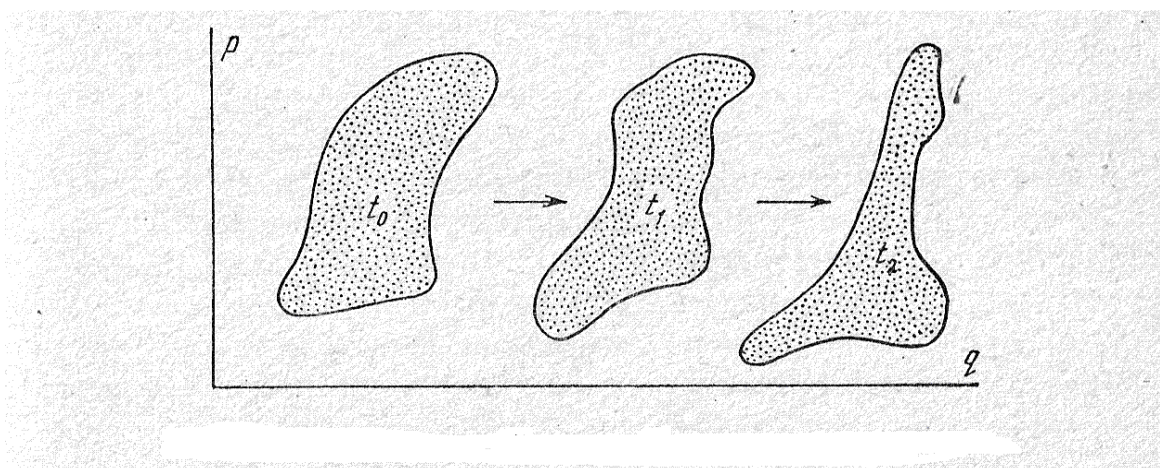

**Figure 6.** Сохранение объема в фазовом пространстве. (Рис. из [14])

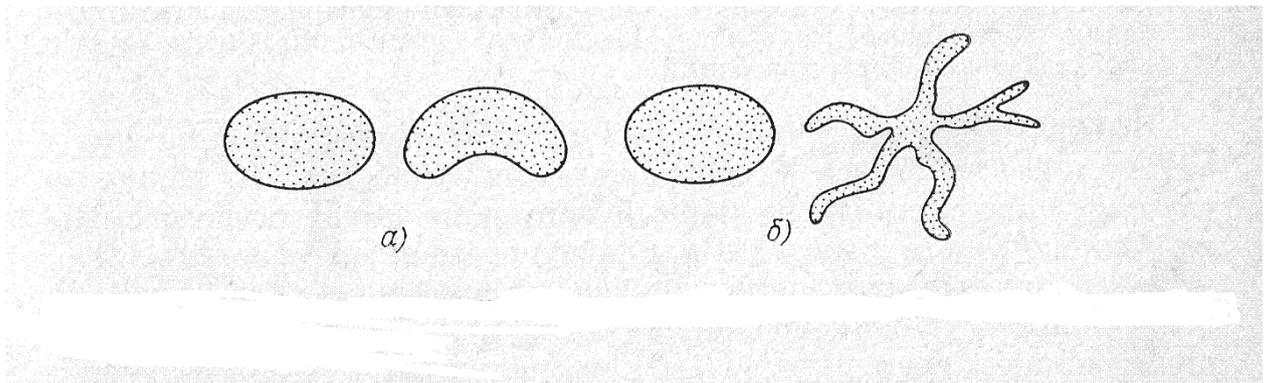

**Figure 7.** Изменение элемента фазового объема в устойчивом a) и неустойчивом b) случаях. (Рис. из [13])

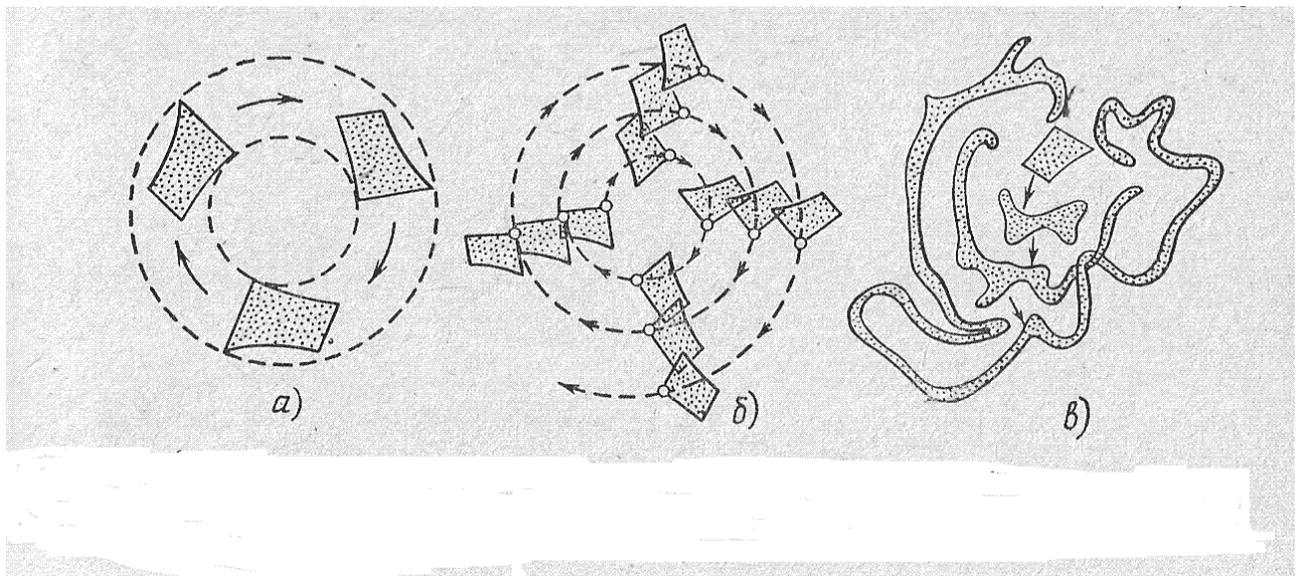

**Figure 8.** Различные типы потоков в фазовых пространствах: а) неэргодический поток b) эргодический поток без перемешивания c) поток с перемешиванием. (Рис. из [14])

## 1.6 Эволюция макроскопической энтропии для хаотических систем.

Из свойств эргодичности следует, что система, изображаемая точкой в фазовом пространстве, из почти любого первоначального состояния спустя некоторое время перейдёт в термодинамически равновесное состояние, и будет находиться в этом состоянии *большинство* времени. Это связано с тем, что большинство микросостояний системы, через которые она проходит в процессе эволюции, отвечает термодинамически равновесноому состоянию.

Термодинамически равновесное состояние обладает максимальной макроскопической энтропией. Даже если в начальном состоянии макроскопическая энтропия была мала, при переходе к термодинамическому равновесию она очень сильно возрастает. Это свойство противоположно свойству энтропии ансамбля, которая остается постоянной. Эта разница связанна с тем, что энтропия ансамбля определяется начальным набором микросостояний. В то время как макроскопическая энтропия определяется набором микросостояний, которые соответствуют текущему макроскопическому состоянию. Для термодинамического равновесия это число микросостояний очень велико.

Для хаотических систем (систем с перемешиванием) верна следующая теорема:

**Процессы изменения** *макропараметров*, **протекающие с уменьшением макроскопической энтропии, сильно неустойчивы по отношению к малому внешнему шуму. При этом процессы изменения** *макропараметров*, **связанные с ростом макроскопической энтропии, являются наоборот устойчивыми**[2].

Докажем это. Рассмотрим вначале процесс с ростом энтропии. Пусть начальное состояние системы описывается неким макроскопическим состоянием, далеким от термодинамического равновесия. Такое состояние характеризуется компактным (замкнутым и ограниченным) и выпуклым (содержащим отрезок, соединяющий любые две его точки) фазовым объемом. Поскольку систем хаотическая, в окрестности каждой точки будет и другая, с которой она экспоненциально удаляется. Из-за сохранения фазового объема (Смотри **Приложение F .**) в окрестности каждой фазовой точки всегда будет и другая, с которой она экспоненциально сближается, а не только удаляется. В результате, из малого компактного начального фазового объема образуется размазанный по всему фазовому пространству объем, который является уже не выпуклым множеством. Он обладает огромным количеством «рукавов» или «ветвей». При этом сам полный объем фазовой «капли» сохраняется. «Рукава» экспоненциально расширяются вдоль их длины и экспоненциально сжимаются вдоль их ширины. С течением времени число «рукавов» или «ветвей» растет, они причудливо изгибаются и покрывают своей «сетью» весь фазовый объем. Этот процесс называют **растеканием фазовой «капли»** [9], [35]. Пусть малый внешний шум выбросил фазовую точку за пределы «рукава» фазовой капли. Но процесс сжатия идет перпендикулярно «рукаву» и фазовая точка будет приближаться к «рукаву», а не удаляться от него. Это означает, что процесс растекания фазовой капли устойчив к шуму.

К тому же, шум может влияния сильно на микросостояние, но не макросостояние. Макросостояние соответствует огромному числу микросостояний молекул. Хотя внешний шум может сильно изменить состояние каждой одиночной молекулы, но полный вклад всех молекул в макросостояние остается неизменным. Это связано с "законом больших чисел" в теории вероятности [56]. Большинство микросостояний, соответствующих некоторому текущему макросостоянию, приводят к росту энтропии, потому что вероятность такого развития намного больше. Когда фазовая капля растечется почти вдоль всей поверхности постоянной энергии, её макросотояние будет соответствовать просто обычному термодинамическому равновесию. При этом даже уже не малый шум уже не может заметно повлиять на ее макросостояние, поскольку подавляющее большинство микросостояний в системе отвечают именно этому состоянию.

Теперь рассмотрим обратный процесс, идущий с уменьшением энтропии. Начальное состояние определяется набором точек фазового пространства, полученных из конечного состояния прямого процесса (растекшейся «фазовой капли») путем обращением скоростей всех молекул**.** При обращении скоростей начальная форма растекшейся «фазовой капли» не меняется. Но направление сжатия из-за обращения скоростей уже не перпендикулярно, а параллельно её «веткам». Вместо растекания фазовой капли будет происходить её сжатие. Пусть малый внешний шум выбросил фазовую точку за пределы «рукава» фазовой капли. Но процесс расширения идет уже перпендикулярно «рукаву» и фазовая точка будет удаляться от него, а не приближаться. Это означает, что процесс сжатия фазовой капли неустойчив к шуму(Рис. 9).

---

[2] На самом деле, рассмотрим простой пример роста энтропии идеального газа, расширяющегося из малого объема ящика до полного заполнения этого ящика. Если процесс расширения устойчив к малому шуму, то обратный процесс сжатия легко предотвращается таким малым внешним шумом, рассеивающим молекулы.

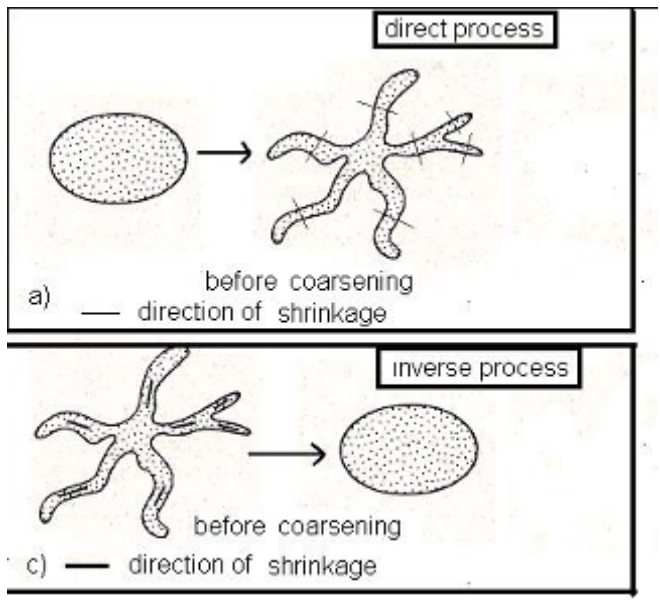

**Figure 9.** Direct process with macroscopic entropy increasing and its inverse process. Directions of shrinkage are denoted.

### 1.7 Второе начало термодинамики и связанные с ним парадоксы.

Теперь мы готовы определить второе начало термодинамики и вносимые им парадоксы. **Второе начало гласит: в замкнутых системах конечного объема макроскопическая энтропия не может уменьшаться, а может только лишь расти или не меняться. Со временем макроскопическая энтропия достигает максимума в состоянии термодинамического равновесия [1], [2].**

Главный парадокс состоит в противоречии этого закона **возрастания энтропии** с описанными выше базисными свойствами статистической физики. Действительно, из обратимости следует, что на каждый процесс с увеличением энтропии есть обратный процесс с таким же её уменьшением. Это **парадокс Лошмидта**. Кроме того, из теоремы Пуанкаре о возвратах следует, что рано или поздно система вернется в прежнее состояние и, следовательно, и ее энтропия также вернётся к начальному значению! Это **парадокс Пуанкаре.**

С этими двумя парадоксами тесно связано понятие **корреляций** скоростей и положений молекул, составляющих физические тела.

### 1.8 Дополнительные микроскопические корреляции и их связь с парадоксами статистической физики.

**Корреляция** является количественной мерой зависимости переменных (В нашем случае это мера зависимости скоростей и положений молекул). Наиболее известна **корреляция Пирсона**. Она является мерой **линейной** зависимости двух переменных. (Смотри **Приложение D.**) Очевидно, что существуют гораздо более сложные зависимости и соответствующие им более сложные корреляции. Корреляции между различными переменными приводят к **ограничениям** на возможность выбора тех или иных значений этих переменных.

Одним из факторов, обеспечивающих корреляцию, является знание макроскопического состояния системы. Действительно, теперь уже не все, а лишь некоторые микросотояния могут соответствовать данному макроскопическому состоянию. Таким образом, их набор уже ограничен, что приведет к возникновению ограничения на скорости и положения молекул, определяющих возможные микросостояние системы. Следует отметить, что все такие **корреляции являются макроскопическими и описываются как зависимость между макроскопическими параметрами системы**. У макроскопических состояний с низкой энтропией ограничения на выбор возможных микросостояний велики и, соответственно, велико число возможных макроскопических параметров и корреляций между ними. У системы находящейся при термодинамическом равновесии энтропия достигает максимума, а число возможных макроскопических параметров и корреляций между ними мало.

**Дополнительные или микроскопические корреляции [33]**, определяются знанием не только текущего макроскопического состояния, но и макроскопической истории системы. Пусть физическая система эволюционировала из начального макроскопического состояния в некое другое текущее макроскопическое состояние. При этом не все микроскопические состояния, отвечающие текущему макроскопическому состоянию возможны. Только такие состояния, которые при обращении скоростей молекул приводят к начальному состоянию, могут быть рассмотрены (свойство обратимости движения.). Это накладывает дополнительные ограничения (корреляции) на набор микросостояний, соответствующих текущему макроскопическому состоянию. **Дополнительную корреляцию** можно определить по-другому, не через знание прошлого, а через знание будущего. Согласно теореме Пуанкаре система должна вернуться в известное начальное макроскопическое исходное состояние через некоторое, заранее известное время. Зная некое текущее макроскопическое состояние и зная, когда в будущем произойдет возврат, мы можем наложить дополнительные ограничения (корреляции) на набор микросостояний, соответствующих этому текущему макроскопическому состоянию. Из определения этих **дополнительных корреляций** видно, что они тесно связаны с парадоксами Пуанкаре и Лошмидта. Именно существование этих **дополнительных корреляций (дополнительных** по отношению к **макроскопическим корреляциям**, определяемых заданием макросостояния) приводит к нарушению закона возрастания энтропии и обеспечивает возможность возвратов и обратимости, наблюдаемых в парадоксах Пуанкаре и Лошмидта.

Одним из основных свойств **дополнительных или микроскопических корреляций** является **неустойчивость** динамики этих дополнительных корреляций. При взаимодействии разных частей системы или взаимодействии этой системы с окружающими системами (в том числе и наблюдателем) дополнительные корреляции исчезают, а, точнее говоря, «растекаются» между частями системы и/или между самой системой окружающими ее системами. Пусть имеется некое начальное состояние с низкой энтропией. После некоторого малого времени происходят первые столкновения между молекулами. Их положения и скорости становятся коррелированными (это можно проверить, обратив скорости). Коррелированны пока что лишь близкие пары столкнувшихся молекул. Однако, по мере роста числа столкновений, возникающие корреляции будут включать все больше молекул и распространятся по все большему объему системы. Происходит "растекание" корреляций по системе (Рис. 10) **[33]**.

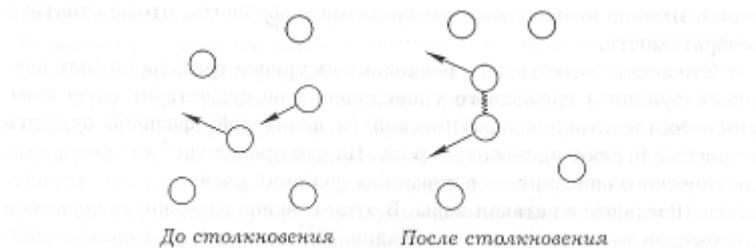

Рис. 3.2. Столкновения и корреляции. Столкновение двух частиц создает корреляцию между ними (условно изображенную волнистой линией).

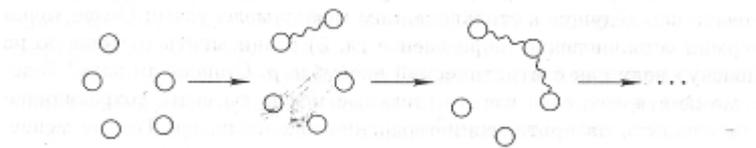

Рис. 3.3. Поток корреляций. Последовательные столкновения приводят к возникновению парных, тройных, ... корреляций.

**Figure 10.** Рассеяние и корреляции. Поток корреляций. (Рис. из [17])

Аналогично, если система состоит из двух не взаимодействующих систем, корреляции будут существовать только внутри каждой системы. Возвраты и обратимость возможны для каждой из таких подсистем. Пусть есть хотя бы малое взаимодействие между этими подсистемами. Тогда корреляции "перетекут" из одной подсистемы в другую, и эти две системы станут зависимыми. Соответственно, будет возможен только их совместный возврат или обратимость.

# 2. Главные парадоксы квантовой механики.

## 2.1 Основные понятия квантовой механики – волновая функция, уравнение Шредингера, амплитуда вероятности, измеряемые величины, соотношение неопределенности Гейзенберга.[3], [36]

Чтобы дальнейшее изложение было яснее, определим вначале основные понятия квантовой механики.

Движение в квантовой механике описывается не траекторией, а **волновой функцией**. Это волна вероятности, а точнее волна "**амплитуды вероятности**". Это означает, что квадрат амплитуды волновой функции в некоторой точке дает вероятность нахождения частицы в этой точке. Движение этой волны вероятности во времени описывается **уравнением Шредингера [36]**. Это линейное уравнение, т.е. сумма двух его решений также являются решением. Таким образом, амплитуды вероятностей суммируются, но не сами вероятности, поскольку они получаются из амплитуд возведением в квадрат. Это и вносит элемент нелинейности в процесс распространения волновой функции.

Любая измеряемая величина (например, импульс) описывается ортонормальным, полным набором функций (набор собственных функций измеряемой величины). Волновая функция может быть разложена по этому набору функций. Каждая из функций набора соответствует некому определенному значению (собственному значению) измеряемой величины. Коэффициенты разложения дают амплитуду вероятности для каждого такого значения. Если волновая функция совпадает с собственной функцией измеряемой величины, то эта величина имеет в этом состоянии одно определенное значение. Если же

нет – мы можем указать лишь вероятности для различных значений измеряемой величины.

Понятие скорость частицы теперь не имеет никакого физического смысла, поскольку нет траектории, а есть лишь «волна вероятности» [36]. Импульс теперь определяется не через скорость, а через коэффициенты разложения волны вероятности на набор собственный функций импульса. Этот набор функций аналогичен ортонормальному, полному набору Фурье функций в Фурье анализе.

Собственные функции координаты пропорциональны Дельта-функциям Дирака. Коэффициент разложения волновой функции для такой функции дается значением волновой функции в точке, где Дельта-функция Дирака обращается в бесконечность. Это соответствует приведенному ранее смыслу волновой функции как амплитуды вероятности.

Поскольку импульс и координата соответствуют различным наборам собственных функций [36], никакая волновая функция не может иметь одновременно точные значения импульса и координаты, в противоположность классической механике. Причина связана с различием определений этих величин в квантовой и классической механике. Это и есть знаменитая **неопределенность Гейзенберга. [36] (**Смотри **Приложение G.**).

## 2.2 Чистые и смешанные состояния. Матрица плотности. [3], [15], [29]

Наиболее полное описание квантовой системы дает её волновая функция. Это, так называемое, **чистое состояние**. Для классической механики это была точка в фазовом пространстве. Каков аналог ансамбля систем в классической статистической физике (облако точек в фазовом пространстве)? Это набор волновых функций, где каждой функции соответствует её вероятность (а не «амплитуда вероятности»!). Так определяется **смешанное состояние** квантовой системы.

Пусть некоторая система является частью некоторой большей системы. Тогда, даже если большая система описывается чистым состоянием, для описания только меньшей подсистемы в общем случае необходимо использовать смешанное состояние. Пусть, например, малая квантовая система взаимодействует с прибором, который находится в чистом состоянии. Тогда большая система (прибор – малая квантовая система) описывается чистым состоянием, но сама малая квантовая система после измерения описывается уже смешанным состоянием.

Для единообразного описания смешанных и чистых состояний используют матрицы плотности. [15] Выберем некоторую измеряемую величину и соответствующей ей набор ортонормальных функций. Тогда матрица плотности в представлении этих функций описывается квадратной матрицей. Каждой такой функции соответствует диагональный элемент матрицы плотности. Он равен вероятности найти систему в состоянии, определяемой этой функцией.

Недиагональные элементы матрицы плотности определяют разницу между чистым и смешанным состояниями. В чистом состоянии недиагональные элементы матрицы плотности имеют максимальное значение, для смешанных состояний их величина уменьшается и может стать равной нулю. Величина недиагональных элементов является мерой корреляции между состояниями системы, определяемыми базисными функциями матрицы плотности. Матрицу плотности всегда можно переписать в другом ортонормальном базисе функций, соответствующем какой-либо другой измеряемой величине. Эволюция во времени матрицы плотности заменяет эволюцию волновой функции в чистом состоянии или ансамбля (набора) волновых функций в смешанном состоянии. Вид этих уравнений в обоих случаях одинаков, что и обуславливает удобство в использовании матрицы плотности вместо набора волновых функций. (Смотри **Приложение I. )**

**2.3 Свойства изолированной квантовой системы с конечным объемом и конечным числом частиц. [29]**

Так же как мы сделали и для классических систем, рассмотрим здесь свойства изолированной (замкнутой) квантовой системы в с конечным объёмом с конечным числом частиц..

1) Такие квантовые системы описываются обратимыми уравнениями движения Шредингера.

2) Для таких систем верна теорема Пуанкаре. Более того, по своим свойствам квантовые системы подобны классическим интегрируемым системам (которые составляют лишь незначительную часть классических систем). Их возвраты происходят почти периодическим образом. Кроме того, период этого возврата очень слабо зависит от начальных условий.

3) Для квантовых систем можно определить энтропию ансамбля. Энтропия – это мера неопределенности, а чистое состояние единственно и дает максимально полное описание квантовой системы. Поэтому для любого чистого состояния энтропия принимается равной нулю по определению. Для смешанного состояния энтропия уже больше нуля, и такому состоянию соответствует уже набор чистых состояний. Пусть вероятность какого-либо одного из чистых состояний, описывающих смешанное состояние, близка к единице. Тогда же смешанное состояние почти чистое и его энтропия почти равна нулю. Когда же все чистые состояния смешанного состояния равновероятны, энтропия достигает своего максимума.

4) При эволюции квантовой системы чистое состояние переходит в чистое. В смешанном состоянии вероятности чистых состояний тоже постоянны. А значит и микроскопическая энтропия не меняется в процессе эволюции квантовой системы.

5) Мы можем описать большую квантовую систему с большим числом квантовых параметров лишь небольшим числом параметров, которые будем называть макроскопическими. Данному макроскопическому состоянию соответствует целый набор чистых состояний, которые будем называть микроскопическими. Соответственно на основе этого набора можно определить энтропию макроскопического состояния, которую будем называть макроскопической энтропией. В отличие от энтропии ансамбля она не должна сохраняться в процессе эволюции квантовой системы.

6) В процессе измерения квантовой системы, она перестает быть замкнутой, поскольку взаимодействует с прибором. Соответственно, ее чистое состояние переходит в смешанное, а микроскопическая энтропия возрастает. Такая эволюция необратима без обращения во времени также и измерительного прибора.

.

**2.4 Теория измерений в квантовой механике.[3], [29] (Приложение J, O, P)**

Для того чтобы проверить научную теорию, необходимо проводить измерения с помощью измерительных приборов. Это, как минимум, два измерения: для начального и конечного состояния. Зная начальное состояние и сравнивая измеренное конечное состояние с предсказанием теории, можно проверить ее достоверность.

В классической механике измерение - это простой процесс определения текущих параметров системы, не влияющий на её динамику. В классической механике полное описание её начального состояния дает однозначный результат измерения.

В квантовой механике всё много сложнее. Измерение влияет на динамику квантовой системы. Кроме того, в квантовой механике мы можем предсказать лишь вероятность того или иного результата измерения, даже если начальное состояние является чистым, т.е. дает наиболее полное описание квантовой системы.

Опишем процесс измерения в квантовой механике подробнее. Пусть система в начале описывается некоторой волновой функцией. Измерение некоторой величины приводит к тому, что волновая функция в момент измерения переходит скачком в одну из собственных волновых функций измеряемой величины. Это происходит, поскольку только для такой собственной функции измеряемая величина имеет определенное значение, получаемое в результате измерения. Как мы уже писали выше, вероятность такого измерения пропорциональна квадрату амплитуды разложения на собственные функции. Таким образом, после измерения система переходит из чистого состояния в смешанное состояние. Этот процесс носит название **редукции волновой функции**. Он не описывается уравнением Шредингера. Действительно, уравнение Шредингера описывает лишь переходы из чистого состояния в чистое. В результате редукции мы получаем из чистого состояния смешанное состояние. Кроме того, уравнение Шредингера обратимо. Процесс редукции необратим. Наличие этих **двух различных типов эволюции** квантовой системы объясняется тем, что в момент измерения квантовая система не замкнута – она взаимодействует с **макроскопическим классическим прибором**.

Макроскопический прибор, чтобы быть непротиворечиво описанным квантовой механикой, должен на самом деле быть **идеально макроскопическим**, т.е. или находиться в бесконечном пространстве, или состоять из бесконечного числа частиц. Идеальный макроскопический прибор не удовлетворяет условиям теоремы Пуанкаре о возвратах и имеет вполне определенное макроскопическое состояние во все моменты измерения. Для идеального макроскопического прибора квантовые законы в течение любого конечного времени дают тот же результат, что и классические законы. Следует отметить, что реальный измерительный прибор в ограниченном замкнутом объеме является макроскопическим лишь приближенно. Это замечание очень существенно для нашего анализа. Оно и является основным источником обсуждающихся в дальнейшем парадоксов.

Таким образом, эволюция квантовой системы разбивается на два вида. Первый это обратимая эволюция Шредингера. Второй это необратимая редукция волновой функции, происходящая при взаимодействии с макроскопическим классическим прибором.

Поскольку мы непосредственно наблюдаем лишь классические приборы, нет необходимости "наглядно" представлять себе "загадочные" квантовые объекты. По сути, квантовые объекты - это всего лишь математические абстракции, позволяющие описать связи между результатами показаний измерительных приборов. Сами приборы вполне классические и представимые. У них нет, например, параметров, которые нельзя померить одновременно, подобно координате и импульсу частицы в квантовой механике. "Наглядное", "физическое", "интуитивное" представление квантовой механики необходимо лишь для облегчения понимания самого сложного математического аппарата квантовой механики. Такое понимание до конца не возможно, поскольку наша интуиция основывается на классическом окружающем мире. Но, как мы уже написали выше, и нет такой практической необходимости. Однако эта невозможность и является основой всем известной **"магии" и "загадочности" квантовой механики**.

## 2.5 Сложность в попытке «классической» интерпретации квантовой механики: введение скрытых параметров и парадокс ЭПР.[3]

Законы квантовой механики носят вероятностный характер, причем многие величины не могут быть измерены одновременно. Однако вероятностны и многие законы классической статистической механики. Там их вероятностный характер объяснялся наличием **скрытых микроскопических параметров (Приложение U)**: скоростей и положений всех молекул. Любое классическое макроскопическое состояние описывается набором возможных микросостояний, соответствующих ему. Мы можем сделать попытку наглядной интерпретации вероятностных квантовых параметров, введя (по аналогии с классической статистической механикой) также и в квантовую механику свои скрытые параметры. Знание всех этих скрытых параметров позволяет однозначно определить все измеряемые в квантовой системе величины. Как и макроскопическое состояние, реальное наблюдаемое состояние в квантовой механике соответствует некому набору возможных значений скрытых параметров, а не их точным значениям. Однако, в отличие от классической статистической механики, в квантовой механике введение таких скрытых параметров возможно лишь при следующих очень серьезных допущениях:

   1) Измерение (за исключением особых случаев совпадения одной из собственных функций измеряемой величины с волновой функцией наблюдаемой системы) неизбежно меняет состояние измеряемой системы. В классическом случае мы полагаем возможность любого измерения без возмущения.

   2) Все скрытые параметры не могут быть померены одновременно. Так, например, мы ранее уже говорили о принципе неопределенности и невозможности померить одновременно координату и импульс частицы. Измерение меняет состояние системы (редукция волновой функции), и, следовательно, все скрытые параметры не могут быть померены также и в результате серии последовательных измерений. Мы можем предположить, что все скрытые параметры имеют некое вполне определенное значение. Но не существует такого реального и наблюдаемого физического состояния, в котором все скрытые параметры имеют некое вполне определенное значение. В любом реальном эксперименте мы можем напрямую найти определенные однозначные значения лишь для части из этих параметров. В то же время это же измерение приведем к неконтролируемому изменению и статистической неопределенности всех других скрытых параметров.

   3) **Введение скрытых параметров невозможно без введения дальнодействия [3]** между ними. Дальнодействие – это взаимодействие, которое распространяется мгновенно на любые расстояния. Это, однако, не противоречит из теории относительности факту, что никакое взаимодействие, связанное с передачей информации или массы, не может распространяться быстрее скорости света. Ведь это дальнодействие не переносит информацию или массу и его, соответственно, невозможно напрямую экспериментально наблюдать! Перенос информации возможен лишь для экспериментально наблюдаемого взаимодействия. Причиной этого служит свойство, описанное в предыдущем пункте. А именно, невозможность одновременно померить все скрытые параметры и обнаружить это взаимодействие.

   Эта необходимость введения ненаблюдаемого дальнодействия является слишком высокой ценой за классическую "наглядность", и поэтому скрытые параметры обычно не используются в квантовой механике. Если же рассматривать законы квантовой механики просто лишь как некую математическую взаимосвязь между результатами реально наблюдаемых показаний макроскопических приборов, то никаких противоречий или необходимости вводить дальнодействие не возникает.

   Эта необходимость введения дальнодействия скрытых параметров иллюстрируется знаменитым **«парадоксом» Энштейна-Подольского-Розена (ЭПР).[3], (Приложение R)** Заметим, что этот «парадокс» на самом деле фиктивный. Он возникает только тогда, когда законы квантовой механики пытаются проинтерпретировать классически и одновременно "легкой ценой", т.е. введением скрытых параметров, но без дальнодействия. Он основывается на анализе состояний пары электрон-позитрон, которые в начале были вместе, а затем разлетелись на большое расстояние.

   Каждый электрон (или позитрон) обладает собственным вращательным моментом, который носит название спина. Классический аналог собственного момента – момент волчка, вращающегося вокруг собственной оси. В отличие от классического момента, абсолютная величина проекции собственного момента электрона имеет неизменную величину (1/2), а её проекция на любую ось имеет только два возможных значения: вдоль оси и против оси (+1/2 и -1/2). Если мы выберем иную ось, то она будет обладать тем же свойством. Однако проекции на две разных оси не могут быть измерены одновременно. Не существует квантового состояния, в котором проекции спина на две разных оси имеют определенное значение.

Пусть рождается электрон-позитрон с суммарным спином ноль. Предположим теперь, что электрон и позитрон движутся в противоположных направлениях, пока не удаляться друг от друга на большое расстояние. Величины их спинов при этом неизвестны. Пусть мы померили спин электрона вдоль некой оси (обозначим ее как ось Z) и получили значение +1/2. Отсюда можно сделать однозначный вывод о результате измерения проекции спина позитрона на туже ось. Действительно, из закона сохранения полного спина (вначале он был равен нулю) эта проекция равна -1/2. Мы можем вместо этого провести измерение проекции спина позитрона вдоль любой другой оси. Если эта ось перпендикулярна оси Z, то равной вероятностью спин позитрона равен как +1/2, так и -1/2. Для других положений оси квантовая механика также позволяет точно рассчитать условные вероятности измерения проекции спина позитрона при известной данной проекции спина электрона на ось Z (Рис. 11).

Предположим, что все возможные проекции спина можно проинтерпретировать «классически». Это означает, что все они имеют некое определенное известное значение, а измерение лишь «проявляет», делает наблюдаемым его. Это означает, что за скрытые параметры принимаются проекции спина частиц на все возможные оси.

Чем тогда объясняется связь спинов электрона и позитрона, описанная выше? Тем, что существует статистическая связь (корреляция) между возможными значениями скрытых параметров? Или измерение спина электрона за счет некого дальнодействия скрытых параметров влияет также на спин позитрона? С помощью так называемого **неравенства Белла (Приложение Т)** можно доказать следующую теорему. **Не существует такого набора скрытых параметров (т.е. проекций спина) и распределения вероятности, ему соответствующего, который определял бы все найденные из квантовой теории условные вероятности только за счет статистических корреляций между скрытыми параметрами (проекциями спина) этого набора.** Т.е. для объяснения связи спинов необходимо введение дальнодействия.

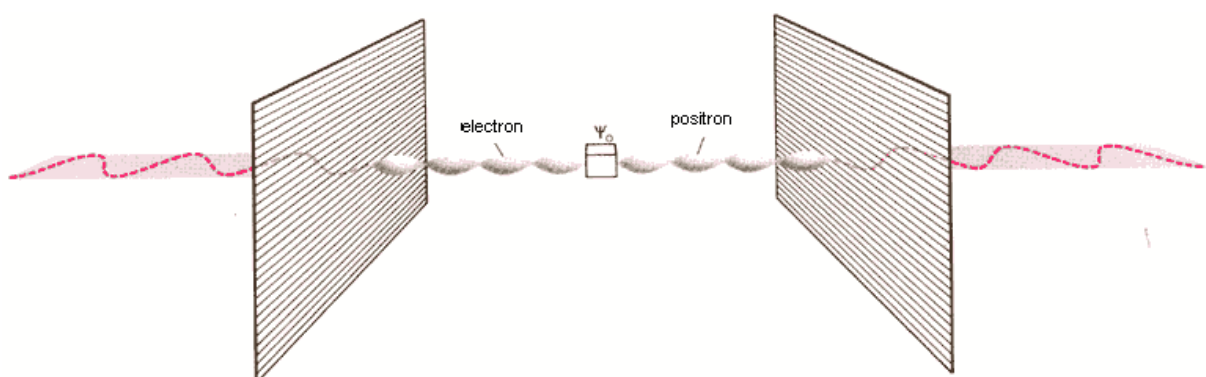

**Рис 11.** Разлетающиеся электрон и позитрон в ЭПР парадоксе [103].

Однако, можно рассматривать не все скрытые параметры, а лишь результаты измерения. В этом случае связь между спинами электрона и позитрона легко объясняется обычными статистическими корреляциями между измеренными параметрами. Это связанно с тем, что что мы способны померить только две проекции спина (одну для электрона и одну для протона), а все.

**Корреляции удаленных объектов,** существующие до измерения, являются **квантовыми,** если попытка их толкования "классически" (т.е. на основе набора скрытых параметров) приводит к необходимости введения дальнодействия скрытых параметров. Это и отличает их от **классических корреляций**, которые подобных допущений не требуют**. После же осуществления измерения, квантовый характер корреляций необратимо исчезает, и они переходит в самые обычные классические корреляции.

Подведем итоги рассмотрения. Мы либо должны отказаться от необходимости вводить скрытые параметры, либо согласиться с возможностью дальнодействия для этих параметров. В случае отказа от скрытых параметров и рассмотрении квантовой механики лишь как некого математического аппарата, дающего связь между результатами измерения, никакого дальнодействия уже не требуется. Существует связь между результатами измерений проекций спинов электрона и позитрона, однако она носит характер обычной статистической корреляции двух вероятностных величин (проекций спина). Она объясняется тем, что в прошлом эти частицы были вместе. Так, например, для проекции спина +1/2 для электрона мы всегда получаем проекцию спина на ту же ось -1/2 для позитрона.

## 2.6 Задача двух щелей как иллюстрация сложностей квантовой механики.

Именно из-за невозможности "легкой" классической интерпретации квантовой механики известный американский физик Ричард Фейнман полагал, что никто не понимает квантовую механику. Однажды он заметил, что «единственная тайна квантовой механики заключена в интерпретации одного-единственного эксперимента — эксперимента с двойной щелью и электронами, современной версии классического опыта, выполненного в 1801 году английским исследователем Томасом Юнгом для демонстрации волновой природы света». Опыт этот был на удивление простым. В опыте Юнга свет от источника, в качестве которого служила узкая щель $S$, падал на экран с двумя близко расположенными щелями $S_1$ и $S_2$. Проходя через каждую из щелей, световой пучок уширялся вследствие дифракции, поэтому на белом экране Э световые пучки, прошедшие через щели $S_1$ и $S_2$, перекрывались. В области перекрытия световых пучков образуется ряд чередующихся светлых и темных полос— то, что мы теперь называем интерференционной картиной. Юнг интерпретировал темные линии как места, где «гребни» световых волн от одной щели встречаются с «впадинами» волн от другой щели, гася друг друга. Яркие же линии получаются в местах, где гребни или впадины от обеих щелей совпадают, вызывая усиление света. На протяжении почти двухсот лет различные варианты двухщелевого опыта Юнга рассматривались как доказательство того, что волны на воде, радиосигналы, рентгеновские лучи, звук и тепловое излучение распространяются в виде волн (Рис. 12).

Определим понятие **разность хода** волн от щелей. Пусть имеется некоторая точка на конечном экране. Разница расстояний от двух щелей, измеренной в единицах длины волны, называется разницей хода волны для этой точки. Если разница хода целое число мы имеем максимум волны в этой точке. Если это целое число плюс половина, то имеем минимум.

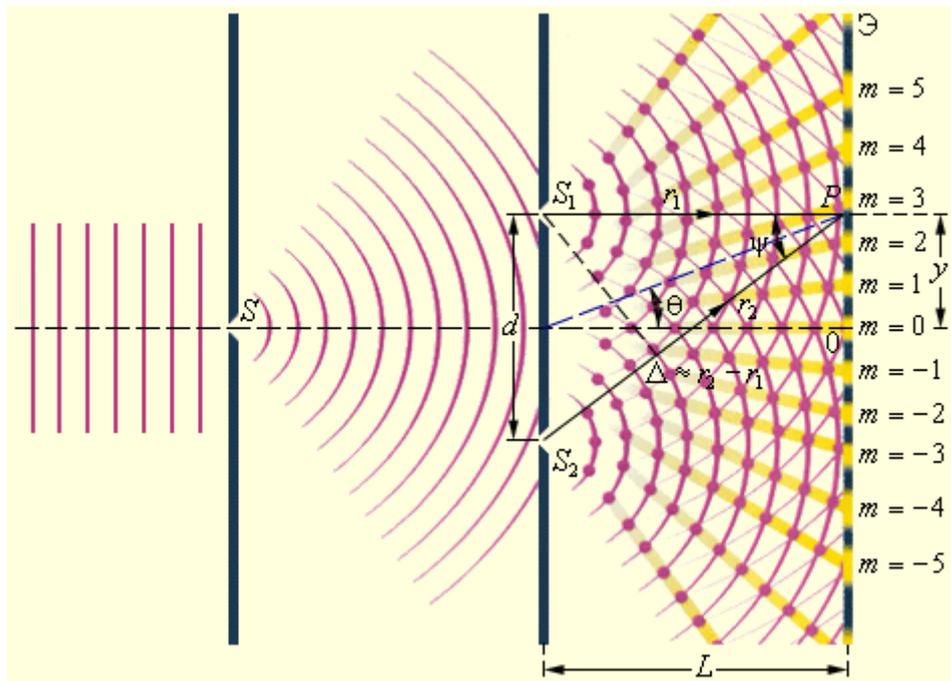

**Figure 12.** Схема интерференционного опыта Юнга. (Рис. из [96])

Замечательно то, что опыт Юнга (Рис. 13, 14) можно выполнить и с электронами. Вместо пучка солнечного света через параллельные щели проходит поток электронов, экраном служит покрытая люминофором пластина (подобная экрану телевизионной трубки). Каждый электрон при ударе о люминофор оставляет на нем светящуюся точку, фиксируя тем самым свое прибытие в виде частицы. Но изображение, сформированное всеми электронами, производит удивительное впечатление. Оно принимает вид интерференционного рисунка: яркие полосы, образованные множеством светящихся точек, чередуются с темными областями, где они почти отсутствуют. Картина в целом подобна той, которая получается в случае света, и совершенно непохожа на ту, которую мы получили бы, бросая, скажем, мячики в изгородь с двумя вынутыми из нее досками, то есть с двумя щелями. Двухщелевой эксперимент с электронами, демонстрируя, что частицы могут вести себя как волны.

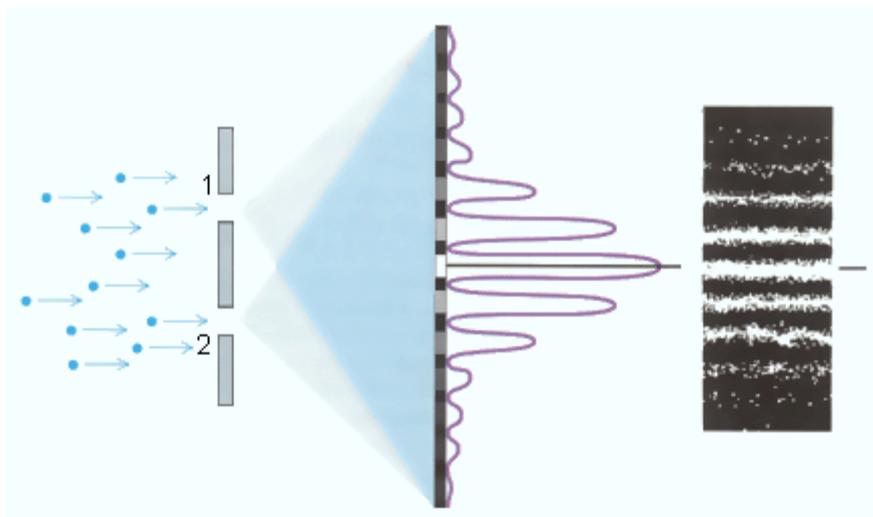

**Figure 13.** Дифракция электронов на двух щелях. (Рис. из [97])

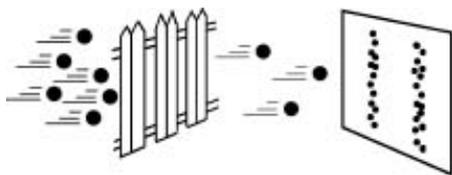

**Рис. 14** Мячики, пролетающие сквозь щель в заборе, оставляют на стене следы в виде двух полос — «изображения» щелей. Так же вел бы себя и свет, будь он просто потоком частиц [104].

Если в опыте по наблюдению дифракции электронов на двух щелях закрыть одну из щелей, то интерференционные полосы исчезнут, и фотопластинка зарегистрирует полосу электронов, продифрагировавших на одной щели. Если затем открыть вторую щель мы получим вторую полосу. Итоговая картина в целом подобна той, которую мы получили бы, бросая, скажем, мячики в изгородь с двумя вынутыми из нее досками, то есть с двумя щелями. Если же открыты одновременно обе щели, то наблюдаемое интерференционное распределение не является простым наложением этих двух независимых распределений от каждой из щелей в отдельности. Результаты этого эксперимента не могут быть объяснены взаимодействием электронов – тот же результат получается, если выпускать электроны по одному. Причина в том, что позиция электрона определяется не некоторой траекторией, а волной вероятности. Две волны от двух щелей складываются и дают интерференционную картину. Квадрат амплитуды этой суммарной волны на экране даёт вероятность попадания туда электрона.

Предположим, мы располагаем детектором, который показывает, через какую из щелей проходит электрон. Итоговая картина в этом случае подобна той, которая получается в результате опыта попеременного закрытия двух щелей, описанного выше, г.е. интерференционная картина исчезает. Этот результат объясняется влиянием измерительного прибора – детектора. Происходит описанная ваше редукция волновой функции и чистое состояние переходит в смешанное. Таким образом, вместо суммы амплитуд волновых функций от двух щелей, суммируются вероятности, определяемые каждой из этих волн, и интерференционная картина исчезает.

Этот эксперимент демонстрирует два важнейших свойства квантовой механики. Во-первых, **мы не можем предсказать точное место попадания электрона в экран**, а можем лишь найти вероятность его попадания в ту или иную точку. Лишь большое число электронов дает вполне определенную и предсказуемую картину распределения их совместных попаданий на экран. В классической механике результат предсказаний был однозначен даже для одной частицы. Во-вторых, **мы не можем произвести** *любое* **измерение промежуточного состояния электрона, не исказив результата последующих измерений**. Так, проверив через какую из щелей прошёл электрон, мы разрушим интерференционную картину. В классической механике теоретически всегда было возможно провести измерение, не меняя последующей динамики системы. В квантовой механике так можно измерить лишь величну, одна из собственных функций которой совпадает с текущей волновой функцией системы.

Опыт с двумя щелями также позволяет объяснить механизм **исчезновение квантовых интерференционных эффектов для макроскопических систем.** Он происходит при следующих трех условиях:

1) взаимодействие падающей на экран когерентной**,** "монохроматической" волны с окружением или источником излучения приводит к **переходу** ее **чистого состояния в смешанное состояние**. Как следствие, волна вероятности не является бесконечной синусоидой, а разбивается на отдельные отрезки синусоид. Такие отрезки синусоид носят название волновых пакетов. Соотношение фаз между волновыми пакетами – случайная величина. Длина волнового пакета около 10-20 длин волн. Она имеет порядок радиуса взаимодействия волны с атомами окружающей среды (атомами источника излучения).

2) **переход к макроскопическим размерам**. Расстояниям между щелями (D) велико и много больше длины волнового пакета (nλ) и расстояния от щелей до экрана (L). Точнее: D>>√(L·nλ), где λ-длинна волны.
3) **рассмотрение макроскопических параметров** (усреднение интенсивности волны на отрезке, много большем длины волнового пакета, и в течение периода времени, много большем времени прохождения волнового пакета через неподвижную точку)

Когда расстояние между щелями растет (при постоянной L), разница хода становится много больше размера волнового пакета для большинства точек экрана. Как результат, разница фаз волн, приходящих от щелей становиться переменной, случайной величиной. Следовательно, складываются уже не амлпитуды волн от щелей, а их огрубленные макроскопические интенсивности. В итоге, на большей части экрана интерференция исчезает. Когда расстояние между щелями становится больше расстояния до экрана, интерференция остается лишь в малой окрестности точки экрана, находящейся точно посередине между щелями. Ее размер равен ширине волнового пакета. При дальнейшем увеличении расстояний между щелями интенсивность волны в интерференционной области начинает уменьшаться и стремится к нулю. Столь малые интерференционные эффекты не наблюдаются при огрубленном (т.е. макроскопическом) описании[3].

Все эти эффекты исчезновения интерференции связанны с макроскопичностью системы и ее параметров, а также смешанным, не чистым начальным состоянием. Такой переход из чистого состояния в хаотическое смешанное состояние из-за взаимодействия с окружением носит название **декогеренции** (от латинского *cohaerentio* - сцепление, связь) [31], [39], [16], [16], [47], [48] , (**Приложение Р**). Система перемешивается или запутывается с окружающей средой. Для макроскопических (т.е. очень больших) систем декогеренция приводит к исчезновению квантовой интерференции, как мы видели выше в опыте с двумя щелями. Теория декогеренции имеет важное следствие: для макросостояния предсказания <u>квантовой теории</u> практически совпадают с предсказаниями классической теории. Но ценой за это совпадение является необратимость, как мы увидим далее.

## 2.7 Парадокс Шредингеровского Кота [51] и спонтанная редукция.[3]

Полное, сто процентное нарушение принципа суперпозиции (т.е. полное исчезновение интерференции) и редукция волновой функции происходит лишь при взаимодействии квантовой системы с идеальным макроскопическим объектом или прибором. Идеальный макроскопический объект имеет либо бесконечный объем, либо состоит из бесконечно числа частиц. Такой идеальный макроскопический объект может быть непротиворечиво описан как квантовой, так классической механикой[4].

---

[3] Следует отметить, что эта система имеет **бесконечные размеры**, поскольку волны бегут к бесконечности в направлении, перпендикулярном экрану, а не отражаются обратно. Поэтому, в отличие от конечных систем, обсуждаемых далее, интерференция исчезая, не появляется вдруг снова вновь. С другой стороны, для расстояния между щелями **стремящимся к бесконечности** (при постоянной L) квантовые интерференционные эффекты стремятся к нулю для любой конечной ширины волновых пактов и любой конечной степени огрубления параметров.

[4] *Так, наблюдая свет удаленной звезды, мы её изучаем, но не оказываем при этом на неё никакого влияния, как ожидалось бы из квантовой теории измерений. Мы изменяем лишь состояние дошедших до нас и наблюдаемых нами фотонов света звезды. Происходит это потому, что мы считаем пространство Вселенной бесконечным. Наблюдаемые фотоны не имеют шанса вернуться к наблюдаемой звезде и изменить ее состояние. В случае же конечной Вселенной, наблюдаемые фотоны могут вернуться к звезде, и таким путем наше наблюдение окажет влияние на звезду. Правда, в случае большого объема Вселенной ждать этого придется очень долго.*

В дальнейшем, если не оговорено иное, мы рассматриваем, как и в случае классической статистической механики, лишь системы конечного объема с конечным числом частиц. Для таких систем приборы или объекты могут считаться лишь приближенно макроскопическими[5].

Тем не менее, реальный эксперимент показывает, что и для таких неидеальных макроскопических объектов может происходить разрушение суперпозиции и редукция волновой функции. Будем называть такую редукцию неидеальных макроскопических объектов **спонтанной редукцией**. Спонтанная редукция приводит к многочисленным парадоксам, которые заставляют сомневаться в полноте квантовой механики, несмотря на все её огромные успехи. Приведем самый впечатляющий парадокс из этой серии – **парадокс «кота Шредингера». [51]**(Эрвин Шредингер ,1935) (Рис. 15)

Это умозрительный эксперимент, проясняющий принцип суперпозиции и редукции волновой функции. Кота помещают в коробку. В ней, кроме кота, находится капсула с ядовитым газом (или бомба), капсула может взорваться с 50-процентной вероятностью благодаря радиоактивному распаду атома плутония или случайно залетевшему кванту света. Через некоторое время коробка открывается и выясняется, жив кот или нет. До тех пор пока коробка не открыта (не произведено измерение), кот пребывает в очень странной суперпозиции двух состояний: «живой» и «мертвый». Применительно к окружающим нас макрообъектам такая ситуация выглядит странновато[6].( Для элементарных частиц нахождение одновременно в двух, казалось бы, взаимоисключающих состояниях совершенно естественно. ) Тем не менее, никакого принципиального запрета на квантовую суперпозицию состояний для макрообъектов нет.

Редукция этих состояний при открытии коробки и внешнем наблюдении не приводит к противоречию с квантовой механикой. Она легко объясняется воздействием внешнего наблюдателя на кота при измерении.

Парадокс возникает при закрытой коробке, когда наблюдателем является сам Кот. Действительно, Кот обладает сознанием и способен наблюдать как самого себя, так и свое окружение. При реальном самонаблюдении кот не может быть одновременно жив и мертв, а находятся лишь в одном из этих двух состояний. Опыт показывает, что любое сознающее существо либо ощущает себя живым, либо оно мертво. Одновременно и тем и другим оно чувствовать себя не может![7] Кот даже вместе со всем содержимым ящика - это не *идеальный* макроскопический объект. Поэтому наблюдаемая и необратимая спонтанная редукция на состояния живой и мертвый, приводящая к отклонению динамики этой системы от идеальной обратимой Шредингеровской эволюции, казалось бы, уже не может быть ничем объяснена. В этом и состоит суть парадокса «кота Шредингера»**[3], [6], [7].**

---

[5] *Например, описанное выше поведение системы звезда-наблюдатель для конечной, но очень большой Вселенной совпадает с поведением в бесконечной Вселенной лишь в течение очень большого, но конечного промежутка времени.*

[6] *Иногда пытаются изобразить такие ситуации средствами искусства через «парадоксальные» картины.* ***[8], [58], (Приложение V).***

[7] *Хотя и существуют экзотические попытки представить себе, как сознание может воспринимать такие смешанные состояния макрообъектов **[8 ], [58], (Приложение V).**.Смотри также предыдущую сноску.*

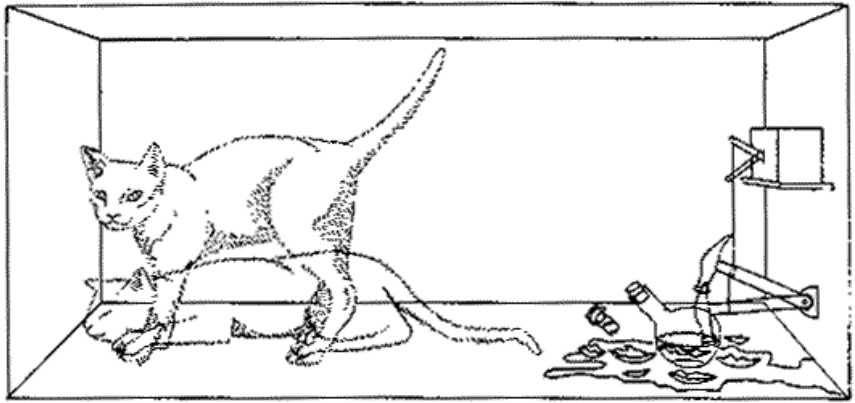

**Рис. 15** Опыт со Шредингеровским Котом. Суперпозиция живого и мертвого Кота. (Рис. из [98])

Возникает масса вопросов, связанных со спонтанной редукцией. Противоречит ли она на самом деле Шредингеровской динамике? Насколько макроскопической должна быть система, чтобы спонтанная редукция происходила? Должна ли такая почти макроскопическая система обладать сознанием, подобно коту? С какой частотой по времени происходит спонтанная редукция?

## 2.8 Парадокс Зенона или «парадокс котелка, который никогда не закипит».

С последним из этих вопросов связан **«парадокс котелка, который никогда не закипит».** На самом деле, здесь возникает не один, а два парадокса.
Пусть происходит во времени некий квантовый процесс. Например, распад частицы или перехода частицы с одного энергетического уровня на другой.

Первый парадокс заключается в том, что если устремить временные интервалы между актами регистрации этого процесса к нулю, указанный процесс вообще никогда не происходит за любой заранее выбранный конечный промежуток времени! Это объясняется неустранимым влиянием квантового измерения. Измерение приводит к редукции смешанного состояния на распавшуюся и не распавшуюся частицу. Кроме того, относительная (на одну частицу) скорость процесса при уменьшении интервала между измерениями стремиться к нулю. Эти два факта приводят к тому, что процесс прекращается при черезчур частых измерениях.

Второй парадокс закючается в следующем. В реальности распад вещества, содержащего большое число частиц, всегда описывается экспоненциальным законом. Эо не случайно. Относительная скорость такого распада постоянна во времени. Соответственно, невозможно экспериментально определить "возраст" такого вещества, если мы не знаем начальное число нераспавшихся частиц, а продукты распада удаляются из системы.

Но квантовый распад, согласно уравнениям квантовой механики, описывается **не экспоненциальным** законом. Поэтому относительная скорость распада в самом начале процесса равна к нулю, а затем растет. Мы приходим к парадоксальному выводу, что можно ввести нефизическое понятие «возраста» системы, который может быть легко определен через текущую относительную скорость распада системы.

Мы разрешим этот второй парадокс в следующей части статьи, посвященной Наблюдаемой Динамике.

## 2.9 Квантовые корреляции состояний системы и их связь с парадоксом Шредингеровсого Кота.

С парадоксом Шредингеровского Кота тесно связано понятие **квантовой корреляции состояний системы**. Пусть происходит спонтанная редукция состояний живого или мертвого кота. Тогда любое дальнейшее измерение будет зависеть от того, в каком состоянии был перед этим Кот. Результаты измерений можно разбить на две группы: одна группа будет соответствовать живому Коту, другая мертвому. Если Кот находится в квантовой суперпозиции, то результат последующего измерения будет зависеть от обоих состояний Кота, его невозможно уже разбить на две несвязанных группы. Эта связь между текущими состояниями, проявляющаяся в невозможности получения независимых результатов измерений в будущем, называется квантовой корреляцией состояний системы.

На языке математики, этот результат объясняется нелинейностью связи между вероятностью результата измерений и волновой функцией. Квадрат суммы не равен сумме квадратов. Появляющиеся дополнительные члены (или интерференционные члены) определяют квантовую корреляцию.

Квантовая корреляция определяется недиагональными членами матрицы плотности. Для смешанного состояния, получающегося в результате измерения, все недиагональные члены нулевые.

Выразим парадокс Шредингеровского Кота на языке квантовых корреляций:

С одной стороны, результат самонаблюдения кота однозначно дает только один из результатов: кот жив или мертв. Таким образом происходит **спонтанная редукция,** и **квантовая корреляция** между этими состояниями **исчезает**. Это означает, что результаты всех последующих измерений можно разбивать на две независимые группы, соответствовать живому и мертвому коту.

С другой стороны, согласно уравнению Шредингера, квантовая корреляция не может сама исчезать, без наличия внешних сил. Это означает, что результаты последующих измерений нельзя разбивать на две независимые группы.

Это противоречие между Шредингеровской динамикой и наблюдаемой спонтанной редукцией приводит к парадоксу.

# 3. Интерпретации квантовой механики. Их нерелевантность для решения парадоксов.

Одной из проблем, о которой мы писали выше, является трудность понимания квантовой механики на основе нашей интуиции, почерпнутой из повседневного мира, описываемого классической механикой. Для облегчения такого понимания и служат различные интерпретации квантовой механики [3] . Надо особо подчеркнуть, что ни одна из интерпретаций квантовой механики не приводит к разрешению описанных выше парадоксов, а лишь позволяет наглядно и понятно для нашей интуиции понять саму квантовую механику. Приведем лишь три из широкого списка возможных интерпретаций. Наиболее популярной ныне и также очень наглядной является многомировая интерпретация.

## 3.1 Многомировая интерпретация. [37], [38],[3].

Опишем её подробнее. На примере Шредингеровского кота мы видели, что квантовая эволюция может приводить к различным, макроскопически различимым состояниям. Мы наблюдаем лишь одно из них. Многомировая интерпретация утверждает, что все эти состояния существуют одновременно в неких «параллельных мирах», но мы (или сам кот в мысленном эксперименте) можем наблюдать лишь одну из макроскопических альтернатив.

Подобный подход следующим образом иллюстрирует понятие спонтанной редукции. Поскольку все миры существуют одновременно, все они могут оказывать влияние на результат последующего измерения. В общем случае результаты измерения невозможно разбить на две несвязанные группы, связанные с живым и мертвым котом. Значит эти миры коррелируют друг с другом и влияют совместно на последующие измерения. Наличие же спонтанной редукции при измерении приводит к потере этой корреляции. Результаты всех последующих измерений распадаются на независимые группы, соответствующие различным мирам (Рис. 16).

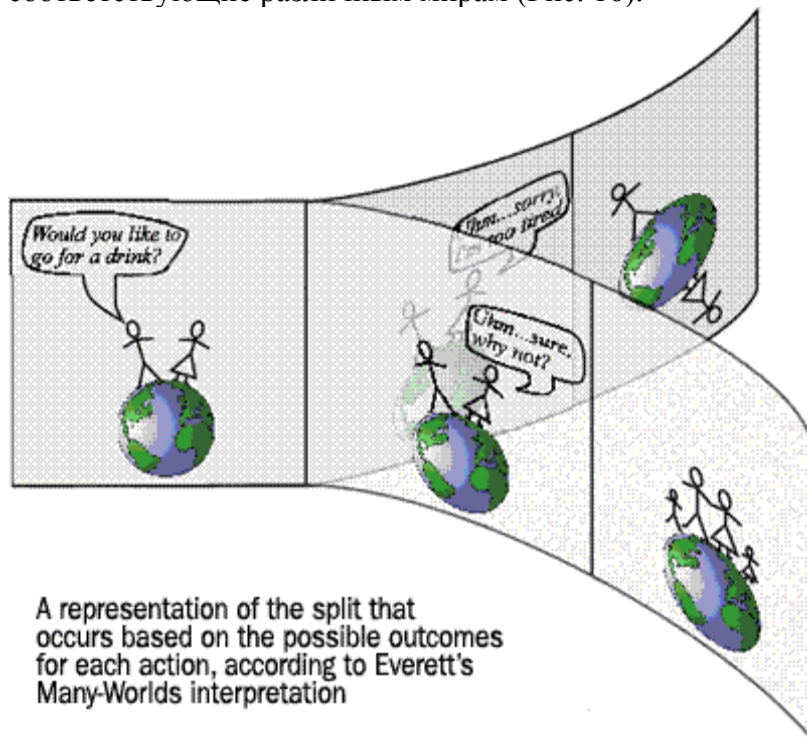

**Рис. 16.** Многомировая интерпретация. (Рис. Max Tegmark из [99])

Многомировая интерпретация не объясняет сама по себе парадокса Шредингеровсого Кота. Действительно, Кот наблюдает лишь один из существующих миров. Результаты же дальнейших измерений зависят от корреляций между мирами. Но ни эти миры, ни эти корреляции не являются наблюдаемыми. Всегда могут существовать параллельные миры, о которых мы ничего не знаем и которые могут повлиять на результат будущего эксперимента. Т.е. знание лишь текущего состояния и законов квантовой механики не позволяет нам предвидеть будущее даже вероятностно! Но именно для такого предсказания и была изобретена квантовая механика. Лишь приняв существование необъяснимой спонтанной редукции, уничтожающей квантовые корреляции между мирами, мы можем предсказать будущее на основе знания лишь текущего (и реально наблюдаемого) состояния нашего мира. Парадокс Шредингеровского Кота возвращается, просто изменив форму.

Попрежнему неясны вопросы, как однозначно определить макроскопические состояния, соответствующие «разделению» на разные миры. (Ведь разложение волновой функции неоднозначно и возможно по различным наборам ортогональных функций.) Непонятно и в какие точно моменты времени, это «разделение» происходит. Но разрешение парадоксов, вопреки очень широко распространенному заблуждению, и не является целью интерпретаций квантовой механики.

## 3.2 Копенгагенская интерпретация.

Другой интерпретацией является Копенгагенская интерпретация. С этой точки зрения, как наиболее стандартной и используемой в литературе, и ведется изложение в данной работе. Она утверждает, что в момент наблюдения макроскопических состояний происходит спонтанная редукция и квантовые корреляции исчезают. Это ведет к появлению парадоксов, изложенных выше.

Следует отметить, что редукция в Копенгагенской интерпретации происходит лишь для *конечного* наблюдателя в цепочки измерений. С его точки зрения и должен описываться весь эксперимент. Редукция, подобно скорости системы, зависит от выбора конкретной системы наблюдения.

Пусть внешний наблюдатель исследует другого наблюдателя, например Шредингеровского Кота. Никакой спонтанной редукции, наблюдаемой Котом, для внешнего наблюдателя не происходит (Рис. 17).

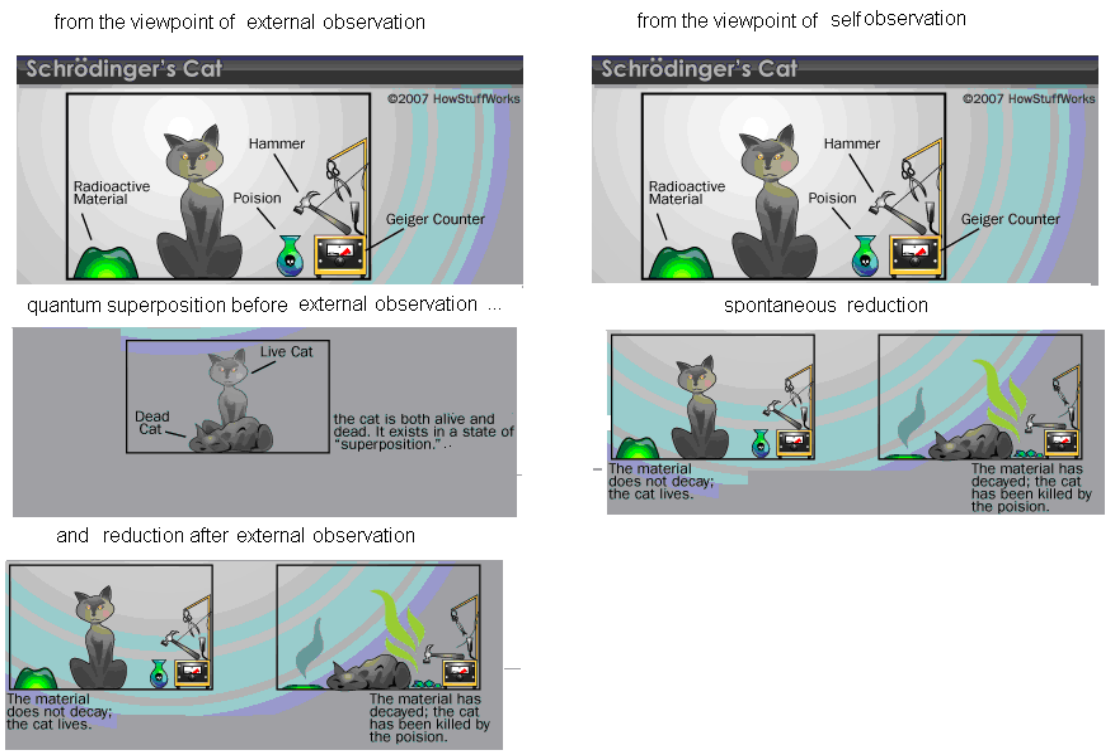

**Рис 17.** Опыт со Шредингеровским Котом с точки зрения внешнего наблюдателя и с точки зрения самого Кота (самонаблюдение). (Рис. из [100])

Редукция происходит лишь в момент внешнего наблюдения над Котом, когда экспериментатор открывает ящик. Значит для внешнего наблюдателя нет и парадокса.

Только, если в качестве конечного наблюдателя эксперимента рассматривается сам Кот, возникает спонтанная редукция, и появляется описанный выше парадокс. Ведь Кот может ощущать себя лишь живым или мертвым, но не как и то и другое одновременно!

Это замечание очень важно, поскольку непонимание его приводит к совершенно ошибочным утверждениям **[37], [38],** что Копенгагенская интерпретация якобы несовместима с Многомировой интерпретацией. На самом деле, как мы увидим далее, разница между этими интерпретациями экспериментально не наблюдаема и обе они вполне законны.

### 3.3 Интерпретация через скрытые параметры.[3], (Приложение S, T, U)

Еще одной интерпретацией, связанной с парадоксом ЭПР и затронутой в этой работе, было введение скрытых параметров. Это, например, теория волны-пилота де Бройля - Бома. [3]. Она включает как скрытые параметры координату, скорость, спин и волновую функцию (волна - пилот), изменяющуюся во времени в соответствии с уравнением Шредингера. Квантовые корреляции при этом, как мы видели в парадоксе ЭПР, приводили к нарушению локальности, т.е. дальнодействию между скрытыми параметрами. Для объяснения связи между реально измеренными, а не скрытыми параметрами, такие дальнодействия уже не нужны. Эти связи описываются обычной корреляцией случайных величин. Таким образом, редукция макроскопического состояния (или происходящая при измерении, или спонтанная) приводит к исчезновению квантового характера корреляций, которые становятся классическими.

Отличие квантовых корреляций от классических корреляций, возникающих после редукции, проявляется не только в дальнодействии. Пусть разбежавшиеся на большое расстояние части системы (находящейся в начале в чистом состоянии) спустя некоторое время снова оказываются в одном месте. В квантовом случае мы получаем при этом чистое состояние, а в классическом случае, сопровождающемся редукцией, - смешанное состояние. В случае спонтанной редукции это приводит к противоречию с Шредингеровской эволюцией. Парадоксы снова не исчезают, а лишь принимают другое обличие.

# 4. Определение полной физической системы в теории измерений.

В теории измерений необходимо включать наблюдателя и окружающую среду в полную систему, поскольку во многих случаях даже их малым влиянием невозможно пренебречь. Как мы увидим в дальнейшем это верно не только для квантовой, но и классической механики. В общем случае полная система, к которой прикладываются законы физики, состоит из трех частей: **наблюдаемая система**, **окружающая среда**, и сам **наблюдатель**. Наблюдатель также состоит из трех частей: **измерительный прибор**, сама **личность наблюдателя** и **память** наблюдателя, необходимая для запоминания последовательности наблюдений для их дальнейшего сравнения с теорией. Необходимо ответить, что **память должна быть изолирована от всего ее окружения**, кроме самого канала получения информации. Если внешние факторы могут влиять на неё, менять или стирать её содержимое, никакие эксперименты с ее участием для проверки теории не являются релевантными. Последнее положение очень важно. Оно разрешает многие парадоксы, связанные с мысленными экспериментами, включающими измерения, которые мы рассмотрим в следующей главе.

Конечным пунктом полной физической системы является память наблюдателя. Система включает только одного наблюдателя с его памятью. Конечно, наблюдателей может быть несколько, но для описания эксперимента мы должны выбрать лишь одного, остальные будут просто частью наблюдаемой системы или окружения. Какого из них выбрать? Вопрос решается аналогично теории относительности – можно выбрать любого. Важно лишь все факты и парадоксы толковать с его точки зрения и не смешивать их, чтобы не возникала путаница. В случае с парадоксом Шредингеровского Кота наблюдателем может быть как сам Кот, так и внешний наблюдатель-экспериментатор.

# 5. Разрешение парадокса Шредингеровского кота.

Напомним, что парадокс Шредингеровского кота заключается в противоречии между спонтанной редукцией, наблюдаемой котом и Шредингеровской эволюцией, запрещающей подобную редукцию. Чтобы правильно понять парадокс Шредингеровского кота нужно рассмотреть его с точки зрения двух наблюдателей: внешнего наблюдателя-экспериментатора или самого Кота, т.е. **самонаблюдение**.

   В случае внешнего наблюдателя-экспериментатора парадокса не возникает. Если экспериментатор пытается посмотреть, жив кот или нет, он оказывает неизбежное в квантовой механике воздействие на наблюдаемую систему, что приводит к редукции. Система не замкнута и, следовательно, не может описываться уравнением Шредингера. Редуцирующую роль наблюдателя может играть и окружающая среда. Этот случай описывает декогеренция. Здесь роль наблюдателя более естественна и сводится лишь к фиксации результата декогеренции. В обоих случаях происходит запутывание состояния измеряемой системы с окружением или наблюдателем, т.е. появляются корреляции измеряемой системы с окружением или наблюдателем.

   Что будет, если мы рассмотрим замкнутую полную физическую систему, включающую наблюдателя и окружение? Такой случай самонаблюдения – это сам Кот с его окружением в ящике. Следует отметить, что **полное самонаблюдение (полное в смысле квантовой механики) и полная проверка законов квантовой механики невозможны в замкнутой системе, в которую включен наблюдатель**. Мы можем в принципе сколь угодно точно измерить и проанализировать состояние внешней системы. Но если мы включаем и себя в описание, то возникают естественные ограничения, связанные с необходимостью записи в память и анализа состояний молекул с помощью самих, же этих молекул. Предположение о такой возможности приводит к противоречиям. (**Приложение М**) Поэтому и возможности найти в реальном эксперименте противоречие между Шредингеровской эволюцией и спонтанной редукцией при самонаблюдении для замкнутых систем ограничены.
   Проверим, можно ли поставить мысленный эксперимент, приводящий к противоречию между Шредингеровской эволюцией и спонтанной редукцией.

   1) Первый пример связан с обратимостью квантовой эволюции. Пусть мы ввели гамильтониан, способный обратить квантовую эволюцию системы кот-ящик.**[37], [38]** Хотя практически это почти невозможно, теоретического запрета нет. Если спонтанная редукция происходит, то процесс был бы необратим. Если спонтанной редукции нет, то система Кот-ящик вернётся в исходное чистое состояние. Однако, подобную проверку может сделать лишь внешний наблюдатель. Сам же Кот не сможет это сделать путем самонаблюдения, поскольку память Кота сотрётся, вернувшись в исходное состояние.
   С точки зрения внешнего наблюдателя никакого парадокса тоже не было, поскольку он спонтанной редукции, приводящей к парадоксу,  не наблюдал.
   2) Второй пример связан с необходимостью возврата квантовой системы к исходному состоянию, согласно теореме Пуанкаре, вытекающей из уравнения Шредингера. Пусть исходное состояние было чистое, а затем наблюдался процесс спонтанной редукции в процессе самонаблюдения Кота, приводящий к смешанному состоянию. Тогда возврат был бы уже невозможен – смешанное состояние уже не может перейти в чистое состояние согласно уравнению Шредингера. Таким образом, зафиксировав возврат, Кот пришел бы к противоречию между спонтанной редукцией и наличием возврата. Но Кот не сможет зафиксировать возврат (в случае верности квантовой механики), поскольку процесс возврата сотрет память Кота. А значит, нет и парадокса.

С точки зрения внешнего наблюдателя (который на самом деле может наблюдать этот возврат, измерив начальное и  конечное состояние этой системы) никакого парадокса тоже нет. На самом деле, внешний наблюдатель никакой спонтанной редукции, приводящей к парадоксу, и не наблюдает (Рис. 18).

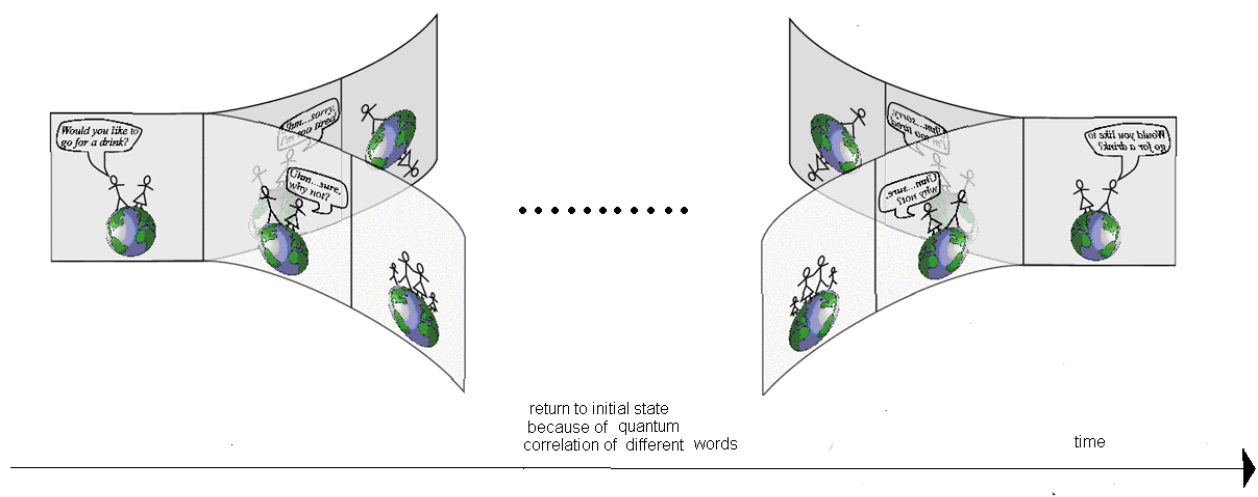

**Рис 18.** Возврат системы к исходному состоянию из-за квантовых корреляций между «Мирами» в Многомировой интерпретации. Этот возврат не может быть зафиксирован при самонаблюдении из-за стирания памяти.

Здесь следует отметить, что противоречие между спонтанной редукцией и Шредингеровской эволюцией может быть экспериментально наблюдаемо только в тех случаях, когда спонтанная редукция зарегистрирована в памяти наблюдателя и эта память не стерта и не повреждена. Все описанные выше эксперименты не попадают под это условие. Таким образом, эти примеры ясно показывают, что хотя спонтанная редукция и приводит к нарушению Шредингеровской эволюции,  это нарушение экспериментально не наблюдаемо.

3)  Приведем третий пример. Предположим, мы имеем суперпозицию живого и мертвого кота в ящике. Теоретически внешний наблюдатель всегда может провести измерение, одной из базисных функций которого будет суперпозиция живого и мертвого кота. Такое измерение не изменит состояния ящик-кот, в отличие от измерения, базисными функциями которого будет живой и мертвый кот. Сообщив коту о результатах измерения, мы войдем в противоречие с наблюдаемой котом спонтанной редукцией, которая имела место как до, так и после измерения. **[40], [41]** Такое рассуждение содержит двойную ошибку.

**Во-первых**, подобный эксперимент ставиться для *проверки* наличия спонтанной редукции Кота, когда наблюдателем является сам Кот. Он не должен влиять на память Кота в любом случае, включая возможность спонтанной редукции. Но внешний наблюдатель не влияет на память Кота только лишь тогда, когда спонтанной редукции нет, и состояние Кота является суперпозицией живого и мертвого состояний. Как мы уже упоминали выше, при возможности влиянии эксперимента на память наблюдателя он не легитимен для проверки достоверности теории.

С точки же зрения внешнего наблюдения (который не видит спонтанной редукции и не наблюдает, следовательно, и парадокса) подобная проверка вполне возможна и легитимна. Ведь она не влияет на память самого внешнего наблюдателя. Более того, подобные проверки, не нарушающие эволюцию наблюдаемой системы, позволяют измерять не только начальное и  конечное состояние системы, но и все промежуточные, т.е. вести непрерывное наблюдение системы, не оказывая на нее влияния!  Правда, создание в реальности системы способной проводить такие измерения – почти не выполнимая задача.

**Во-вторых**, передача данных коту о его состоянии, записывается в его память и таким образом меняет как состояние, так и всю дальнейшую эволюцию наблюдаемой системы, включающей Кота, т.е. система перестает быть изолированной.

Следует отметить, что для внешнего наблюдателя реальное наблюдение суперпозиции живого и мертвого кота теоретически возможно, но практически почти невыполнимо. В то же время для малых квантовых систем наблюдение суперпозиции вполне посильная задача. Это приводит к тому, что квантовую механику обычно рассматривают как теорию малых систем. Но для небольших макроскопических (**мезоскопических**) объектов такое наблюдение также возможно. Примером может быть большой коллектив частиц при низких температурах или некоторые фотонные состояния [43].

.

# 6. Разрешение парадоксов Лошмидта и Пуанкаре в классической механике. Объяснение закона роста макроскопической энтропии.

Мы также рассмотрим тут два случая – когда наблюдатель включен в наблюдаемую систему, и когда он находиться вне нее.

Основное противоречие классической статистической механики – это противоречие между законом возрастания энтропии и обратимыми классическими законами движения. Оно находит выражение в парадоксах Пуанкаре и Лошмидта.

В случае классической механики, в отличие от квантовой механики, более простой случай самонаблюдения, когда наблюдатель включен в описываемую систему. Возврат Пуанкаре системы в исходное состояние приводит к стиранию памяти, аналогично тому, как мы описали в предыдущей главе, и делает невозможным экспериментальное наблюдение парадокса Пуанкаре. Обращение скоростей действительно приводит к уменьшению энтропии, однако само направление времени относительное понятие, и  мы должны логично определить положительное направление стрелы времени. Разумно выбрать ее в направлении роста энтропии. Назовем ее **собственной стрелой времени** системы. Относительно такого направления стрелы времени энтропия растет, и парадокс Лошмидта исчезает. Следует отметить, что в разрешении обих парадоксов происходит как стирание памяти в конечном состоянии системы, так и рост энтропии в направлении *собственной* стрелы времени системы. При приближении к конечному состоянию направление собственной стрелы времени меняется (реверсируется) по отношению к стреле времени вблизи начального состояния. Основной причиной, делающей парадоксы ненаблюдаемыми, является **невозможность полного знания состояния системы для случая самонаблюдения**.

Для внешнего наблюдателя ситуация сложнее. Теоретически, взаимодействие между наблюдателем и наблюдаемой системой может быть сделано сколь угодно малым  в классической механике. Следовательно, ничто не препятствует наблюдать уменьшение энтропии. В этом случае направления собственных стрел времени наблюдателя и наблюдаемой системы противоположны.

Но возможно ли это на самом деле? Теоретически да, но практически это почти невозможно. Все дело в том, что  подавляющее большинство реальных физических систем являются  системами с перемешиванием (хаотическими системами). Это означает, что их фазовые траектории экспоненциально неустойчивы по отношению к малому шуму. Малое же взаимодействие наблюдателя или окружающей среды с наблюдаемой системой почти всегда неизбежно присутствует. Напомним, что для хаотических систем верна следующая теорема:

**Для внешнего наблюдателя процессы изменеия *макропараметров*, протекающие с уменьшением макроскопической энтропии, сильно неустойчивы по отношению к малому внешнему шуму. При этом процессы изменеия *макропараметров*, связанные с ростом макроскопической энтропии, являются наоборот устойчивыми.**

Поэтому неизбежное малое взаимодействие приводит к прекращению процесса с убыванием энтропии и *синхронизации направлений собственных стрел времени* наблюдателя и наблюдаемой системы. Система оказывается незамкнутой, и парадоксы верные для классической механики замкнутых систем недействительны. Малое взаимодействие со стороны окружающей среды на наблюдателя и наблюдаемую систему будет иметь тот же эффект, что и взаимодействие наблюдателя и наблюдаемой системы – синхронизация направлений всех собственных стрел времени подсистем. В этом случае роль наблюдателя более пассивна и естественна (Рис. 19).

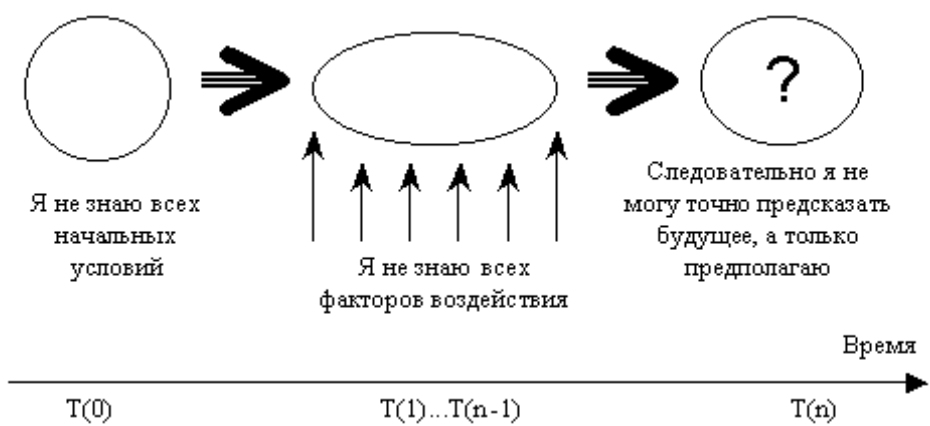

**Рис. 19** Две причины, приводящие к несоответствию реальных законов движения законам классической и квантовой динамики. Это внешний шум и погрешность начальных условий. Эти факторы приводят к разрешению парадоксов в квантовой механике и классической статистической физике. Они объясняют отклонение от законов идеальной динамики в этих парадоксах [105].

В литературе **[2], [28], [44], [45]** часто с недоумением рассматривается синхронность стрел времени в окружающем нас мире. Почему мы никогда не встречаем процессов с уменьшением энтропии, хотя их вероятность обнаружения равна вероятность обнаружения процессов, идущим с увеличением энтропии? Часто при этом пытаются найти объяснения в особенностях происхождения нашей Вселенной[8]. **[2], [28], [44], [45]** Ничего подобного не нужно. Достаточным объяснением является неизбежное очень малое взаимодействие между системами, синхронизирующее все собственные стрелы подсистем.

Таким образом, пути, ведущие к уменьшению энтропии, просто не устойчивы к малому шуму со стороны внешнего наблюдателя (Рис. 20). В случае квантовой механики такой шум даже теоретически неизбежен при измерении, если мы не знаем истиного начального состояния измеряемой системы. Измерение состояния приводит к неизбежному нарушению этого измеряемого состояния. В классической механике измерение может быть теоретически проведено сколь угодно точно. Поэтому мы должны вводить малый внешний шум и/или погрешности начального состояния «руками», чтобы объяснить рост энтропии для наблюдателя.

---

[8] Почему наша нынешняя Вселенная не была создана неким «Создателем» из Хаоса путем процесса с направлением времени, обратного теперешнему времени! **[44], [45]** Это было бы чем-то вроде гигантской флуктуации! Почему ее началом стало очень низкоэнтропийное состояние, приведшее к Большому Взрыву? Ответ дает статья Элицура **[60].** В ней он показывает, что оптимальным процессом для создания низкоэнтропийного состояния с точки зрения некого «Создателя» является создание начального еще более низкоэнтропийного состояния. Этот метод он называет «лыжный лифт». Он проводит аналогию с лыжным подъемником на гору, на которую стоит вначале полностью подняться, прежде чем частично спуститься.

В реальных измерениях такой малый внешний шум всегда присутствует. И требуются огромные усилия, приводящие к росту энтропии окружения, чтобы от этого шума избавиться. При этом такой рост энтропии намного превышает выигрыш от уменьшения энтропии наблюдаемой системы вследствие исчезновения шума. Таким образом, закон роста энтропии опять будет выполняться. Здесь есть аналогия с разрешением парадокса Демона Максвелла [49], [18], который использует сортировку молекул для уменьшения энтропии. Получение информации необходимой для сортировки приводит к увеличению энтропии, компенсирующее ее уменьшение в сортируемой системе.

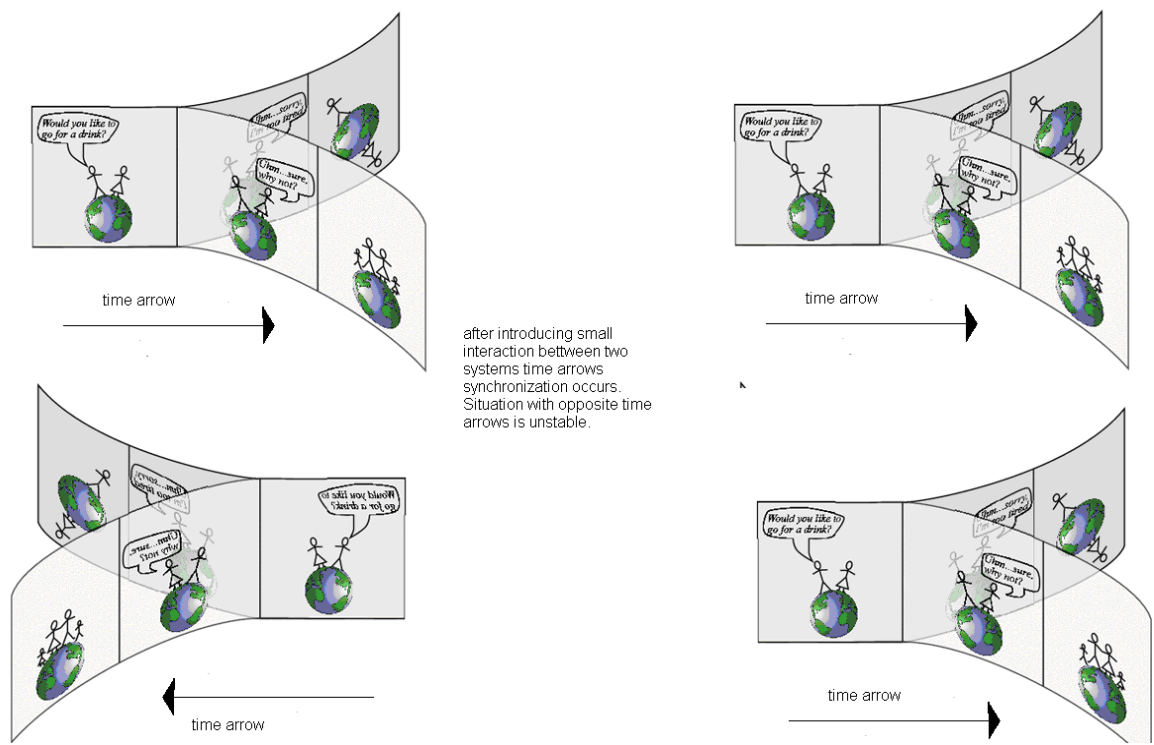

**Рис. 20** Состояние с противоположными стрелами времени у двух подсистем неустойчиво. Малое взаимодействие между системами приводит к синхронизации стрел времени. Поэтому всюду во Вселенной стрелы времени сонаправлены. Положительное направление стрелы времени идет в направлении роста энтропии. Поэтому энтропия, повсюду во Вселенной только растет.

Хотя борьба с малым и всепроникающим взаимодействием систем очень трудная задача, но для очень малых систем она легко выполнима. И мы повсеместно наблюдаем малые флюктуации, которые и являются наблюдаемым отклонением от закона роста энтропии. Также вполне возможно наблюдать обратимость и возвраты для динамики небольшого количества тел, если пренебречь трением. Это происходит в небесной механике.

# 7. Аналогия между квантовой механикой (КМ) и классической статистической механикой (КСМ). (Приложение N)

Из приведенных выше соображений можно догадаться, что существует почти полная аналогия между свойствами и парадоксами КСМ и КМ конечных замкнутых систем, а также между методами их разрешения. Приведем эти аналогии более подробно

1) Обе механики обратимы во времени.
2) Для обоих механик применима теорема Пуанкаре о возвратах. Однако если для КСМ системы с почти периодическими возвратами являются довольно узким классом систем, для КМ все системы в ограниченном объеме являются почти периодическими.
3) В КСМ возможны корреляции, вытекающие из знания макросостояния, и скрытые дополнительные микроскопические корреляции, связанные со знанием ее «истории». В квантовой механике также возможны два типа корреляции: классические корреляции, определяемые диагональными элементами матрицы плотности и сохраняющиеся при редукции и скрытые квантовые корреляции, определяющиеся недиагональными элементами и приводящие к парадоксам. Малый внешний шум со стороны наблюдателя или окружения уничтожает дополнительные корреляции в классических системах и приводит к огрублению функции фазовой плотности. Аналогично этому, запутывание наблюдаемой системы с наблюдателем (взаимодействие при измерении) или окружением (декогренция засчет взаимодействия) приводят к исчезновению квантовых корреляций (недиагональных членов матрицы плотности) и редукции волновой функции.
4) В случае самонаблюдения регистрация возвратов Пуанкаре или Лошмидта невозможна из-за стирания памяти. Из-за этого при самонаблюдении невозможна и регистрация дополнительных (в КСМ) или квантовых (в КМ) корреляций, приводящих к парадоксам.
5) В КМ и КСМ можно определяются два вида энтропии - энтропия ансамблей (или функции фазовой плотности) и макроскопическая энтропию. Энтропия ансамблей при обратимой эволюции остается неизменной, макроэнтропия может как возрастать, так и убывать. При самонаблюдении убывание энтропии становиться ненаблюдаемым. Для внешнего наблюдателя малое взаимодействие этого наблюдателя с наблюдаемой системой или системы с окружением также делает невозможным (или очень труднодостижимым) убывание энтропии. Для изоляции от окружения и наблюдения систем с убыванием энтропии требует увеличение энтропии большее, чем выигрывается от ее убывания в результате изоляции. Это связано с отсутствием в реальном мире абсолютно изолированных и непроницаемых полостей и бесконечно легких регистрирующих частиц, а также наличием малых, но всепроникающих взаимодействий.
6) Процесс спонтанной редукции в КМ связан с отбросом квантовых корреляций и переходом из чистого состояния в смешанное что, таким образом, приводит к увеличению макроскопической энтропии. Интересно, что подобным образом водиться и рост макроскопической энтропии для уравнения Больцмана в КСМ. Это достигается путем введением «гипотезы молекулярного хаоса», когда отбрасывается корреляции между парами частиц (т.е. их положения и координаты рассматриваются как независимые). При этом функция распределения двух частиц рассматривается как произведение одночастичных функций. Таким образом, введение спонтанной редукции в уравнения КМ эквивалентно введению закона возрастания макроскопической энтропии в уравнения КСМ.

7) Законы КМ статистические. Наблюдения в КМ при условии, что состояние системы нам неизвестно, неизбежно ведет к воздействию на эволюцию наблюдаемой системы. Поскольку большинство систем в КСМ являются системами с перемешиванием, в реальности их поведение также случайно. Это связано, во-первых, с наличием малого, но конечного взаимодействия с наблюдателем или окружением. Во-вторых, с конечной точностью знания начального состояния. Однако в пределе очень высокой точности измерений и изоляции их поведение можно предсказать сколь угодно точно. Достижение такой точности в реальности требует огромных затрат энтропии.

8) В обоих случаях парадоксы возникают лишь для макросистем. Законы поведения микросистем оказываются не применимыми к ним в полной мере из-за малого внешнего шума и конечной точности знания начального состояния. Единственная серьёзная разница между КМ и КСМ заключается в том, что в КСМ малое, но конечное взаимодействие при измерении (наблюдении) или малые погрешности начального состояния приходится вводить «руками», а в КМ они возникают естественно, само собой[9].

Подводя итог мы приходим к неожиданному выводу: парадокс Шредингеровского Кота в КМ является по сути дела квантовым аналогом закона возрастания энтропии в КСМ. Поскольку в КМ обычно исследуются микросистемы, а в КСМ обычно макросистемы, эти, по сути, эквивалентные парадоксы, имеют столь разную внешнюю форму (Рис. 21).

---

[9] Очень часто приводят примеры «чисто квантовых парадоксов», якобы не имеющих аналогии в классической статистической механике. Примером является парадокс Элитцура-Вайдмана [61] с бомбой, которую можно обнаружить без взрыва:

1) Пусть волновая функция одного кванта света разветвляется по двум каналам некого устройства. В конце эти каналы снова объединяются, и происходит интерференция двух волн вероятности. Внесение в один из каналов бомбы нарушит процесс интерференции и позволит таким образом обнаружить бомбу, даже если квант света не подорвет ее. ( Квант света считается способным взорвать бомбу)

2) Полной аналогией является следующий эксперимент КСМ. В один из рукавов, где нет бомбы, пустим макроскопический поток многих легких частиц. В другой канал, где может быть есть бомба – только одну легкую частицу. Эта частица не способна взорвать ее, но бомба может отклонить ее назад. Эта частица не обнаружима макроскопически из-за конечной чувствительности приборов. Но если поток частиц в конце канала имеет хаотическую динамику, даже наличие одной дополнительной частицы может сильно ее изменить (так называемый «эффект бабочки»). Это позволит зарегистрировать новую частицу, которая пройдет через второй канал, если бомбы нет. Полная аналогия КМ и КСМ!

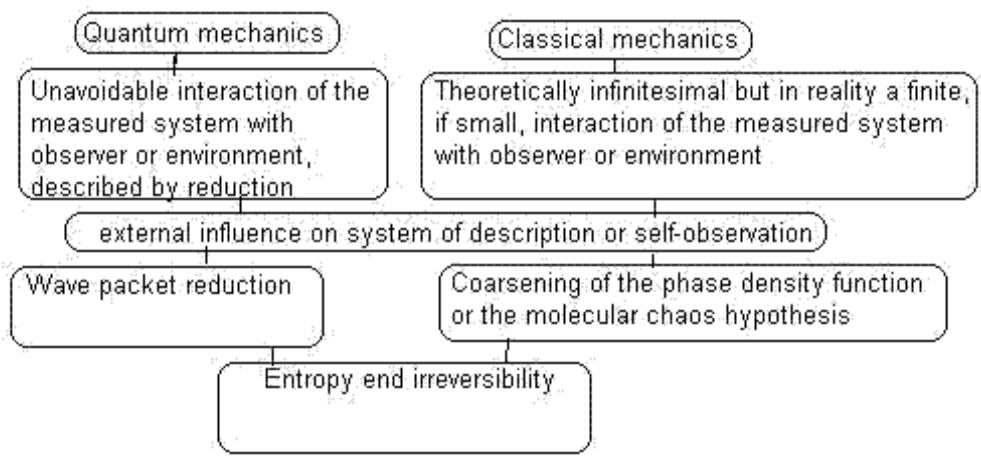

Figure 21. Sources of irreversibility and entropy in physics.

# 8. Синхронизация стрел времени/декогеренция [88-94].

Может появиться следующий вопрос. Предположим, что существует процесс, в котором уменьшается энтропия. Для определенности, позвольте нам рассмотреть спонтанную реконструкцию здания (ранее разрушенного землетрясением). Позвольте также взять простой пример газа, расширяющегося из маленькой области пространства в центре полости. Если после некоторого времени все скорости молекул газа будут полностью обращены, то газ вернется в стартовый маленький объем. Пусть мы включаем камеру, чтобы получить серию кадров, записывающих различные стадии спонтанной реконструкции здания/ (сжатия газа). Мы ожидаем, что камера сделает эту запись. Почему же камера не будет в состоянии сделать это? Что точно помешает камере?

Ответ этот вопрос следующий: даже очень маленькое взаимодействие между камерой и наблюдаемой системой разрушает обратный процесс уменьшения энтропии и приводит к синхронизации стрел времени наблюдателя и наблюдаемой системы. (Стрела времени определена в направлении увеличения энтропии). Это очень маленькое взаимодействие появляется из-за света, испущенного наблюдаемым объектом и отраженного камерой (и также из-за света, испущенного камерой). В отсутствии камеры роль наблюдателя может играть  среда, испускающая и отражающая свет. (Процесса без наблюдателя не имеет никакого физического смысла. Он должен появиться на некоторой стадии процесса. Но его влияние намного меньше чем влияние среды). Внешний шум (воздействие) от среды или наблюдателя разрушает корреляцию между молекулами наблюдаемой системы. Это приводит к прекращению обратного процесса с уменьшением энтропии. В квантовой механике такой процесс определен как "декогеренция". Здание не будет восстановлено / (газ не будет сжиматься). В противоположность этому процессы увеличения энтропии устойчивы.

Декогеренция (синхронизация стрел времени и перепутывание) и релаксация (во время релаксации система переходит в термодинамически равновесное состояние) являются совершенно различными физическими процессами! Во время релаксации макроскопические переменные (энтропия, температура, давление) сильно изменяются до равновесных величин, а невидимые микроскопические корреляции между частями системы увеличиваются. Во время декогеренции макроскопические переменные (энтропия, температура, давление) являются почти постоянными. Невидимые микроскопические корреляции внутри подсистем (среда, наблюдатель, наблюдаемая система) сильно разрушаются, а новые корреляции между этими подсистемами появляются. Это называют "перепутыванием" в квантовой механике. Во время этого процесса также происходит синхронизация стрел времени. Время релаксации намного больше чем время декогеренции.

Позвольте взять простой пример газа, расширяющегося из маленькой области пространства в большой объем. В этом процессе с увеличением энтропии изменение макроскопических параметров во времени устойчиво к маленьким внешним возмущениям. Если после некоторого времени все скорости молекул газа будут полностью обращены, то газ вернется в стартовый маленький объем; но это верно только в отсутствии любого внешнего возмущения. Этот процесс, идущий с уменьшением энтропии, с очевидностью нестабилен, и даже маленькое внешнее возмущение вызвало бы его переключение на процесс с ростом энтропии. Таким образом, процессы, идущие с увеличением энтропии устойчивы, а с ее уменьшением - нет.

Следующий пример - цитата из статьи Маккони [69], [70]:
"Однако, наблюдатель является макроскопическим по определению, и все удаленно взаимодействующие макроскопические системы становятся коррелированными очень быстро (Например, Борель сделал замечательное вычисление, что перемещение грамма материала на звезде Сириус на 1 м может влиять на траектории частиц в газе на Земле на временных масштабах порядка секунды [20])"

Но нет никакой проблемы полностью обратить во времени как наблюдателя (камера) так и наблюдаемую систему совместно. Из-за теоремы о возвращении Пуанкаре для замкнутой системы (включающей как наблюдателя, так и наблюдаемую систему) это должно произойти автоматически после очень большого времени. Однако стирание памяти наблюдателя не позволит ему зарегистрировать этот процесс.

Большинство реальных систем является *хаотическими*, т.е. слабое возмущение может привести к значительному расхождению траекторий, и также между двумя всегда есть ненезначительное взаимодействие между наблюдаемой системой и наблюдателем/средой. Но *в принципе* и в квантовой и в классической механике мы можем сделать невозмущающее наблюдение за процессом, идущим с уменьшением энтропии. Хороший пример такого мезоскопического устройства - это квантовый компьютер: никакого закона увеличения энтропии не существует для такой системы. Это устройство очень хорошо изолировано от среды и от наблюдателя. Но *на практике* невозмущающее наблюдение почти невозможно для макроскопических систем. Мы можем заключить, что закон возрастания энтропия - это *FAPP закон* (т.е. закон, применимый во всех *практических* случаях).

Позвольте рассмотреть синхронизацию стрел времени для двух систем, не взаимодействующих ранее некоторого начального момента времени. У этих систем стрелы времени были первоначально противоположны.
Это означает, что существует две невзаимодействующие системы таким образом, что в одной из них потоки времени (то есть, увеличение энтропии) идет в одном направлении, в то время как в другой время течет в другом (противоположном) направлении. Однако, когда они входят во взаимодействие друг с другом, тогда одна из них ("более сильная") перенаправит другую ("более слабую") таким образом, чтобы в конечном счете у них обоих было время, текущее в том же самом направлении, что и в "более сильной".

Что точно это означает быть "более сильным"? Это что-то, что увеличивается с числом степеней свободы системы? Это не верно. "Более сильная" или "более слабая" не зависит от числа степеней свободы систем. Для первой системы взаимодействие появляется в ее *будущем* после начального момента времени. (В начальный момент системы имеют *противоположно направленные* стрелы времени). Во второй системе взаимодействие было в ее *прошлом*. Таким образом ситуация не *симметрична во времени,* и первая система всегда "более сильная". Это происходит из-за неустойчивости процессов с уменьшением энтропии и стабильности процессов с увеличением энтропии, что было описано выше.

Действительно, предположим, что у нас есть два первоначально изолированных сосуда с газом. В первом газ расширяется (увеличение энтропии). Во втором газ сжимается (уменьшение энтропии).

В первом сосуде газ расширяется из маленького объема в центре сосуда. Скорости молекул направлены от центра сосуда к его границе. Физически очевидно, что маленькое изменение скоростей не может остановить расширение газа. Действительно, скорости после случайного маленького изменения продолжат быть направлены от центра сосуда к его границе. Возмущение может даже усилить расширение . Таким образом, процесс расширения устойчив.

Во втором сосуде газ сжимается из полного объема сосуда к его центру. Скорости всех молекул направлены к центру сосуда. Физически очевидно, что маленькое случайное изменение скоростей может легко остановить сжатие газа. Действительно, скорости даже после маленького изменения не будут направлены к центру сосуда. Таким образом, процесс сокращения будет остановлен. Следовательно, мы можем заключить, что процесс сжатия неустойчив. Этот процесс сжатия может быть получен путем обращения расширения газа во времени. Если мы полностью обращаем скорости молекул расширяющегося газа *перед* столкновениями молекул друг с другом и границей сосуда, такая неустойчивость будет линейна и слаба. Но при обращении скоростей уже *после* столкновений эта неустойчивость становиться экспоненциальной и значительно более сильной.

Оба направления времени играют одинаковую роль. Но маленькое случайное взаимодействие нарушает эту симметрию для описанных выше двух систем из-за неустойчивости процессов с уменьшением энтропии. Симметрия времени существует только для *полной* системы, включающей эти две, определенные выше подсистемы. Но стрелы времени взаимодействующих подсистем должны быть сонаправлены.

В действительности, взаимодействие в течение бесконечного временем может быть заменено взаимодействием в течение большого, но конечного времени Т, которое выберем много меньше, чем время возвращения Пуанкаре. Тогда в первой системе взаимодействие происходит во время [0, Т], а во второй - во время [-Т, 0]. Остаются ли наши аргументы по-прежнему релевантными? Вместо асимметрии сил в этом случае мы получаем асимметрию начальных условий: Для первой системы координат ([0, Т]) у этих двух сосудов различные собственные стрелы времени в начальный момент 0. Однако, для второй системы координат [-Т, 0] у этих двух сосудов одинаковые собственные стрелы времени в отрицательном направлении в начальный момент -Т. Только если время Т точно равно времени возврата Пуанкаре, ситуация будет действительно симметрична. Для такой ситуации две собственные стрелы времени являются также различными в момент Т, но противоположны их направлениям в момент времени 0. Снова у "более сильной" системы силы взаимодействия находятся в ее будущем относительно собственной стрелы времени.

Это теория объясняет, почему энтропия растет во всех частях Вселенной в одном направлении. Однако она не может объяснить низкую начальную величину энтропии Вселенной. Думается, что это результат антропного принципа **[66].**

# 9. Закон возрастания энтропии и синхронизация стрел времени/декогеренция в теории гравитации.

В общей теории относительности Эйнштейна движение так же, как и в классической механике обратимо. Но имеется и важное отличие от классической механики. Это *неоднозначность* решения задачи Коши: получения конечного состояния системы из полного набора начальных и граничных условий. В общей теории относительности два различных состояния за *конечное* время могут дать бесконечно близкие состояния. Это происходит, например, при образовании черной дыры в результате коллапса. Рассмотрим обратный процесс, описывающий белую дыру. В этом процессе бесконечно близкие начальные состояния за *конечное* время могут дать разные конечные состояния. Это означает, что наблюдатель/окружение, даже бесконечно слабо взаимодействующий с белой дырой может значительно повлиять на ее эволюцию за конечное время. Это свойство приводит к тому, что закон возрастания энтропии превращается из приближенного (FAPP, для всех практических целей) закона в точный, а энтропия становится фундаментальным понятием. Действительно, появляется такая фундаментальная величина, как энтропия черной дыры. Появление этой энтропии также можно объяснить пертурбацией (создаваемой наблюдателем), которая в отличие от классической механики теперь может быть даже бесконечно малой. Образование черной дыры идет с увеличением энтропии. Обращение времени приводит к появлению белой дыры и ведет к уменьшению энтропии.

Белая дыра не может существовать в реальности по тем же причинам, что невозможны процессы с уменьшением энтропии в классической механике: из-за ее неустойчивости (много более сильной, чем в классической механике) и вытекающей из этого синхронизации собственных стрел времени белой дыры и наблюдателя/окружения. Направление собственной стрелы времени белой дыры меняется на противоположное, совпадающее со стрелой времени наблюдателя/окружения. Белая дыра превращается в черную.

Здесь же возникает и знаменитый информационный парадокс [71] – информация, которая в классической и квантовой механике сохраняется, в черной дыре исчезает навсегда. Казалось бы здесь нет никакой проблемы – возможно внутри черной дыры она храниться в какой-либо форме. Однако хаотическое излучение Хокинга делает этот процесс потери информации явным – черная дыра испаряется, а информация не восстанавливается.

Излучение Хокинга относится к квазиклассической гравитации. Однако парадокс может быть сформулирован и в рамках общей теории относительности. Сферическая черная дыру можно превратить в белую дыру в некоторый момент и процесс обращается во времени (Это трудно осуществимо в реальности, но похожее преобразование рассмотрено в [95]. Там обращение во времени заменено «кротовой норой» из черной дыры в белую дыру, находящуюся в другой Вселенной) Информация при этом может не восстанавливаться из-за неоднозначности (бесконечно большой нестабильности) эволюции белой дыры.

Обычно рассматривают только два ответа на этот вопрос. Либо информация действительно пропадает, либо из-за внутренних корреляций излучения Хокинга (или точного обращения сжатия черной дыры при ее превращении в белую дыру) информация сохраняется. Но, скорее всего, верным является третий ответ. Из-за неизбежного влияния наблюдателя/окружения экспериментально различить эти две ситуации просто невозможно! А что нельзя проверить экспериментально, не является предметом науки.

Как для общей теории относительности, так и для квазиклассической гравитации разрешение парадокса находится с помощью учета влияния наблюдателя/окружения. Даже если бы излучение Хокинга было коррелированным, а не случайным (или белая дыра была в точности обратна черной дыре) бесконечно малое влияние наблюдателя/окружения приводило бы к нарушению этих корреляций за конечное время и делало бы потерю информации неизбежной. Включение наблюдателя в систему

бессмысленно – полное самоописание и самоанализ невозможен. Закон сохранения информации не может быть проверен для такого случая, даже если он выполняется.

Мы не имеем сейчас законченную теорию квантовой гравитации. Однако для частного случая 5 мерного анти-де-Ситтеровского мира этот парадокс ныне многими учеными считается разрешенным в пользу сохранения информации, вследствие гипотезы о AdS/CFT дуальности, т. е. гипотезы о том, что квантовая гравитация в анти-де-ситтеровском (то есть с отрицательным космологическим членом) 5-мерном пространстве математически эквивалентна конформной теории поля на 4-поверхности этого мира. Она была проверена в некоторых частных случаях, но пока не доказана в общем виде. Полагают, что если эта гипотеза действительно верна, то это автоматически влечёт за собой разрешение проблемы об исчезновении информации. Дело в том, что конформная теория поля, по построению, унитарна. Если она дуальна квантовой гравитации, то значит и соответствующая квантовогравитационная теория тоже унитарна, а значит, информация в этом случае не теряется.. Отметим, что это не так. Процесс образования черной дыры и ее дальнейшее испарение происходит на *всей* поверхности анти-де-Ситтеровского мира (описываемого квантовой теорией поля), который включает также и наблюдателя/окружение. Но наблюдатель не может точно знать начальное состояние и анализировать систему, частью которой он сам и является! А значит не может и сделать свое влияние на нее пренебрежимо малой даже теоретически. Таким образом, экспериментальная проверка информационного парадокса снова становится невозможной!

Рассмотрим с точки зрения энтропии и такой парадоксальный объект общей теории относительности, как кротовая нора [72] (червоточина). Выберем ее вариант, предложенный Торном [73]. Путем очень простой процедуры (погружение одного из концов на космический корабль, его движение со скоростями сравнимыми со световыми, а затем возвращение этого конца на прежнее место) пространственная кротовая нора может быть преобразована во временную (wormhole traversing space into one traversing time). В том числе она может быть использована как машина времени, приводя к знаменитому парадоксу дедушки. Как же может быть разрешен этот парадокс?

Для макроскопических кротовых нор разрешение может быть найдено с помощью закона возрастания энтропии, обеспечиваемого неустойчивостью процессов с убыванием энтропии и вытекающей из этого синхронизацией стрел времени.

Действительно, пространственная кротовая дыра не приводит к парадоксу. Объекты поглощенные ее одним концом выходят из другого конца в более позднее время. Таким образом, объекты из более упорядоченного низкоэнтропийного прошлого попадают в менее упорядоченное высокоэнтропийные будущее. При движении вдоль кротовой норы они также переходят из более упорядоченного состояния в менее упорядоченное. Таким образом, собственные стрелы времени путешествующего в кротовой норе объекта и окружающего мира сонаправлены. Тоже верно для путешествия по временной кротовой норе из прошлого в будущее.

Однако для путешествия из будущего в прошлое стрелы времени путешественника в кротовой норе и окружающего мира будут уже противоположны. Действительно, сам объект путешествует из менее упорядоченного будущего в более упорядоченное прошлое, но при этом его собственная энтропия растет, а не убывает! Как мы говорили ранее, такой процесс неустойчив и будет предотвращен (принудительно обращен) процессом синхронизации стрел времени в тот момент, когда движущийся вход кротовой норы вернется в начальное положение. «Свобода воли» позволяет нам инициировать лишь устойчивые процессы с ростом энтропии, но не с ее убыванием. Таким образом, мы не сможем послать объект из будущего в прошлое. Процесс синхронизации стрел времени и вытекающий из него закон роста энтропии запрещает начальные условия, необходимые для путешествия макроскопических объектов в прошлое и реализацию парадокса дедушки.

В работе [74] доказывается, что собственная термодинамическая стрела времени не может все время иметь одинаковую ориентацию с собственной координатной стрелой времени при путешествии по замкнутой временноподобной траектории (closed timelike curve) вследствие закона роста энтропии. Процесс синхронизации стрел времени (связанный с бесконечно большой неустойчивостью, неоднозначностью процессов с убыванием энтропии) и является тем самым *физическим механизмом*, который фактически обеспечивает как эту невозможность, так и выполнение закона роста энтропии для всех подсистем вдоль их единой собственной стрелы времени.

Для микроскопических кротовых нор ситуация совершенно иная. Если начальные условия совместимы с путешествием в прошлое по кротовой норе, нет никаких причин, которые могут помешать ему. Если сколь угодно малое (даже бесконечно малое малое!) изменение начальных условий приводит к противоречию с существованием кротовой норы, она может всегда быть легко разрушена [75]. Действительно, здесь реализуется уже ранее отмеченное замечательное свойство экстремальной неустойчивости общей теории относительности: бесконечно малое изменение начальных условий может повлечь значительное изменение конечного состояния за конечное время!

Однако, это не является решением парадокса дедушки, который является макроскопическим, а не микроскопическим явлением. Действительно, предположим, что существуют два процесса с противоположными стрелами времени: весь Космос и космонавт, путешествующий по червоточине из будущего Космоса в его прошлое. Тогда для собственной стрелы времени космонавта это будет путешествием из его прошлого в его будущее. Для общей теории относительности описанная выше ситуация невозможна даже в принципе (в противоположность с классической механикой): даже бесконечно малое взаимодействие приводит к синхронизации стрел времени из-за бесконечно большой неустойчивости (неоднозначности) процессов с убыванием энтропии (в данном случае процесс с убыванием энтропии - это космонавт, путешествующий из будущего в прошлое). Эта синхронизация стрел времени может сопровождаться как разрушением червоточины [75], так и сохранением червоточины и изменением только начальных условий [74]. Но в действительности закон о росте энтропии (и соответствующая синхронизация стрел времени) *не позволяет даже появление* таких ситуаций с противоречием между макроскопическими начальными условиями и первоначально определенной (заданной и неизменной) макроскопической топологией пространства-времени (включающей ряд червоточин) [74] . Сформулируем окончательный вывод: *для макроскопических процессов* бесконечно большая неустойчивость (неоднозначность) процессов с убыванием энтропии и сопутствующая ей синхронизация стрел времени делает невозможным появление начальных условий несовместимых с существованием заданных червоточин, и тем самым предотвращает как их разрушение, так и путешествия по ним макроскопических тел в прошлое, приводящее к «парадоксу дедушки»

Подведем и общие итоги. Мы видим необыкновенную вещь. Те же самые соображения, которые нам позволили разрешить парадокс редукции, парадоксы Лошмидта (Loshmidt) и Пуанкаре позволяют разрешить информационный парадокс черных дыр и парадокс дедушки для кротовых нор. Замечательная универсальность!

# 10. Идеальная и Наблюдаемая Динамика.

## 10.1 Определение Идеальной и Наблюдаемой Динамики. Почему необходима Наблюдаемая Динамика?

Мы видим, что точные уравнения квантовой и классической механики описывают **ИДЕАЛЬНУЮ** динамику, которая обратима и приводят к возвратам Пуанкаре. Уравнения физики, описывающие **НАБЛЮДАЕМУЮ** динамику, например, уравнения гидродинамики вязкой жидкости, уравнение Больцмана в термодинамике, закон роста энтропии в изолированных системах (*master equations*) - необратимы и исключают возвраты Пуанкаре в исходное состояние (Рис. 22).

. Дадим определения Наблюдаемой и Идеальной Динамик, а также объясним необходимость введения Наблюдаемой Динамики. **Идеальной Динамикой** мы будем называть точные законы квантовой или классической механики. Почему мы назвали их идеальными? Потому что для большинства реальных систем выполняется закон возрастания энтропия или спонтанная редукция, противоречащие законом Идеальной Динамики. Как видно из объяснения этих противоречий, нарушения Идеальной Динамики связано или с незамкнутостью измеряемых систем (т.е. объясняется влиянием внешней среды или наблюдателя), или невозможностью полного самоизмерения и самоанализа для замкнутых и полных физических систем, включающих как внешнюю среду, так и наблюдателя. Что же делать в таких случаях? Реальная система незамкнута или неполна, т.е. Идеальная Динамика невозможна и мы должны отказаться от использования физики? Отнюдь нет! Очень многие такие системы могут быть описаны уравнениями точной (или вероятностной) динамики, несмотря на незамкнутость или неполноту описания. Мы будем называть её **Наблюдаемой Динамикой**. Большинство уравнений физики, называемые *master equations* (такие как, например, как уравнения гидродинамики вязкой жидкости, уравнение Больцмана в термодинамике, закон роста энтропии в замкнутых системах) являются уравнениями Наблюдаемой Динамики.

Для того чтобы обладать указанным выше свойством Наблюдаемая Динамика должна отвечать определенным условиям. Она не может оперировать полным набором **микропеременных**. В Наблюдаемой Динамике мы определяем лишь много меньшее число **макропеременных**, которые являются некими функциями микропеременных. Это делает ее много устойчивее по отношению к ошибкам в задании начальных условий и шуму. Действительно,изменение микросостояния не приводит неизбежно к изменению макросостояния, поскольку одному макросостоянию отвечает большой набор микросостояний. Для газа макропеременными являются, например, плотность, давление, температура и энтропия. Микропеременными же являются скорости и координаты всех его молекул.

Как из Идеальныой Динамики получается Наблюдаемая Динамика? Они получаются или введением в идеальные уравнения малого, но конечного внешнего шума или же введением погрешностей начального состояния. Погрешности и/или шумы должны быть достаточно большими, чтобы нарушить ненаблюдаемую реально обратимость движения или возвраты Пуанкаре. С другой стороны они должны быть достаточно малы, чтобы не влиять на протекание реально наблюдаемых процессов с ростом энтропии.

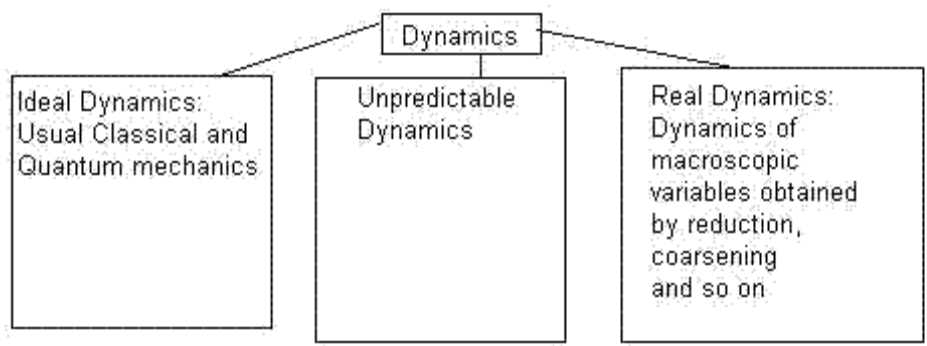

Figure 22   Three types of Dynamics.

Наблюдаемая Динамика для процессов с ростом энтропии дает результат совпадающий с Идеальной Динамикой. Однако обратные процессы, идущие с уменьшением энтропии, и возвраты Пуанкаре в ней невозможны.

**Возможность введения Наблюдаемой Динамики связана с указанной выше устойчивостью процессов, идущих с ростом энтропии, по отношению к ошибкам в задании начальных условий и внешнему шуму.** С другой стороны, **процессы, идущие с уменьшением энтропии**, и возвраты Пуанкаре **неустойчивы** даже к малому шуму и погрешностям начальных условий. Но эти явления и не наблюдаются в реальных экспериментах.

С чем связана (при удачном выборе макропеременных) в Наблюдаемой Динамике устойчивость по отношению к шуму для процессов, идущих с ростом энтропии? На две основные причины указал ещё Шредингер [11].

Первая уже нами была указана. Она заключается в том, что макросостояние определяется параметрами огромного числа молекул. Хотя внешний шум может сильно менять состояние отдельной молекулы, суммарный их вклад в макросостояние остается неизменным. Это связано с законом больших чисел в теории вероятности [56]. Примером термодинамического закона, связанного с эти, может служить макроскопические законы движения жидкости или газа.

Вторая связана с дискретностью состояний в квантовой механике. Поскольку разные квантовые состояния дискретны и сильно отличаются энергией, малый шум не может их изменить. Это приводит, например, к устойчивости химической связи и позволяет рассматривать, например, термодинамику макромолекулы

Зачем нужно использовать именно Наблюдаемую, а не Идеальную динамику, если они дают одинаковый результат во всех важных экспериментальных ситуациях, связанных с ростом энтропии? Потому что **описание, даваемое Наблюдаемой Динамикой, много проще, чем даваемое Идеальной**. Наблюдаемая Динамика отбрасывает ненаблюдаемые процессы (вроде обратных процессов с убыванием энтропии или возвраты), оперирует с меньшим числом переменных и описывается более простыми уравнениями. Кроме того, она позволяет отвлечься от малого внешнего шума или неполноты описания системы, позволяя сделать точное или вероятностное описание системы.   .

Мы знаем, что истинной теорией является Идеальная Динамика. Наблюдаемая Динамика дает отличные от нее результаты. Можно ли экспериментально обнаружить эту разницу между теориями, при условии истинности Идеальной Динамики? Та или иная теория может быть верной или неверной только в том случае если может быть поставлен реальный эксперимент, способной опровергнуть теорию. Такая теория называется **фальсифицируемой** в смысле определенном **Карлом Поппером [42].** Предположим, что Иеальная Динамика верна. Является ли Наблюдаемая Динамика фальсифицируемой в определенном выше смысле?

**Для полной физической системы, включающей наблюдателя, наблюдаемую систему и окружающую среду наблюдаемая динамика не фальсифицируема в смысле Поппера (при условии верности Идеальной Динамики).** Т.е. разницу между Идеальной и Наблюдаемой Динамикой в этом случае невозможно наблюдать в эксперименте.[10]

Для измеряемой системы без включения в нее наблюдателя и окружения Наблюдаемая Динамика **в принципе фальсифицируема**. Для этого надо исключить даже очень малое шумовое взаимодействие с окружающей средой и подготовить некое точно заданное начальное состояние, а затем измерить конечное и сравнить полученный результат с теорией. Тут возникает соблазн спросить: а может на самом деле исходно верной является не Идеальная, а некая Наблюдаемая Динамика **[10], [44], [45]**? Например, предположим, что спонтанная редукция происходит для достаточно больших макроскопических систем. То-есть, спонтанная редукция наблюдается не только при самонаблюдении, но также и для внешнего наблюдателя при полной изоляции макросистемы от шума окружения. Однако в случаях небольших изолированных систем, когда Идеальная Динамика экспериментально проверяема, она всегда оказывается верной.[11]

Однако, поскольку для макроскопических систем исключить это очень малое шумовое взаимодействие практически невероятно трудно, то выполняется **практическая нефальсифицируемость** Наблюдаемой Динамики для измеряемой системы. Т.е.

---

[10] *Тем не менее, она может иметь самостоятельную ценность, позволяя более просто проводить вычисления или понять физику явления.*

*Так, например, Галлилеевская система отсчета, связанная с солнцем, позволила гораздо проще и точнее рассчитать динамику планет, чем Птолемеевская, связанная с Землей. Хотя за начало отсчета мы вправе выбрать любую из них. Соответственно выбор «что вокруг чего вращается» остается за нами и определяется лишь красотой описания и нашим удобством.*

*Аналогично этому, в математической науке выбор определений и аксиом определяется лишь нашим удобством и требованием непротиворечивости аксиом. Теория, объясняющая как делать тот или иной выбор, отсутствует (в отличие от теоремы Геделя о неполноте) Обычно прибегают к аргументу «красоты» и «всеобщности» теорем, получающихся при подходящим задании аксиом и определений. Однако эти вещи требуют более точных определений.[13]*

[11] *Подобные эксперименты существуют и проводятся на системах промежуточного размера между макро и микро, так называемых мезоскопических системах. Все эти эксперименты подтверждают Идеальную, а не Наблюдаемую Динамику. В этих экспериментах наблюдается квантовая интерференция (при отсутствии спонтанной редукции) и происходят флюктуации энтропии.[43],[10].*

*Однако для действительно больших макроскопических систем аналогичный эксперимент поставить в ближайшее время вряд ли будет возможно. В фундаментальной физике аналогичная ситуация существует для Струнных Теорий и Великих Объединений. Эксперимента, который сможет подтвердить или опровергнуть их , нужно ждать сотни лет, если не случиться чуда. Впрочем, и в теории гравитации Эништейна, которая проверена точно лишь для не слишком большой силы гравитации, ситуация похожа. (вспомним, например, загадочное темное вещество и энергию, а также новые теории гравитации Мильгрома [53] и Логунова [54]).*

теоретически опровергнуть её можно, но в реальном эксперименте сделать это очень трудно

Сделаем тут очень важное примечание. Вполне возможны и случаи, когда введение такой динамики невозможно и система остается все-таки непредсказуемой, вследствие незамкнутости или неполноты описания. Это случай **Непредсказуемой Динамики**, обсуждаемой в следующей главе.

## 10.2 Чем ограничивается выбор макроскопических переменных Наблюдаемой динамики?

Очень важно отметить, что **выбор макропараметров не может быть произвольным** (Рис. 23). Наблюдаемая динамика должна приводить к закону возрастания энтропии и необратимости. Следовательно, макропеременные должны вводиться таким образом, чтобы макросостояния и описываемые ими процессы, ведущие к росту энтропии (определяемой этими макропеременными), были устойчивы по отношению к внешнему шуму. Они также должны происходить аналогично процессам, описываемых идеальной динамкой изолированных систем при возрастании энтропии. Процессы идеальной динамики, связанные с уменьшением макроскопической энтропии, должны быть наоборот неустойчивыми и нарушаться. Это условие накладывает серьезные ограничения на выбор возможных макроскопических состояний. Так в классической статистической механике, если мы рассмотрим набор всех микросостояний соответствующих некому макросостоянию, то в фазовом пространстве они имеют вид компактной, выпуклой капли. Это макросостояние и его динамика, идущая с ростом энтропии, будут устойчивыми по отношению к малому внешнему шуму.

Возьмем набору точек, соответствующему расплывшейся фазовой капле с множеством узких ответвлений («рукавов») и обратим скорости всех молекул. Такому ансамблю не соответствует никакого макросостояния. Хотя ансамбли (т.е. наборы точек в фазовом пространстве), обладающие таким свойством существуют и соответствуют начальным состояниям процессов с убыванием энтропии. Невозможность введения такого макросостояния связана с тем, что подобное макросостояние было бы неустойчивым к малому шуму. Неустойчивостью (для систем с перемешиванием) объясняется и невозможность выбора в качестве макропараметров (даже в качестве некого предельного случая) самих микропараметров , т.е. скоростей и координат всех молекул. Конечно, еще одной причиной для этого является и огромное число таких параметров.

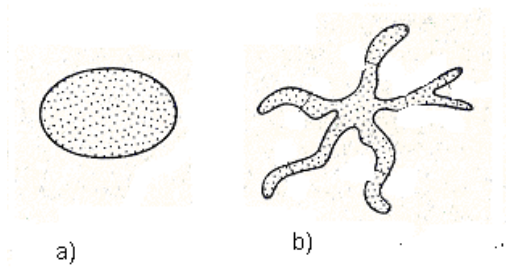

a)                    b)

**Рис 23** a) возможное макросостояние    b)невозможное макросостояние

Аналогично этому в КМ часто возникает вопрос: почему для основных макросостояний Шредингеровского кота мы выбираем состояния живой или мертвый кот, а не их разницу и сумму **[14]**? Связано это с тем, что состояния живой или мертвый кот устойчивы к флюктуациям внешней среды. В тоже время, их сумма или разность в очень короткое время запутываются с окружением и переходят в смесь живой и мертвый кот. Мы уже раньше назвали этот процесс декогеренцией. Он происходит много быстрее, чем все иные термодинамические релаксационные изменения. **[15], [16], [47], [48]** Т.е. выбор состояний как именно живой или мертвый кот диктуется необходимостью устойчивости к внешнему шуму.

Какое же свойство уравнений идеальной динамики приводит к приоритету подобных состояний? Это свойство **локальности взаимодействия**. Сильно взаимодействуют друг с другом лишь близкие в пространстве молекулы. Поскольку состояния живой и мертвый кот пространственно сильно разделены их суперпозиция легко редуцируется в их смесь. **Определение таких приоритетных макроскопических состояний (*pointer states*) в случае квантовой механики описано в работах Зурека [39], [31].** Если бы сила взаимодействия между молекулами или атомами определялась бы не близостью координат, а, например, близостью импульсов частиц ситуация была бы совсем другая.

Для систем близких к состоянию термодинамического равновесия приоритетные макроскопические состояния (*pointer states*) соответствуют собственным функциям энергии. В энергетическом представлении при термодинамическом равновесии матрица плотности диагональна.

Отметим тут, что выбор, как уравнений Наблюдаемой Динамики, так и набора макропеременных неоднозначен. Существует огромное количество различных и непротиворечивых между собой Наблюдаемых Динамик. По сути дела, все **master equations** термодинамики (например, уравнения гидродинамики вязкой жидкости, уравнение Больцмана в термодинамике, закон роста энтропии в изолированных системах ) являются уравнениями Наблюдаемой Динамики. Различные Наблюдаемые Динамики отличаются как по степени их «макроскопичности», так и по выбору конкретных наборов макропеременных.

Равновесный ансамбль, находящийся в равновесии с термостатом, в квантовой механике в энергетическом представлении описывается диагональной матрицей плотности. Аналогично этому, в классической механике, в равновесии отсутствуют корреляции между молекулами, аналогами неdiagональных элементов матрицы плотности. Нарушение равновесия проявляется **двояко**. **Во-первых**, через наличие макроскопических корреляций макропараметров. Они определяются неравновесными значениями диагональных элементов матрицы плотности в энергетическом представлении для КМ. Эти корреляции исчезают при переходе к равновесному состоянию в процессе, называемом релаксацией. **Во-вторых**, через неустойчивые микроскопические корреляции (квантовые корреляции в КМ или дополнительные корреляции в КСМ). Квантовые корреляции соответствуют не равным нулю неdiagональным элементам матрицы плотности в энергетическом представлении. Эти микроскопические корреляции много более неустойчивы и гораздо быстрее затухают, чем макроскопические. Процесс их исчезновения был назван выше декогеренцией. Время декогеренции много меньше времени релаксации.

## 10.3 Методы получения Наблюдаемой Динамики.(Master equations)

Рассмотрим методы получения Наблюдаемой Динамики. В соответствии с двумя основными причинами, приводящими к необходимости применения Наблюдаемой Динамики (внешний шум или неполнота знаний о состоянии системы) мы можем разбить все методы получения Наблюдаемой Динамики на две группы.

Первый метод связан с введением малого неконтролируемого шума от внешнего большого термостата (например, вакуум является термостатом с нулевой температурой). **[15]** Этот шум уничтожает дополнительные корреляции, приводящие к возвратам и обратимости.

Второй метод связан с неполнотой знания о состоянии полной физической системы для самонаблюдения, что позволяет провести огрубление («размазывание») функции, описывающей состояния системы **[35]**, **[2]**, (**Приложение К**). Для КСМ такой функцией является фазовая плотность, для КМ – матрица плотности. Если мы рассмотрим случай КСМ, то огрублением («размазыванием») может служить сглаживание (усреднение) функции фазовой плотности в окрестности каждой точки с некоторым периодом времени. Между сглаживаниями эволюция описывается обычными уравнениями Идеальной Динамики. Для КМ аналогичная процедура связана с периодической редукцией волновой функции **[3]** и использования уравнения Шредингера во временных промежутках между редукциями. Подобные методы огрубления уничтожают дополнительные корреляции, приводящие к возвратам и обратимости. Как мы уже отмечали выше, эти корреляции эспериментально не наблюдаемы для случая самонаблюдения.

Наблюдаемая динамика (при самонаблюдении) должна описывать поведение системы (с ростом энтропии) и давать результаты, совпадающие с Идеальной Динамикой, только в течение некоторого ограниченного интервала времени.Это время много меньшего времени возврата. Ведь система просто не может быть наблюдаема экспериментально в течение большего времени из-за стирания памяти наблюдателя при возвратах.

## 10.4 Разрешение парадокса Зенона с точки зрения Наблюдаемой Динамики. Экспоненциальный распад частиц– закон Наблюдаемой, а не Идеальной Динамики.

В связи с необходимостью введения Наблюдаемой Динамики, мы можем тут вспомнить ранее уже упомянутый квантовый **парадокс Зенона**. Разрешение второй части этого парадокса (о не экспоненциальном характере распада) мы находим в рамках Наблюдаемой Динамики.

Пусть количество нераспавшихся частиц N (с небольшой возможной ошибкой) является измеряемым макропараметром. В начальный момент $t_0$ было $N_0$ нераспавшихся частиц. Идеальная квантовая динамика дальнейшего распада не экспоненциальная. Но квантовое измерение распада неизбежно вносит искажение в Идеальную динамику, делая ее неприменимой. Мы можем лишь свести помехи, вносимые измерением, к минимуму, выбрав большие интервалы между измерениями. Временной интервал между измерениями (редукциями) с одной стороны должен быть достаточно большим, чтобы не влиять существенно на динамику распада. ($\Delta t \gg \hbar/\Delta E$, где $\Delta t$ – временной интервал между редукциями, $\hbar$ –постоянная планка, $\Delta E$ – разница энергий между состояниями распавшаяся и не распавшаяся частицы ) С другой стороны, его нужно выбрать меньше среднего времени жизни частицы.( $\hbar/\Delta E \ll \Delta t < \tau$, где $\tau$ – среднее время жизни распадающейся частицы). Общее время наблюдения процесса должно быть много меньше времени возврата Пуанкаре для замкнутых систем конечного объема (включающей наблюдателя), поскольку подобные возвраты не наблюдаемы из-за стирания памяти наблюдателя. ($n \cdot \Delta t \ll T_{return}$ , где n – число наблюдений [редукций] , $T_{return}$ – время возврата Пуанкаре) Пусть временной интервал между измерениями распавшихся частиц и общее время наблюдения выбраны корректно, т.е. отвечают всем этим условиям. Для такой ситуации полученный закон распада является уже строго экспоненциальным и не зависит от конкретной точной величины выбранного временного интервала между измерениями (редукциями). Этот **экспоненциальный закон распада ( N=$N_0$·exp(-(t-$t_0$)/τ) ) является уже законом Наблюдаемой динамики, а не Идеальной динамики**, согласно самому определению понятия Наблюдаемой динамики.

## 10.5 Примеры различных методов получения Наблюдаемых Динамик путем «огрубления» («размазывания»): уравнение Больцмана и Новая Динамика Пригожина.

Примером Наблюдаемой Динамики является **уравнение Больцмана**.[1],[2] Огрубление («размазывание») (**Приложение K**) проводиться при его получении в два этапа. Вначале фазовая функция заменяется функцией распределения для одной частицы. Это, по сути, соответствует усреднению фазовой плотности по всем частицам, кроме одной. Полученное уравнение для одночастичной функции остается обратимым уравнением Идеальной динамики и зависит от двухчастичной функции распределения. Уравнение Наблюдаемой Динамики получаются путем дальнейшего огрубления. Вводится «гипотеза молекулярного хаоса». Корреляции между любыми двумя частицами принимаются нулевыми, и двучастичная функция распределения заменяется произведением двух одночастичных функций. Подставляя полученную двучастичную функцию в полученное ранее уравнение, приходим к необратимому и нелинейному уравнению Больцмана. Это очень напоминает редукцию в КМ, когда все корреляции между возможными результатами измерения отбрасываются (т.е. обнуляются недиагональные элементы матрицы плотности).

Еще одним примером получения Наблюдаемой динамики методом огрубления может служить «**Новая Динамика» Пригожина. [9], [12], (Приложение L)** Она дает очень красивые методы получения уравнений. Как само огрубление («размазывание»), так и уравнения движения (получающиеся простой подстановкой функции, обратной огрубленной, в уравнения Идеальной динамики) являются линейными. При этом в КСМ фазовая функция распределения в окрестности каждой точки огрубляется анизотропно. Как мы описывали ранее для систем с перемешиванием, в окрестности каждой точки есть направление, вдоль которого траектории расходятся экспоненциально (**направление расширения**). Также есть и направление, вдоль которого траектории сходятся экспоненциально (**направление сжатия**). Только вдоль направления сжатия и производиться огрубление («размазывание») фазовой функции.

Рассмотрим макроскопическое состояние, которому соответствует некая компактная выпуклая «фазовая капля». Первоначальная компактная «фазовая капля», растекаясь по фазовому пространству, приобретает много «веток». Направление сжатия перпендикулярно этим «веткам». Поэтому огрубление функции вдоль них приводит к росту площади растекшейся «фазовой капли», и, соответственно, росту числа микросостояний и энтропии.

Теперь рассмотрим обратный процесс. Начальное состояние определяется набором точек фазового пространства, полученных из конечного состояния прямого процесса (растекшейся «фазовой капли») путем обращением скоростей всех молекул. При обращении скоростей форма растекшейся «фазовой капли» не меняется. Но направление сжатия из-за обращения скоростей уже не перпендикулярно, а параллельно её «веткам». Поэтому при огрублении этой вдоль направления сжатия площадь растекшейся фазовой капли» почти не меняется. Соответственно, число микросостояний и энтропия тоже почти не меняются в противоположность с увеличением энтропии растекшейся «фазовой капли» при прямом процессе (Рис. 24).

Таким образом, **анизотропное огрубление нарушает симметрию по отношению к обращению времени**, и полученные уравнения оказываются также необратимыми.

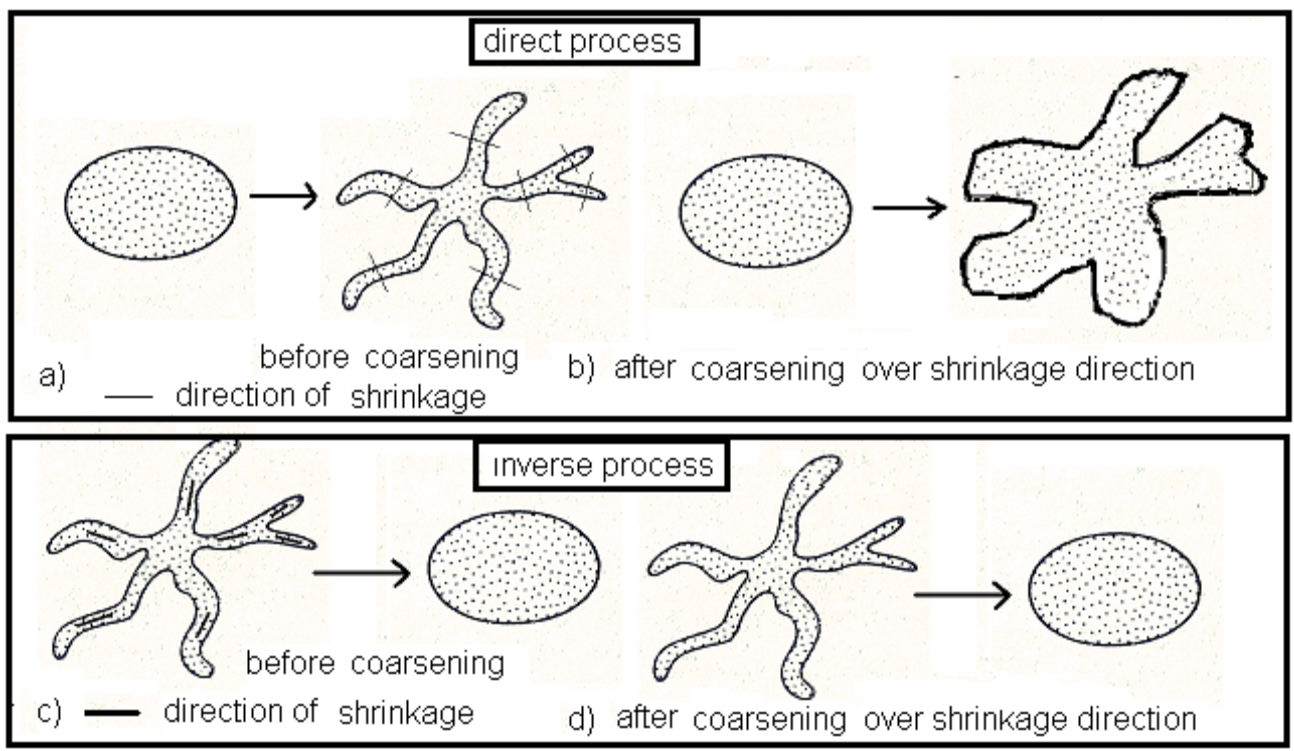

**Рис. 24** Огрубление («размазывание») фазовых капель в направлении сжатия приводит к нарушению обратимости во времени. Прямой и обратный процессы после огрубления протекают уже не симметрично. («Новая динамика» Пригожина.)

Рассмотрим «Новую Динамику» для квантовых систем. Подобную линейную анизотропную процедуру огрубления можно провести и в случае КМ, но лишь для бесконечно больших квантовых систем ( в бесконечном объеме или бесконечным числом частиц). Конечные квантовые системы с конечным числом частиц являются почти периодическими, и имеют конечное время возврата. Для них «Новая Динамика» уже не применима[12].

Пригожин предлагает рассматривать все реальные квантовые системы как бесконечно большие. Решение, однако, может быть много проще. Наблюдаемая динамика имеет смысл только в течение времени, много меньше времени возврата. Описание в течение большего времени не имеет смысла. Ведь система не может быть наблюдаема экспериментально в течение большего времени из-за стирания памяти при возвратах (при самонаблюдении). В течение такого времени в термодинамическом пределе (объем стремиться к бесконечности, но отношение числа частиц к объему остается постоянной) большие конечные системы, для которых существует такой предел, почти не отличаются от бесконечных систем. Если же у нас имеется система, для которой термодинамический предел вообще не существует (например, макромолекула) ее можно просто погрузить в резервуар (окружение) для которого такой предел есть. Все эти приемы позволяют использовать уравнения «Новой Динамики» и для конечных систем квантовой механики.

Часто «Новую Динамику» подвергают критике.[28] Основной аргумент при этом заключается в том, что все парадоксы КМ и КСМ можно объяснить, не прибегая к «Новой Динамике». Это верно, и мы проделали это в данной работе. Соответственно, говорят эти критики, «Новая Динамика» не нужна и избыточна! Однако нужно рассматривать «Новую Динамику» не как замену КМ или КСМ (что, к сожалению, пытается сделать и сам Пригожин), а лишь как одну из полезных форм Наблюдаемой Динамики. Она позволяет

---

[12] *Также Новая Динамика не применима к почти периодическим системам КСМ. Впрочем, большинство реальных систем КСМ является системами с перемешиванием (хаотическими). Поэтому для КСМ эта проблема менее важна*

удобно описать физические системы, исключая ненаблюдаемые в реальности обратимость и возвраты. В этом и состоит ее преимущество.

Обсудим теперь ситуацию, когда Наблюдаемую Динамику ввести не возможно. Это случай мы назвали выше Непредсказуемой Динамикой.

# 11. Непредсказуемая Динамика.

Не во всех случаях нарушенная внешним шумом (или неполная при самонаблюдении) Идеальная Динамика может быть заменена предсказуемой Наблюдаемой Динамикой. Для ряда систем их динамика становиться в принципе непредсказуемой. Назовем динамику, описывающую такую систему, **Непредсказуемой Динамикой**. Как следует из самого определения таких систем, для них невозможно ввести макропараметры, характерные для Наблюдаемой Динамики и предсказывать их поведение. Их динамика не описывается и не предсказывается научными методами. Таким образом, **наука сама ставит границы своей применимости.**

Мы не сомневаемся в верности и универсальности основных законов физики. Но невозможность полного знания состояния или законов динамики системы (из-за взаимодействия с наблюдателем и окружением или неполноты при самонаблюдении) делает в некоторых случаях невозможным их полную экспериментальную проверку. Это дает нам некоторую свободу их произвольно изменять, не входя в противоречие с экспериментом. Когда эти изменения приводят к предсказуемой динамике системы, она называется Наблюдаемой Динамикой. В тех же случаях, когда никакая предсказуемая динамика становиться вообще невозможна, мы говорим уже о Непредсказуемой Динамике.

Приведем несколько примеров Непредсказуемой Динамики.

1) Точки фазовых переходов или точки бифуркаций. В этих точках макроскопическая система, описываемая Наблюдаемой Динамикой, в процессе эволюции во времени или в процессе изменения какого-либо внешнего параметра может перейти не в одно, а в несколько различных макроскопических состояний. То есть, в этих точках Наблюдаемая Динамика теряет свою однозначность. В этих точках возникают огромные макроскопические флюктуации и использование макропараметров становиться бессмысленным. Эволюция также становиться непредсказуемой, т.е. возникает Непредсказуемая Динамика.

2) В современных космологических моделях есть дополнительные явления, кроме уже описанных выше явлений. Они связаны с потерей информации и ведут к неполноте нашего знания о состоянии системы и, следовательно, непредсказуемости. Так, за конечное время разные начальные состояния вследствие коллапса приводят к одинаковому конечному состоянию - Черной Дыре. Это приводит к потере информации в Черной Дыре и непредсказуемости динамики системы, ее включающей. (Таким, например, однозначно непредсказуемым явлениям, например, как Хокингское испарение Черной Дыры.) Тот же эффект потери информации имеет и ускоренное расширение Вселенной – появляются ненаблюдаемые области, откуда до нас не доходит даже свет. Следовательно, они ненаблюдаемые, и содержащаяся в них информация потеряна. Это опять ведет к непредсказуемости.

3) Возьмем микроскопическую или мезоскопическую систему, описываемую Идеальной Динамикой, изолированную от внешней декогеренции. Ее динамика зависит от неконтролируемых микроскопических **квантовых корреляций**. Эти корреляции очень неустойчивы и вследствие декогеренции (т.е. запутывания с окружением или наблюдателем) исчезают. Пусть некий первый наблюдатель фиксирует лишь начальное и конечное состояние системы. В промежутке времени между ними система полностью или почти изолирована от окружения или этого наблюдателя. В таком случае эти микроскопические корреляции не исчезают и влияют на динамику. Рассмотрим другого внешнего наблюдателя, не знающего начального состояния системы. В отличие от первого наблюдателя, знающего начального состояния системы, поведение системы для второго наблюдателя становиться непредсказуемым! Т.е. с точки зрения такого наблюдателя возникает Непредсказуемая Динамика. В квантовой области примерами таких систем являются **квантовые компьютеры** (Рис. 25) и **квантовые криптографические передающие системы** (Рис. 26) **[16], [47], [48]** .

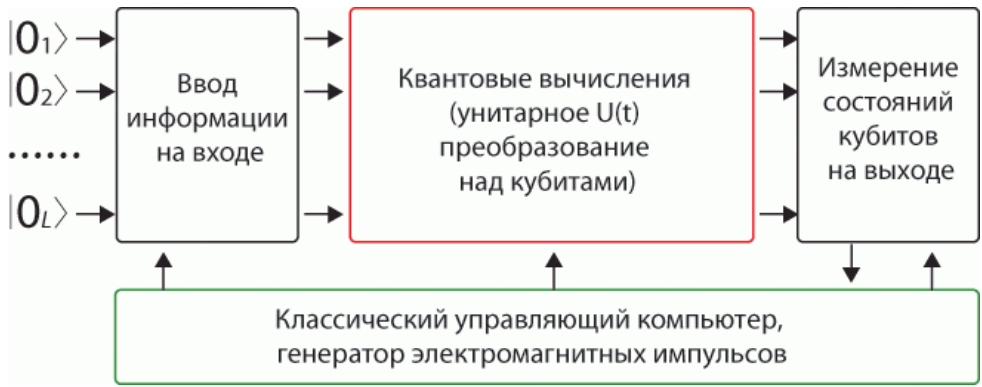

**Рис. 25** Схема квантового компьютера

Квантовые компьютеры имеют не только свойство непредсказуемости для наблюдателя, не информированного об их состоянии при запуске вычислений. Другим важным свойством является их высокая параллельность вычисления. Оно достигается за счет того, что начальное состояние является суперпозицией многих возможных начальных состояний «квантовых битов информации». За счет линейности уравнений квантовой механики эта суперпозиция сохраняется и «обработка» всех состояний, входящих в суперпозицию, происходит одновременно (параллельно). Эта параллельность приводит к тому, что многие задачи, которые обычный компьютер решает очень медленно из-за того, что рассматривает все случаи последовательно, квантовый решает очень быстро. С этим свойством и связаны надежды на практическую пользу квантовых компьютеров.

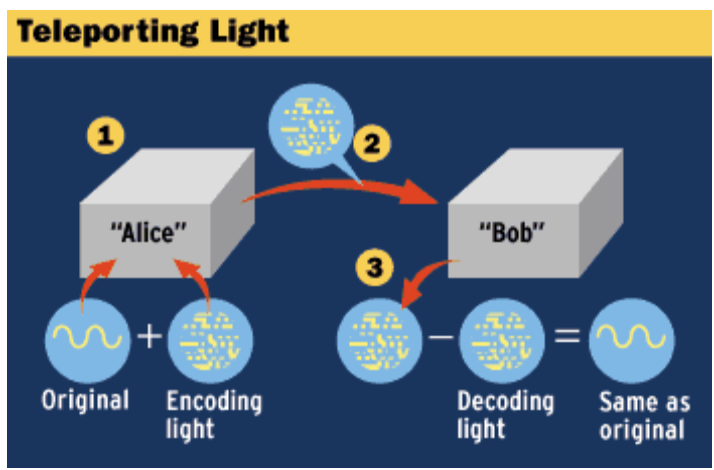

**Рис 26.** Квантовое кодирование в квантовых криптографических передающих системах. (Рис. из [102])

Квантовые криптографические передающие системы используют в первую очередь свойство своей непредсказуемости «передаваемых сообщений» (на самом деле, просто неких специальных систем, описываемых квантовыми законами) для внешнего наблюдателя, не информированного об их состоянии при начале передачи. Любая попытка прочесть передаваемое сообщение приводит к его взаимодействию с этим наблюдателем и, следовательно, «разрушению» передаваемого сообщения и невозможности прочесть это сообщение. Таким образом, перехват сообщений оказывается **в принципе невозможным** по законам физики.

Следует особо отметить, что вопреки широко распространенному мнению, как квантовые компьютеры так и квантовая криптография **[16], [47], [48]** имеют классические аналоги. Действительно, в классических системах в отличие от квантовых систем измерение можно провести абсолютно точно, не искажая измеряемое состояние. Однако и в классических хаотических системах имеются **неконтролируемые и неустойчивые микроскопические дополнительные корреляции**, обеспечивающие обратимость и возвраты системы. Введём «руками» конечное, но малое взаимодействие в классическое измерение или конечную погрешность в начальные условия, которые в реальных ситуациях, и на самом деле, всегда существуют. Они стирают разницу между классической и квантовой системой, как мы уже отмечали выше. В реальных системах всегда присутствует млый внешний шум, выполняющий эту роль. Изолируя хаотическую классическую систему от этого шума, ма получаем **классические аналоги изолированных квантовых устройств с квантовыми корреляциями [86-87]**.

Аналогом квантовых компьютеров являются **молекулярные компьютеры [52, 86-87]**. Большое количество молекул обеспечивает параллельность вычислений. Неконтролируемые и неустойчивые микроскопические дополнительные корреляции, обеспечивающие обратимость и возвраты системы, делают динамику неопределенной для наблюдателя, не информированного о состоянии компьютера в момент старта. Введение конечного, но малого взаимодействия приведет к тому, что такой наблюдатель нарушит нормальный запланированный ход вычислений при попытке чересчур точно померить координаты и скорости молекул, чтобы предсказать результат работы компьютера.

Аналогичные аргументы могут быть использованы для создания **классических криптографических передающих систем**, использующих явление классических неустойчивых микроскопических дополнительных корреляции. Неустранимое малое взаимодействия с перехватчиком сообщений разрушает эти корреляции. Тем самым оно делает ненаблюдаемый перехват принципиально невозможным также и в классическом случае.

4) Сохранение неустойчивых микроскопических корреляций может быть обеспечено не только за счет пассивной изоляции от внешней среды и наблюдателя, но и за счет динамического, компенсирующего помехи механизма. Это происходит в так называемых физических **стационарных системах**, в которых равновесие системы поддерживается за счет непрерывного **потока энергии или вещества через систему**. Примером могут служить микромазеры **[25]** – маленькие и хорошо проводящие полости с электромагнитным излучением внутри. Размер полостей настолько мал, что излучение уже необходимо описывать квантово. Оно постепенно затухает из-за взаимодействия со стенками. Эту систему оптимально описывать матрицей плотности в базисе состояний, соответствующих различным собственным энергиям системы. Этот базис наиболее устойчив к внешним шумам для любой системы близкой к термодинамическому равновесию и, следовательно, наиболее подходит для Наблюдаемой Динамики. Микроскопические корреляции соответствуют недиагональным элементам матрицы плотности и стремятся к нулю много быстрее, чем диагональные элементы при затухании излучения. (Иными словами, время декогеренции много меньше времени релаксации.) Однако пропускание через микромазер пучка возбужденных частиц приводит к сильному замедлению затухания недиагональных элементов матрицы плотности (иными словами микрокорреляций) и отличному от нуля стационарному излучению.

5) Примером **очень сложных стационарных** систем являются **живые системы**. Они находятся в состоянии очень далеком от термодинамического равновесия и крайне сложны. Они упорядочены, хоть эта упорядоченность сильно отличается от периодичного неживого кристалла. Низкоэнтропийное неравновесное состояние живого поддерживается за счет роста энтропии в окружении[13]. Неравновесное состояние поддерживается за счет метаболизма – непрерывного потока вещества и энергии через живой организм. С другой стороны само это неравновесное состояние является катализатором метаболического процесса, т.е. создает и поддерживает его на необходимом уровне. Поскольку состояние живых систем является сильно неравновесным, оно может поддерживать и существующие неустойчивые микрокорреляции, препятствуя процессу декогеренции. Эти корреляции могут быть как между частями самой живой системы, так и между живой системой и другими (живыми или неживыми) системами. Если это происходит, то динамику живой системы можно отнести к Непредсказуемой Динамике. Несомненные успехи молекулярной биологии позволяют предсказать и описать многие черты динамики живых систем. Но нет никаких фактов, свидетельствующих, что она будет способна **полностью** описать всю сложность процессов в живой системе, даже с учетом ее дальнейших достижений. Довольно трудно проанализировать реальные живые системы в рамках концепций Идеальной, Наблюдаемой и Непредсказуемой Динамик из-за их огромной сложности. Но возможно построить математические модели гораздо менее сложных неравновесных стационарных систем с метаболизмом и понять возможную роль всех трех Динамик для таких систем. Эта важная задача для дальнейшей работы физиков и математиков в этой области. Некоторые шаги в этом направлении сделаны в ниже следующей главе о Синергетике.

Следует отметить очень важное обстоятельство. Неустойчивые микрокорреляции существуют не только в квантовой, но и в классической механике. Например, классические системы с перемешиванием. Следовательно, подобные модели не должны носить только квантовый характер. Они могут быть и классическими! Очень многие допускают эту ошибку, считая, что только квантовая механика может описывать подобные явления.**[44], [45]** Как мы уже много раз указывали выше, введение «руками» малого, но конечного взаимодействия при классическом измерении и погрешностей начального состояния стирает разницу между свойствами квантовой и классической механики (при наличии неустойчивых микрокорреляций).

6) Описанными выше случаями не описывается все многообразие Непредсказуемых Динамик. Нахождение точных условий, при которых Идеальная Динамика переходит в Наблюдаемую и Непредсказуемую Динамику – еще полностью не решенная задача для математики и физики. Также такой еще полностью не решенной проблемой (и, по-видимому, связанной с предыдущей задачей) является роль этих трех динамик в сложных стационарных системах. Решение этой проблемы позволит глубже понять физические принципы, лежащие в основе жизни. Этому вопросу мы посвятим следующую главу.

---

[13] Так, например, растет энтропия Солнца, служащего источником энергии для жизни на Земле.

# 12. Жизнь и смерть.

Отметим с самого начала, что если предыдущие главы носили более или менее строгий характер, то данная глава в силу очевидных причин носит более гипотетический характер и является скорее набором гипотез.

Будем исходить в данной статье из положения, что жизнь полностью соответствует законам физики.

Следующие вопросы должны быть обсуждены ниже:

Что такое жизнь и смерть с точки зрения физики?

Есть ли у живой материи некие свойства не совместимые с физикой?

Чем живые системы отличаются от неживых систем с точки зрения физики?

Когда у живых систем появляется сознание и свобода воли с точки зрения физики?

Жизнь определяется, обычно, как особая высокоорганизованная форма существования органических молекул, обладающая способностью к обмену веществ, размножению, адаптации, движению, реакцией на внешние раздражители, способностью к самосохранению в течение долгого времени или даже повышению уровня самоорганизации. Это верное, но слишком узкое определение: многие из живых систем обладают лишь частью из этих свойств, некоторые из них присущи и неживой материи, вполне возможны и неорганические формы жизни.

## 12.1 Жизнь с точки зрения физики – предыдущие работы.

Первую попытку описать жизнь с точки зрения физики дал Шредингер [11]. В своей работе он определил жизнь как апериодический кристалл, т.е. высокоупорядоченную[14], но не основанную на простом повторении, в отличие от кристалла, форму материи. Он также привел две причины, делающие Наблюдаемую Динамику живых систем устойчивой к их внутреннему и внешнему шуму: статистический закон больших чисел и дискретность квантовых переходов, обеспечивающую устойчивость химических связей. Сам принцип действия живых организмов он уподобляет часам: и там и там возникает «порядок из порядка» несмотря на высокую температуру.

Следующий вклад в понимание жизни сделал Бор [26]. Он обратил внимание, что полное измерение состояния системы вносит в квантовой механике неизбежные искажения в поведение системы, чем возможно и объясняется принципиальная непознаваемость жизни. Критика этих взглядов Бора Шредингером [19] не состоятельна. Он пишет, что полное знание состояния квантовой системы возможно. Просто в отличие от классического случая полное знание состояния квантовой системы позволяет предсказать будущее лишь вероятностно. Проблема, однако, состоит в том, что измерение вносит неизбежные искажение в поведение системы, более сильные, чем вытекающие из вероятностного характера КМ. Т.е. предсказание нельзя делать даже вероятностно! В отсутствии измерение поведение было бы иным, чем при его наличии [20]. Измерение нарушает дополнительные (или квантовые) корреляции между частями системы, меняя ее поведение. Таким образом, измерение делает поведение системы в принципе непредсказуемым, а не только вероятностным. Это относится не только к квантовой механике, но и к классической механике, где между реальными системами существует малое конечное взаимодействие.

---

[14] Т.е. обладающую низкой энтропией. Живая система «питается» негоэнтропией из окружения. Таким образом, это принципиально открытая система.

В своей работе Бауэр[15] [17] определил, что не только высокая упорядоченность (и, соответственно, низкая энтропия) проявляются не только в неравновесности распределения веществ в живой материи, но и сама структура живой материи является низкоэнтропийной и сильно неустойчивой. Эта неустойчивая структура не только поддерживается за счет процесса обмена веществ, но и сама является катализатором этого обмена. Казалось бы, белки или вирусы имеют упорядоченную структуру и в кристаллической форме. Однако внутри живой материи можно встретить их более высокоупорядоченные и низкоэнтропийные модификации. С течением времени, тем не менее, происходит постепенная деградация структуры. Это приводит к неизбежности смерти и необходимости размножения для самого сохранения жизни. Т.е. процесс обмена веществ лишь очень сильно замедляет распад сложной структуры живой материи, а не поддерживает ее все время неизменной. Экспериментальные результаты, приведенные Бауэром, подтверждают выделение энергии и соответственного увеличение энтропии в процессе автолиза. Автолиз - это распад живой материи в отсутствие поддерживающего ее обмена веществ. На первой стадии процесса энергия выделяется из-за разрушения сильно неустойчивой исходной структуры материи, а на второй стадии процесса - в результате действии протолитических (разлагающих) ферментов, освобождающихся или появляющихся при автолизе. Наличие этой избыточной *структурной* энергии Бауэр и считал неотъемлемой характеристикой жизни.

Во всех указанных выше работах дано рассмотрение отдельного живого организма, в то время как жизнь, как совокупность всех организмов в целом (биосфера) также может быть рассмотрено. Сюда же относится и вопрос о происхождении и источнике жизни. Наиболее полный и современный ответ на эти вопросы с точки зрения физики был дан в работе Элицура [18]. В ней он рассматривает источник жизни как ансамбль саморазмножающихся молекул. Проходя через сито Дарвиновского естественного отбора, жизнь накапливает в своих генах информацию (точнее *полезную* информацию (знания) в определении, данном Элицуром) об окружающей среде, повышая, тем самым, уровень своей организации (негоэнтропии) в соответствии со вторым началом термодинамики. Ламаркизм в его слишком прямолинейной формулировке приведен в противоречие с этим законом физики. Обзор широкого спектра работ в этой области также сделан в этой статье [18]. К ее недостаткам относятся:

1) Рассмотрение верно для жизни в целом, как явления, но не для отдельно взятого живого организма.

2) Предложенное доказательство отвергает лишь слишком грубую, прямолинейную модель Ламаркизма, в то время как есть много гипотез и опытов, иллюстрирующих возможность реализации его элементов даже в реальной жизни [32]. Идея Ламаркизма заключается в том, что полезные изменения, происходящие в организме в течение его жизни, запоминаются и передаются потомству. Это противоречит Дарвиновской концепции отбора, когда случайные изменения происходят в наследственном веществе и удачные из них закрепляются у потомков естественным отбором, а неудачные отбрасываются.

3) За самоорганизующимися диссипативными системами, предложенными Пригожиным, например, ячейками Бернара [50], отрицается свойство адаптации, в отличие от живых организмов. Естественно, их адаптивные способности несравнимы с живыми системами, но в зачаточной форме, тем не менее, существуют. Так, например, ячейки Бернара [50] меняют свою геометрию или даже исчезают, как функция разницы температур между нижним и верхним слоем жидкости. Это и есть примитивная форма адаптации.

---

[15] Талантливый советский биофизик, трагически погибший в сталинских лагерях. Он намного опередил свое время.

## 12.2 Жизнь как процесс сопротивления процессу релаксации и декогеренции. Сохранение при этом как макроскопических корреляций, так и дополнительных классических (или квантовых) корреляций.

Жизнь была определена Бауэром как самоподдерживающаяся за счет движения и обмена веществ сильная неустойчивость. Живые системы сопротивляются процессу перехода из неустойчивого состояния в более устойчивое. Но мы можем предположить, что большая часть этой неустойчивости проистекает также из сильно неустойчивых (дополнительных в КСМ или квантовых в КМ) микроскопических корреляций. Живые системы стремятся поддержать и сохранить эти корреляции, сопротивляясь также процессу декогеренции. Следует отметить, что живые системы могут поддерживать такие корреляции как между своими внутренними элементами, так и с окружением.

Напомним, что в физических системах мы выделяем два типа корреляций. Первый – это устойчивые к малому внешнему шуму макроскопические корреляции между параметрами системы. Например, связь между давлением, плотностью и температурой для идеального газа. Второй тип – это неустойчивые микроскопические корреляции, которые приводят к обратимости и возвратам Пуанкаре, как в квантовых, так и в классических системах. Декогеренция разрушает эти корреляции, нарушая обратимость и предотвращая возвраты. Это проявляется в законе возрастания энтропии. Мы предполагаем, что живые системы обладают способностью сохранять эти неустойчивые корреляции, тормозя или предотвращая процесс декогеренции.

Проведем сравнение живых и неживых систем по их свойству сопротивляться разрушению и переходу к термодинамическому равновесию. Системы могут активно сопротивляться процессу релаксации. Вспомним живые или неживые стационарные системы, обменивающиеся энергией или веществом с окружением. Однако система может сопротивляться процессу не только процессу релаксации, но и декогеренции. В неживых системах это достигается пассивным путем, т.е. просто изоляцией системы от окружения. В живых открытых системах это достигается путем активного взаимодействия с окружением, внешним и внутренним движением, метаболизмом.

Способность жизни поддерживать не только макроскопические, но дополнительные (или квантовые) микроскопические корреляции делает жизнь в принципе непредсказуемой, как и предполагал Бор. Причем здесь необязательна квантовая механика, подобные корреляции присущи и классической механике. Аналогом квантовых корреляций являются дополнительные корреляции в КСМ.

Успехи молекулярной генетики не опровергали наличие неустойчивых и непредсказуемых корреляций, возможно свойственных жизни. Построение Наблюдаемой Динамики жизни в принципе возможно. Действительно, живые системы - это открытые системы, активно взаимодействующие со случайным окружением. Внешний наблюдатель взаимодействует с ними обычно много слабее и не может вызвать принципиальное изменение в их поведении. Однако попытка понять и предсказать жизнь слишком подробно и детально может нарушить сложные и тонкие корреляции, сохраняемые жизнью. Она может привести к Непредсказуемой Динамике живых систем и, т.е. дать предсказанный Бором эффект.

## 12.3 Синергетические системы - модели физических свойств сложных (живых) систем.

Введем понятие синергетических физических систем. Будем называть таковыми простые физические или математические системы, иллюстрирующие в зачаточной форме некие действительные или предполагаемые свойства сложных (живых) систем. Для нас важно создание в первую очередь синергетических моделей систем, способных противостоять шуму (декогеренции в КМ). Они сохраняют свои внутренние корреляции (квантовые или классические), приводящие к обратимости движения или возвратам Пуанкаре.

Существуют три метода сделать это

1) Пассивный метод – создание неких «стенок» непроницаемых для шума. Примером могут служить модели современных квантовых компьютеров.

2) Активный метод , обратный пассивному– подобно диссипативным или живым системам, сохраняющими своя неравновесное состояние благодаря активному взаимодействию и обмену энергией и веществом с окружением (метаболизмом). Думается, что будущие модели квантовых компьютеров должны браться из этой области.

3) Когда корреляции охватывают ВСЮ Вселенную. Внешний источник шума здесь просто отсутствует. Источник корреляций Вселенной в том, что Вселенная произошла из малой области путем Большого Взрыва. Назовем это явление глобальными корреляциями.

Следует отметить два обстоятельства:

1) Многие сложные системы в своем развитии проходят динамические точки бифуркации – когда существуют несколько альтернативных путей развития и выбор конкретного из них зависит от малейших изменений состояния системы в точке динамической бифуркации [62-64]. Тут даже слабые (и сохраненные, указанными выше путями) корреляции могут оказать огромное влияние. Наличие подобных корреляций ограничивает предсказательную силу науки, но отнюдь не ограничивает нашу личную интуицию. Поскольку мы являемся неотделимой частью нашего мира, то мы вполне способны на субъективном уровне «ощущать» эти корреляции, недоступные научному предсказанию (Но, ни в кое мере не противоречащие самим законам физики!)

2) Как пассивная, так и активная защита требуют огромных затрат негоэнтропии, которая черпается из окружения, поэтому суммарная энтропия системы и окружения только растет. Закон возрастания энтропии остается незыблемым для «большой» системы (наблюдаемая система + окружение + наблюдатель), хотя неверен для самой наблюдаемой системы. Уменьшения энтропии в большой системе согласно уравнениям Идеальной динамики происходят, но являются ненаблюдаемыми, как объяснено выше. Поэтому они могут просто игнорироваться.

Приведем примеры синергетических моделей физических процессов.

Так, рост кристаллов моделирует способность живых систем к размножению. Кстати, анализ этих систем позволяет найти слабое место [27] в аргументах Вигнера [6] , видящем противоречие между способностью к размножению и квантовой механикой. Положим, что взаимодействие между размножающейся **квантовой** системой и окружением **случайно**. В этом случае Вигнер показывает, что вероятность размножения почти нулевая. Однако, на самом деле, это размножение не случайно, а задается кристаллической решеткой уже образовавшегося кристаллика. В случае же живых организмов оно определяется последовательностью нуклеотидов в ДНК и тоже не случайно. [27]

Активное замедление процессов декогеренции, более близкое к методам поддержания корреляций в живых системах, достигается в таких открытых системах, как микромазеры, уже описанные выше [25], которые являются еще одним примером синергетических систем.

Еще одним активным механизмом предотвращения декогеренции является, так называемая «**квантовая телепортация**»[47]. В результате этого процесса создается точная копия произвольного квантового состояния, сопровождающаяся разрушением самого копируемого состояния. (Создавать копии любого квантового состояния без разрушения его самого невозможно.**[47])** Проводя подобный процесс много раз с малыми интервалами времени, можно долгое время сохранять исходное состояние квантовой системы, препятствуя декогеренции. Этот процесс эквивалентен сохранению исходного измеренного состояния в парадоксе «котелка, который никогда не закипит», наблюдаемого при многократном и частом измерении текущего состояния. Отличие состоит в том, что в процессе «квантовой телепортации» копируемое состояние остается не только неизменным, но и неизвестным.

Диссипативные системы тоже являются активными синергетическими системами и иллюстрируют свойства открытых живых систем к замедлению релаксации, поддержанию низкой энтропии и примитивной адаптации к изменению условий окружающей среды.

Другой пример - это квантовые изолированные системы (например, современные квантовые компьютеры при низких температурах). Они демонстрируют свойство сохранения неустойчивых квантовых корреляций. Это похоже на  поддержание сильной неустойчивости в живых системах, связанной с сохранением подобных квантовых или классический корреляций. Однако, в отличие от живых систем, это сохранение пассивно. Пример таких систем, возможно используемых мозгом для вычислений, дает Пенроуз.**[44],[45]** Это система молекул - димеров тубулинов, из которых состоят микротрубочки цитоскелета нейронов (клетки, составляющие мозг). Система молекул тубулинов рассматривается как некий квантовый компьютер. .**[16], [47], [48]**  Даже если эта гипотетическая модель неверна, она иллюстрирует **принципиальную возможность** существования квантовых корреляций в мозгу. Аналогом квантовых корреляций в КСМ являются дополнительные корреляции, а аналогом квантового компьютера – молекулярный компьютер с дополнительными корреляциями между молекулами. Подобные корреляции возникают в хаотических или почти хаотических классических системах с перемешиванием. Возможно построить модель мозга и на основе таких вполне классических систем. При этом они будут обладать всеми свойствами квантовых компьютеров – непредсказуемость, параллельность вычислений. Поскольку Пенроуз ошибочно считает классические хаотические системы непригодными для моделирования живых систем, он даже не рассматривает такую возможность.

Другим примером синергетических систем, иллюстрирующих свойства квантовых корреляций, являются квантовые осциллирующие, почти изолированные от окружения системы.**[10]** Пусть имеется некое сверхпроводящее кольцо, и состояние А, например, соответствует движению тока по часовой стрелке, а состояние В – против часовой стрелке. Тогда  эта осциллирующая квантовая почти изолированная система  изменяются по следующей схеме: А -> сумма А и В -> В -> разница А и В-> А. «Сумма А и В» или «разница А и В» - это квантовые суперпозиции состояний А и В.Пусть мы хотим измерить направления тока в кольце. Это измерение нарушает состояния суперпозиции, если система в момент измерения находилось в таком состоянии. Оно, таким образом, меняет динамику системы и уничтожает корреляции между состояниями суперпозиции**[10].**

Можно придумать и другой, интересный для биологов и химиков пример таких чувствительных к измерениям колебательных систем. Они активно защищаются от влияния внешнего шума. Пусть у нас имеется процесс, состоящий из трех стадий. На первой стадии появляется белок (фермент) с неустойчивой конформацией А. На второй стадии он катализирует некий химический процесс, который, в свою очередь, препятствует разрушению конформации А. Усложним немного процесс. Пусть существуют две неустойчивые конформации А и В, которые способны катализировать процесс на второй стадии, и происходит непрерывный переход конформации из состояния А в В, из В в А и т.д..  На третьей стадии белок попадает в третью реакцию. Если в этот момент он был в конформации А он катализирует реакцию и сохраняется, если же в конформации В то не катализирует реакцию и разрушается. Таким образом, сохранение белка зависит в каком состоянии он покинул вторую стадию в А или В. Пусть во время второй стадии мы решили с помощью некого метода (ядерный магнитный резонанс, например) померить текущую коформацию белка. Поскольку переход из А в В является неустойчивым, измерение нарушит фазу перехода из состояния А в В. Как результат, на третьей стадии белок может оказаться в конформации В вместо А и, таким образом, разрушиться, вместо того, чтобы сохраниться. То есть, измерение промежуточного состояния белка нарушит ход самого процесса, меняя его результат. Причем это может быть верно, как для квантовой, так и для классической механики.

Перейдем к синергетическим моделям глобальных корреляций, охватывающих всю вселенную.

С помощью синергетических «игрушечных» моделей можно понять синхронистичность (одновременность) причинно не связанных процессов, а также явление глобальных корреляций.

Примером являются нестационарные системы с «обострением» (blow up) **[62-64]**, рассмотренные школой Курдюмова. В этих процессах определяется некая функция на плоскости. Ее динамика описывается нелинейными уравнениями, подобными уравнению горения. Для решения с «обострением» значение функции может стремиться к бесконечности за *конечное* время в одной точке или нескольких изолированных точках на плоскости. Интересно, что функция достигает бесконечности во всех этих точках в один и тот же момент времени, то есть синхронно.

С помощью таких моделей иллюстрируют рост населения (или уровня технического развития цивилизаций) в мегаполисах нашей планеты **[65].** Точки бесконечного роста – это мегаполисы, а плотность населения – это значение самой функции.

Усложним задачу. Пусть в некий момент этого процесса происходит очень быстрое расширение («инфляция») плоскости, в которой протекает процесс с «обострением». Тем не менее, процессы достижения бесконечности в изолированных точках остаются синхронными, несмотря на то, что они уже разделены большим расстоянием.

Этой более сложной моделью можно качественно объяснить синхронность развития процессов в очень далеких частях нашей резко расширившейся Вселенной в результате «инфляция» после Большого Взрыва. То, что эти процессы с «обострением» появляются лишь при некотором узком наборе коэффициентов уравнения горения, позволяет провести аналогию с «антропным принципом»**[66].** Антропный принцип утверждает, что фундаментальные постоянные имеют именно такие значения, чтобы в итоге могла возникнуть именно наша наблюдаемая Вселенной с «антропными» существами, способными ее наблюдать.

Также интересно проиллюстрировать сложные процессы с помощью «клеточных» моделей. Хорошей базой служат дискретная модель Хопфилда [67-68]. Эту систему можно описать как квадратную двухмерную решетку ячеек, которые могут быть либо черными, либо белыми. Зададим некоторое начальное состояние решетки. Коэффициенты линейного взаимодействия между ячейками неравны. Их можно выбрать так, что в процессе дискретной эволюции начальное состояние переходит в одно из возможных конечных состояний, из заранее заданного набора состояний (аттракторов) . Пусть эти аттракторы - буквы А или В.

Существуют такие начальные неустойчивые состояния которые отличаются лишь на одну ячейку (критический элемент). При этом одно из них имеет в качестве аттрактора состояние А, а другое – состояние В. Подобная синергетическая модель хорошо иллюстрирует свойство *глобальной неустойчивости* сложных систем. Оно также показывает, что эта неустойчивость присуща всей системе в целом, а не ее части. Лишь внешний наблюдатель может привести к изменению значения критического элемента и изменить эволюцию системы. Внутренняя динамика самой системы сделать это не может.

*Глобальная корреляция* между ячейками неустойчивого начального состояния определяет к какому именно аттрактору эволюционирует эта решетка (либо А, либо В).

Эта модель может интерпретироваться как нейронная сеть с обратной связью или как спиновая решетка (спиновое стекло) с неодинаковыми взаимодействиями между спинами. Подобная система используется для целей распознавания образов.

Можно несколько усложить модель. Пусть каждая ячейка в описанной выше решетке сама является аналогичной подрешеткой. Пусть процесс происходит в два этапа.

На первом этапе крупные ячейки не взаимодействуют, взаимодействие есть лишь в подрешетках, которое идет по обычной схеме. Начальные состояния всех подрешеток можно выбрать неустойчивыми. Итоговое состояние А подрешетки будем ассоциировать как черную ячейку крупной решетки, а В – белую.

Второй этап эволюции определяется как обычную эволюция уже этой крупной решетки, без изменений в подрешетках. Ее возникшее на первом этапе начальное состояние тоже может быть неустойчивым.

Пусть итоговое состояние крупнозернистой решетки будет буква А, а каждой ее крупной ячейки - тоже буква А. Назовем его состоянием «А-А». Тогда наличие именно такого, а не иного финального состояния можно объяснить глобальными корреляциями неустойчивого начального состоянии и конкретным выбором всех коэффициентов взаимодействия между ячейками.

Будем считать, что до начала описанного выше двухэтапного процесса, наша крупнозернистая решетка занимала очень малую область пространства, но в результате расширения («инфляции») расширилась до больших размеров, после чего и начался описанный выше процесс. Тогда наличие коррелированного неустойчивого начального состояния составной решетки, приводящего именно к итоговому состоянию «А-А» можно объяснить тем, что до «инфляции» все ячейки находились близко друг от друга. Специфический выбор коэффициентов взаимодействия ячеек, приводящий к итоговой асимптотике (состоянию «А-А»), можно объяснять по аналогии с «антропным принципом».

Действительно, всю эти крупнозернистую решетку в целом можно сравнить с нашей «Вселенной», а ее крупные ячейки (подрешетки) с «организмами», препятствующие (активно или пассивно) «декогеренции». «Декогеренции» - это влияние «окружения» некоторой крупной ячейки (т.е. влияние других ячеек) на процессы внутри этой ячейки. Тогда глобальные корреляции неустойчивого начального состояний решетки могут служить аналогами возможных глобальных корреляций неустойчивого начального состоянии нашей Вселенной, а коэффициенты взаимодействия ячеек соответствуют фундаментальным константам. Начальный процесс расширения решетки соответствует Большому Взрыву.

## 12.4 Гипотетические последствия взгляда на жизнь как средство сохранения корреляций.

Кстати, определение жизни, как систем, способствующих сохранению корреляций в противовес внешнему шуму, хорошо объясняет загадочное молчание КОСМОСА, т.е. отсутствие сигналов от других разумных миров. Вселенная произошла из единого центра (Большой Взрыв) и все ее части коррелированны, жизнь лишь поддерживает эти корреляции и существует на их основе. Поэтому процессы возникновения жизни в различных частях скоррелированны и находятся на одном уровне развития, т.е. сверхцивилизаций, способных достичь Земли пока просто нет.

Эффектами дальних корреляций можно объяснить и часть поистине чудесных проявлений человеческой интуиции и парапсихологических эффектов. Возможно, особо тонкая человеческая интуиция и некоторые парапсихологические эффекты и лежат в этой области Непредсказуемости. Об этом подробно писал известный психиатр Карл Юнг в свое статье «О «синхронистичности»» **[55].** Приведем наиболее интересные отрывки из этой статьи. Вот отрывок, дающий определение "синхронистичности":

*« ... концепцией "синхронистичности". С точки зрения этимологии, этот термин каким-то образом связан со временем, или, если точнее, с чем-то вроде одновременности. Вместо "одновременности" мы можем также использовать концепцию "смыслового совпадения" двух или более событий, когда речь идет не о вероятности случая, а о чем-то другом. Статистическое - то есть вероятностное - совпадение событий, типа иногда имеющего место в больницах "дублирования случаев", относится к категории случайности....*
*Пространство и время, а значит и причинность, являются факторами, которые можно исключить, в результате чего становится возможным беспричинный феномен, еще называемый "чудом". Все природные явления такого рода являются уникальными и чрезвычайно любопытными комбинациями случайностей, образовавших единое целое благодаря общему для всех них значению. Хотя "смысловые совпадения" бесконечно разнообразны в своей фено-менологии, будучи беспричинными событиями, они, тем не менее, образуют элемент, который является частью научной картины мира. Причинность - это способ, каким мы объясняем связь между двумя последовательными событиями. Синхронистичность указывает на параллельность времени и смысла между психическими и психофизическими событиями, которую наука пока что не способна свести к общему принципу. Сам этот термин ничего не объясняет, он просто указывает на существование "смысловых совпадений", которые сами по себе являются случайными происшествиями, не настолько невероятными, что мы вынуждены предположить, - они основаны на некоем принципе или на каком-то свойстве эмпирического мира. Между параллельными событиями нельзя проследить никакой взаимной причинной связи, и именно это и придает им характер случайности. Единственной заметной и доказуемой связью между ними является общность смысла или эквивалентность. Древняя теория соответствия была основана на ощущении таких связей - теория, венец такой точкой, а заодно и временным концом которой стала идея Лейбница о заранее установленной гармонии. После чего эту теорию заменили причинностью. Синхронистичность - это современный и модернизированный вариант устаревшей концепции соответствия, взаимопонимания и гармонии. Он основан не на философских предположениях, а на эмпирических ощущениях и экспериментальной работе»*

А вот цитата, дающая пример "синхронистичности" из личной практики Юнга:

*«Поэтому я сосредоточил свое внимание на определенных наблюдениях и ощущениях, которые, как я могу с уверенностью сказать, "навязались" мне за время моей долгой врачебной практики. Они относятся к спонтанным "смысловым совпадениям" такой низкой степени вероятности, что поначалу в них просто невозможно поверить. Поэтому я опишу вам только один случай такого рода, просто в качестве примера, характеризующего всю категорию этих явлений. Совершенно все равно, откажитесь ли вы поверить в этот конкретный случай или отмахнетесь от него с объяснением ad hoc(Для данного случая (лат.)). Я мог бы рассказать вам великое множество таких историй, которые, в равной степени или невероятны, чем неопровержимые результаты, полученные Рейном, и вы вскоре поймете, что почти каждый случай требует индивидуального объяснения. Но причинное объяснение, единственно возможное с точки зрения естественной науки, оказывается несостоятельным из-за психической относительности пространства и времени, которые являются обязательными условиями причинно-следственных связей.*

*Героиней этой истории является молодая пациентка, которая, несмотря на обоюдные усилия, оказалась психологически закрытой. Трудность заключалась в том, что она считала себя самой сведущей по любому вопросу. Ее великолепное образование дало ей в руки идеально подходящее для этой цели "оружие", а именно, слегка облагороженный картезианский рационализм с его безупречно "геометрической" идеей реальности. После нескольких бесплодных попыток "разбавить" ее рационализм несколько более человечным мышлением, я „был вынужден ограничиться надеждой на какое-нибудь неожиданное и иррациональное событие, на что-то, что разнесет интеллектуальную реторту, в которой она сама себя запечатала. И вот, однажды, я сидел напротив нее, спиной к окну, слушая поток ее риторики. Этой ночью ее посетило впечатляющее сновидение, в котором кто-то дал ей золотого скарабея - ценное произведение ювелирного искусства. Она все еще рассказывала мне этот сон, когда я услышал тихий стук в*

*окно. Я обернулся и увидел довольно большое насекомое, которое билось о стекло, явно пытаясь проникнуть с улицы в тёмную комнату. Мне это показалось очень странным. Я тут же открыл окно и поймал насекомое, как только оно залетело в комнату. Это был скарабеевидный жук или хрущ обыкновенный (Cetonia aurata), желто-зеленая окраска которого очень сильно напоминала цвет золотого скарабея. Я протянул жука моей пациентке со словами: "Вот ваш скарабей". Это событие пробило желаемую брешь в её рационализме и сломало лед её интеллектуального сопротивления. Теперь лечение могло принести удовлетворительные результаты.*

*Эта история - не более чем парадигма бесчисленных случаев "смысловых совпадений", которые наблюдались не только мною, но и многими другими, и обильно задокументированы. Они включают в себя все, что относится к категории ясновидения, телепатии и т. д., от подтвержденного очевидцами видения Сведенборгом большого пожара в Стокгольме до недавнего рассказа маршала авиации сэра Виктора Годара о сновидении неизвестного офицера, в котором была предсказана действительно имевшая впоследствии место авария самолета Годара.*

*Все упомянутые мною феномены можно разделить на три категории:*

*1) Совпадение психического состояния наблюдателя с происходящим в момент этого состояния, объективным внешним событием, которое соответствует психическому состоянию или его содержимому (например, скарабей), в котором не прослеживается причинная связь между психическим состоянием и внешним событием, и в котором, учитывая психическую относительность времени и пространства, такой связи не может и быть.*

*2) Совпадение психического состояния с соответствующим (происходящим более-менее в то же время) внешним событием, имеющим место за пределами восприятия наблюдателя, то есть на расстоянии, удостовериться в котором можно только впоследствии (например, стокгольмский пожар).*

*3) Совпадение психического состояния с соответствующим, но еще не существующим будущим событием, которое значительно отдалено во времени и реальность которого тоже может быть установлена только впоследствии.*

*События групп 2 и 3 еще не присутствуют в поле зрения наблюдателя, но уже ему известны, если, конечно, их реальность будет подтверждена. Поэтому я назвал эти события "синхронистическими", что не следует путать с "синхронными"»*

Эту, указанную Юнгом «синхронистичность» можно объяснить с научной точки зрения не только лишь случайным совпадением, но и проявлениям неустойчивых скрытых корреляций между живыми организмами и окружающими объектами. Как мы уже говорили выше, процессы метаболизма могут поддерживать такие корреляции и препятствовать их «перепутыванию» с окружением в процессе декогеренции. Поскольку эти корреляции неустойчивы, то они не наблюдаемы (т.е. соответствуют Непредсказуемой динамике). Причем здесь необязательна квантовая механика, подобные корреляции присущи и классической механике, имеющей аналоги квантовых корреляций. Подобные корреляции часто ошибочно увязывают лишь с квантовой механикой.

Ненаблюдаемость этих корреляций для внешнего наблюдателя отнюдь не означает, что они не могут регистрироваться нашим субъективным опытом в форме некого «предчувствия нечто». Так, внешний наблюдатель не может померить или предсказать результат вычислений квантового компьютера, не исказив его. Предположим, что у квантового компьютера существует своё «сознание». Тогда он может «предчувствовать» свое будущее в отличие от внешнего наблюдателя.

Данные рассуждения **не «доказывают»** однако, что «синхроничность» на самом деле связана с неустойчивыми корреляциями. Они лишь говорят о том, что подобные явления не противоречат физике, даже если их рассматривать не только как случайные совпадения. Экспериментальная проверка этой гипотезы вообще вряд ли возможна **в принципе.** Причиной является **принципиальная ненаблюдаемость** самих этих **корреляций**.

Так, например, пусть некий человек, запустил два изначально коррелированных квантовых компьютера и знал их начальное состояние. В дальнейшем он погиб или исчез. Тогда мы никогда не сможем заранее предсказать результаты их совместной работы. Мы не сможем определить, случайна ли корреляция между результатами их работы или она бала заложена изначально. Ведь любая попытка определить внутреннее состояние квантового компьютера приводит неизбежно к нарушению его работы! Подобные рассуждения верны не только для квантовых, но и для хаотических классических систем с неустойчивыми дополнительными микроскопическими корреляциями.

Может весь наш мир живых существ на самом деле и есть набор таких коррелированных компьютеров с ненаблюдаемыми неустойчивыми корреляциями между ними? И роль живой материи состоит лишь в поддержании этих корреляций? Ведь их точное начальное состояние мог бы знать лишь Бог, если допустить его существование. Возможность таких корреляций вполне возможна, поскольку весь наш мир произошел из одной точки в результате «Большого Взрыва». Да и все живые организмы на нашей планете, возможно, являются потомком одной единственной «протоклетки».

Особо тонкая человеческая интуиция и некоторые парапсихологические эффекты могут лежать в узкой области на грани постижимости точной наукой. Эта область Непредсказуемой динамики. Их принципиальная неуловимость и неустойчивость не дает естественному отбору усилить эти свойства [21], [22]. По причине неустойчивости и непредсказуемости также нет возможности постичь полностью эти явления средствами науки. Хотя эти явления не противоречат законам физики.

Менский [46] в своей книге также пытается обосновать некоторые тонкости человеческой интуиции и парапсихологические эффекты через специфические особенности квантовой механики. Однако он допускает несколько очень характерных ошибок

1) Для обоснования этих эффектов нет необходимости в «нарушениях» квантовой механики, например, перехода с помощью «силы сознания» «медиума» в другие Миры, описываемых Многомировой интерпретацией. Достаточно предположить лишь корреляцию между желанием «медиума» и происходящими событиями. За счет этих корреляций окружение может «подыгрывать» нашим желаниям. Сознание, в свою очередь, за счет этих корреляций может «предчувствовать» будущее. Попытка же заранее «помелить» или «обнаружить» эти неустойчивые квантовые корреляции приведет лишь к их исчезновению и необратимому изменению дальнейшей динамики этих событий. Никакого «нарушения» обычных законов квантовой механики «медиумом» тут не нужно.
2) Эффекты, подобные квантовым, происходят и в классических системах близким к хаотическим. Соответственно, все эти эффекты могут быть промоделированы и классически, без обращения к квантовой механике.
3) Менский пишет о сложности и даже невозможности проверки индивидуальной истории системы с научной точки зрения. Ведь в процессе такой истории происходят разнородные события. Они характеризующиеся различными распределениями вероятности. Действительно, обычно, в науке для проверки вероятностной теории используют ансамбль однородных событий и используют «закон больших чисел».[56] Тем не менее, существует формулировка «закона больших чисел» и для разнородных событий! (Обобщенная теорема Чебышева, Теорема Маркова) [56] И с помощью такой формулировки может быть проверена на соответствие законам физики также и индивидуальная история системы.

# 13. Заключение.

Приведенная статья **не является** лишь **философским** абстрактным построением. Не понимание ее основ приводит к конкретным физическим ошибкам. Подавляющее большинство реальных систем не описывается идеальными уравнениями квантовой или классической механики. Из-за всегда существующего (и неизбежно существующего для измерений в квантовой механике) воздействия измерительной аппаратуры и окружения системы на эту систему эти уравнения нарушаются. Попытка включить измерительные приборы и окружение в описываемую систему приводит к самонаблюдаемой системе. (**Приложение М**)  Такая система не может измерить и запомнить в полной мере собственные состояния и ее даже приближенное самоописание имеет смысл лишь для промежутков времени много меньших времени возврата, определяемого согласно теореме Пуанкаре, после которого, вся память о прошлом неизбежно стирается. Однако систему можно описать в этой ситуации не Идеальной, а Наблюдаемой динамикой. Наблюдаемая динамика может быть получена и на основе огрубления фазовой функции распределения или редукции матрицы плотности, поскольку начальное состояние точно не определено при самонаблюдении. Наблюдаемая динамика может быть получена и введением малых внешних шумов, источником которых является внешний наблюдатель или окружение.  Это  возможно, поскольку, для широкого интервала величин внешних шумов свойства этой динамики не зависят от величины и вида этих шумов, а определяются лишь свойствами самой системы. Введение такой Наблюдаемой динамики разрешает все известные парадоксы классической и  квантовой механики.

Приведем несколько конкретных примеров ошибок, сделанных из-за  непонимания этих основ сделанных, например,  в  теории полюсов  для задачи движения фронта пламени и роста «пальца» на поверхности раздела жидкостей.

Севашинский[23] утверждал, что Идеальная Динамика полюсов приводит к ускорению фронта пламени, и это ускорение не вызвано шумом, поскольку оно не меняется  при уменьшении шума и зависит лишь от свойств самой системы. Но ведь также Наблюдаемая Динамика, связанная с шумом, не зависит от него в широком интервале значений.

Танвир [24] нашел различие в росте «пальца» в теории и численных экспериментах, не поняв, что эта разница связана с численным шумом, приводящим к новой Наблюдаемой Динамике. Это лишь два рядовых примера, взятых из повседневной практики автора статьи, а встретить их можно много. Изложенные в этой статье результаты необходимы для понимания основ нелинейной динамики, термодинамики и квантовой механики.

В обычных задачах физики обычно не возникает необходимости в глубоком анализе ситуации, сделанном в этой статье. Связано это с тем, что обычные физические  системы – это либо системы с малым числом частиц, либо системы многих частиц близких к состоянию термодинамического равновесия. В таких системах можно либо использовать достаточно точно Идеальную динамику, либо использовать упрощенные «приближенные» методы получения уравнений Наблюдаемой Динамики. Это, например, редукция в КМ или уравнение Больцмана в КСМ. Поэтому интерес многих физиков к статьям подобной этой довольно мал. Однако достаточно много физических систем не входит в класс, описываемой Идеальной или Наблюдаемой Динамикой. Их поведение в принципе непредсказуемо, даже вероятностно. Мы определяем их поведение как Неопределенную Динамику. К таким системам относятся, например, квантовые компьютеры с точки зрения наблюдателя, не присутствовавшего при их «запуске». К системам с Неопределенной Динамикой могут относиться и некоторые стационарные системы, далекие от термодинамического равновесия. Живые организмы тоже могут являться подобными стационарными системами. Даже если подобные системы могут  быть частично или полностью описаны Наблюдаемой Динамикой, ее получение становиться нетривиальной задачей. Когда физика попытается описать подобные сложные системы, понимание методов разрешения ее парадоксов, изложенных в этой статье, становиться необходимостью. На многие вопросы необходимо найти еще ответы физикам и математикам:

1) Каковы методы получения  Наблюдаемой Динамики сложных систем, которая полностью или частично их описывает?

2) Когда получение Наблюдаемой Динамики невозможно и система описывается Неопределенной Динамикой?

3) Можно ли создать «на бумаге» «синергетичекие» системы (подобные, например, тубулинам Пенроуза), которые бы иллюстрировали саму **принципиальную возможность** возникновения и существования сложных стационарных систем (как классических, так и квантовых), описываемых Неопределенной Динамикой? Мы приведем здесь лишь некоторые догадки, как можно получить подобные системы.

**a)** Системы, описываемые Неопределенной Динамикой, должны быть способны сопротивляться декогеренции и сохранять неустойчивые дополнительные классические (или квантовые) корреляции как внутри сложных систем, так и между самими системами.

**b)** Такие системы могут обладать несколькими неустойчивыми состояниями, возможно способными переходить друг в друга с течением времени. Поток негоэнтропии, вещества или энергии (т.е. метаболизм) позволит поддерживать и сохранять эти неустойчивые состояния или процессы, не влияя на них, но защищая их от влияния внешнего шума. С другой стороны, сами неустойчивые системы могут служить катализаторами этого метаболизма. В таких системах будут возможны как обратные процессы, так и возвраты Пуанкаре. Ведь они защищены от внешнего шума (декогеренции) метаболизмом. Внешний шум должен стать настолько малым, чтобы стать неспособным уничтожить эти процессы или состояния. Попытка измерить текущее состояние или процесс в неустойчивой системе приведёт к нарушению ее динамики. Таким образом, их динамика будет ненаблюдаемой. Такие системы могут быть не только квантовыми, но и классическими.

**c)** В физике обычно макросостояние рассматривается как некая пассивная функция микросостояния. Однако можно рассмотреть случай, когда система **сама** измеряет как свое макросостояние, так и макросостояние окружения. Таким образом **обратная связь макросостояний через микросостояние** появляется. [58] (**Приложения M и V**)

**d)** Это могут быть саморазмножающиеся клеточные автоматы. [57]

Последние годы были опубликованы очень интересные работы в направлении создания таких «синергетичеких» систем, возможно подобных живым организмам [76], [77], [78].

**Следует отметить, что построение таких моделей является задачей физики и математики, а не философии.**

## Приложение А Функция фазовой плотности. [1], [2], [9]

В данный момент времени t состояние системы N одинаковых частиц с точки зрения классической механики определяется заданием значений координат $\mathbf{r}_1$, …, $\mathbf{r}_N$ и импульсов $\mathbf{p}_1$, …, $\mathbf{p}_N$ всех N частиц системы. Для краткости будем использовать обозначение

$$x_i=(\mathbf{r}_i, \mathbf{p}_i) \quad (i=1, 2, …, N)$$

для совокупности значений координат и компонент импульса отдельной частицы и обозначению

$$X= (x_1, …, x_N)=(\mathbf{r}_1, …, \mathbf{r}_N, \mathbf{p}_1, …, \mathbf{p}_N)$$

для совокупности значений координат и значений импульсов всех частиц системы. Такое пространство 6N переменных называют 6N-мерным фазовым пространством.

Чтобы определить понятие функции распределения состояний системы, рассмотрим набор одинаковых макроскопических систем - ансамбль Гиббса. Для всех этих систем условия опыта одинаковы. Поскольку, однако, эти условия определяют состояние системы неоднозначно, то разным состояниям ансамбля в данный момент времени t будут соответствовать разные значения X.

Выделим в фазовом пространстве объем dX около точки X. Пусть в данный момент времени t в этом объеме заключены точки, характеризующие состояния dM систем ансамбля из их полного числа M. Тогда предел отношения этих величин

$$\lim_{m \to \infty} dM/M = f_N(X,t)dX$$

и определяет фазовую функцию плотности распределения
в момент времени t.

$$\int f_N(X,t)dX = 1$$

Уравнение Лиувилля для функции фазовой плотности
Можно записать в виде

$$i\frac{\partial f_N}{\partial t} = L f_N = \{H, f_N\}$$

где L - линейный оператор:

$$L = -i\frac{\partial H}{\partial p}\frac{\partial}{\partial x} + i\frac{\partial H}{\partial x}\frac{\partial}{\partial p},$$

где H - энергия системы.

## Приложение B Определения энтропии. [1], [2], [9], [29]

В качестве определения энтропии укажем выражение

$$S = -k\int_{(X)} f_N(X,t) \ln f_N(X,t)$$

Для квантовой механики через матрицу плотности [29]

$$S = -k \, \mathrm{tr}\, \rho \ln \rho$$

где tr - след матрицы.

Эти энтропии ансамбля не меняются при обратимой эволюции. Чтобы получить меняющиеся энтропии в качестве $f_N$ или $\rho$ используются их огрубленные величины. [35]

## Приложение C. Доказательство теоремы Пуанкаре о возвратах. [2]

*Число фазовых точек, покидающих при своем движении заданный фазовый объем g и не возвращающихся в него, с течением времени будет меньше любой сколько-нибудь заметной доли полного числа фазовых точек.* Докажем это положение.

Рассмотрим систему, имеющую конечный общий фазовый объем *G*. Выделим внутри этого объема некоторую *фиксированную* поверхность σ, ограничивающую малый объем *g*. Рассмотрим фазовые точки, вытекающие через поверхность σ из объема *g*. Скорость перемещения фазовой точки по фазовой траектории зависит только от фазовых координат, поэтому число точек, вытекающих в единицу времени через фиксированную поверхность σ, не зависит от времени. Обозначим через *g'* объем, занимаемый фазовыми точками, которые вытекают в единицу времени из фазового объема *g*, не возвращаясь в него вновь. За время *T* из объема *g* вытекает *g'T* объемов фазовой жидкости. Поскольку вытекший объем *g'T*,. по предположению, не возвращается более в объем *g*, поскольку он должен заполнять остальную часть полного фазового объема $G$. Фазовая жидкость несжимаема, поэтому вытекший из $g$ объем $g'T$ не должен превышать объем, в который он вытечет, т. е.

$$Tg' < G-g < G. \qquad (1)$$

Объем $G$ конечен, поэтому при конечном $g'$ это неравенство может быть удовлетворено только для конечного времени $T$. Если же $T \to \infty$, то неравенство (1) удовлетворяется лишь при $g' \to 0$, что и требовалось доказать.

## Приложение D . Корреляция. [56]

Рассмотрим следующую задачу. Была проведена серия измерений двух случайных величин $X$ и $Y$, причем измерения проводились попарно: т.е. за одно измерение мы получали два значения - $x_i$ и $y_i$. Имея выборку, состоящую из пар $(x_i, y_i)$, мы хотим определить, имеется ли между этими двумя переменными зависимость. Эта зависимость носит название **корреляции**. Корреляция может существовать не только между двумя, но и большим числом величин.

Зависимость между случайными величинами может иметь функциональный характер, т.е. быть строгим функциональным отношением, связывающим их значения. Однако при обработке экспериментальных данных гораздо чаще встречаются зависимости другого рода: статистические зависимости. Различие между двумя видами зависимостей состоит в том, что функциональная зависимость устанавливает строгую взаимосвязь между переменными, а статистическая зависимость лишь говорит о том, что распределение случайной величины $Y$ зависит от того, какое значение принимает случайная величина $X$.

### Коэффициент корреляции Пирсона

Существует несколько различных коэффициентов корреляции, к каждому из которых относится сказанное выше. Наиболее широко известен коэффициент корреляции Пирсона, характеризующий степень линейной зависимости между переменными. Он определяется, как

$$r = \frac{\sum_i (x_i - \bar{x})(y_i - \bar{y})}{\sqrt{\sum_i (x_i - \bar{x})^2} \sqrt{\sum_i (y_i - \bar{y})^2}}$$

## Приложение E. Термодинамическое равновесие изолированной системы. Микроканоническое распределение. [2], [9]

Нас будут интересовать *адиабатическая система* — система, изолированная от внешних тел и имеющая определенную, строго заданную энергию E.

### МИКРОКАНОНИЧЕСКОЕ РАСПРЕДЕЛЕНИЕ

Рассмотрим адиабатическую систему, т. е. систему, которая при неизменных внешних параметрах не может обмениваться энергией с внешними телами. Для такой системы, очевидно,

$$H(X, a) = E = const \qquad (1)$$

и функция фазового распределения φ(ε) должна иметь вид острого максимума, поскольку энергия системы может быть практически точно фиксирована и не будет изменяться с течением времени.

Но фазового распределения φ(ε) в пределе при ΔE —>0 превращается, с точностью до постоянного множителя, в дельта-функцию δ {ε — E}. Таким образом, для адиабатически изолированной системы можно положить:

$$\omega(X) = \frac{1}{\Omega(E, a)} \, \delta\{E - H(X, \alpha)\}, \qquad (2)$$

где $1/\Omega(E, \alpha)$ — нормирующий множитель, определяемый из условия нормировки, т. е.

$$\Omega(E, \alpha) = \int\limits_{(X)} \delta \{E - H(X, a)\} dX \qquad (3)$$

Выражение (2) называется микроканоническим распределением Гиббса, при помощи которого можно вычислять фазовые средние любых физических величин для адиабатически изолированной системы по формуле

$$\overline{F} = \int\limits_{(X)} F(X) \frac{1}{\Omega(E, a)} \delta\{E - H(X, a)\} dX \qquad (4)$$

Величина $\Omega(E, \alpha)$ имеет наглядный геометрический смысл. $\Omega(E, \alpha)dE$ имеет смысл фазового объема бесконечно тонкого слоя, заключенного между гиперповерхностями $H(X, a) = E$ и $H(X, a) = E + dE$..

## Приложение F .Теорема о неизменности объема фазовой «жидкости». [1], [2]

Пусть каждая материальная точка системы описывается декартовыми координатами

$$x_k, \ y_k, \ z_k \quad (k = l, \ 2, \ \ldots, N).$$

Совокупность этих трех координат мы будем также иногда обозначать вектором $\vec{r}_k$ . Эта система $N$ материальных точек может описываться также $3N$ обобщенными координатами:

$$q_n(x_l, \ldots, z_N) \quad (n=1, 2, \ \ldots, 3N).$$

Уравнениями движения такой консервативной системы являются уравнения Лагранжа:

$$\frac{d}{dt} \frac{\partial L}{\partial \dot{q}_k} - \frac{\partial L}{\partial q_k} = 0 \quad (k = l, \ 2, \ \ldots, 3N). \quad (1)$$

где $L=K—U$—функция Лагранжа, или лагранжиан; К — кинетическая энергия; $U$—потенциальная энергия системы. Однако в статистической физике удобнее пользоваться уравнениями движения в гамильтоновой форме:

$$\left.\begin{array}{l} \dot{q}_k = \dfrac{\partial H}{\partial p_k} \\[2mm] \dot{p}_k = -\dfrac{\partial H}{\partial q_k} \end{array}\right\} \quad k = 1,2,...,3N, \\[4mm] H = \sum_{k=1}^{3N} p_k \dot{q}_k - L, \quad p_k = \dfrac{\partial L}{\partial \dot{q}_k} \right\} \tag{2}$$

— функция Гамильтона, или гамильтониан, а *($q_1$, $q_2$ ,..., $q_{3N}$ ; $p_1$ , $p_2$, ..., $p_{3N}$) — совокупность канонических переменных.* В общих выводах все канонические переменные мы будем обозначать буквой *X&,* полагая:

$$q_k = X_k , \; p_k = X_{k+3N} \quad \{k=1, 2, \; ..., 3N). \tag{3}$$

Для сокращения формул вся совокупность переменных *($X_1$, $X_2$ ..., $X_{6N}$)* часто будет обозначаться одной буквой (X), а произведение всех дифференциалов *$dX_1$ $dX_2$, ..., $dX_{6N}$* будет обозначаться через *dX.*

Уравнения Гамильтона представляют систему дифференциальных уравнений первого порядка, в силу чего значения всех переменных *X* в момент *t* полностью определяются, если известны значения этих переменных *$X^0$* в момент t = 0. Это свойство гамильтоновой формы механически позволяет ввести геометрически наглядное изображение системы и ее движения в фазовом пространстве. Движение фазового ансамбля в фазовом пространстве можно рассматривать как движение *фазовой жидкости,* по аналогии с движением обычной жидкости в трехмерном пространстве. Иначе говоря, точки фазового пространства отождествляются с точками воображаемой фазовой жидкости, заполняющей пространство.

Легко доказать, что для систем, удовлетворяющих уравнениям Гамильтониана, фазовая жидкость несжимаема. Действительно, плотность обычной трехмерной несжимаемой жидкости постоянна. Следовательно, в силу уравнения непрерывности

$$-\frac{\partial \rho}{\partial t} = div \, \vec{\upsilon} \; \rho = \rho \; div \, \vec{\upsilon} + \vec{\upsilon} \cdot \nabla \rho \tag{4}$$

и условия ρ = const для несжимаемой жидкости имеем:

$$div \, \vec{\upsilon} = \frac{\partial \dot{x}}{\partial x} + \frac{\partial \dot{y}}{\partial y} + \frac{\partial \dot{z}}{\partial z} \tag{5}$$

Для жидкости в многомерном пространстве эта теорема без труда обобщается, а следовательно, для несжимаемой фазовой жидкости должно удовлетворяться условие равенства нулю многомерной дивергенции, т. е.

$$\sum_{k=1}^{6N} \frac{\partial \dot{X}_k}{\partial X_k} = 0 \tag{6}$$

Легко, однако, видеть, что в силу уравнений Гамильтона (2)

$$\sum_{k=1}^{6N} \frac{\partial \dot{X}_k}{\partial X_k} = \sum_{k=1}^{6N} \left( \frac{\partial \dot{q}_k}{\partial q_k} + \frac{\partial \dot{p}_k}{\partial p_k} \right) = \sum_{k=1}^{6N} \left( \frac{\partial^2 H}{\partial q_k \partial p_k} - \frac{\partial^2 H}{\partial p_k \partial q_k} \right) \equiv 0 \tag{7}$$

что и требовалось доказать.

Поскольку фазовая жидкость несжимаема, поскольку при ее движении остается неизменным фазовый объем, занимаемый любой частью этой жидкости.

## Приложение G. Основные понятия квантовой механики. [36], [3], [15]

**Волновая функция.**

Основу математического аппарата квантовой механики составляет утверждение, что состояние системы может быть описано определенной (вообще говоря, комплексной) функцией координат $\Psi(q)$, причем квадрат модуля этой функции определяет распределение вероятностей значений координаты q: $|\Psi|^2 dq$ есть вероятность того, что произведенное над системой измерение обнаружит значения координат в элементе $dq$ пространства. Функция $\Psi$ называется *волновой функцией* системы.

**Наблюдаемые величины.**

Рассмотрим некоторую физическую величину *f*, характеризующую состояние квантовой системы. Строго говоря, в нижеследующих рассуждениях следовало бы говорить не об одной величине, а сразу о целом полном их наборе. Однако все рассуждения от этого по существу не меняются, и в целях краткости и простоты мы говорим ниже всего лишь об одной физической величине.

Значения, которые может принимать данная физическая величина, называют в квантовой механике ее *собственными значениями*, а об их совокупности говорят как о *спектре* собственных значений данной величины. В классической механике величины пробегают, вообще говоря, непрерывный ряд значений. В квантовой механике тоже существуют физические величины (например, координаты), собственные значения которых заполняют непрерывный ряд; в таких случаях говорят о *непрерывном спектре,* собственных значений. Наряду с такими величинами в квантовой механике существуют, однако, и другие, собственные значения которых образуют некоторый дискретный набор; в таких случаях говорят о *дискретном спектре.*

Будем считать сначала для простоты, что рассматриваемая величина *f* обладает дискретным спектром. Собственные значения величины *f* обозначим через $f_n$, где индекс *n* пробегает значения 1, 2, 3, …. Обозначим волновую функцию системы в состоянии ,в котором величина *f* имеет значение $f_n$ посредством $\Psi_n$. Волновые функции $\Psi_n$ называют *собственными функциями* данной физической величины *f*. Каждая из этих функций предполагается нормированной, так что

$$\int |\Psi_n|^2 dq = 1. \qquad (1)$$

Если система находится в некотором произвольном состоянии с волновой функцией $\Psi$, то произведенное над нею измерение величины *f* даст в результате одно из собственных значений $f_n$. В соответствии с принципом суперпозиции можно утверждать, что волновая функция $\psi$ должна представлять собой линейную комбинацию тех из собственных функций $\psi_n$ которые соответствуют значениям $f_n$ , могущим быть обнаруженными с отличной от нуля вероятностью при измерении, произведенном над системой, находящейся в рассматриваемом состоянии. Поэтому в общем случае произвольного состояния функция $\Psi$ может быть представлена в виде ряда

$$\Psi = \Sigma \, a_n \Psi_n, \qquad (2)$$

где суммирование производится по всем *n*, а $a_n$ — некоторые постоянные коэффициенты.

Таким образом, мы приходим к выводу, что всякая волновая функция может быть, как говорят, разложена по собственным функциям любой физической величины. О системе функций, по которым можно произвести такое разложение, говорят как о *полной системе функций.*

Разложение (2) дает возможность определить вероятности обнаружения (путем измерений) у системы в состоянии с волновой функцией $\psi$ того или иного значения $f_n$ величины *f*. А именно, квадрат модуля $|a_n|^2$ каждого из коэффициентов разложения (2) определяет вероятность соответствующего значения $f_n$ величины $f$ в состоянии с волновой функцией Ψ. Сумма вероятностей всех возможных значений $f_n$ должна быть равна единице; другими словами,  должно  иметь  место  соотношение

$$\sum_n |a_n|^2 = 1 \qquad (3)$$

Наблюдаемые величины могут быть определены с помощью операторов, действующих над пространством функций. Результатом действия оператора на функцию также является функция. Тогда собственные функции наблюдаемой $\psi_k$ и её собственные значения $\lambda_k$ являются просто решением операторного уравнения на собственные функции и собственные значения:

$$A \psi_k = \lambda_k \psi_k \qquad (4)$$

Где А – оператор наблюдаемой величины.

Обозначение Н обычно используют для Гамильтониана - оператора энергии.
Для одной частицы во внешнем поле $U(x, y, z)$ он определяется следующей формулой:

$$H \equiv -\frac{\hbar^2}{2m}\Delta + U(x, y, z) \qquad (5)$$

Где $\Delta = \dfrac{\partial^2}{\partial x^2} + \dfrac{\partial^2}{\partial y^2} + \dfrac{\partial^2}{\partial z^2}$  - оператор Лапласа

**Матричная форма квантовой механики**.

Давайте разложим функцию, на которую действует оператор, и функцию, являющуюся результатом действия оператора, по собственным функциям некоторой наблюдаемой. Тогда обе эти функции могут быть записаны как столбцы коэффициентов этого разложения. Сам оператор наблюдаемой может быть записан в виде квадратной матрицы. Произведение этой матрицы на столбец коэффициентов разложения первой функции даст коэффициенты разложения второй функции.  Подобная форма записи операторов и функций носит название **матричной формой квантовой механики**.

**Уравнение Шредингера.**

Выпишем здесь волновое уравнение движения для частицы во внешнем поле

$$i\hbar\frac{\partial \Psi}{\partial t} = -\frac{\hbar^2}{2m}\Delta\Psi + U(x, y, z)\Psi. \qquad (6)$$

Уравнение (6) было установлено *Шредингером* в 1926 г. и   называются  *уравнение Шредингера.*

**Соотношение неопределенности Гейзенберга.**

Если характеризовать неопределенности координат и импульсов средними   квадратичными флуктуациями

$$\delta x = \sqrt{\overline{(x - \bar{x})^2}}, \quad \delta p_x = \sqrt{\overline{(p_x - \bar{p}_x)^2}},$$

то можно дать точную оценку наименьшего возможного значения их произведения.

Рассмотрим одномерный случай — пакет с волновой функцией $\psi$ (x), зависящей только от одной координаты; предположим для простоты, что средние значения $x$ и $p_x$ в этом состоянии равны нулю. Исходим из очевидного неравенства

$$\int_{-\infty}^{+\infty} \left| \alpha \ x\varphi + \frac{d\varphi}{dx} \right|^2 dx \geq 0$$

где $\alpha$ — произвольная вещественная постоянная. При вычислении этого интеграла замечаем, что

$$\int x^2 |\varphi|^2 dx = (\delta x)^2 \ ,$$

$$\int \left( x\frac{d\varphi^*}{dx}\varphi + x\varphi^*\frac{d\varphi}{dx} \right)dx = \int x\frac{d|\varphi|^2}{dx}dx = -\int |\varphi|^2 dx = -1,$$

$$\int \frac{d\varphi^*}{dx}\frac{d\varphi}{dx}dx = -\int \varphi^*\frac{d^2\varphi}{dx^2}dx = \frac{1}{\hbar^2}\int \varphi^* \hat{p}_x^2\varphi \ dx = \frac{1}{\hbar^2}(\delta p_x)^2,$$

и получаем

$$\alpha^2(\delta x)^2 - \alpha + \frac{(\delta p_x)^2}{\hbar^2} \geq 0$$

Для того чтобы этот квадратичный (по $\alpha$) трехчлен был положительным при любых значениях $\alpha$, его дискриминант должен быть отрицательным. Отсюда получаем неравенство

$$\delta x \ \delta p_x \geq \hbar/2 \tag{7}$$

Наименьшее возможное значение произведения равно $\hbar/2$.

Эти *соотношения неопределенности* (7) были установлены *Гейзенбергом* в 1927 г.

Мы видим, что чем с большей точностью известна координата частицы (т. е. чем меньше $\delta x$), тем больше неопределенность $\delta p_x$ в значении компоненты импульса вдоль той же оси, и наоборот. В частности, если частица находится в некоторой строго определенной точке пространства ($\delta x = \delta y = \delta z = 0$), то $\delta p_x = \delta p_y = = \delta p_z = \infty$. Это значит, что все значения импульса при этом равновероятны. Наоборот, если частица имеет строго определенный импульс p, то равновероятны все ее положения в пространстве.

### Приложение I. Матрица плотности. Чистое и смешанное состояние. [15], [29]

**Определение чистых и смешанных состояний, а также матрице плотности.**
Рассмотрим для простоты случай двух базисных ортонормальных состояний частицы, которые могут определяться, например, значением проекции спина (собственного момента частицы) вдоль некой оси. При этом частица может находиться в **чистых состояниях** с волновыми функциями $|\chi_a\rangle$, $|\chi_b\rangle$ или их суперпозиции. Выберем набор базисных ортонормальных состояний, соответствующих спину направленному вверх и вниз вдоль оси z ( обычно $|+1/2\rangle$ и $|-1/2\rangle$) и разложим состояния $|\chi_a\rangle$ и $|\chi_b\rangle$ по этому набору согласно соотношению

$|\chi_a\rangle = a_1^{(a)} |+1/2\rangle + a_2^{(a)} |-1/2\rangle,$
$|\chi_b\rangle = a_1^{(b)} |+1/2\rangle + a_2^{(b)} |-1/2\rangle.$

Применяя правила умножения матриц, получим для матрицы плотности $\rho_a$ чистого состояния $|\chi_a\rangle$:

$$\rho_a = \begin{pmatrix} a_1^{(a)} \\ a_2^{(a)} \end{pmatrix} (a_1^{(a)*}, a_2^{(a)*}) = \begin{pmatrix} |a_1^{(a)}|^2 & a_1^{(a)} a_2^{(a)*} \\ a_1^{(a)*} a_2^{(a)} & |a_2^{(a)}|^2 \end{pmatrix}$$

и аналогичное выражение для для матрицы плотности $\rho_b$ чистого состояния $|\chi_b\rangle$:

$$\rho_b = \begin{pmatrix} a_1^{(b)} \\ a_2^{(b)} \end{pmatrix} (a_1^{(b)*}, a_2^{(b)*}) = \begin{pmatrix} |a_1^{(b)}|^2 & a_1^{(b)} a_2^{(b)*} \\ a_1^{(b)*} a_2^{(b)} & |a_2^{(b)}|^2 \end{pmatrix}$$

Наблюдаемые в квантовой механике можно рассматривать как в матричной, так и в операторной форме. Подобно этому, матрицу плотности в квантовой механике можно записать не только в матричной форме, но и в форме оператора над пространством функций. Оператор матрицы плотности, соответствующий состоянию $|\chi_a\rangle$:

$$\rho_a = |\chi_a\rangle\langle\chi_a|$$

**Результат действия оператора $\rho_a$ на функцию $|\psi\rangle$ определяется следующим выражением:**

$$\rho_a |\psi\rangle = \langle \chi_a|\psi\rangle |\chi_a\rangle,$$

**где $\langle \chi_a|\psi\rangle$ число, определяющееся следующей формулой:** $\langle \chi_a|\psi\rangle = \int \psi \, \chi_a^* \, dV$

**Возможно и смешанное состояние: рассмотрим пучок из $N_a$ частиц, приготовленных в состоянии с волновой функцией $|\chi_a\rangle$, и еще один независимый от первого пучок из $N_b$ частиц, приготовленных в состоянии с волновой функцией $|\chi_b\rangle$. Для описания объединенного пучка введем матрицу плотности $\rho$ смешанного состояния.**

. Матрица плотности смешанного состояния будет определяться как взвешенная сумма этих двух матриц плотности;

$$\rho = W_a \rho_a + W_b \rho_b = \begin{pmatrix} W_a|a_1^{(a)}|^2 + W_b|a_1^{(b)}|^2 & W_a a_1^{(a)} a_2^{(a)*} + W_b a_1^{(b)} a_2^{(b)*} \\ W_a a_1^{(a)*} a_2^{(a)} + W_b a_1^{(b)*} a_2^{(b)} & W_a|a_2^{(a)}|^2 + W_b|a_2^{(b)}|^2 \end{pmatrix}$$

где $W_a = N_a/N$, $W_b = N_b/N$, $N = N_a + N_b$.

Поскольку при выводе этого выражения были использованы базисные состояния $|\pm 1/2\rangle$, полученное выражение называют матрицей плотности в $\{|\pm 1/2\rangle\}$-представлении. Можно написать матрицы плотности всех этих состояний и в представлении функций $|\chi_a\rangle$, $|\chi_b\rangle$, **если они являются ортонормальными**. В этом представлении

$$\rho_a = \begin{pmatrix} 1 & 0 \\ 0 & 0 \end{pmatrix}$$

$$\rho_b = \begin{pmatrix} 1 & 0 \\ 0 & 0 \end{pmatrix}$$

а матрица плотности смешанного состояния будет иметь очень простой диагональный вид:

$$\rho = W_a\,\rho_a + W_b\,\rho_b = \begin{pmatrix} W_a & 0 \\ 0 & W_b \end{pmatrix}$$

**Статистическая матрица плотности $P_0$,**

Мы знаем, что в классической статистической термодинамике все возможные макроскопические состояния системы рассматриваются как априори равновероятными (другим словами, они считаются равновероятными, если нет каких-либо сведений о значении полной энергии или о контакте с термостатом, поддерживающим постоянную температуру системы, и т.д.). По аналогии с этим в волновой механике все состояния системы, определяемые различными функциями, образующие полную систему ортонормированных функций, можно априори предполагать равновероятными. Пусть $\varphi_1,\ldots,\varphi_k$,- такая система базисных функций $\varphi_k$; зная, что система характеризуется смесью состояний $\varphi_k$ в отсутствие какой-либо другой информации можно считать, что статистическая матрица системы имеет вид

$$P_0 = \sum_k p P_{\varphi_k}\,,\ \text{где}\ \sum_k p = 1\,,$$

т.е., что $P_0$ - статистическая диагональная матрица такого смешанного состояния, для которого все диагональные веса равны между собой. Принимаем $\varphi_k$ за базисные функции, матрицу $P_0$ можно представить в виде

$$(P_0)_{kl}=p\delta_{kl}$$

Эта матрица обладает замечательным свойством. Пусть статистическое состояние ансамбля систем в начальный момент времени характеризуется матрицей $P_0$. Если во всех системах ансамбля провести измерение одной и той же величины A, то статистическое состояние ансамбля будет по-прежнему характеризоваться той же матрицей $P_0$.

**Уравнения движения для матрицы плотности $\rho$.**

$$i\frac{\partial \rho_N}{\partial t} = L\rho_N$$

где L - линейный оператор:

$L\rho = H\rho - \rho H = [H,\rho]$,

где H - оператор энергии системы.

Если A – оператор некой наблюдаемой величины,

Среднее значение этой величины может быть найдено следующим образом:
$<A>=\mathrm{tr}A\rho$
Где tr – оператор следа.

**Приложение J. Редукция матрицы плотности и теория измерения. [6], [3]**

Пусть при измерении некоторого объекта мы «четко различаем» состояния $\sigma^{(1)}$, $\sigma^{(2)}$, ... . Производя измерения над объектом, находящимся в этих состояниях, мы получаем числа $\lambda_1$, $\lambda_2$, ... . Начальное состояние измерительного прибора обозначим через а. Если измеряемая система сначала находилась в состоянии $\sigma^{(v)}$, то состояние полной системы «объект измерения плюс измерительный прибор» до того, как они вступили во взаимодействие будет определяться прямым произведением а $\times$ $\sigma^{(v)}$ . После измерения:

$$a \times \sigma^{(v)} \rightarrow a^{(v)} \times \sigma^{(v)}$$

Пусть теперь начальное состояние будет не «четко различимым», а произвольной комбинацией $\alpha_1 \sigma^{(1)} + \alpha_2 \sigma^{(2)} + ...$
таких состояний. В этом случае из линейности квантовых уравнений следует

$$a \times [\Sigma \alpha_v \sigma^{(v)}] \rightarrow \Sigma \alpha_v [a^{(v)} \times \sigma^{(v)}]$$

В итоговом состоянии, возникающем в результате измерения,
существует статистическая корреляция между состоянием объекта и состоянием прибора: одновременное измерение у системы «объект измерения плюс измерительный прибор» двух величин ( первой- подлежащий измерению характеристики исследуемого объекта и второй- положение стрелки прибора) всегда приводит к согласующимся результатам. Вследствие этого одно из названных измерений становится излишним - к заключению о состоянии объекта измерения можно прийти на основании наблюдения за прибором. Результатом измерения является не вектор состояния, представляемый в виде суммы, стоящей в правой части найденного соотношения, а так называемая смесь, т.е. один из векторов состояния вида

$$a^{(v)} \times \sigma^{(v)}$$

и что при взаимодействии между объектом измерения и измерительным прибором это состояние возникает с вероятностью $|\alpha_v|^2$. Данный переход называется редукцией волнового пакета и соответствует переходу матрицы плотности из недиагонального $\alpha_v \alpha_\mu^*$ в диагональный вид $|\alpha_v|^2 \delta_{v\mu}$. Этот переход не описывается квантовыми уравнениями движения.

**Приложение К. Огрубление функция фазовой плотности и гипотеза молекулярного хаоса**. [33], [2]

Огрублением функции плотности будем называть ее замену на приближенную величину. Например

$$f_N^*(X,t) = \int_{(y)} g(X-Y) f_N(Y,t)$$

где

$$g(X) = 1/\Delta \; D(X/\Delta)$$

$$D(x) = 1 \; for \; |X| < 1$$
$$D(x) = 0 \; for \; |X| \geq 1$$

Другой пример огрубления - «гипотеза молекулярного хаоса»

Замена двухчастичной функции распределения на произведение одночастичных функций
$$f(x_1, x_2, t) \rightarrow f(x_1, t) f(x_2, t)$$

## Приложение L. «Новая динамика» Пригожина. [9], [12]

В статье неоднократно упоминается "Новая Динамика" Пригожина. Дадим здесь ее очень краткое изложение, опираясь на книги [9, 12]

Позвольте ввести линейный оператор $\Lambda$ на функции $\rho$ фазовой плотности или матрицы плотности

$$\widetilde{\rho} = \Lambda^{-1} \rho$$
$$\Lambda^{-1} \, 1 = 1$$
$$\int \widetilde{\rho} = \int \rho$$

$\Lambda^{-1}$ сохраняет положительность.

Такой, что функция $\widetilde{\rho}$ обладает свойством:для функции $\Omega$ определенной как

$$\Omega = tr \widetilde{\rho}^{+} \widetilde{\rho} \quad \text{или} \quad \Omega = tr \widetilde{\rho} \ \ln \widetilde{\rho}$$

имеем $d\Omega/dt \leq 0$

Уравнение движения для $\rho$

$$\frac{\partial \widetilde{\rho}}{\partial t} = \Phi \widetilde{\rho}$$

где $\Phi = \Lambda^{-1} L \Lambda$

$\Phi$-необратимая марковская полугруппа

$\Lambda^{-1}(L) = \Lambda^{+}(-L)$

Для функции фазовой плотности оператор $\Lambda^{-1}$ соответствует огрублению в направлении сжатия фазового объема. Для квантовой механики такой оператор может быть найден лишь для бесконечного объема или бесконечного числа частиц. Введем в квантовой механике проекционный оператор P, обнуляющий недиагональные элементы матрицы плотности. Оператор $\Phi$ и базисные вектора матрицы плотности выбираются таким образом, чтобы сделать оператор $\Phi$ перестановочным с оператором P:

$$\frac{\partial P \widetilde{\rho}}{\partial t} = P \Phi \widetilde{\rho} = \Phi P \widetilde{\rho}$$

## Приложение М. Невозможность самопредсказания системы.

Предположим, что существует мощная вычислительная машина способная предсказать будущее свое и окружения на основе расчета движения всех молекул. Пусть ее предсказания - выкатывание черного или белого шара из некого устройства, входящего в одну систему с машиной и описываемого ею. Устройство выкатывает белый шар, если машина предсказывает черный и наоборот черный шар при предсказании белого. Ясно, что предсказания машины всегда не верны. Так как выбор окружающей среды произволен, следовательно, этот контрпример доказывает невозможность точного самонаблюдения и самовычисления. Так как результат действия устройства всегда противоречит предсказаниям компьютера, следовательно, полное самопредсказание системы, включая и компьютер, и устройство невозможно (Рис. 27).

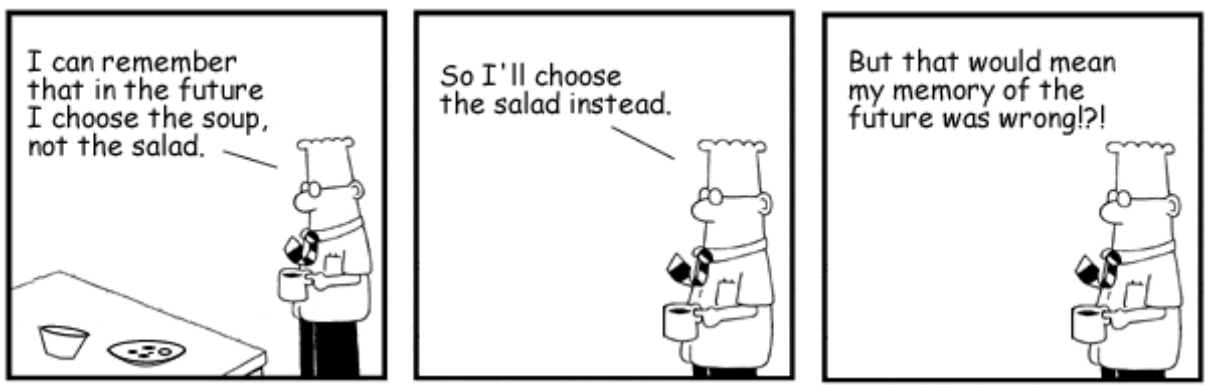

**Рис. 27.** Невозможность самопредсказания. (Рис. из [101])

**Приложение N. Соответствие квантовой и классической механик. Таблицы.**

| *Квантовая механика* | *Классическая механика* |
|---|---|
| Матрица плотности | Функция фазовой плотности |
| Уравнение движения для матрицы плотности | Уравнение Лиувилля |
| Редукция волнового пакета | Огрубление функции фазовой плотности или гипотеза молекулярного хаоса |
| Неустранимое воздействие наблюдателя или окружения, описываемое редукцией | Теоретически бесконечно малое, но в реальности конечное взаимодействие системы с наблюдателем или окружением |
| Ненулевые недиагональные элементы матрицы плотности | Корреляции между скоростями и положениями молекул в разных частях системы |

*Таблица.* Вероятностные формулировки классической и квантовой механик.
**[3]**

| | Классическая механика | Квантовая механика |
|---|---|---|
| Чистое состояние | Точка *(q, ρ)* фазового пространства | Вектор состояния $|\psi>$ |
| Общее метасостояние | Плотность вероятности $\rho(q, p)$ | Положительный эрмитов оператор $\rho$ |
| Условие нормировки | $\int \rho dq dp = l$ | $\mathrm{tr}\,\rho = 1$ |
| Условие того, что состояние чистое | $\rho = \delta$-функции | $\rho = |\psi> <\psi|$ (ранг оператора $\rho$ равен 1) |
| Уравнение движения | $\dfrac{\partial \rho}{\partial t} = \{H, \rho\}$ | $i\hbar \dfrac{\partial \rho}{\partial t} = [H, \rho]$ |
| Наблюдаемая | Функция $A(q, p)$ | Эрмитов оператор $A$ |

**Приложение О. Редукция системы при измерении**. [29], [3]

Рассмотрим ситуацию, когда прибор вначале находился в состоянии $|\alpha_0\rangle$, а объект — в суперпозиционном состоянии $|\psi\rangle = \sum c_i |\psi_i\rangle$, где $|\psi_i\rangle^{-}$ собственные состояния эксперимента. Начальный статистический оператор дается выражением

$$\rho_0 = |\psi\rangle\, |\alpha_0\rangle\langle\alpha_0|\, \langle\psi|. \tag{1}$$

Парциальный след этого оператора, совпадающий со статистическим оператором системы, составленной из одного объекта, имеет вид

$$\mathrm{tr}_A(\rho_0) = \sum_n \langle\varphi_n| \rho_0 |\varphi_n\rangle$$

где $|\varphi_n\rangle$ — какая-то полная система состояний прибора. Таким образом,

$$\mathrm{tr}_A(\rho_0) = \sum |\psi\rangle\, \langle\varphi_n|\alpha_0\rangle\langle\alpha_0|\varphi_n\rangle\langle\psi| = |\psi\rangle\langle\psi|, \tag{2}$$

где использовано соотношение $\sum |\varphi_n\rangle\langle\varphi_n| = 1$ и тот факт, что состояние $|\alpha_0\rangle$ нормировано. Мы получили в точности тот статистический оператор, который должны были приписать объекту, если бы он находился в состоянии $|\psi\rangle$. После акта измерения возникает корреляция между состояниями прибора и состояниями объекта, так что состояние комбинированной системы, составленной из прибора и объекта, описывается вектором состояния

$$|\Psi\rangle = \sum c_i e^{i\theta_i} |\psi_i\rangle |\alpha_i\rangle. \tag{3}$$

а статистический оператор дается выражением

$$\rho = |\Psi\rangle\langle\Psi| = \sum c_i c_j^{*} e^{i(\theta_i - \theta_j)} |\psi_i\rangle|\alpha_i\rangle\langle\alpha_j|\langle\psi_j|. \tag{4}$$

Парциальный след этого оператора равен

$$\mathrm{tr}_A(\rho)=\sum_n \langle\varphi_n| \,\rho\, |\varphi_n\rangle =$$
$$=\sum_{(ij)}c_i c_j^* \,e^{i(\theta i - \theta j)} \,|\psi_i\rangle \{\sum_n \langle\varphi_n |\alpha_i\rangle\langle\alpha_j|\varphi_n\rangle\} \langle\psi_j|=$$
$$=\sum_{(ij)}c_i c_j^* \,\delta_{ij} \,|\psi_i\rangle\langle\psi_j| \tag{5}$$

(так как различные состояния $|\alpha_i\rangle$ прибора ортогональны друг другу); таким образом,

$$\mathrm{tr}_A(\rho)=\sum|c_i|^2|\psi_i\rangle\langle\psi_i|. \tag{6}$$

Мы получили статистический оператор для системы, состоящей из одного объекта, описывающий ситуацию, когда имеются вероятности $|c_i|^2$ пребывать объекту в состояниях $|\psi_i\rangle$. Итак приходим к формулировке следующей теоремы.

**Теорема 5.5** (об измерении). Если две системы $S$ и $A$ взаимодействуют таким образом, что каждому состоянию $|\psi_i\rangle$ системы $S$ соответствует определенное состояние $|\alpha_i\rangle$ системы $A$, то статистический оператор $\mathrm{tr}_A(\rho)$ над полной системой ($S$ и $A$) воспроизводит действие редукции, применяемого к акту измерения, производимого над системой $S$, находившейся до измерения в состоянии $|\psi\rangle=\sum c_i|\psi_i\rangle$. ∎

Метасостояние системы, находясь в котором она не имеет определенного состояния, но является частью большой системы, которая находится в чистом состоянии, называется *несобственным смешанным состоянием*.

## Приложение Р. Теорема о декогеренции при взаимодействии с макроскопическим прибором.[3], [30]

Учтем теперь, что прибор является макроскопической системой. Это означает, что каждая различимая конфигурация прибора (например, положение его стрелки) не является чистым квантовым состоянием, никоим образом ничего не утверждая о состоянии движения каждой отдельной молекулы стрелки. Таким образом, в вышеприведенном рассуждении начальное состояние прибора $|\alpha 0\rangle$ следует заменить некоторым статистическим распределением по микроскопическим квантовым состояниям $|\alpha 0,s\rangle$; начальный статистический оператор не дается выражением (1), а равен

$$\rho_0 = \sum_s p_s \,|\psi\rangle|\alpha_{0,s}\rangle \langle\alpha_{0,s}|\langle\psi|. \tag{7}$$

Каждое состояние прибора $|\alpha_{0,s}\rangle$ будет реагировать на каждое собственное состояние $|\psi_i\rangle$ объекта тем, что превратится в некоторое другое состояние $|\alpha_{i,s}\rangle$, которое является одним из квантовых состояний, макроскопическое описание которого состоит в указании, что стрелка занимает положение i; точнее имеем формулу

$$e^{iH\tau/\hbar}(|\psi\rangle|\alpha_{0,s}\rangle) = e^{i\theta i,s}|\psi\rangle|\alpha_{i,s}\rangle. \tag{8}$$

Обратим внимание на появление фазового множителя, который зависит от индекса $s$. Разности энергий квантовых состояний $|\alpha_{0,s}\rangle$ с учетом времени $\tau$ должны быть такими, чтобы фазы $\theta_{i,s}(\mathrm{mod}\,2\pi)$ были случайно распределены между 0 и $2\pi$.

Из формул (7) и (8) следует, что при $|\psi\rangle=\sum c_i|\psi_i\rangle$ статистический оператор после измерения будет даваться следующим выражением:

$$\rho = \sum_{(s,\,i,\,j)} p_s c_i c_j^* \,e^{i(\theta i,s - \theta j,s)} \,|\psi_i\rangle|\alpha_{i,s}\rangle \langle\alpha_{j,s}|\langle\psi_j| \tag{9}$$

Так как из (9) получаем тот же результат (6), то видим, что статистический оператор (9) воспроизводит действие редукции, примененной к данному объекту. Он также практически воспроизводит действие редукции, примененной к одному прибору («практически» в том смысле, что речь идет о «макроскопической» наблюдаемой). Такая наблюдаемая не различает разные квантовые состояния прибора, соответствующие одному и тому же макроскопическому описанию, т. е. матричные элементы этой наблюдаемой между состояниями $|\psi_i\rangle|\alpha_{i,s}\rangle$ и $|\psi_i\rangle|\alpha_{i,s}\rangle$не зависят от $r$ и $s$. Среднее значение такой макроскопической наблюдаемой $A$ равно

$$\mathrm{tr}\,(\rho A) = \sum_{(s,\,i,\,j)} p_s c_i c_j^* \, e^{i\,(\theta_{i,\,s} - \theta_{j,\,s})} \, \langle\alpha_{j,\,s}|\langle\psi_j|A\,|\,\psi_i\rangle|\,\alpha_{i,\,s}\rangle =$$
$$= \sum_{(i,\,j)} c_i c_j^* \, a_{i,\,j} \sum_s p_s e^{i\,(\theta_{i,\,s-}\,\theta_{j,\,s})} \tag{10}$$

Так как фазы $\theta_{i,s}$ распределены случайным образом, суммы по s обращаются в нуль при i≠j; следовательно,

$$\mathrm{tr}\,(\rho A) = \sum |c_i|^2 a_{ii} = \mathrm{tr}\,(\rho\,'\,A). \tag{11}$$
где
$$\rho\,' = \sum |c_i|^2\, p_s\,|\,\psi_i\rangle|\,\alpha_{i,s}\rangle\,\langle\alpha_{j,s}|\langle\psi_j| \tag{12}$$

Получаем статистический оператор, который воспроизводит действие редукции прибора. Если стрелка прибора наблюдается в положении i, состояние прибора при некотором s будет $|\alpha_{i,s}\rangle$, причем вероятность того, что оно будет именно состоянием $|\alpha_{i,s}\rangle$, равна вероятности того, что до акта измерения было состояние $|\alpha_{i,s}\rangle$. Таким образом, приходим к формулировке следующей теоремы.

**Теорема 5.6. О декогеренции макроскопического прибора**. Пусть квантовая система взаимодействует с макроскопическим прибором таким образом, что возникает хаотическое распределение фаз состояний прибора. Пусть $\rho$ — статистический оператор прибора после измерения, рассчитанный с использованием уравнения Шредингера, а $\rho'$ — статистический оператор, полученный в результате применения редукции к оператору $\rho$. Тогда невозможно произвести такой эксперимент с макроскопическим прибором, который зарегистрировал бы различие между $\rho$ и $\rho'$. ∎

Для широкого класса приборов доказано, что хаотичность в распределении фаз, о которой идет речь в теореме 5.6, действительно имеет место, если при проведении очередного измерения прибор эволюционирует необратимым образом.

Это так называемая теорема *Данери — Лойнжера — Проспери* **[30].**

**Приложение R. Парадокс Зенона. Теорема о непрерывно наблюдаемом котелке, который никак не закипает. [3]**

*Теорема:*

Пусть $A$ — наблюдаемая квантовой системы, имеющая собственные значения 0 и 1. Предположим, что измерения наблюдаемой $A$ производятся в моменты времени $0 = t_0,\ t_1,$ ..., $t_N = T$ на временном интервале $[0,\ T]$ и редукция применяется после каждого такого измерения. Пусть $p_n$ — вероятность того, что измерение в момент $t_n$ дает результат 0. Тогда, если $N \to \infty$, причем так, что $\max(p_{n+1} - p_n) \to 0$, то

$$p_N - p_0 \to 0 \tag{5.52}$$

(так что, если система находилась в собственном состоянии наблюдаемой $A$ в момент времени $t = 0$, она будет иметь то же значение наблюдаемой $A$ и в момент времени $t = T$).

*Доказательство.* Пусть $P_0$ обозначает проекционный оператор на собственное пространство наблюдаемой $A$, соответствующее собственному значению 0, и пусть $P_1 = 1$ — $P_0$ обозначает проекционный оператор на собственное пространство с собственным значением 1. Пусть $\rho_n$ — статистический оператор, характеризующий состояние системы непосредственно перед измерением в момент времени $t_n$. Тогда, статистический оператор после измерения дается выражением

$$\rho_n' = P_0\,\rho_n\,P_0 + P_1\,\rho_n\,P_1 \tag{5.53}$$

так что статистический оператор, характеризующий состояние системы непосредственно перед измерением, производимым в момент времени $t_{n+1}$, будет равен

$$\rho_{n+1} = e^{-iH\tau_n}\,\rho_n{}'\,e^{iH\tau_n} \qquad (5.54)$$

где $H$ — гамильтониан системы, а $\tau_n = t_{n+1} - t_n$. Заметим, что если оператор $\rho_n$ равен сумме $k$ слагаемых вида $|\psi\rangle\langle\psi|$, то оператор $\rho_{n+1}$ будет суммой не более чем $2k$ таких слагаемых; поскольку $\rho_n = |\psi_0\rangle\langle\psi_0|$, отсюда следует, что оператор $\rho_n$ является суммой конечного числа указанных слагаемых. Согласно (5.54), имеем

$$\rho_{n+1} = \rho_n{}' - i\tau_n[H, \rho_n{}'] + O(\tau_n^2).\qquad (5.55)$$

Так как $P_0{}^2 = P_0$, $P_0\,P_1 = 0$, то

$$P_0\,\rho_{n+1}\,P_0 = P_0\,\rho_n\,P_0 - i\tau_n[P_0\,H\,P_0,\ P_0\,\rho_n\,P_0] + O(\tau_n^2). \qquad (5.56)$$

Следовательно, вероятность того, что измерение в момент времени $t_n$ даст результат 0, равна

$$p_{n+1} = \mathrm{tr}\,(\rho_{n+1}\,P_0) = \mathrm{tr}(P_0\,\rho_{n+1}\,P_0) =$$
$$= \mathrm{tr}(P_0\,\rho_n\,P_0) - i\tau_n\,\mathrm{tr}[P_0\,H\,P_0,\ P_0\,\rho_n P_0] + O(\tau_n^2). \quad (5.57)$$

причем второе равенство справедливо потому, что $P_0{}^2 = P_0$. Учтем теперь, что оператор $P_0\rho_n P_0$ равен сумме конечного числа слагаемых вида $|\psi\rangle\langle\psi|$ и для любого оператора $X$ справедливо соотношение

$$\mathrm{tr}\,(X\,|\psi\rangle\langle\psi|) = \langle\psi|X|\psi\rangle = \mathrm{tr}\,(|\psi\rangle\langle\psi|\,X). \qquad (5.58)$$

Следовательно, след коммутатора в (5.57) обращается в нуль, поэтому

$$p_{n+1} = p_n + O(\tau_n^2). \qquad (5.59)$$

Обозначим максимальное значение $\tau_n$ через $\tau$ ($\tau = \max \tau_n$); тогда существует такая постоянная $k$, что

$$p_{n+1} - p_n \le k\tau_n{}^2 \le k\tau\tau_n \qquad (5.60)$$

поэтому

$$p_N - p_0 = \sum_{n=0}^{N-1}(p_{n+1} - p_n) \le k\tau\sum_{n=0}^{N-1}\tau_n = k\tau T \to 0$$

при $\tau\to 0$.∎

## Приложение S. Парадокс Эйнштейна — Подольского — Розена. [3]

Можно встать на точку зрения, что обсуждавшиеся в квантовой механике трудности связаны исключительно с тем, что квантово-механический вектор состояния не дает нам всех сведений о состоянии системы: что существуют другие переменные, скрытые от нас в настоящее время, называемые скрытыми переменными, значения которых полностью характеризуют состояние системы и определяют ее будущее поведение более полно, чем это позволяет сделать квантовая механика. Серьезный аргумент в пользу существования таких дополнительных скрытых переменных выдвинули Эйнштейн, Подольский и Розен в 1935 г. Рассмотрим электрон и позитрон, рожденные одновременно в состоянии с полным спином 0. Это спиновое состояние представляет собой антисимметричную комбинацию спиновых состояний двух частиц со спином 1/2, т. е. будет иметь вид

$$| \Psi \rangle = \frac{1}{\sqrt{2}} \left( | \uparrow \rangle \, | \downarrow \rangle - | \downarrow \rangle \, | \uparrow \rangle \right) \qquad (5.62)$$

где $| \uparrow \rangle$ и $| \downarrow \rangle$ — одночастичные собственные состояния компоненты спина $s_z$ с собственными значениями +1/2 и —1/2 соответственно, причем в двухчастичном спиновом состоянии (5.62) спиновое состояние электрона записано первым множителем.

Так как состояние с нулевым угловым моментом инвариантно относительно вращений, оно должно иметь вид (5.62) независимо от того, какая ось использована для определения базиса одночастичных спиновых состояний. Таким образом, можно также записать

$$| \Psi \rangle = \frac{1}{\sqrt{2}} \left( | \rightarrow \rangle \, | \leftarrow \rangle - | \leftarrow \rangle \, | \rightarrow \rangle \right) \qquad (5.63)$$

где $| \leftarrow \rangle$ и $| \rightarrow \rangle$ — одночастичные собственные состояния компоненты спина $s_x$.

Предположим теперь, что электрон и позитрон движутся в противоположных направлениях, пока не удалятся друг от друга на большое расстояние, а затем производится измерение z - компоненты спина электрона. Таким образом, измеряется наблюдаемая $s_z(e^-)$ полной системы; после такого измерения состояние системы спроектируется на некоторое собственное состояние данной наблюдаемой: если измерение даст значение +1/2, то после акта измерения система перейдет в состояние $| \uparrow \rangle | \downarrow \rangle$. Это означает, что позитрон будет находиться в состоянии $| \downarrow \rangle$ и измерение z-компоненты его спина $s_z(e^+)$ с полной определенностью даст значение —1/2. Заметим, что эта информация о позитроне получена посредством эксперимента, произведенного над электроном, находящимся на большом расстоянии от позитрона, без какой-либо возможности физически влиять на него. Эйнштейн, Подольский и Розен заключили поэтому, что обнаруженный в описываемом эксперименте результат в отношении позитрона (а именно что $s_z(e^+)= $ —1/2) должен быть реальным объективным фактом, который имел место и до проведения эксперимента с электроном.

Предположим теперь, что в эксперименте над электроном измеряли не z-компоненту, а x-компоненту его спина. Тогда из (5.63) следует, что состояние системы будет спроектировано либо на состояние $| \leftarrow \rangle | \rightarrow \rangle$, либо на состояние $| \rightarrow \rangle | \leftarrow \rangle$, так что позитрон будет иметь теперь определенное значение x-компоненты $s_x(e+)$. Такое состояние позитрона также должно было существовать до проведения последнего эксперимента. Следовательно, до эксперимента позитрон имел определенные значения и $s_z(e+)$, и $s_x(e+)$. Но это несовместные наблюдаемые, и они не имеют одновременных собственных состояний: нет такого квантово-механического состояния, в котором они обе могли бы иметь определенные значения. Эйнштейн, Подольский и Розен сделали отсюда заключение, что квантово-механическое описание неполное и имеются «элементы/реальности», которые квантовая механика не учитывает. Прежде чем приступить к описанию попыток превращения квантовой механики в более полную теорию, как того требует приведенный аргумент, рассмотрим более подробно, как ситуацию Эйнштейна, Подольского, Розена (ЭПР) объясняет ортодоксальная квантовая механика. После эксперимента, проведенного над электроном, вся система действительно перешла в собственное состояние $| \uparrow \rangle | \downarrow \rangle$, если измерялась наблюдаемая $s_z(e^-)$ и было получено ее значение +1/2, или в собственное состояние $| \rightarrow \rangle | \leftarrow \rangle$, если измерялась наблюдаемая $s_x(e^-)$ и было получено ее значение +1/2. Это означает, что после эксперимента позитрон находится в определенном состоянии, либо $| \downarrow \rangle$, либо $| \leftarrow \rangle$ соответственно, и это состояние отлично от того, в котором позитрон находился до эксперимента. Но это не означает, что состояние позитрона было изменено экспериментом, проведенным над электроном, так как позитрон вообще не имел какого-либо определенного состояния до проведения эксперимента. Если все же настаивать на том, что позитрон должен как-то описываться отдельно, то следует обратиться к его статистическому оператору. Согласно (5.62) и (5.63), этот оператор (до эксперимента) дается выражением

$$\rho_{_{\ldots}} = tr_{_{\ldots}} \mid \Psi \rangle \langle \Psi \mid = \frac{1}{2} \left( \mid\uparrow\rangle \; \langle\uparrow\mid + \mid\downarrow\rangle \; \langle\downarrow\mid \right) = \tag{5.64}$$

$$= \frac{1}{2} \left( \mid\rightarrow\rangle \; \langle\rightarrow\mid + \mid\leftarrow\rangle \; \langle\leftarrow\mid \right) \tag{5.65}$$

т. е. равен единичному оператору, умноженному на 1/2, в двумерном спиновом пространстве позитрона. Рассмотрим теперь статистический оператор позитрона непосредственно после проведения эксперимента, но до того, как информация о его результате дойдет до позитрона. Если в эксперименте с электроном измерялась компонента $s_z$, то состояние позитрона было бы либо $\mid\uparrow\rangle$, либо $\mid\downarrow\rangle$ с равной вероятностью, и статистический оператор имел бы вид (5.64). Если же в эксперименте с электроном измерялась компонента $s_x$, то состояние позитрона было бы либо $\mid\rightarrow\rangle$, либо $\mid\leftarrow\rangle$ с равной вероятностью, и статистический оператор имел бы вид (5.65), т. е. в точности тот же, что в предыдущем случае и до проведения эксперимента с электроном. Хотя эти три ситуации (до эксперимента, после эксперимента по измерению $s_z$ и после эксперимента по измерению $s_x$) по-разному описываются с использованием состояний позитрона, но они все соответствуют одному и тому же статистическому оператору, и между ними нет никакого экспериментально наблюдаемого различия. Таким образом, нет никакого экспериментально обнаруживаемого действия на расстоянии между электроном и удаленным позитроном, т. е. эксперимент ЭПР нельзя использовать для передачи информации со скоростью больше скорости света.

### Приложение Т. Неравенство Белла.[3]

Покажем теперь, что мгновенное действие на расстоянии неизбежно для всякой теории со скрытыми переменными, которая приводит к тем же следствиям, что и квантовая механика.

Рассмотрим ситуацию, в которой эксперименты проводятся над двумя разделенными в пространстве частицами, и выведем следствия из допущения, что результаты эксперимента, проведенного над одной из частиц, определяются только самим этим экспериментом и не зависят от результатов эксперимента, который может проводиться над другой частицей. В этом состоит предположение о *локальности* теории. Ниже будет показано, что требование локальности накладывает такие ограничения на корреляции между экспериментами, производимыми над разными частицами, которые противоречат предсказаниям квантовой механики.

В принципе между локальностью и детерминизмом связи нет. Свойством локальности может обладать теория, которая делает только вероятностные высказывания в отношении результатов экспериментов. Ее можно развить следующим образом. Предположим, что вероятности определяются некоторым числом переменных, совокупность которых обозначим символом $\lambda$ (в случае двух разделенных в пространстве частиц эти переменные могут состоять из переменных, описывающих индивидуально обе частицы, и переменных, описывающих общие устройства, оказывающие одновременное действие на обе частицы). Тогда для каждого эксперимента $E$ можно указать вероятность $p_E(\alpha\mid\lambda)$ получения результата $\alpha$, когда переменные имеют значения $\lambda$. Теория будет *локальной,* если эксперименты $E$ и $F$, которые разделены в пространстве, независимы в смысле теории вероятностей. Отсюда заключаем, что

$$p_{E \oplus F}(\alpha \oplus \beta \mid \lambda) = p_E(\alpha\mid\lambda) p_F(\beta\mid\lambda). \tag{5.75}$$

Любая локальная теория, которая воспроизводит все предсказания квантовой механики в отношении эксперимента ЭПР для разделенных двух частиц со спином 1/2, будет эквивалентна детерминированной теории. Пусть в эксперименте $E$ измеряется компонента спина электрона в определенном направлении, а в эксперименте $F$ измеряется компонента спина удаленного на большое расстояние позитрона в том же направлении. Обозначим стрелками $\uparrow$ и $\downarrow$ два возможных результата измерений. Тогда, поскольку полный спин равен нулю, нам известно, что эксперименты $E$ и $F$ всегда будут давать противоположные результаты; согласно теории вероятностей,

$$p_{E \oplus F}(\uparrow \oplus \uparrow) = p_{E \oplus F}(\downarrow \oplus \downarrow) = 0 \quad (5.76)$$

Пусть $\rho$ $(\lambda)$ — плотность вероятностей, характеризующая вероятность того, что переменные имеют значения $\lambda$; тогда полная вероятность (5.76) равна

$$p_{E \oplus F}(\uparrow \oplus \uparrow) = \int p_{E \oplus F}(\uparrow \oplus \uparrow)\rho(\lambda)d\lambda =$$
$$= \int p_{E}(\uparrow | \lambda) p_{F}(\uparrow | \lambda)\rho(\lambda)d\lambda \qquad (5.77)$$

Так как полная вероятность равна нулю, подынтегральное выражение, будучи положительным, должно всюду обращаться в нуль. Следовательно,

либо $\rho(\lambda) = 0$, либо $p_{E}(\uparrow|\lambda) = 0$, либо $p_{F}(\uparrow|\lambda)=0$. (5.78)

Аналогично заключаем, что

либо $\rho(\lambda) = 0$, либо $p_{E}(\downarrow|\lambda) = 0$, либо $p_{F}(\downarrow|\lambda)=0$. (5.79)

Поскольку эксперимент $E$ имеет только два результата $\downarrow$ и $\uparrow$, имеем эквивалентные утверждения

$$p_{E}(\uparrow | \lambda) = 0 \Leftrightarrow p_{E}(\downarrow | \lambda) = 1 \qquad (5.80)$$

Из (5.78) — (5.80) следует, что если $\rho(\lambda) \neq 0$, то все четыре вероятности должны быть равны либо 0, либо 1. Следовательно, для всех значений $\lambda$, которые возможны в действительности, результаты экспериментов полностью определяются значением $\lambda$.

Таким образом, если считать, что на распределение вероятностей скрытых переменных не оказывает влияние то, какой эксперимент производится над частицами, то мы придем к выводу, что должны рассматриваться только детерминированные теории.

Предположим, что каждая из двух удаленных друг от друга частиц может быть подвергнута одному из трех экспериментов $A$, $B$, $C$, каждый из которых может дать только два результата (скажем «да» или «нет»). Тогда в детерминированной локальной теории результат эксперимента $A$ над частицей 1 определяется свойством системы, которое обозначим $a_1$: это переменная, которая может принимать значения $+$ и $—$. Имеем также аналогичные переменные $b_1$, $c_1$, $a_2$, $b_2$, $c_2$. Предположим теперь, что эксперимент $A$ всегда дает противоположные значения для двух частиц; тогда $a_1=—a_2$. Аналогично будем считать, что эксперименты $B$ v. $C$ тоже дают противоположные результаты для обеих частиц, т. е. $b_1=—b_2$ и $c_1=—c_2$.

Рассмотрим теперь частицы, которые приготовляются с фиксированной вероятностью наборов значений $a$, $b$ и $c$.

Пусть $P(a =1, b =1)$ обозначает вероятность того, что частица имеет указанные значения $a$ и $b$. Тогда

$P(b = 1, c = —1) = P(a =1, b = 1, c = —1) + P(a = —1, b = 1, c = —1) \leq P(a =1, b = 1) + P(a = —1, c = —1).$ (5.81)

Следовательно, когда пары частиц приготовляются с противоположными значениями $a$, $b$ и $c$, имеем

$P(b_1 = 1, c_2 = 1) \leq P(a_1 = 1, b_2 = —1) + P(a_1 = —1, c_2 = 1).$ (5.82)

Каждое слагаемое в правой части этого неравенства дает вероятность результата эксперимента, проведенного над различными частицами; поэтому неравенство можно проверить даже в том случае, когда эксперименты *A, B, C* не могут быть проведены одновременно над одной частицей.

Неравенству (5.82) не удовлетворяют вероятности, рассчитанные согласно правилам квантовой механики, в следующем случае. Предположим, что две частицы со спином 1/2 приготовляются в состоянии с полным спином, равным 0, как рассмотренные в начале этого параграфа электрон и позитрон; тогда мы знаем, что измерение компоненты спина в любом данном направлении даст противоположные результаты для обеих частиц. Пусть *A, B, C* обозначают эксперименты по измерению компонент спина вдоль трех осей, лежащих в одной плоскости, причем пусть угол между осями *A* и *B* равен θ, а угол между осями *B* и *C* равен φ. Вычислим вероятность $P(b_1 = 1, c_2 = 1)$, входящую в левую часть неравенства (5.82); ее следует интерпретировать как вероятность того, что оба измерения компонент спинов частиц 1 и 2 вдоль осей, угол между которыми равен φ, дадут одни и тот же результат +1/2. Возьмем в качестве оси для частицы 1 ось z; тогда если при измерении компоненты спина частицы 1 вдоль указанной оси получим значение 1/2, то после измерения частица 1 перейдет в собственное состояние |↑ >, а частица 2 — в собственное состояние |↓>. Собственные состояния измерения, произведенного над частицей 2, получаются поворотом состояний |↑> и |↓> на угол φ (скажем вокруг оси x); таким образом, собственное состояние, соответствующее собственному значению + 1/2, имеет вид

$$| + (\varphi) \rangle = e^{-i\varphi J_x} |\uparrow\rangle = [\cos\left(\tfrac{1}{2}\varphi\right) + 2iJ_x \sin\left(\tfrac{1}{2}\varphi\right)]|\uparrow\rangle =$$
$$= \cos\left(\tfrac{1}{2}\varphi\right)|\uparrow\rangle + i\sin\left(\tfrac{1}{2}\varphi\right)|\downarrow\rangle \qquad (5.83)$$

Следовательно, искомая вероятность равна

$$P(b_1 = 1, c_2 = 1) = {}^1/_2 |\langle +(\varphi) | \downarrow\rangle|^2 = {}^1/_2 \sin^2 \left({}^1/_2 \varphi\right) \qquad (5.84)$$

(так как вероятность результата +1/2 для частицы 1 равна 1/2). Аналогично можно вычислить входящие в (5.82) вероятности

$$P(a_1 = 1, b_2 = -1) = {}^1/_2 \cos^2 \left({}^1/_2 \theta\right)$$

и

$$P(a_1 = -1, c_2 = 1) = {}^1/_2 \cos^2 [{}^1/_2 (\theta + \varphi)].$$

Таким образом, неравенство (5.82) сводится к неравенству
$$\sin^2 \left({}^1/_2\varphi\right) \leq \cos^2({}^1/_2\theta) + \cos^2 \left[{}^1/_2 (\theta + \varphi)\right],$$
или к неравенству
$$\cos\theta + \cos\varphi + \cos(\theta + \varphi) \geq -1, \qquad (5.85)$$
которые не выполняются при θ = φ = 3π/4. В результате приходим к следующей теореме.

**Теорема 5.8 (Белла).** Пусть две удаленные одна от другой частицы могут быть подвержены одному из трех двузначных экспериментов, и один и тот же эксперимент, производимый над обеими частицами, всегда дает противоположные результаты. Если частицы описываются локальной теорией и тип экспериментов, которые мы собираемся произвести над ними, не влияет на вероятности обнаружения тех или иных свойств, то вероятности результатов экспериментов удовлетворяют неравенству (5.82).

Это неравенство не выполняется в квантовой механике для системы двух частиц со спином 1/2, имеющих полный спин, равный 0.

## Приложение U. Теория волны-пилота де Бройля — Бома. [3]

Характерные особенности квантовой механики (в частности, интерференционные эффекты) затрудняют построение теории скрытых переменных в описанном выше смысле. Долгое время думали даже, что, доказана теорема (фон Неймана) **[29]**, что никакая теория такого типа не может воспроизвести все следствия квантовой механики. Но это доказательство оказалось ошибочным, как показывает следующий контрпример. Рассмотрим отдельную простую частицу, движущуюся в потенциале V(**r**). Предположим, что частица описывается в момент времени *t* не только волновой функцией ψ (**r**, *t*), но также некоторым вектором q(t), и волновая функция удовлетворяет обычному уравнению Шредингера

$$i\hbar \frac{\partial \Psi}{\partial t} = -\frac{\hbar^2}{2m}\Delta\Psi + U(x, y, z)\Psi. \qquad (5.69)$$

а вектор q удовлетворяет уравнению

$$\frac{dq}{dt} = \frac{j(q,t)}{\rho(q,t)}$$

где j и ρ — плотность тока вероятности и плотность самой вероятности:

$$j = \frac{\hbar}{m}\mathrm{Im}\left[\overline{\psi}\nabla\psi\right], \quad \rho = |\psi|^2. \qquad (5.71)$$

Предположим теперь, что в момент времени *t* = 0 имеем большое число таких частиц, каждая из которых описывается одной и той же волновой функцией ψ (**r**, 0), но своим вектором q.

Пусть доля частиц, для которых значение этого вектора лежит в объеме dV, содержащем точку q, равна σ(q, 0)dV; пусть эта доля в момент времени *t* равна σ(q, *t*)dV. Тогда, если считать q координатой частицы, можно рассматривать весь коллектив частиц как жидкость с плотностью *σ* и полем скоростей u = j/ρ согласно (5.70). Последние величины должны удовлетворять уравнению неразрывности

$$\frac{\partial \sigma}{\partial t} + \nabla \cdot (\sigma u) = 0 \qquad (5.72)$$

т.е.

$$\frac{\partial \sigma}{\partial t} = -\nabla \cdot \left(\frac{\sigma j}{\rho}\right) \qquad (5.73)$$

Уравнение неразрывности имеет единственное решение σ(**r**, *t*), если заданы σ(**r**, 0), j(**r**, *t*) и ρ(**r**, *t*). Этому уравнению удовлетворяет также решение *σ* = ρ, так как при этом уравнение (5.73) превращается в уравнение неразрывности (3.42), которое, как было показано, является следствием уравнения Шредингера (5.69). Следовательно, если распределение значений q по частицам описывается в момент *t* = 0 функцией ρ, то оно будет характеризоваться этой функцией и во все последующие моменты времени.

Таким образом, мы можем предположить, что каждая частица, волновая функция которой удовлетворяет уравнению Шредингера (5.69), имеет определенную координату q в пространстве и любое наше экспериментальное устройство, создающее частицы с волновой функцией *ψ*, производит частицы с определенным распределением их координат; доля этих частиц |ψ(q)|²dV находится в объеме dV около точки q. Это справедливо, если экспериментальное устройство, создающее частицы с волновой функцией ψ, с вероятностью |ψ(q)|²dV производит частицы в объеме dV. Так как величины ψ и q эволюционируют во времени, согласно детерминированным уравнениям (5.69), (5.70), такое распределение сохранится во все моменты времени, если оно имелось в какой-то начальный момент.

Можно также такую теорию обобщить на системы из нескольких частиц, но при этом возникает одна очевидная трудность. Рассмотрим, например, систему двух частиц. Переменными будут q₁ и q₂, а двухчастичная волновая функция имеет вид ψ (r₁, r₂). В качестве уравнений движения следует рассмотреть двухчастичное уравнение Шредингера и два уравнения

$$\frac{dq_1}{dt} = \frac{j_1}{\rho}, \qquad \frac{dq_2}{dt} = \frac{j_2}{\rho},$$

Где

$$j = \frac{\hbar}{m}\operatorname{Im}[\overline{\psi}\nabla_1\psi], \quad j = \frac{\hbar}{m}\operatorname{Im}[\overline{\psi}\nabla_2\psi], \rho = |\psi|^2.$$

Здесь $j_1$, а следовательно, $dq_1/dt$ может быть функцией $q_2$: движение первой частицы зависит от положения второй частицы. Таким образом, имеется мгновенное действие на расстоянии между двумя частицами, и оно должно наблюдаться даже в том случае, когда вообще не будет никакого потенциала $V(r_1, r_2)$ сил взаимодействия между частицами. Это является отражением корреляций между частицами, возникающих в формализме квантовой механики, оперирующем с волновыми функциями. В частности, волновая функция ЭПР демонстрирует такого рода корреляции между отдельными частицами.

Приложение V. **Водоворот Эшера, где скрещиваются все уровни. [58]**

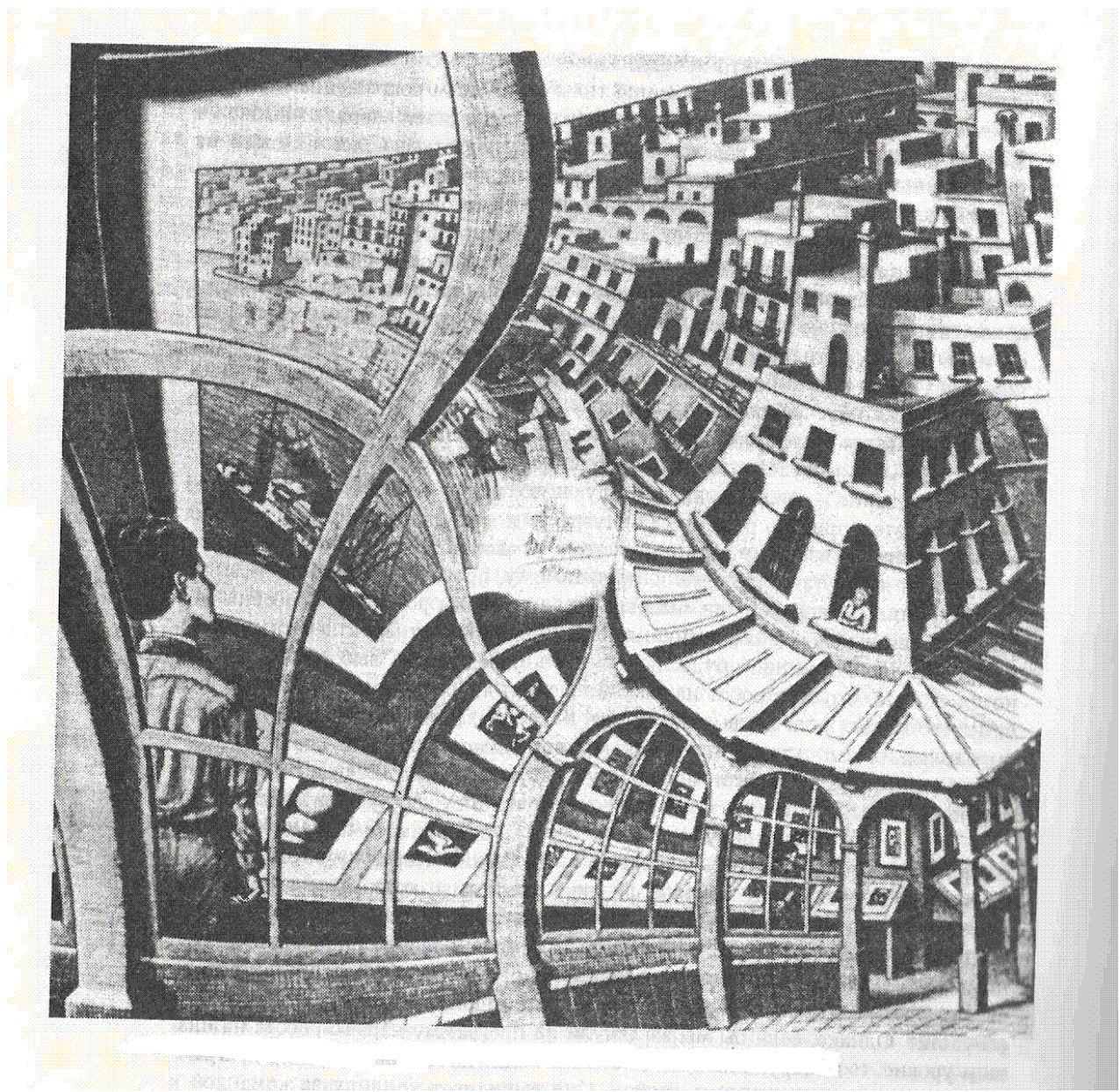

**Figure 28. М. К. Эшер Картинная галерея (литография 1956)**

Поразительно красивая и в то же время странно тревожащая иллюстрация "глаза" циклона, порожденного Запутанной Иерархией, дана нам Эшером в его "Картинной галерее (рис. 28). На этой литографии изображена картинная галерея, где стоит молодой человек, глядя на картину корабля в гавани небольшого городка, может быть, мальтийского, судя по архитектуре, с его башенками, куполами и плоскими каменными крышами, на одной из которых сидит на солнце мальчишка; а двумя этажами ниже какая-то женщина — может быть, мать этого мальчишки — глядит из окна квартиры, расположенной прямо над картинной галереей, где стоит молодой человек, глядя на картину корабля в гавани небольшого городка, может быть, мальтийского — Но что это!? Мы вернулись к тому же уровню, с которого начинали, хотя логически этого никак не могло случиться. Давайте нарисуем диаграмму того, что мы видим на этой картине (рис. 29):

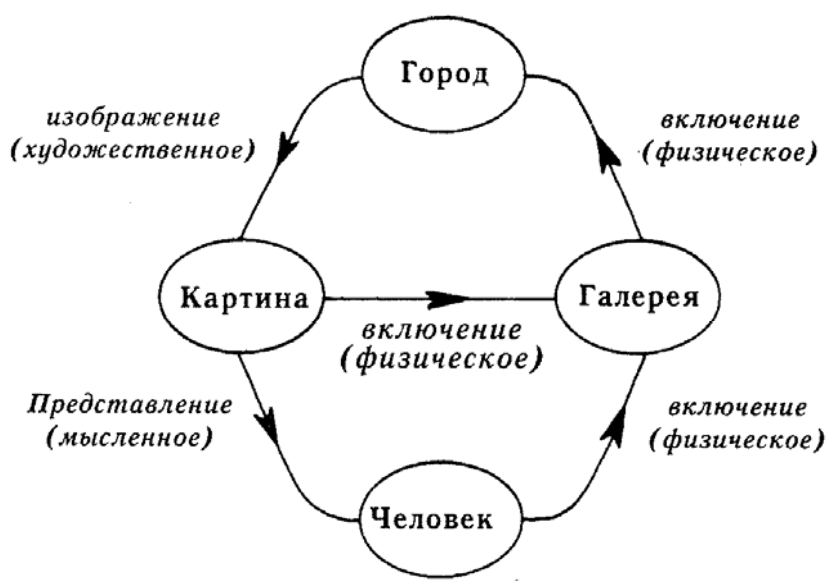

**Figure 29.** Абстрактня диаграмма «Картинная галерея» М. К. Эшера

На этой диаграмме показаны три вида включения. Галерея физически включена в город ("включение"); город художественно включен в картину ("изображение"); картина мысленно включена в человека ("представление") Хотя эта диаграмма может показаться точной, на самом деле она произвольна, поскольку произвольно количество показанных на ней уровней. Ниже представлен другой вариант верхней половины диаграммы (рис 30):

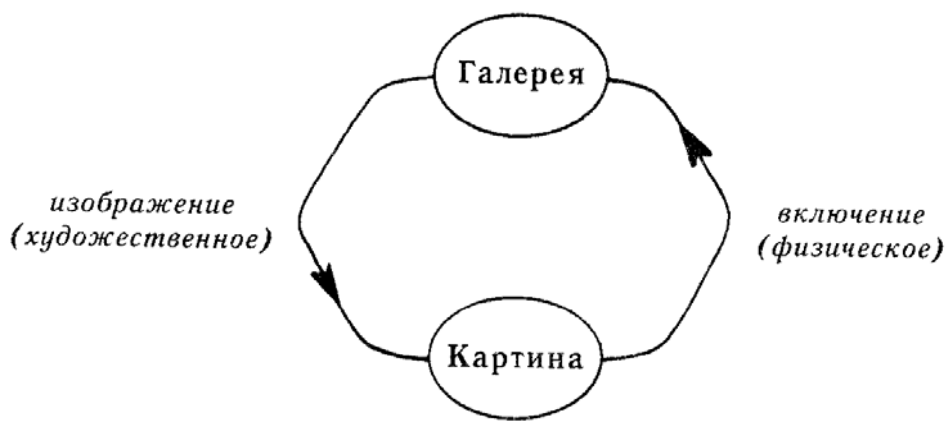

**Figure 30.** Сокращенная версия предыдущей диаграммы.

Мы убрали уровень города: хотя концептуально он полезен, без него можно вполне обойтись. Рис. 30 выглядит так же, как диаграмма "Рисующих рук": это двухступенчатая Странная Петля. Разделительные знаки произвольны, хотя и кажутся нам естественными. Это видно яснее из еще более упрощенной диаграммы "Картинной галереи":

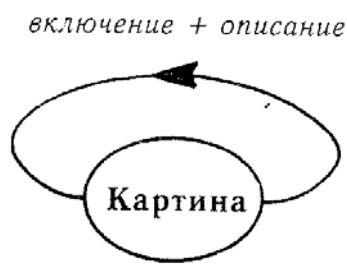

**Figure 31.** Дальнейшее сокращение рис 29.

Парадокс картины выражен здесь в крайней форме. Но если картина "включена в саму себя", то молодой человек тоже включен сам в себя? На этот вопрос отвечает рис. 32.

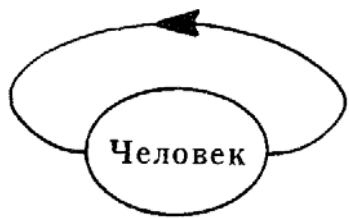

**Figure 32.** Другой способ сокращения рис 29.

Здесь мы видим молодого человека внутри самого себя , в том смысле, какой получается от соединения трех аспектов "внутренности".

Оказываются ли зрители, глядящие на "Картинную галерею", затянутыми "в самих себя"? На самом деле, этого не происходит. Нам удается избежать этого водоворота благодаря тому, что мы находимся вне системы. Глядя на картину, мы видим то, что незаметно молодому человеку, — например, подпись Эшера "MCE" в центральном "слепом пятне". Хотя это пятно кажется дефектом, скорее всего, дефект заключается в наших ожиданиях, поскольку Эшер не мог бы закончить этот фрагмент картины без того, чтобы не вступить в противоречие с правилами, по которым он ее создавал. Центр водоворота остается — и должен оставаться — неполным. Эшер мог бы сделать его сколь угодно малым, но избавиться от него совсем он не мог. Таким образом, мы, глядя снаружи, видим, что "Картинная галерея" неполна, чего молодой человек на картине заметить не в состоянии. Здесь Эшер дал художественную метафору Теоремы Гёделя о неполноте. Поэтому Эшер и Гёдель так тесно переплетены в моей книге.

# Литература.